\documentclass[12pt,titlepage]{article}%
\usepackage{amsrefs}
\usepackage{amssymb}
\usepackage{amsmath}
\usepackage{amsfonts}
\usepackage{graphicx}
\usepackage{color, colortbl}
\usepackage{setspace}
\providecommand{\U}[1]{\protect\rule{.1in}{.1in}}
\newtheorem{theorem}{Theorem}

\newtheorem{lemma}[theorem]{Lemma}

\newtheorem{proposition}[theorem]{Proposition}

\renewcommand{\theequation}{\thesection.\arabic{equation}}

\newfont{\bbf}{cmbx12 scaled 1435}

\newtheorem{hp}{Assumption}
\renewcommand{\thehp}{\Alph{hp}}

\renewcommand{\theequation}{\thesection.\arabic{equation}}

\ifx\pdfoutput\relax\let\pdfoutput=\undefined\fi
\newcount\msipdfoutput
\ifx\pdfoutput\undefined\else
\ifcase\pdfoutput\else
\ifx\paperwidth\undefined\else
\ifdim\paperheight=0pt\relax\else\pdfpageheight\paperheight\fi
\ifdim\paperwidth=0pt\relax\else\pdfpagewidth\paperwidth\fi
\fi\fi\fi
\begin{document}

\title{{\LARGE Quantile-regression methods for first-price auctions}}
\author{Nathalie Gimenes\\Department of Economics\\PUC-Rio\\Brazil\\\texttt{ngimenes@econ.puc-rio.br}
\and Emmanuel Guerre\\School of Economics and Finance\\Queen Mary University of London\\United Kingdom\\\texttt{eoguerre@gmail.com}}
\date{September 2020 }
\maketitle

\newpage\thispagestyle{empty}

\begin{center}
\bigskip\textbf{Abstract}
\end{center}

The paper proposes a quantile-regression inference framework for first-price
auctions with symmetric risk-neutral bidders under the independent private-value paradigm. It is first shown that a private-value quantile regression
 generates a quantile regression for the bids. The private-value quantile
regression can be easily estimated from the bid quantile regression and its
derivative with respect to the quantile level. 
This also allows to  test for various specification  or exogeneity  null hypothesis using the observed bids in a simple way. 
A new local polynomial
technique is proposed to estimate the latter over the whole quantile level
interval. Plug-in estimation of functionals is also considered, as needed for
the expected revenue or the case of CRRA risk-averse bidders, which is
amenable to our framework. A quantile-regression analysis to USFS timber is
found more appropriate than the homogenized-bid methodology and illustrates
the contribution of each explanatory variables to the private-value distribution. 
Linear interactive sieve extensions are proposed and studied in the Appendices.

\bigskip

\textit{JEL}:\textit{\ }C14, L70

\bigskip

\textit{Keywords}: First-price auction; independent private values; dimension
reduction; quantile regression; local polynomial estimation; specification testing;
boundary correction; sieve estimation.

\bigskip

\vfill

{\footnotesize A previous version of this paper has been circulated under the
title "Quantile regression methods for first-price auction:a signal approach".
The authors acknowledge useful discussions and comments from Xiaohong Chen,
Valentina Corradi, Yanqin Fan, Phil Haile, Xavier d'Haultfoeuille, Vadim
Marmer, Isabelle Perrigne, Martin Pesendorfer and Quang Vuong, and the
audience of many conferences and seminars. Nathalie Gimenes also thanks Ying
Fan and Ginger Jin for encouragements. Many thanks to Elie Tamer and three anonymous referees,  who all  have been extremely stimulating and helpful to enrich the paper. 
All remaining errors are our
responsibility. Both authors would like to thank the School of Economics and
Finance, Queen Mary University of London, for generous funding. Nathalie Gimenes would like to express her gratitude to PUC Rio for generous support.}

\bigskip

\pagebreak

\setcounter{page}{1}

\section{Introduction}
Since Paarsch (1992), many parametric methods have been proposed to estimate first-price auction models under the independent private-value paradigm. See Laffont, Ossard and Vuong (1995), Athey and Levin (2001), Hirano and Porter (2003), Li and Zheng (2012), Paarsch and Hong (2012) and the references therein  to name just a few. Validating specification choice is difficult and seldom attempted. 

On the other hand, the nonparametric approach is very flexible and less subject to misspecification of functional form, so that it is commonly considered in applications and theoretical studies. See Guerre, Perrigne and Vuong (2000, hereafter GPV),  Lu and Perrigne (2008), Krasnokutskaya (2011), Marmer and Shneyerov (2012), Hubbard, Paarsch and Li (2012), Campo, Guerre, Perrigne and Vuong (2013), Marmer, Shneyerov and Xu (2013a,b), Hickman and Hubbard (2015), Enache and Florens (2017), Liu and Luo (2017), Liu and Vuong (2018), Luo and Wan (2018), Zincenko (2018) and Ma, Marmer and Shneyerov (2019) among others. But the nonparametric approach comes with the burden of the curse of dimensionality, which  considerably limits its scope of applications.

Haile, Hong and Shum (2003, HSS hereafter) and Rezende (2008) have proposed to circumvent the curse of dimensionality using a regression specification that purges the bids from the covariate effects. The resulting homogenized bids are then used as in GPV to backup the density of their private-value counterparts. This approach can tackle linear dependence, but is not appropriate to capture more complex interactions. The present paper proposes to use instead a more flexible quantile-regression specification.

The use of quantile in first-price auctions is not new. Milgrom (2001, Theorem 4.7) reformulates the identification relation of Guerre, Perrigne and Vuong (2000, GPV afterwards) using quantile
function. See Guerre, Perrigne, Vuong (2009) and Campo et al. (2013) for the use of quantile in risk-aversion identification and, for related estimation methods, Menzel and Morganti (2013), Enache and Florens (2017), Liu and Vuong (2018), Luo and Wan (2018). Marmer and Shneyerov (2012) have proposed a
quantile-based estimator of the private-value probability density function
(pdf), which is an alternative to the two-step GPV method. See also Marmer, Shneyerov and Xu (2013b) who consider a  nonparametric single-index quantile model.  
Guerre and Sabbah
(2012) have noted that the private-value quantile function can be estimated
using a one-step procedure from the estimation of the bid quantile function
and its first derivative. Gimenes (2017) has developed a flexible but parsimonious quantile-regression estimation strategy for ascending auction. The present paper is however the first to develop a quantile inference framework in a first-price auction setting allowing for many covariates.

Using Koenker and Bassett (1978) quantile regression framework is appealing for several reasons. 
First, the quantile-regression specification is flexible enough to capture economically relevant effects as in Gimenes (2017), which could be ignored using parametric ones or less interpretable nonparametric models. These parsimonious specifications can be estimated with reasonable nonparametric rates, allowing implementation in small samples with rich covariate environment. Compared to GPV, this estimation method is one-step and only requests one bandwidth parameter, which theoretical choice follows from standard bias variance expansion. As detailed in \ref{App:Sieve}, quantile-regression specification can be enriched to include more nonparametric features using sieve extensions ranging from the additive  specification of Horowitz and Lee (2005) to fully nonparametric one as in Belloni,  Chernozhukov, Chetverikov and
Fern\'{a}ndez-Val (2019).
Second, the quantile approach comes with a stability property 
of linear specifications, which ensures that a private-value quantile regression generates an bid quantile-regression. This is key for our estimation procedure and also for testing, as
it transfers many null hypotheses of interest for the latent private-value distribution to the bid quantile-regression slopes. Tests derived from Koenker and Xiao (2002), Escanciano and Velasco (2010), Rothe and Wied (2013), Escanciano and Goh (2014) or Liu and Luo (2017) can be used to test correct specification of the quantile-regression or homogenized-bid models, or exogeneity of the auction format and of entry.
Third, the quantile representation used in the paper can play the role of a reduced form generated by  a more complex  model, such as the random-coefficient model considered in Berry, Levinsohn and Pakes (1995), Hoderlein, Klemel\"{a} and Mammen (2010),  or Backus and Lewis (2019) to name just a few.
In particular, random coefficients drawn from an elliptical distribution generates a quantile specification which extends homogenized bid and can be easily estimated.

Fourth, the proposed augmented quantile-regression estimation methodology is based upon local polynomial for quantile levels, and  is therefore not affected by asymptotic boundary bias. This permits better estimation of the upper tail distribution than most  nonparametric methods, which is important as the winner's private value is high for a large number of bidders. 
Fifth, it can also be used to recover important parameters such as the probability or cumulative density functions (pdf and cdf hereafter), mitigating the curse of dimensionality that affects most nonparametric methods. Plug in estimation of the seller expected revenue, optimal reserve price and of  agent constant relative risk-aversion parameter are also considered.

The rest of the paper is organized as follows. The next section \ref{model} introduces our stability result for linear quantile  specification. Section \ref{Alter} considers the homogenized-bid and random-coefficient specifications. Section \ref{Tests} reviews some testing strategies based upon the bid quantile regression. 
Section \ref{Func} explains how to use our quantile specification for estimating agent's risk-aversion, seller's expected revenue, and the cdf and pdf of the private values.
A difficulty of the quantile approach for first-price auction is the need to estimate the bid quantile derivative with respect to quantile levels, see Guerre and Sabbah (2012) and the reference therein for related approaches.
Section \ref{ASQRE} introduces our new augmented quantile regression estimators, which use a quantile-level local-polynomial approach to jointly estimate the bid quantile regression and its higher-order derivatives. 
Sections \ref{Res} and \ref{Funcest} group our main theoretical results, including Integral Mean Squared Error (IMSE), optimal bandwidth choice, optimal uniform convergence rate and Central Limit Theorem for the proposed private-value quantile-regression estimators.

Our theoretical results are illustrated with a simulation experiment and an
application to USFS first-price auctions in Sections \ref{Simexp} and \ref{Appli}. 
Some simulation experiments illustrate how the new estimation procedure improves on the GPV two-step density estimator and homogenized bids.
A preliminary quantile-regression
analysis of the bid quantile function suggests that the homogenized-bid
technique should not be applied here because the quantile-regression slopes
are not constant. The private-value quantile-regression slope functions reveal
the covariates impact, and how strongly bidders in the top of the
distribution can differ from the bottom. 
 Section \ref{conclusion} concludes the paper.

\ref{App:Sieve} details an interactive localized sieve quantile extension and related theoretical results which are the counterparts of the ones obtained for the quantile-regression specification. \ref{App:Prelims} briefly sketches the main proof arguments and states some preliminary lemmas used for the proofs of the two key bias and linearization results in \ref{App:Bias} and \ref{App:Bahadur}, from which our main results follow. The two remaining Appendices group the proofs of our main and intermediary results.

%\texttt{La hache est ici}

%TCIMACRO{\TeXButton{TeX field}{\setcounter{equation}{0}}}%
%BeginExpansion
\setcounter{equation}{0}%
%EndExpansion

\section{First-price auction and quantile specification\label{model}}

A single and indivisible object with some characteristic $X\in\mathbb{R}^{D}$
is auctioned to $I\geq2$ buyers. The potential number of bidders $I$ and $X$ 
are known to the bidders and the econometrician. Bids $B_i$ are sealed so that a
bidder does not know the other bids when forming his own bid. The object is sold
to the highest bidder who pays his bid  to the seller, and all the bids $B_i$ are then observed by the econometrician. Under the
symmetric IPV paradigm, each potential bidder is assumed to have a private
value $V_{i}$, $i=1,\ldots,I$ for the auctioned object. A buyer knows his
private value but not the other  ones,  the common
distribution of the independent $V_{i}$ being common knowledge. The private-value conditional cdf $F\left(  \cdot|X,I\right)$ has a bounded support, or
equivalently the conditional private-value quantile function
\[
V\left(  \alpha|X,I\right)  =F^{-1}\left(  \alpha|X,I\right)  ,\quad
\alpha\text{ in }\left[  0,1\right] ,
\]
is  finite for $\alpha=0$ and $\alpha=1$.

The private-value quantile function $V(\alpha|x,I)$ plays an important economic role. The bidder's rent at quantile level $\alpha$ is $V(\alpha|x,I)-B(\alpha|x,I)$  where $B(\cdot|x,I)$ is the bid conditional quantile function, and assuming bids depend in a monotonous way on private values as considered below. The private-value quantile conditional function is important to compute counterfactuals, such as the bid quantile function in an alternative auction mechanism. In particular, it can be used to compute the seller expected revenue achieved with any reserve price, see (\ref{ER}) below. It allows, as a consequence, to compute an optimal reserve price, or more generally to propose suitable auction designs.

\subsection{Private value quantile identification}

It is well-known that the bidder $i$ private-value rank
\[
A_{i}=F\left(  V_{i}|X,I\right)
\]
has a uniform distribution over $\left[  0,1\right]  $ and is independent of
$X$ and $I$. It also follows from the IPV paradigm that the private-value
ranks $A_{i}=1,\ldots,I$ are independent. The dependence between the private
value $V_{i}$ and the auction covariates $X$ and $I$ is therefore fully
captured by the non separable quantile representation%
\begin{equation}
V_{i}=V\left(  A_{i}|X,I\right)  ,\quad A_{i}\overset{\text{iid}}{\sim
}\mathcal{U}_{\left[  0,1\right]  }\perp\left(  X,I\right)  ,
\label{Vquant}
\end{equation}
which, when the private values are generated by an economic structural model, can be also  viewed as a nonparametric reduced form.
Following Milgrom and Weber (1982) or Milgrom (2001), $V\left(  \cdot
|X,I\right)  $ can be also interpreted as a \textit{valuation function,} the
private-value rank $A_{i}$ being the associated signal. In what follows,
$G\left(  \cdot|X,I\right)  $ and $g\left(  \cdot|X,I\right)  $ stand for
respectively the bid conditional cdf and pdf.

Maskin and Riley (1984) have shown that Bayesian Nash Equilibrium bids
$B_{i}=\sigma\left(  V_{i};X,I\right)  $ of symmetric risk-averse or risk-neutral bidders are strictly increasing and continuous in $V_i$. It follows that $B_{i}=B\left(  A_{i}%
|X,i\right)  $, where $B\left(  \cdot;X,i\right)  =\sigma\left(  F\left(
\cdot|X,I\right)  ;X,I\right)  $ can be viewed as a bidding strategy depending
upon the rank $A_{i}$. If $F\left(  \cdot|X,I\right)  $ is also strictly
increasing, so is $B\left(  \cdot|X,I\right)  $ and since $A_{i}$ is uniform
it holds
\[
G\left(  b|X,I\right)  =\mathbb{P}\left[  B\left(  A_{i}|X,I\right)  \leq
b|X,I\right]  =\mathbb{P}\left[  A_{i}\leq B^{-1}\left(  b|X,I\right)
|X,I\right]  =B^{-1}\left(  b|X,I\right)
\]
showing that the bidding strategy\ $B\left(  \cdot|X,I\right)  $ is also the
bid quantile function.

A standard best response argument will show how to identify the private-value
quantile function $V\left(  \cdot|X,I\right)  $ from $B\left(  \cdot
|X,I\right)  $. Suppose bidder $i$ signal $A_{i}$ is equal to $\alpha$, but
that her bid is a suboptimal $B\left(  a|X,I\right)  $, all other bidders
bidding $B\left(  A_{j}|X,I\right)  $. Then the probability that bidder $i$
wins the auction is%
\begin{align}
\mathbb{P}\left[  \left.  B\left(  a|X,I\right)  >\max_{1\leq j\neq i\leq
I}B\left(  A_{j}|X,I\right)  \right\vert A_{i}=\alpha,X,I\right]   &
=\mathbb{P}\left[  \left.  a>\max_{1\leq j\neq i\leq I}A_{j}\right\vert
A_{i}=\alpha,X,I\right] \nonumber\\
&  =a^{I-1} \label{Winprob}%
\end{align}
because the $A_{j}$'s are independent $\mathcal{U}_{\left[  0,1\right]  }$
independent of $X$ and $I$. It follows that the expected revenue of such a bid
is, for a risk-neutral bidder, $\left(  V\left(  \alpha|X,I\right)  -B\left(
a|X,I\right)  \right)  a^{I-1}$. If $B\left(  \cdot|X,I\right)  $ is a
best-response bidding strategy, the optimal bid of a bidder with signal
$\alpha$ is $B\left(  \alpha|X,I\right)  $, that is%
\[
\alpha=\arg\max_{a}\left\{  \left(  V\left(  \alpha|X,I\right)  -B\left(
a|X,I\right)  \right)  a^{I-1}\right\}  .
\]
As $B\left(  \cdot|X,I\right)  $ is continuously differentiable, it follows
that%
\begin{equation}
\left.  \frac{\partial}{\partial a}\left\{  \left(  V\left(  \alpha
|X,I\right)  -B\left(  a|X,I\right)  \right)  a^{I-1}\right\}  \right\vert
_{a=\alpha}=0 \label{BR}%
\end{equation}
or equivalently 
\begin{align*}
&
\frac{d}{d\alpha}
\left[  
\alpha^{I-1}B\left(  \alpha
|X,I\right)  
\right]  
=
\left(  I-1\right)  \alpha^{I-2}V\left(  \alpha
|X,I\right) 
\\
& 
\text{ with }
\frac{d}{d\alpha}
\left[  
\alpha^{I-1}B\left(  \alpha
|X,I\right)  
\right]
=
\left(  I-1\right)  \alpha^{I-2}V\left(  \alpha
|X,I\right)
+
\alpha^{I-1}
\frac{d}{d\alpha}B\left(
\alpha|X,I\right).
\end{align*} 
Solving with the initial condition $B\left(  0|X,I\right)
=V\left(  0|X,I\right)  $ and rearranging the equation above gives Proposition
\ref{Ident}, which is the cornerstone of our estimation method. From now on
$B^{\left(  1\right)  }\left(  \alpha|X,I\right)  =\frac{d}{d\alpha}B\left(
\alpha|X,I\right)  $.

\begin{proposition}
\label{Ident}Consider a given $\left(  X,I\right)  $, $I\geq2$, for which
$\alpha\in\left[  0,1\right]  \mapsto V\left(  \alpha|X,I\right)  $ is
continuously differentiable with a derivative $V^{\left(  1\right)  }\left(
\cdot|X,I\right)  >0$. Suppose the bids are drawn from the symmetric
differential Bayesian Nash equilibrium. Then,
\begin{enumerate}
\item The conditional equilibrium quantile function $B\left(  \cdot
|X,I\right)  $ of the $I$ iid optimal bids $B_{i}$ satisfies%
\begin{equation}
B\left(  \alpha|X,I\right)  =\frac{I-1}{\alpha^{I-1}}\int_{0}^{\alpha}%
a^{I-2}V\left(  a|X,I\right)  da \text{ with }
\lim_{\alpha \downarrow 0} 
B\left(  \alpha|X,I\right)
=
V\left(  0 |X,I\right).
\label{V2B}%
\end{equation}

\item The bid quantile function $B\left(  \alpha|X,I\right)  $ is continuously
differentiable over $\left[  0,1\right]  $ and it holds%
\begin{equation}
V\left(  \alpha|X,I\right)  =B\left(  \alpha|X,I\right)  +\frac{\alpha
B^{\left(  1\right)  }\left(  \alpha|X,I\right)  }{I-1}. \label{B2V}%
\end{equation}
\bigskip
\end{enumerate}
\end{proposition}

A key feature is the linearity with respect to $V\left(  \cdot|X,I\right)$  of the private-value to bid quantile functions
mapping (\ref{V2B}), which implies that a private value quantile linear model
is mapped into a similar bid linear model, as detailed below for the well-known quantile regression. Proposition \ref{Ident}-(ii) shows that the private-value quantile function is identified from the bid quantile function and its
derivative. It is a quantile version of
the identification strategy of GPV, which is based on the identity\footnote{This can be recovered from (\ref{B2V}) taking
$\alpha=A_{i}$ as $V_{i}=V\left(  A_{i}|X,I\right)  $, $B_{i}=B\left(
A_{i}|X,I\right)  $ implying that $A_{i}=G\left(  A_{i}|X,I\right)  $ and
$B^{\left(  1\right)  }\left(  A_{i}|X,I\right)  =1/g\left(  B\left(
A_{i}|X,I\right)  |X,I\right)  =1/g(B_{i}|X,I)$.}%
\begin{equation}
V_{i}=B_{i}+\frac{1}{I-1}\frac{G\left(  B_{i}|X,I\right)  }{g\left(
B_{i}|X,I\right)  }.
\label{GPV}
\end{equation}
Versions of (\ref{B2V}) with $B^{\left(  1\right)  }\left(  \alpha|X,I\right)
$ changed into $1/g\left(  B\left(  \alpha|X,I\right)  |X,I\right)  $ can be
found in Milgrom (2001, Theorem 4.7), Liu and Luo (2014), Liu and Vuong (2016), Luo and Wan (2016), Enache and Florens
(2017) and, under risk-aversion,
in Guerre et al. (2009) and Campo et al. (2011). 

\subsection{Private-value quantile regression \label{Qrhmg}}

%\paragraph{Private value quantile regression.}

The linearity of (\ref{V2B}) has
important model stability implications useful for practical implementation.
Consider a private-value quantile given by the quantile-regression specification%
\begin{equation}
V\left(  \alpha|X,I\right)  =\gamma_{0}\left(  \alpha|I\right)  +X^{\prime
}\gamma_{1}\left(  \alpha|I\right)  =X_{1}^{\prime}
\gamma\left(  \alpha|I\right),
\quad
X_{1} = \left[  1,X^{\prime}\right]^{\prime}.
\label{Vqr}%
\end{equation}
As a linear regression  is often viewed as an alternative to a nonparametric one which is difficult to estimate, this quantile regression is  simpler to estimate than a general quantile function which must be estimated nonparametrically. As pointed by a Referee, the quantile level $\alpha$ can be viewed as a measure of the bidder efficiency and the slope function $\gamma(\cdot|I)$ indicates how this efficiency affects valuation in the covariate dimension. While more flexible than the homogenized bid specification detailed in Section \ref{HHSmodel}, the quantile approach only involves a unique signal: in particular, if each slope entries are increasing, then each covariate contribution to the value increases with efficiency. More flexibility is possible with the random coefficient model of Section \ref{Randcoeffmodel}, which attaches a specific signal to each auction covariate.

Proposition \ref{Ident}-(i) implies that the conditional bid quantile function
satisfies,%
\begin{equation}
B\left(  \alpha|X,I\right)  =X_{1}^{\prime}  \beta\left(
\alpha|I\right)  \text{ with }\beta\left(  \alpha|I\right)  =\frac{I-1}%
{\alpha^{I-1}}\int_{0}^{\alpha}a^{I-2}\gamma\left(  a|I\right)  da,
\label{Bqr}%
\end{equation}
showing that $B\left(  \alpha|X,I\right)  $ belongs to the quantile-regression
specification. Hence (\ref{B2V}) gives
\begin{equation}
\gamma\left(  \alpha|I\right)  =\beta\left(  \alpha|I\right)  +\frac
{\alpha\beta^{\left(  1\right)  }\left(  \alpha|I\right)  }{I-1},
\label{Bqr2Vqr}%
\end{equation}
so that estimating $\gamma\left(  \alpha|I\right)  $ amounts to estimate $\beta\left(  \alpha|I\right)  $ and $\beta^{\left(  1\right)
}\left(  \alpha|I\right)  $. 

This approach  extends to more flexible nonparametric linear specifications, as developed in \ref{App:Sieve} which considers a sieve extension
\begin{equation}
\left\{
\begin{array}{l}
V(\alpha|x,I) = P(x)^{\prime} \gamma(\alpha|I) + \text{approx. error},
\\
B(\alpha|x,I) = P(x)^{\prime} \frac{I-1}{\alpha^{I-1}} \int_0^{\alpha} a^{I-2} \gamma (a|I) da + \text{approx. error} 
\end{array}
\right.
\label{Sieve}
\end{equation}
where $P(\cdot)$ is a localized sieve vector whose dimension grows with a smoothing parameter $h$. The choice of $P(\cdot)$ can be tailored to cover additivity or less stringent interaction restrictions. As for the quantile-regression estimators proposed below, the sieve approach developed in \ref{App:Sieve} is not affected by asymptotic boundary issues.

\subsection{Alternative models and specification testing strategies \label{Alter}}

\subsubsection{Homogenized bids \label{HHSmodel}}
HHS and Rezende (2008) consider a regression specification
\begin{equation}
V_i = X^{\prime} \gamma_1 + v_i,
\quad
i=1,\ldots,I,
\label{HHS}
\end{equation}
where the iid  $v_i$, the ``homogenized'' private values, are independent of $X$ and not centered.\footnote{Centering the $v_i$'s would amount to introduce an intercept parameter $\gamma_0$, which would be changed to a new intercept $\beta_0 (I)$ when turning to the bid regression when the $v_i$'s are independent of $I$. In contrast, the bid regression slope is $\gamma_1$, therefore unchanged. Hence (\ref{HHS}) does not include an intercept to better focus on the invariant parameter, the purpose being to estimate $\gamma_1$ and the distribution of $v_i$. Estimating an intercept in the bid regression is however necessary to consistently estimate $\gamma_1$ using OLS because the regression error term in (\ref{Homobid}) is not centered.} The corresponding homogenized-bid quantile-regression specification is the following restriction of (\ref{Vqr})
\[
V(\alpha|X,I) = X^{\prime} \gamma_1 + v(\alpha|I)
\]
where $v(\cdot|I)$ is the quantile function of the $v_i$'s.
Since
$\frac{I-1}{\alpha^{I-1}}\int_{0}^{\alpha}a^{I-2}da=1$, it follows that the
associated bid quantile function is, by (\ref{V2B})
\[
B\left(  \alpha|X,I\right)  =X^{\prime}\gamma_{1}+b\left(
\alpha|I\right)  ,\text{ where }b\left(  \alpha|I\right)  =\frac{I-1}%
{\alpha^{I-1}}\int_{0}^{\alpha}a^{I-2}v\left(  a|I\right)  da.
\]
This gives the bid regression model%
\begin{equation}
B_{i}=X^{\prime}\gamma_{1}+b_{i},
\quad 
b_i = b\left(  A_{i}|I\right)
\text{ and } i=1,\ldots,I
\label{Homobid}
\end{equation}
where the  $b_{i}$ are the homogenized bids
of HHS, which are independent of $X$ but depend upon $I$.
Given a sample $X_{\ell},I_{\ell}, B_{1 \ell}, \ldots, B_{I_{\ell}\ell}$ of $\ell=1,\ldots,L$ first-price auctions,
HSS and Rezende (2008) propose to backup the homogenized bids by regressing the bids $B_{i\ell}$ on $X_{1\ell}=[1, X_{\ell}^{\prime}]^{\prime}$, so that the estimation of the homogenized bids are $\widehat{b}_{i\ell} =B_{i\ell} - X_{\ell}^{\prime} \widehat{\gamma}_1$, where $\widehat{\gamma}_1$ is the OLS slope estimator. The pdf of $v_{i}$ can be estimated applying the GPV two-step method to the homogenized-bid estimates. 
An important feature of this model is that the dependence of the private values to the covariate is simple enough to allow for  accurate estimation of $\gamma_1$.
As noted in Paarsch and
Hong (2006), a similar two-step procedure applies for the nonparametric regression model $V_i = m(X|I) +v_i$ where the $v_i$'s are independent of $X,I$, see also Marmer, Shneyerov and Xu (2013b).

However this approach requests independence between the regression error term
$v_{i}$ and the covariate $X$, an assumption which may be too restrictive in
practice as found by Gimenes (2017) and the application below. When
$\gamma_{1}\left(  \cdot\right)  $ is not a constant and $V (\alpha|X,I) = X^{\prime} \gamma_1
(\alpha|I)
+ v(\alpha|I)$, it holds for $\beta_1 (\alpha|I) =\frac{I-1}{\alpha^{I-2}} \int_{0}^{\alpha} a^{I-2} \gamma_{1} (a|I) da$ and the OLS limit $\beta_1 (I) = \mathbb{E} [\beta_1 (A_i|I)]$ obtained when regressing the bids on the constant and $X$,
\[
B_i
=
X^{\prime}
\beta_1 (I)
+
b (A_i|X,I)
\text{ where }
b (A_i|X,I) = 
b(A_i|I) 
+ 
X^{\prime} 
\left[
\beta_1 (\alpha|I)
-
\beta_1 (I)
\right].
\]
As
$b (A_i|X,I)$ depends upon $X$,
the homogenized-bid approach does not apply. As explained below, estimating the slope $\gamma_1 (\cdot)$ involves nonparametric techniques that cannot deliver the parametric rate feasible in the homogenized-bid model.

\subsubsection{Random-coefficient specification} \label{Randcoeffmodel}

 Consider $I$ private values from 
\begin{equation}
V_i = X_{1}^{\prime} \Gamma_i,
\quad
i=1,\ldots,I 
\label{Randcoeff}
\end{equation}
where the random coefficients $\Gamma_i$ are iid $1\times(D+1)$  vectors independent of $X_1$. Since  $V_i = \Gamma_{0i} + X^{\prime} \Gamma_{1i}$, taking $\Gamma_{1i}$ constant across bidders gives a homogenized-bid specification, which is therefore a particular case of random-coefficients regression.
Compared to (\ref{Vquant}) version which involves a unique signal $A_i$, (\ref{Randcoeff}) allows for $D+1$ individual signals $\Gamma_{id}$ which models the impact of the common covariate $X_{d}$ on the private value $V_i$. How a quantile approach can be useful is first discussed when  $\Gamma_i$ is drawn from an elliptical distribution.

\paragraph{Elliptical random coefficient.} $\Gamma_{i}$ is drawn from an elliptical distribution with translation parameter $\gamma(I)$ and symmetric nonnegative dispersion matrix  $\Sigma_{\Gamma}(I)$ if  the characteristic function $\mathbb{E} \left[ \exp \left( \mathsf{i} t^{\prime} (\Gamma_i-\gamma(I))\right)\right]$ only depends upon  $t^{\prime} \Sigma_{\Gamma} (I) t$. Examples include the multivariate normal, lognormal or Student distribution, which can be truncated  to satisfy our finite support restriction. A convenient representation of $\Gamma_i$ involves the Euclidean norm $R_i=\left\| \Sigma_{\Gamma}^{-1/2} (I) \left(\Gamma_i-\gamma(I)\right)\right\|$ and independent draws $\mathcal{S}_i$ from the uniform distribution over the $D+1$ dimensional unit sphere. Let $\mathcal{C}_i$ be the first coordinate of $\mathcal{S}_i$, noticing that $t'\mathcal{S}_i$ is distributed as $\| t \| \mathcal{C}_i$ for any $(D+1) \times 1$ vector $t$. Then by Fang, Kotz and Ng (1990, p.29), $\Gamma_i$ and $\gamma(I) + R_i \Sigma_{\Gamma}^{1/2} (I)\mathcal{S}_i$ have the same distribution, for independent $R_i$ and $\mathcal{S}_i$. It then follows by (\ref{Randcoeff}), $\stackrel{d}{=}$ indicating random variables with identical distribution
\[
V_i \stackrel{d}{=} X_1^{\prime} \gamma(I) + \left( \Sigma_{\Gamma}^{1/2} (I)X_1 \right)^{\prime} R_i \mathcal{S}_i
\stackrel{d}{=} X_1^{\prime} \gamma (I)+ \left\| \Sigma_{\Gamma}^{1/2} (I) X_1 \right\| R_i \mathcal{C}_i.
\] 
Hence the quantile specification generated by (\ref{Randcoeff}) is
\begin{equation}
V(\alpha|X,I) = X_1^{\prime} \gamma (I)+ \left\| \Sigma_{\Gamma}^{1/2} (I)X_1 \right\| v(\alpha|I)
\label{Vquant_rc}
\end{equation}
where  the unknown quantile function $v(\alpha|I)$ is the one of $R_i \mathcal{C}_i$ given $I$. The generated bids have a common quantile function
\begin{equation}
B(\alpha|X,I) = X_1^{\prime} \gamma(I) + \left\| \Sigma_{\Gamma}^{1/2} (I)X_1 \right\| b(\alpha|I),
\quad
b(\alpha|I)
=
\frac{I-1}{\alpha^{I-1}}
\int_{0}^{\alpha}
a^{I-2} v(a|I) da
\label{Bquant_rc}
\end{equation}
by (\ref{V2B}). Using the normalization $b(1/2|I)=1$ for identification purpose gives that
$
B(1/2|X,I) = X_1^{\prime} \gamma (I)+ \left\| \Sigma_{\Gamma}^{1/2} (I) X_1 \right\|
$, so that the conditional bid median can be used to identify $\gamma(I)$ and $\Sigma_{\Gamma} (I)$. Identification of $v(\cdot|I)$ works as in Proposition \ref{Ident} as $v(\alpha|I) = b(\alpha|I) + \alpha b^{(1)} (\alpha|I)/(I-1)$, observing that $v(\cdot|I)$ identifies the common distribution of the $R_i$'s.

\paragraph{The general case.}  Hoderlein et al. (2010) propose a nonparametric  method that could be used to estimate the distribution of the random slope $\Gamma_i$ of (\ref{Randcoeff}) if the private values were observed. This suggests to implement a two-step method using estimated private values. \ref{App:Sieve} proposes a sieve method to estimate $V(\cdot|x,I)$, which is not subject to asymptotic boundary bias. Consider $S$ estimated private values $\widehat{V} (A_s|X_s,I)$ for arbitrary values $X_s$ of the covariate and independent uniform draws $A_s$, $s=1,\ldots,S$. Assuming that the $X_s/\| X_s \|$ are drawn from the uniform distribution on the unit sphere suggests to estimate the density $f_{\Gamma} (\gamma|I)$ of $\Gamma_{i}$ given $I$ using in the second step the Hoderlein et al (2010) kernel estimator
\begin{align}
\widehat{f}_{\Gamma} (\gamma|I)
& =
\frac{1}{n}
\sum_{s=1}^{S}
K_{HKM,h}
\left(
\frac{\widehat{V} (A_s|X_s,I)-X_s^{\prime}\gamma}{\| X_s\|}
\right)
\text{ with }
\label{HKM}
\\
&
K_{HKM,h} (u)
=
\frac{1}{(2 \pi)^{D+1}}
\int_{0}^{1/h}
\cos (tu)
t^D
\left(1-(ht)^r\right)
dt,
\nonumber
\end{align}
where $h>0$ is a bandwidth parameter and $0<r\leq \infty$.

\subsection{Specification testing strategies \label{Tests}}

The stability of private-value quantile-regression specification allows to use the bid one to test many hypothesis of interest, see Liu and Luo (2017) for a related point of view. This can be useful to obtain better performing tests as the presence of the derivative $\widehat{B}^{(1)} (\alpha|x,I)$ in the implementable private-value expression (\ref{B2V}) makes its use for testing harder. Examples of tests based on this idea are as follows.

\paragraph{Quantile-regression goodness of fit.}
There is a recent literature  that considers the null hypothesis of correct specification  of a quantile-regression model over an subinterval $\mathcal{A}$ of $(0,1)$. See Escanciano and Velasco (2010), Rothe and Wied (2013), Escanciano and Goh (2014) and the references therein. These three papers propose test statistics of the form $\widehat{T} (\widehat{\beta}(\cdot|I))$, where $\widehat{\beta}(\cdot|I)$ is a quantile-regression estimator which converges to the true slope over $\mathcal{A}$ with a parametric rate, such as the standard quantile-regression estimator or the augmented ones proposed in Section \ref{ASQRE}. See the Application Section 7 for the $\widehat{T}(\cdot)$ used by Rothe and Wied (2013). Liu and Luo (2017) based an entry exogeneity test on the integral of the squared difference of two quantile estimators, see (\ref{Liuluo}) below.

\paragraph{Homogenized bid and elliptical random coefficient.}
The correct specification of (\ref{HHS}) or (\ref{Bquant_rc}) can be tested using  Rothe and Wied (2013) without any restriction on the quantile alternative. 
If the alternative is restricted to a quantile-regression model, Koenker and Xiao (2002) or Escanciano and Goh (2014) can be used to test the homogenized-bid null hypothesis, as this specification coincides with the location-shift model considered by these authors.
Following Liu and Luo (2017) suggests to consider, for the same null, a test statistic
\begin{equation}
L
\int_{0}^{1}
\sum_{\ell=1}^{L}
\left(
X_{1\ell}^{\prime} \widehat{\beta}_{H_0} (\alpha)
-
X_{1\ell}^{\prime} \widehat{\beta} (\alpha)
\right)^{2}
d\alpha
\label{Liuluo}
\end{equation}
where $L$ is the number of auctions in the sample, $X_{\ell}$ the auction covariate, $\widehat{\beta} (\cdot)$ a quantile-regression slope estimator $\sqrt{L}$-consistent over $[0,1]$ as the one proposed in the next section, and for instance $\widehat{\beta}_{H_0} (\cdot)=[\widehat{\beta}_0 (\cdot),\widehat{\beta}_{1,OLS},\ldots,\widehat{\beta}_{1D,OLS}]^{\prime}$.
Confidence bands can also be used, see Gimenes (2017) and the theory developed in Fan, Guerre and Lazarova (2020).

\paragraph{Exogenous auction format.}
Let $V_{j} (\alpha|x,I)=x^{\prime} \gamma_j (\alpha|I)$ be the private-value quantile function conditionally on participation to an ascending auction ($j=asc$) or a first-price one ($j=fp$). A null hypothesis of interest is exogeneity of the auction format, $H_0^{F}: V_{fp} (\cdot|\cdot,I) = V_{asc} (\cdot|\cdot,I)$. Gimenes (2017) gives a consistent quantile-regression estimator $\widehat{\gamma}_{asc} (\cdot|I)$  of $\gamma_{asc} (\cdot|I)$ using ascending auction data. It then follows by (\ref{V2B}) that $\widehat{\beta}_{H_0} (\alpha|I) = (I-1) \alpha^{-(I-1)} \int_0^{\alpha} a^{I-2} \widehat{\gamma}_{asc} (a|I) da$ is consistent under the null but not the alternative.\footnote{As the standard quantile-regression estimator may not be well-defined for quantile levels near $0$, it may be more suitable to use an augmented quantile-regression estimator as in Section \ref{ASQRE} to implement Gimenes (2017).} Then using first-price auction data to compute a test statistic $
\widehat{T} (\widehat{\beta}_{H_0} (\cdot|I))$ from Escanciano and Goh (2014) or Rothe and Wied (2013) for an arbitrary alternative, or using Liu and Luo (2017) statistic (\ref{Liuluo}) with a quantile-regression alternative, allow to test for auction format exogeneity.

\paragraph{Participation exogeneity.}
	The participation exogeneity null hypothesis states that the private values are independent of the number of bidder conditionally on the covariate
	$
	H_0^{E}: V(\cdot|\cdot,I) = V(\cdot|\cdot)
	$ for all $I$,  see also Gimenes (2017) for the ascending auction case. Liu and Luo (2017) use an integral version of $H_0^{E}$ to eliminate the bid quantile derivative in  (\ref{B2V}). In a quantile-regression setup, Proposition \ref{Ident} implies under $H_0^{E}$,
	\begin{align}
	\beta (\alpha|I_2)
	& = 
	\frac{I_2-1}{\alpha^{I_2-1}}
	\int_{0}^{\alpha}
	a^{I_2-2}
	\left[
	\beta (a|I_1) + \frac{a \beta^{(1)}(a|I_1)}{I_1-1}
	\right]
	da
	=
	\beta_{I_1} (\alpha|I_2)
	\label{Betai1i2}
	\\
	&
	\text{ where }
	\beta_{I_1} (\alpha|I_2)
	=
	\frac{I_2-1}{I_1-1}
	\beta (\alpha|I_1)
	+
	\frac{(I_2-1)(I_1-I_2)}{(I_1-1)\alpha^{I_2-1}}
	\int_{0}^{\alpha}
	a^{I_2-2}
	\beta (a|I_1) da.
	\nonumber
	\end{align}
	Then tests for entry exogeneity can be obtained using the same construction than for the auction format exogeneity null, 
	using a sample of first-price auction with $I_1$ bidders to estimate $\beta_{I_1} (\alpha|I_2)$ and another sample with $I_2$ bidders to compute a test statistic. 
	
	Under participation exogeneity, private value estimates  can be averaged over $I$ to improve accuracy. Another important motivation for exogenous participation is risk-aversion estimation, see Guerre, et al. (2009).  This approach can be modified to cope with an additional risk-aversion parameter which can be estimated with a parametric rate as shown in Section \ref{Funcest}.

\setcounter{equation}{0}

\section{Risk-aversion, expected payoff and other functionals \label{Func}}

Many auction parameters of interest can be written using the private-value
quantile function or, by (\ref{B2V}), the bid quantile function and its quantile
derivative. We focus here on the conditional and unconditional
integral functionals%
\begin{equation}
\theta\left(  x\right)  =\int_{0}^{1}\mathcal{F}\left[  \alpha,x,B\left(
\alpha|x,I\right)  ,B^{\left(  1\right)  }\left(  \alpha|x,I\right)
;I\in\mathcal{I}\right]  d\alpha,\quad\theta=\int_{\mathcal{X}}\theta\left(
x\right)  dx \label{Thetafunc}%
\end{equation}
where $\mathcal{F}\left(  \alpha,x,b_{0I},b_{1I};I\in\mathcal{I}\right)  $ is
a real valued continuous function. Three illustrative examples are as follows.

\paragraph{Example 1: CRRA parameter.}

For symmetric risk-averse bidders with a concave utility function, the best-response condition (\ref{BR}) becomes%
\[
\left.  \frac{\partial}{\partial a}\left\{  U\left(  V\left(  \alpha
|X,I\right)  -B\left(  a|X,I\right)  \right)  a^{I-1}\right\}  \right\vert
_{a=\alpha}=0.
\]
Rearranging as in Guerre et al. (2009) yields that $V\left(  \alpha|X,I\right)  =B\left(
\alpha|X,I\right)  +\lambda^{-1}\left(  \frac{\alpha B^{\left(  1\right)
}\left(  \alpha|X,I\right)  }{I-1}\right)  $ where $\lambda\left(
\cdot\right)  =U\left(  \cdot\right)  /U^{\prime}\left(  \cdot\right)  $. For
risk-averse bidders with a CRRA utility function $U\left(  t\right)
=t^{\nu}$, arguing as for Proposition \ref{Ident} shows%
\begin{align}
V\left(  \alpha|X,I\right)   &  =B\left(  \alpha|X,I\right)  +\nu
\frac{\alpha B^{\left(  1\right)  }\left(  \alpha|X,I\right)  }{I-1}%
,\label{B2Vcrra}\\
B\left(  \alpha|X,I\right)   &  =\frac{\frac{I-1}{\nu}}{\alpha^{\frac{I-1}{\nu}}}\int%
_{0}^{\alpha}a^{\frac{I-1}{\nu}-1}V\left(  a|X,I\right)  da.\nonumber
\end{align}
These two formulas show that the stability implications of Proposition
\ref{Ident} for linear private-value and bid quantile functions are preserved
under CRRA. Assuming as in Guerre et al. (2009) that the number of bidders is exogenous, i.e
$V\left(  \alpha|X,I\right)  =V\left(  \alpha|X\right)  $ for all $I$, gives that the 
	risk-aversion $\nu$ satisfies,
for any pair $I_{0}\neq I_{1}$%
\begin{equation}
\nu
=\
\frac{\theta_{n}}{\theta_{d}}=\frac{
	\int_{\mathcal{X}}\left[  \int%
_{0}^{1}\left(  B\left(  \alpha|x,I_{1}\right)  -B\left(  \alpha
|x,I_{0}\right)  \right)  \left(  \frac{\alpha B^{\left(  1\right)  }\left(
\alpha|x,I_{0}\right)  }{I_{0}-1}-\frac{\alpha B^{\left(  1\right)  }\left(
\alpha|x,I_{1}\right)  }{I_{1}-1}\right)  d\alpha\right]  dx
}{
\int_{\mathcal{X}}\left[  \int_{0}^{1}\left(  \frac{\alpha B^{\left(  1\right)
}\left(  \alpha|x,I_{0}\right)  }{I_{0}-1}-\frac{\alpha B^{\left(  1\right)
}\left(  \alpha|x,I_{1}\right)  }{I_{1}-1}\right)  ^{2}d\alpha\right]  dx},
\label{Crrafp}
\end{equation}
which gives identification of $\nu$. Following Lu and Perrigne (2008), the risk-aversion
parameter $\nu$ can also be identified combining ascending and first-price
auctions data. As seen from Gimenes (2017), the private-value quantile
function $V_{asc}\left(  \alpha|X,I\right)  $ can be easily estimated from
ascending auctions. Equating $V_{asc}\left(  \alpha|X,I\right)  $ to $V\left(
\alpha|X,I\right)  $ in (\ref{B2Vcrra}) gives that $\nu$ satisfies%
\begin{equation}
\nu=\frac{\int_{\mathcal{X}}\left[  \int_{0}^{1}\left(  V_{asc}\left(
\alpha|x,I\right)  -B\left(  \alpha|x,I\right)  \right)  \frac{\alpha
B^{\left(  1\right)  }\left(  \alpha|x,I\right)  }{I-1}d\alpha\right]
dx}{\int_{\mathcal{X}}\left[  \int_{0}^{1}\left(  \frac{\alpha B^{\left(
1\right)  }\left(  \alpha|x,I\right)  }{I-1}\right)  ^{2}d\alpha\right]  dx}.
\label{Crraasc}
\end{equation}

\paragraph{Example 2: Expected revenue.}

Suppose that the seller decides to reject bids lower than a reserve price $R$
and let $\alpha_{R}=\alpha_{R}\left(  X,I\right)  $ be the associated
screening level, i.e. $\alpha_{R}=F\left(  R|X,I\right)  $. For CRRA bidders,
the first-price auction seller expected revenue is\footnote{It is assumed
for the sake of brevity that the seller value for the good is $0$. The expected
revenue formula for the general case follows from Gimenes (2017).}%
\begin{align}
ER_{\nu}\left(  \alpha_{R}|X,I\right)   &  =\frac{\nu\cdot I\cdot
V\left(  \alpha_{R}|X,I\right)  }{\left(  I-1\right)  \left(  \nu-1\right)
+\nu}\alpha_{R}^{^{\frac{I-1}{\nu}}}\left(  1-\alpha_{R}^{^{\left(
I-1\right)  \frac{\nu-1}{\nu}+1}}\right) \nonumber\\
&  +\frac{I\left(  I-1\right)  }{\left(  I-1\right)  \left(  \nu-1\right)
+\nu}\int_{\alpha_{R}}^{1}a^{\frac{I-1}{\nu}-1}\left(  1-a^{\left(
I-1\right)  \frac{\nu-1}{\nu}+1}\right)  V\left(  a|X,I\right)  da.
\label{ER}%
\end{align}
This expression includes an integral item%
\[
\theta\left(  X;\alpha_{R}\right)  =\int_{\alpha_{R}}^{1}a^{\frac{I-1}{\nu
}-1}\left(  1-a^{\left(  I-1\right)  \frac{\nu-1}{\nu}+1}\right)
V\left(  a|X,I\right)  da
\]
which can be estimated by plugging in a risk-aversion estimator
$\widehat{\nu}$ and an estimator $\widehat{V}\left(  \alpha|X,I\right)  $
of the private-value quantile function, or estimators of the bid quantile
function and its derivative by (\ref{B2V}).\footnote{Under risk-neutrality,
integrating by parts gives that%
\[
\int_{\alpha_{R}}^{1}B^{\left(  1\right)  }\left(  \alpha|X,I\right)
\alpha^{I-1}\left(  1-\alpha\right)  d\alpha=B\left(  \left.  \alpha
_{R}\right\vert X,I\right)  \alpha_{R}^{I-1}\left(  1-\alpha_{R}\right)
-\int_{\alpha_{R}}^{1}B\left(  \alpha|X,I\right)  \alpha^{I-1}\left(
I-1-I\alpha\right)  d\alpha,
\]
estimation of $\theta\left(  X;\alpha_{R}\right)  $ can also be done using
only a bid quantile estimator.}

\paragraph{Example 3: Private-value distribution}
Additional examples of conditional parameter $\theta (\cdot)$ are  the private-value conditional cdf and pdf. Note first that (\ref{Vquant}) shows that the conditional private-value cdf is an integral functional of the private-value quantile function 
\begin{equation}
F\left(  v|X,I\right)  =\mathbb{E}\left[  
\left. \mathbb{I}\left[  V\left(
A|x,I\right)  \leq v\right] \right|X,I \right]  =\int_{0}^{1}%
\mathbb{I}\left[  V\left(  \alpha|X,I\right)  \leq v\right]  d\alpha,\quad
A\sim\mathcal{U}_{\left[  0,1\right]  } .
\label{Cdf}
\end{equation}
Dette and Volgushev (2008) have considered a smoothed version $\mathbb{I}%
_{\eta}\left(  \cdot\right)  $ of the indicator function
\[
F_{\eta}\left(  v|X,I\right)  =\int_{0}^{1}\mathbb{I}_{\eta}\left[  v-V\left(
\alpha|X,I\right)  \right]  d\alpha
\]
where $\mathbb{I}_{\eta}\left(  t\right)  =\int_{-\infty}^{t/\eta}k\left(
u\right)  du$, $k\left(  \cdot\right)  $ being a kernel function and $\eta$ a
bandwidth parameter. Differentiating $F_{\eta}\left(  v|X,I\right)  $ gives%
\[
f_{\eta}\left(  v|X,I\right)  =\frac{1}{\eta}\int_{0}^{1}k\left(
\frac{v-V\left(  \alpha|X,I\right)  }{\eta}\right)  d\alpha
\]
which converges to the private-value pdf when $\eta$ goes to $0$. Note that
$F_{\eta}\left(  v|X,I\right)  $ and $f_{\eta}\left(  v|X,I\right)  $ can be
estimated by plugging in an estimator $\widehat{V}\left(  \alpha|X,I\right)  $
of $V\left(  \alpha|X,I\right)  $. The resulting cdf and pdf estimators inherit of the dimension reduction property of $\widehat{V}\left(  \alpha|X,I\right)  $. As
the private-value estimator proposed
in the next section is consistent over the whole $\left[  0,1\right]  $, no boundary
trimming is needed. This contrasts with the GPV pdf estimator.
As noted by Escanciano and Guo (2019) in a general context, the integral in  $f_{\eta}\left(  v|X,I\right)$ can  be replaced by a sample average over iid uniform draws $A_s$, as used for the density estimator (\ref{HKM}). 

%TCIMACRO{\TeXButton{TeX field}{\setcounter{equation}{0}}}%
%BeginExpansion
\setcounter{equation}{0}%
%EndExpansion

\section{Augmented quantile-regression estimation\label{ASQRE}}

Proposition \ref{Ident} suggests to base the estimation of the private-value
quantile function on estimations of $B\left(  \alpha|x,I\right)  $ and of its
derivative $B^{\left(  1\right)  }\left(  \alpha|x,I\right)  $ with respect to
$\alpha$. The \textit{augmented }methodology
applies local polynomial expansion with respect to $\alpha$ for joint
estimation of $B\left(  \alpha|x,I\right)  $ and $B^{\left(  1\right)
}\left(  \alpha|x,I\right)  $.  To
ensure comparability with the auction literature which considers private-value pdf having $s$ continuous derivatives, we assume that the private-value
quantile function $V\left(  \alpha|x,I\right)  $ has $s+1$ continuous
derivatives with respect to $\alpha$. As seen from (\ref{V2B}), this implies
that the bid quantile function $B\left(  \alpha|x,I\right)  $ has $s+2$
continuous derivatives with respect to $\alpha>0$. Let
$\left(X_{\ell}, I_{\ell}, B_{1\ell},\ldots, B_{I_{\ell}\ell}\right)$, $\ell=1,\ldots,L$, be an iid first-price auction sample  with $I_{\ell}$ bids $B_{i \ell}$ and good characteristics $X_{\ell}$.

\subsection{Augmented quantile estimation without covariate}

\paragraph{Estimation.}
Assume
first that $V\left(  \alpha|X,I\right)  =V\left(  \alpha|I\right)  $ so that
$B\left(  \alpha|X,I\right)  =B\left(  \alpha|I\right)  $. Let $\rho_{\alpha
}\left(  \cdot\right)  $ be the check function
\[
\rho_{\alpha}\left(  q\right)  =q\left(  \alpha-\mathbb{I}\left(
q\leq0\right)  \right)  .
\]
It is well known that
\[
B\left(  \alpha|I\right)  =\arg\min_{q}\mathbb{E}\left[  \mathbb{I}\left(
I_{\ell}=I\right)  \rho_{\alpha}\left(  B_{i\ell}-q\right)  \right]
,\quad\alpha\in\left(  0,1\right)  \text{.}%
\]
We now exhibit a functional objective function which achieves its minimum at the restriction of $B(\cdot|I)$ over $\left[  \alpha-h,\alpha+h\right]
\cap\left[  0,1\right]
$. 
It easily follows that, for a non negative kernel function $K\left(  \cdot\right)$ with support $[-1,1]$ and a positive bandwidth $h=h_{L}$,
\begin{align}
&  \left\{  B\left(  \tau |I\right)  ,\tau \in\left[  \alpha-h,\alpha+h\right]
\cap\left[  0,1\right]  \right\} \nonumber\\
&  \quad=\arg\min_{q\left(  \cdot\right)  }\int_{0}^{1}\mathbb{E}\left[
\mathbb{I}\left(  I_{\ell}=I\right)  \rho_{a}\left(  B_{i\ell}-q\left(
a\right)  \right)  \right]  \frac{1}{h}K\left(  \frac{a-\alpha}{h}\right)  da,
\label{Bmin}%
\end{align}
where the minimization is performed over the set of functions $q\left(
\cdot \right)  $ over $\left[  \alpha-h,\alpha+h\right]
\cap\left[  0,1\right]  $.
This  can be used to estimate the derivative $B^{\left(  1\right)  }\left(  \alpha|I\right)  $, using minimization over Taylor polynomial of order $s+1$ instead of $q(\cdot)$. A Taylor expansion of order $s+1$ gives 
\begin{align*}
B\left(  a|I\right)  
& =
B\left(  \alpha|I\right)  +B^{\left(  1\right)  }\left(
\alpha|I\right)  \left(  a-\alpha\right)  +\cdots+\frac{B^{\left(  s+1\right)
}\left(  \alpha|I\right)  }{\left(  s+1\right)
!}
\left(  a-\alpha\right)^{s+1}
+O\left(  h^{s+2}\right)
\\  
& =
\pi \left( a - \alpha \right)^{\prime} b(\alpha|I) 
+O\left(  h^{s+2}\right)
\text{ where }
\\
b(\alpha|I) & = \left[B\left(  \alpha|I\right), \ldots, B^{(s+1)} (\alpha|I)
\right]^{\prime}
\text{ and }
\pi\left(  t\right)  =\left[  1,t,\frac{t^{2}}{2}\ldots,\frac{t^{s+1}}{\left(
	s+1\right)  !}\right]  ^{\prime}.
\end{align*}
The $(s+2) \times 1$ vector $b(\cdot|I)$ stacks the successive bid quantile derivatives, and is the parameter to be estimated.
Let $b=\left[  \beta_{0},\ldots,\beta_{s+1}\right]  ^{\prime} \in \mathbb{R}^{s+2}$ be the generic
coefficients of such a Taylor polynomial function.
The sample version of the objective function (\ref{Bmin}), restricted to
local polynomial functions $\pi (\cdot)^{\prime} b$ instead of $q(\cdot)$, is
\begin{align*}
\widehat{\mathcal{R}}\left(  b;\alpha,I\right)   &  =\frac{1}{LI}\sum_{\ell
=1}^{L}\mathbb{I}\left(  I_{\ell}=I\right)  \sum_{i=1}^{I_{\ell}}\int_{0}%
^{1}\rho_{a}\left(  B_{i\ell}-\mathcal{\pi}\left(  a-\alpha\right)  ^{\prime
}b\right)  \frac{1}{h}K\left(  \frac{a-\alpha}{h}\right)  da\\
&  =\frac{1}{LI}\sum_{\ell=1}^{L}\mathbb{I}\left(  I_{\ell}=I\right)
\sum_{i=1}^{I_{\ell}}\int_{-\frac{\alpha}{h}}^{\frac{1-\alpha}{h}}\rho
_{\alpha+ht}\left(  B_{i\ell}-\pi\left(  ht\right)  ^{\prime}b\right)
K\left(  t\right)  dt.
\end{align*}
The \textit{augmented quantile }estimator is $\widehat{b}\left(
\alpha|I\right)  =\arg\min_{b\in \mathbb{R}^{s+2}}\widehat{\mathcal{R}}\left(  b;\alpha,I\right)
$, $\widehat{\beta}_{0}\left(  \alpha|I\right)  $ and $\widehat{\beta}%
_{1}\left(  \alpha|I\right)  $ being estimators of $B\left(  \alpha|I\right)
$ and its first derivative $B^{\left(  1\right)  }\left(  \alpha|I\right)  $,
respectively.

\paragraph{Homogenized bid and elliptical random coefficients.}
A two-step version of the augmented method presented above can be used to estimate the homogenized private-value quantile function $v(\cdot|I)$ from (\ref{HHS}). Regressing $B_{i \ell}$ on $X_{\ell}$ and an intercept for those auctions with $I_{\ell}=I$ gives a consistent estimator $\widehat{\gamma}_1 (I)$ of $\gamma_1 (I)$. Let $\widehat{B}_{i \ell}= B_{i \ell} - X_{\ell}^{\prime} \widehat{\gamma}_1$ be the estimated homogenized bids. Then replacing $B_{i \ell}$ with $\widehat{B}_{i \ell}$ in the objective function $\widehat{\mathcal{R}}\left(  b;\alpha,I\right)$  gives estimators $\check{\beta} (\cdot|I)$ and $\check{\beta}_1 (\cdot|I)$ of the homogenized-bid quantile function and of its first derivative. The resulting estimator of the private-value quantile function is then
\[
\widehat{V} (\alpha|X,I)
=
X^{\prime} \widehat{\gamma}_{1} (I)
+
\check{\beta} (\alpha|I)
+
\frac{\alpha \check{\beta}_1 (\alpha|I)}{I-1}.
\]   
The elliptical random-coefficient quantile specification (\ref{Bquant_rc}) can be estimated similarly.
Studying the asymptotic properties of this two-step procedures is outside the scope of this paper. Bhattacharya (2019) considers a related two-step procedure that can be useful for ascending auctions, where estimating quantile derivative is not needed.

\subsection{Augmented quantile-regression}

An extension of this procedure is the \textit{augmented quantile-regression }estimator, AQR hereafter, which assumes
$
V\left(  \alpha |x,I\right)  =x_{1}^{\prime}  \gamma\left(
\alpha|I\right)$, recalling $x_1 = [1,x^{\prime}]^{\prime}$.
Proposition \ref{Ident}-(i) then gives $B\left(  \alpha|x,I\right)  =
x_{1}^{\prime} \beta\left(  \alpha|I\right)$. 
Define now 
\begin{align}
P\left(  x,t\right)  
& 
=
\pi\left(  t\right)  
\otimes
x_{1}
=
\left[
1,
x^{\prime},
t,
t\cdot x^{\prime},
\ldots
,
\frac{t^{s+1}}{(s+1)!} 
,
\frac{t^{s+1}}{(s+1)!} \cdot x^{\prime}
\right]^{\prime} 
\in \mathbb{R}^{(s+2)(D+1)}
\label{Pxt}
\\
b(\alpha|I)
&
=
\left[
\beta(\alpha|I)^{\prime}
,
\beta^{(1)}(\alpha|I)^{\prime}
,
\ldots
,
\beta^{(s+1)}(\alpha|I)^{\prime}
\right]
\nonumber
\end{align}
so that the Taylor expansion of $B\left(  \alpha|X,I\right)  $
writes
\[
B\left(  \alpha+ht|x,I\right)  
=
\sum_{k=0}^{s+1}
x_1^{\prime}
\beta^{(k)}(\alpha|I)
\frac{(ht)^k}{k!}
+O\left(  h^{s+2}\right)
=
P\left(  x,ht\right)  ^{\prime}b\left(
\alpha|I\right)  +O\left(  h^{s+2}\right).
\]
The corresponding generic parameter is the $(s+2)(D+1) \times 1$ column vector
$
b=
\left[  
\beta_{0}^{\prime}
,
\beta_{1}^{\prime},
\ldots,
\beta_{s+1}^{\prime}
\right]^{\prime}
$
where  the $\beta_{j}$ are all of dimension $D+1$, and the objective
function  becomes
\begin{align}
\widehat{\mathcal{R}}\left(  b;\alpha,I\right)   &  
=\frac{1}{LI}\sum_{\ell
=1}^{L}\mathbb{I}\left(  I_{\ell}=I\right)  \sum_{i=1}^{I_{\ell}}\int_{0}%
^{1}\rho_{a}\left(  B_{i\ell}-P\left(  X_{\ell},a-\alpha\right)  ^{\prime
}b\right)  \frac{1}{h}K\left(  \frac{a-\alpha}{h}\right)  da
\nonumber\\
&  =\frac{1}{LI}\sum_{\ell=1}^{L}\mathbb{I}\left(  I_{\ell}=I\right)
\sum_{i=1}^{I_{\ell}}\int_{-\frac{\alpha}{h}}^{\frac{1-\alpha}{h}}\rho
_{\alpha+ht}\left(  B_{i\ell}-P\left(  X_{\ell},ht\right)  ^{\prime}b\right)
K\left(  t\right)  da \label{CalR}%
\end{align}
which accounts for the covariate $X_{\ell}$. The estimation of $b\left(
\alpha|I\right)  $ is 
\[
\widehat{b}\left(  \alpha|I\right)  
=
\arg
\min_{b \in \mathbb{R}^{(s+2)(D+1)}}
\widehat{\mathcal{R}}\left(  b;\alpha,I\right)
\] 
and the private-value quantile-regression estimator is
\[
\widehat{V}\left(  \alpha|x,I\right)  
=
 x_{1}^{\prime}
\widehat{\gamma}\left(  \alpha|I\right)  \text{ with }\widehat{\gamma}\left(
\alpha|I\right)  =\widehat{\beta}_{0}\left(  \alpha|I\right)  +\frac
{\alpha\widehat{\beta}_{1}\left(  \alpha|I\right)  }{I-1}.
\]
The bid quantile function and its derivatives can be estimated using
$\widehat{B}\left(  \alpha|x,I\right)  =x_{1}^{\prime}\widehat{\beta}_{0}\left(  \alpha|I\right)  $ and $\widehat{B}^{\left(
1\right)  }\left(  \alpha|x,I\right)  =x_1^{\prime}\widehat{\beta}_{1}\left(  \alpha|I\right)  $, so that
$
\widehat{V} (\alpha|x,I)
=
\widehat{B} (\alpha|x,I)
+
\frac{\alpha \widehat{B}^{(1)} (\alpha|x,I)}{I-1}
$. 
The rearrangement method of
Chernozhukov, Fern\'{a}ndez-Val and Gallichon (2010) can be used to obtain increasing quantile estimators.

\begin{figure}[h]
	\centering
	\includegraphics[width=.7\linewidth]{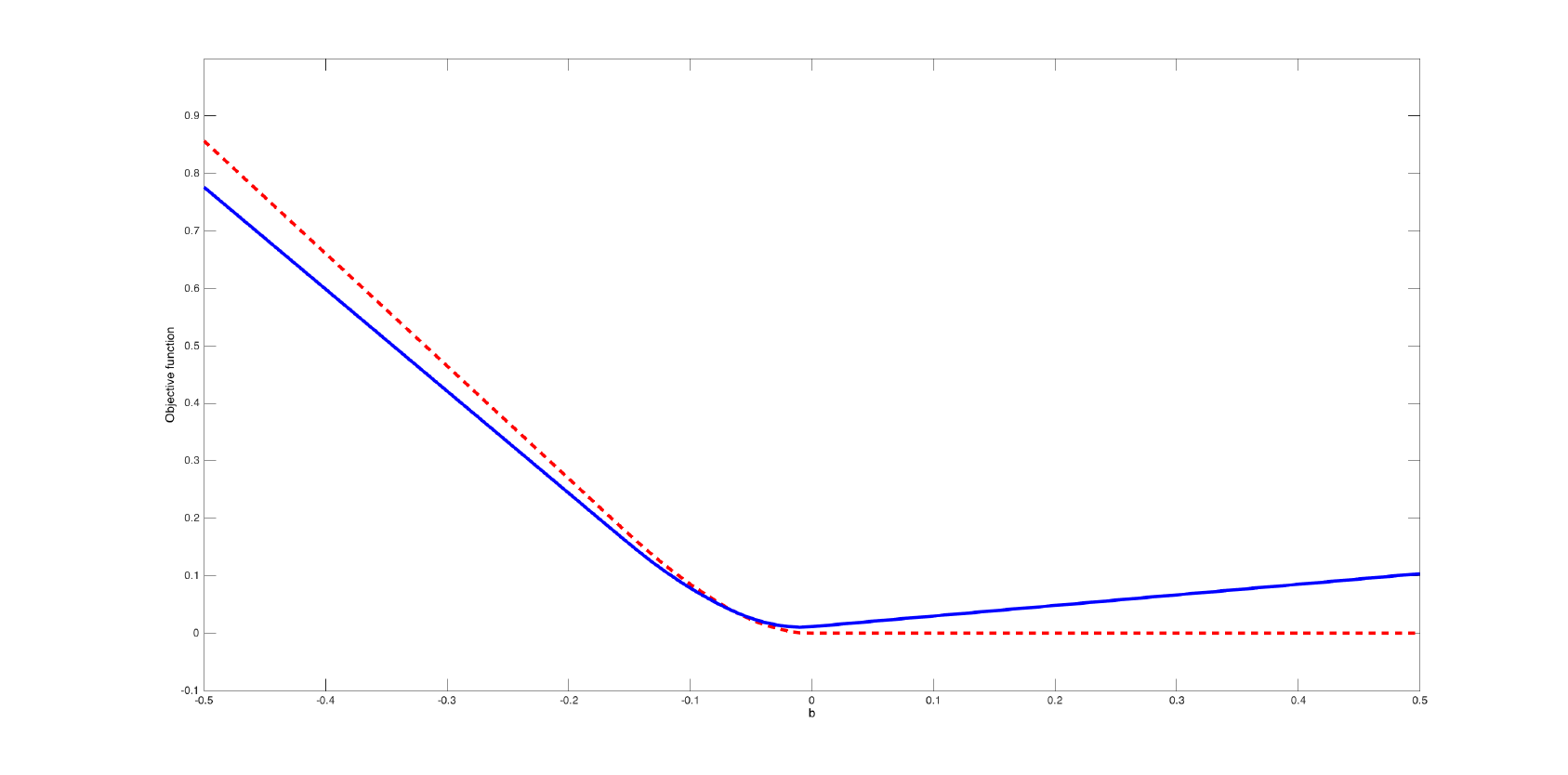}
	\caption{A path of the AQR objective function $\protect\widehat{\mathcal{R}%
		}\left(  b\cdot;1,I\right)  $ (solid line)  and of the objective function of the standard quantile-regression estimator (dotted line) when $b$ varies in the direction $\left[
		1,\ldots,1\right]  ^{\prime}$.}
	\label{Fig1}
\end{figure}

\paragraph{AQR estimator properties.}
Bassett and Koenker (1982) report that standard quantile-regression estimators
are not defined near the extreme quantile levels $\alpha=0$ or $\alpha=1$, mostly because the associated objective function has some flat parts. The AQR is better behaved  because the objective function $\widehat{\mathcal{R}}\left(
b;\alpha,I\right)  $ averages the check function $\rho_{a}\left(
\cdot\right)  $ for quantile levels $a$ in $\left[  \alpha-h,\alpha+h\right]
\cap\left[  0,1\right]  $, ensuring that the AQR objective function is not flat for extreme quantile levels, as illustrated in Figure \ref{Fig1}.\footnote{This averaging effect requests
that $t\mapsto P\left(  X_{\ell},ht\right)  ^{\prime}b$ is not constant
meaning that the derivative components of $b$ should not vanish.} 

%TCIMACRO{\FRAME{fhFU}{6.7481in}{3.3313in}{0pt}{\Qcb{A path of the objective
%function $\protect\widehat{\mathcal{R}}\left(  b\cdot;1,I\right)  $ (solid
%line) of the augmented quantile regression estimator and of the objective
%function of the standard quantile regression estimator (dotted line) when $b$
%diverges in the direction $\left[  1,\ldots,1\right]  ^{\prime}$.}}%
%{\Qlb{Fig1}}{objgraph.bmp}{\special{ language "Scientific Word";
%type "GRAPHIC";  maintain-aspect-ratio TRUE;  display "USEDEF";
%valid_file "F";  width 6.7481in;  height 3.3313in;  depth 0pt;
%original-width 19.1997in;  original-height 9.4403in;  cropleft "0";
%filename 'objgraph.bmp';file-properties "XNPEU";}} }%
%BeginExpansion
Therefore the AQR estimator is easier to define for the extreme
quantile levels $\alpha=0$ and $\alpha=1$ than the standard quantile-regression estimator. This is especially relevant for estimating auction
models as the winner is expected to belong to the upper tail as soon as the
number of bidders is large enough.
It also follows from the theoretical
study of the objective function $\widehat{\mathcal{R}}\left(  \cdot
;\cdot,I\right)  $ that the AQR  estimator is uniquely defined for
all quantile levels with a probability tending to 1.  The bid AQR estimator is also smoother
than the standard quantile-regression one, see  Figure \ref{BidI2} in
the Application Section and \ref{App:Bias} for a formal argument.

\setcounter{equation}{0}

\section{Main results\label{Res}}

Some additional notations are as follows. 
Let $S_{0} = [1,0,\ldots,0]$ and 
$S_{1}=\left[ 0,1,0,\ldots,0\right]$ be $1\times\left(  s+2\right)  $ selection vectors such that $S_0 \pi(t)=1$, $S_1 \pi(t)=t$. Let $\operatorname{Id}_{D+1}$ be the $(D+1)\times (D+1)$ identity matrix and set
$\mathsf{S}_{j}=S_{j}
\otimes\operatorname*{Id}_{D+1}$, $j=0,1$,
so that
$\mathsf{S}_{0}
\widehat{b}\left(  \alpha|I\right)
=\widehat{\beta}_{0}\left(  \alpha|I\right)  $ and $\mathsf{S}_{1}
\widehat{b}\left(  \alpha|I\right)
=\widehat{\beta}_{1}\left(  \alpha|I\right)  $ are respectively estimators of $\beta(\alpha)$ and its first derivative $\beta^{(1)} (\alpha)$. $\mathrm{Tr}(\cdot)$ is the trace of a square matrix and 
	$\partial_u^n$ stands for $\frac{\partial^n}{ \partial u^n}$.
	For two sequences $\{a_L\}$ and $\{b_L\}$, $a_{L}\asymp b_{L}$ means
	that both $a_{L}/b_{L}=O\left(  1\right)  $ and $b_{L}/a_{L}=O\left(
	1\right)  $. The norm $\left\Vert \cdot\right\Vert $ is the Euclidean one,
	i.e. $\left\Vert e\right\Vert =\left(  e^{\prime}e\right)  ^{1/2}$. For a matrix $A$, $\| A\| = \sup_{b:\|b\|=1} \| Ab \|$.
 Convergence in distribution is denoted  as `$\stackrel{d}{\rightarrow}$'.

\subsection{Main assumptions}

\setcounter{hp}{0}

\begin{hp}
\label{Auct}(i) The auction variables $\left(  I_{\ell},X_{\ell},V_{i\ell
},B_{i\ell},i=1,\ldots,I_{\ell}\right)  $ are iid across $\ell$. 
The support $\mathcal{X}$ of $X_{\ell}$ given $I_{\ell}=I$ is independent of $I$, compact with non empty interior. The support $\mathcal{I}$ of  $I_{\ell} \geq 2$ is finite. The matrices $\mathbb{E} \left[\mathbb{I} \left(I_{\ell}=I\right)
	X_{1\ell} X_{1\ell}^{\prime}
	\right]$, where $X_{1\ell} = [1,X_{\ell}^{\prime}]^{\prime}$, have an inverse for all $I$ of $\mathcal{I}$.

(ii) Given $\left(  X_{\ell},I_{\ell}\right)  =\left(  x,I\right)  $, the
$V_{i\ell}$, $i=1,\ldots,I_{\ell}$ are iid with a continuously differentiable conditional quantile
function $V\left(  \alpha|x,I\right)  $ with
$
\min_{\left(  \alpha,x,I\right)  \in\left[  0,1\right]  \times\mathcal{X\times
I}}V^{\left(  1\right)  }\left(  \alpha|x,I\right)  >0$  and $\max_{\left(
\alpha,x,I\right)  \in\left[  0,1\right]  \times\mathcal{X\times I}}V^{\left(
1\right)  }\left(  \alpha|x,I\right)  <\infty$.
\end{hp}

\setcounter{hp}{18}

\begin{hp}
\label{Specification} For some $s\geq1$ and each $I\in\mathcal{I}$, $V\left(
\alpha|X,I\right)=X_{1}^{\prime} \gamma(\alpha|I)$ is as in (\ref{Vqr}), where $X_{1}=[1,X^{\prime}]^{\prime}$ and $\gamma(\cdot|I)$ is $\left(  s+1\right)  -$times continuously
differentiable over $\left[  0,1\right]$.  
\end{hp}

%

%TCIMACRO{\TeXButton{TeX field}{\setcounter{hp}{7}}}%
%BeginExpansion
\setcounter{hp}{7}%
%EndExpansion

\begin{hp}
\label{Kernel}The kernel function $K\left(  \cdot\right)  $ with support
$\left(  -1,1\right)  $ is symmetric, continuously differentiable over the
straight line, and strictly positive over $\left(  -1,1\right)  $. The
positive bandwidth $h$ goes to $0$ with
$
\lim_{L\rightarrow\infty}\frac{\log^2 L}{Lh^{2}}=0
$.
\end{hp}

\setcounter{hp}{5}

\begin{hp}
\label{Funct}
For all $x$ in $\mathcal{X}$ and $\alpha$ in $\left[  0,1\right]
$, the function $\mathcal{F}\left[  \alpha,x,b_{0I},b_{1I};I\in\mathcal{I}%
\right]  $ in (\ref{Thetafunc}) is twice  differentiable with respect to $b_{0I}$ and $b_{1I}$, $I$
in $\mathcal{I}$. For each $I$ in
$\mathcal{I}$, these partial derivatives of order 1 and 2 are continuous with
respect to all the variables. 
\end{hp}

\bigskip

Assumption \ref{Auct}-(i) is standard. Assumption \ref{Auct}-(ii) allows for private values depending on the number of bidders. Recall that
\begin{equation}
	V^{\left(  1\right)  }\left(  \alpha|x,I\right)  =\frac{1}{f\left(  V\left(
		\alpha|x,I\right)  |x,I\right)  }, \label{V2f}%
\end{equation}
where $f\left(  v|x,I\right)  $ is the conditional private-value pdf. Hence
Assumption \ref{Auct}-(ii) amounts to assume that $f\left(  v|x,I\right)  $ is
bounded away from $0$ and infinity on its support $\left[  V\left(
0|x,I\right)  ,V\left(  1|x,I\right)  \right]  $ as assumed for instance in
Riley and Samuelson (1981), Maskin and Riley (1984) or GPV. The condition
$0<f\left(  v|x,I\right)  <\infty$ is also used for asymptotic normality of
quantile-regression estimator, see Koenker (2005). Assumption \ref{Specification} is a standard smoothness condition which, by (\ref{V2f}), parallels GPV who assume that the pdf $f(v|x,I)$ is $s$-times differentiable. 

The bandwidth rate in Assumption \ref{Kernel} is unusual in kernel or local polynomial nonparametric estimation, where rate conditions as $1/(Lh)=o(1)$ are more common. This is due to a key linearization expansion for $\widehat{V} (\alpha|x,I)$, which holds with  an  $O_{\mathbb{P}} \left(\log L/(L \sqrt{h}) \right)$ error term  that must go to $0$, see (\ref{WidetildeV}) and (\ref{IMSEBaha}) in Theorem \ref{IMSE} below.\footnote{See also Theorem \ref{Baha} in \ref{App:Bahadur}, where it is shown more specifically that this bandwidth order is needed for the linearization of $\widehat{B}^{(1)} (\alpha|x,I)$.}

Assumption \ref{Funct} holds for most of the examples of functionals above. A
	notable exception is the cdf $F\left(  v|x,I\right)  $ in Example 3, which
	involves an indicator function which is not smooth. However it holds for the
	smoothed approximation $F_{\eta}\left(  v|x,I\right)  $ of the cdf, although
	Assumption \ref{Funct} implicitly rules out vanishing bandwidth $\eta$ in
	Example 3.

\subsection{Private value quantile estimation results \label{AQR_results}}

The next sections give our theoretical results for integrated mean squared
error, uniform consistency and asymptotic distribution of the augmented estimator $\widehat{V}%
\left(  \cdot|\cdot,I\right)  $.  
These results are derived using a pseudo-true value framework, in which $\widehat{b} (\cdot|I)$ is viewed as an estimator of the minimizer $\overline{b} (\cdot|I)$ of the population counterpart of $\widehat{\mathcal{R}}\left(  b;\alpha,I\right)$
\[
\overline{b}\left(  \alpha|I\right)  
=
\arg
\min_{b \in \mathbb{R}^{(s+2)(D+1)}}
\overline{\mathcal{R}}\left(  b;\alpha,I\right)
\text{ where }
\overline{\mathcal{R}}\left(  b;\alpha,I\right)
=
\mathbb{E}
\left[\widehat{\mathcal{R}}\left(  b;\alpha,I\right) \right]
\]
which asymptotic existence and uniqueness is established for the proofs of our main results.
Define accordingly 
$\overline{\beta}_{0}\left(  \alpha|I\right)=\mathsf{S}_{0}\overline{b}\left(  \alpha|I\right)
$ and $\overline{\beta}_{1}\left(  \alpha|I\right) = \mathsf{S}_{1} \overline{b}\left(  \alpha|I\right)$ and
\begin{equation}
\overline{V} (\alpha|x,I)
=
x_{1}^{\prime}
\left(\overline{\beta}_{0}+\frac{\alpha \overline{\beta}_{1}\left(  \alpha|I\right)}{I-1}\right).
\label{BarV}
\end{equation}
The difference $\overline{V} (\alpha|x,I)- V(\alpha|x,I)$ can be interpreted as a bias term.

Because $\widehat{V} (\cdot|\cdot,I)$ is defined in an implicit way via the minimization of the objective function (\ref{CalR}), 	 
its asymptotic study relies on a linearization of $\widehat{b} (\alpha|I)-\overline{b} (\alpha|I)$ which, in a quantile setup, is called a Bahadur expansion, see Theorem \ref{Baha} in \ref{App:Bahadur} and Koenker (2005, Chap. 4).
It is shown that, in a vicinity of $\overline{b} (\alpha|I)$, $b \mapsto \widehat{\mathcal{R}}\left(  b;\alpha,I\right)$ is twice differentiable with a first derivative  
$\widehat{\mathcal{R}}^{(1)} \left(  b;\alpha,I\right)$
satisfying $\mathbb{E} \left[\widehat{\mathcal{R}}^{(1)} \left(  \overline{b} (\alpha|I);\alpha,I\right)\right] =0$,
and  with a Hessian matrix  $\overline{\mathcal{R}}^{(2)}\left(  \overline{b} (\alpha|I);\alpha,I\right)$ which is  invertible.  The leading term of $\widehat{b}(\alpha|I)-\overline{b}(\alpha|I)$ is 
$
-
\left[
\overline{\mathcal{R}}^{(2)}\left(  \overline{b}(\alpha|I);\alpha,I\right)\right
]^{-1}
\widehat{\mathcal{R}}^{(1)} \left(  \overline{b}(\alpha|I);\alpha,I\right)
$ as shown in Theorem \ref{Baha},
so that the leading term of 
$\widehat{V}(\alpha|x,I)$ is 
\begin{eqnarray}
%\lefteqn{
\widetilde{V}(\alpha|x,I) 
=
\overline{V} (\alpha|x,I)
%}
%\nonumber \\
%&&
-
x_{1}^{\prime}
\left(\mathsf{S}_0+\frac{\alpha \mathsf{S}_1}{I-1} \right)
\left[
\overline{\mathcal{R}}^{(2)}\left(  \overline{\beta}(\alpha|I);\alpha,I\right)\right
]^{-1}
\widehat{\mathcal{R}}^{(1)} \left(  \overline{\beta}(\alpha|I);\alpha,I\right),
\label{WidetildeV}
\end{eqnarray}
see (\ref{IMSEBaha}) below. 
Because direct computations of the moments of $\widehat{V}(\alpha|x,I)$ are difficult due to its implicit definition and to potential nonlinearities,  moments of its linear leading term $\widetilde{V}(\alpha|x,I)$ are used as an approximation.

\subsubsection{Integrated mean squared error and uniform consistency rates}

Let us first introduce some notations for the integrated mean squared error (IMSE). 
Let $\Pi^{1}\left(  \alpha\right)  $ be the second column of the inverse of
$\int\pi\left(  t\right)  \pi\left(  t\right)  ^{\prime}K\left(  t\right)
dt$, i.e.,%
\[
\Pi^{1}\left(  \alpha\right)  =\left(  \int\pi\left(  t\right)  \pi\left(
t\right)  ^{\prime}K\left(  t\right)  dt\right)  ^{-1}S_{1}^{\prime}%
\]
and consider the variance terms
\begin{align*}
v^{2}\left(  \alpha\right)   &  =\Pi^{1}\left(  \alpha\right)  ^{\prime}%
\int\int\pi\left(  t_{1}\right)  \pi\left(  t_{2}\right)  ^{\prime}\min\left(
t_{1},t_{2}\right)  K\left(  t_{1}\right)  K\left(  t_{2}\right)  dt_{1}%
dt_{2}\Pi^{1}\left(  \alpha\right)  ,\\
\Sigma\left(  \alpha|I\right)   &  =\frac{\alpha^{2}v^{2}\left(
\alpha\right)  }{\left(  I-1\right)  ^{2}}\mathbb{E}^{-1}\left[
\frac{  X_{1\ell}    X_{1\ell}^{\prime}
\mathbb{I}\left(  I_{\ell}=I\right)  }{B^{\left(  1\right)  }\left(
\alpha|X_{\ell},I_{\ell}\right)  }\right] 
%\\
%&  
%\times
\mathbb{E}
\left[  X_{1\ell}    X_{1\ell}^{\prime}\mathbb{I}\left(  I_{\ell}=I\right)  \right]  
\mathbb{E}^{-1}
\left[
\frac{X_{1\ell}    X_{1\ell}^{\prime}
\mathbb{I}\left(  I_{\ell}=I\right)  }{B^{\left(  1\right)  }\left(
\alpha|X_{\ell},I_{\ell}\right)  }
\right]  ,\\
\Sigma_{I}  &  =\int_{\mathcal{X}}\int_{0}^{1}x_{1}^{\prime
}\Sigma\left(  \alpha|I\right)  x_{1}  d\alpha dx,
\end{align*}
where $\mathbb{E}^{-1}
\left[
\frac{X_{1\ell}    X_{1\ell}^{\prime}
	\mathbb{I}\left(  I_{\ell}=I\right)  }{B^{\left(  1\right)  }\left(
	\alpha|X_{\ell},I_{\ell}\right)  }
\right] $ is the inverse matrix of 
$
\mathbb{E}
\left[
\frac{X_{1\ell}    X_{1\ell}^{\prime}
	\mathbb{I}\left(  I_{\ell}=I\right)  }{B^{\left(  1\right)  }\left(
	\alpha|X_{\ell},I_{\ell}\right)  }
\right] 
$.
That $v^{2}\left(  \alpha\right)  $, and then $\Sigma_{I}$, is strictly
positive follows from the proof of Theorem \ref{IMSE} below, see in particular
Lemma \ref{Leadterm} in \ref{App:Prelims}. The bias, and integrated squared bias, of the estimator are asymptotically proportional to, respectively\footnote{The expression above depends upon the derivative $\beta^{(s+2)}(\alpha|I)$, which exists by (\ref{V2B}) for all $\alpha \neq 0$ since $\gamma(\cdot|I)$ is $(s+1)$-th differentiable. Proposition \ref{SeriesB2}-(iii) in \ref{App:Sieve} shows that $\alpha \beta^{(s+2)}(\alpha|I)$ can be defined over the whole quantile interval $[0,1]$ since 
	$\lim_{\alpha\downarrow 0} \alpha \beta^{(s+2)}(\alpha|I) =0$. \label{Ders2}}
\begin{align*}
&  \mathsf{Bias}(\alpha|x,I)=\frac{\alpha B^{(s+2)}(\alpha|x,I)}{I-1}
S_{1}\left(  
\int\pi\left(  t\right)  \pi\left(  t\right)  ^{\prime}K\left(  t\right)
dt\right)  ^{-1}\int\frac{t^{s+2}\pi\left(  t\right)  }{\left(  s+2\right)
!}K\left(  t\right)  dt,\\
&  \mathsf{Bias}_{I}^{2}=\int_{\mathcal{X}}\int_{0}^{1}
 \mathsf{Bias}^2 (\alpha|x,I) d\alpha dx.
\end{align*} 
The next Theorem deals with the IMSE of $\widehat{V} (\cdot|\cdot,I)$ and with its difference to its linearization $\widetilde{V} (\cdot|\cdot,I)$ in (\ref{WidetildeV}). 
\begin{theorem}
	\label{IMSE} 
	Under Assumptions \ref{Auct},
	\ref{Kernel}, \ref{Specification} and for $\widetilde{V} (\cdot|\cdot,I)$ as in (\ref{WidetildeV}), it holds for all $I$ in $\mathcal{I}$ 
	\begin{eqnarray}
	\mathbb{E}\left[  \int_{\mathcal{X}}\int_{0}^{1}\left(  \widetilde{V}\left(
	\alpha|x,I\right)  -V\left(  \alpha|x,I\right)  \right)  ^{2}d\alpha
	dx\right]   &  =& h^{2\left(  s+1\right)  }\mathsf{Bias}_{I}^{2}+\frac
	{\Sigma_{I}}{LIh}
	\label{IMSEexp}
	\\
	&  &+o\left(  h^{2\left(  s+1\right)  }+\frac{1}{LIh}\right),  
	\nonumber \\
	\textit{ with }
	\sup_{(\alpha,x) \in [0,1] \times \mathcal{X}}
	\sqrt{LIh}
	\left|
		\widehat{V}\left(  \alpha|x,I\right)
		-\widetilde{V}\left(  \alpha|x,I\right)
	\right|
	& = &
	O_{\mathbb{P}}
	\left(
	\frac{\log L}{h\sqrt{LI}}
	\right)
	=
	o_{\mathbb{P}}(1).
	\label{IMSEBaha}
	\end{eqnarray}
	It also holds, for each $I$ of $\mathcal{I}$,
	\begin{eqnarray}
	\sup_{(\alpha,x) \in [0,1]\times \mathcal{X}}
	\left|
	  \widehat{V} (\alpha|x,I) - V (\alpha|x,I)
	\right|
	& = &
	O_{\mathbb{P}}
	\left(\sqrt{\frac{\log L}{L I h}}\right)
	+
	O(h^{s+1}),
	\label{CVU.V}
	\\
	\sup_{(\alpha,x) \in [0,1]\times \mathcal{X}}
	\left|
	\widehat{B} (\alpha|x,I) - B (\alpha|x,I)
	\right|
	& = &
	O_{\mathbb{P}}
	\left(\sqrt{\frac{\log L}{L I}}\right)
	+
	o(h^{s+1}).
	\label{CVU.B}
	\end{eqnarray}
\end{theorem}

Theorem \ref{IMSE} gives the IMSE of the linearization $\widetilde{V} (\cdot|\cdot,I)$ of $\widehat{V} (\cdot|\cdot,I)$ in (\ref{IMSEexp}). Then (\ref{IMSEBaha}) gives the order of the linearization error in a uniform sense, which is negligible with the order $1/\sqrt{Lh} + O(h^{s+1})$ of the squared root IMSE under Assumption \ref{Kernel}. 
The linearization result (\ref{IMSEBaha}) requests
$\log L / (h \sqrt{LI}) = o(1)$, which
 is the main motivation for the unusual rate of bandwidth rate of Assumption \ref{Kernel}. This condition is driven by the linearization of the bid quantile derivative estimator $\widehat{B}^{(1)} (\cdot|\cdot,I)$.

The bias and variance leading terms in the IMSE expansion (\ref{IMSEexp}) are from the bid quantile derivative estimator 
$\alpha \widehat{B}^{(1)} (\alpha|x,I)/(I-1)$. 
As
\[
B^{\left(  1\right)  }\left(  \alpha|x,I\right)  
=
\frac{1}{g\left[  B\left(\alpha|x,I\right)  |x,I\right]  },
\]
where $g\left(  \cdot|\cdot\right)  $ is the bid conditional pdf, estimation of this item is similar to estimating a pdf. The rate  $1/Lh$ of the variance term $\Sigma_I/(LIh)$  is the rate of a kernel density estimator  in the absence of covariate. This is due to the quantile-regression specification.
 Compared to GPV density estimation rate $1/\sqrt{Lh^{D+1}}$, the rate of the AQR estimator does not suffer from the curse of dimensionality.
The order $O(h^{s+1})$ of the bias is given by (\ref{V2B}), implying that $B^{(1)} (\alpha|x,I)$ has as many derivatives as $V(\alpha|x,I)$, hence the exponent $s+1$.

Minimizing the leading term of the IMSE expansion (\ref{IMSEexp}) yields the optimal bandwidth%
\begin{equation}
h_{\ast}=\left(  \frac{  \Sigma_{I}}{2\left(
	s+1\right)  \mathsf{Bias}_{I}^{2}}\frac{1}{LI}\right)  ^{\frac{1}%
	{2s+3}}.
 \label{Hstar}%
\end{equation}
As in kernel estimation, a pilot bandwidth can be computed using a simple
private-value quantile-regression model to proxy $\Sigma_{I}$ and
$\mathsf{Bias}_{I}^{2}$ in a parametric way. The corresponding square root IMSE rate is
$
L^{\frac{s+1}{2s+3}}%
$
which corresponds to the optimal minimax rate given in GPV in the absence of covariate, up to a logarithmic term and an exponent $s+1$ due to estimation of the private-value quantile function, instead of $s$ appearing for pdf. In particular, it is $L^{-2/5}$ for $s=1$, with an exponent $2/5=.4$ close to $1/2$ suggesting potential good performances in small samples even in the presence of covariate. 

A similar rate can also be derived for the uniform consistency of $\widehat{V}(\cdot|\cdot,I)$ stated in (\ref{CVU.V}). Note also that (\ref{CVU.B}) shows that the bid quantile estimator $\widehat{B} (\cdot|\cdot,I)$ converges uniformly to $B(\cdot|\cdot,I)$ with a rate which is nearly parametric.\footnote{The uniform consistency rate in (\ref{CVU.B}) includes a bias term $o(h^{s+1})$, which is $O(h^{s+2})$ for $\alpha \neq 0$ because $B(\cdot|x,I)=x_{1}^{\prime} \beta (\cdot|I)$ is $(s+2)$-th times continuously differentiable over $(0,1]$. Because $B(\cdot|x,I)$ may have only $(s+1)$ derivatives at $\alpha=0$, the bias order $h^{s+2}$ may not hold uniformly over $[0,1]$. } 
All these convergence results take place over the whole $[0,1] \times \mathcal{X}$, meaning that potential boundary biases disappear asymptotically.

\subsubsection{Bias, variance and Central Limit Theorem}

While Theorem \ref{IMSE} reviews the global performance of the AQR estimator, this section details some of its local features, and in particular its upper boundary behavior. Define 
\[
\Pi_{h}^{1}\left(  \alpha\right)  =\left(  \int_{-\frac{\alpha}{h}}%
^{\frac{1-\alpha}{h}}\pi\left(  t\right)  \pi\left(  t\right)  ^{\prime
}K\left(  t\right)  dt\right)  ^{-1}
S_{1}^{\prime},
\]%
\[
v_{h}^{2}\left(  \alpha\right)  =\Pi_{h}^{1}\left(  \alpha\right)  ^{\prime
}\int_{-\frac{\alpha}{h}}^{\frac{1-\alpha}{h}}\int_{-\frac{\alpha}{h}}%
^{\frac{1-\alpha}{h}}\pi\left(  t_{1}\right)  \pi\left(  t_{2}\right)
^{\prime}\min\left(  t_{1},t_{2}\right)  K\left(  t_{1}\right)  K\left(
t_{2}\right)  dt_{1}dt_{2}\Pi_{h}^{1}\left(  \alpha\right)  ,
\]%
\begin{align}
\Sigma_{h}\left(  \alpha|I\right)   &  =\frac{\alpha^{2}v_{h}^{2}\left(
\alpha\right)  }{\left(  I-1\right)  ^{2}}\mathbb{E}^{-1}\left[
\frac{X_{1\ell} X_{1\ell}^{\prime}%
\mathbb{I}\left(  I_{\ell}=I\right)  }{B^{\left(  1\right)  }\left(
\alpha|X_{\ell},I_{\ell}\right)  }\right] 
%\nonumber
%\\
%&  
%\times
\mathbb{E}\left[  X_{1\ell} X_{1\ell}^{\prime}\mathbb{I}\left(  I_{\ell}=I\right)  \right]  \mathbb{E}^{-1}\left[
\frac{X_{1\ell} X_{1\ell}^{\prime}
\mathbb{I}\left(  I_{\ell}=I\right)  }{B^{\left(  1\right)  }\left(
\alpha|X_{\ell},I_{\ell}\right)  }\right]  , \label{AQR_Var}%
\end{align}%
\begin{align}
&  \mathsf{Bias}_{h}(\alpha|x,I)=\frac{\alpha B^{(s+2)} (\alpha|x,I)}{I-1}
S_{1}
\left(
\int_{-\frac{\alpha}{h}}^{\frac{1-\alpha}{h}}\pi\left(  t\right)  \pi\left(
t\right)  ^{\prime}K\left(  t\right)  dt\right)  ^{-1}\int_{-\frac{\alpha}{h}%
}^{\frac{1-\alpha}{h}}\frac{t^{s+2}\pi\left(  t\right)  }{\left(  s+2\right)
!}K\left(  t\right)  dt. \label{AQR_bias}%
\end{align}
setting $\mathsf{Bias}_{h}(0|x,I)=0$, see Footnote \ref{Ders2}.
The next Theorem gives some variance and bias expansions and the pointwise asymptotic distribution of the estimator. Recall that, in our pseudo true value framework, the bias of $\widehat{V} (\alpha|x,I)$ is
given by $\overline{V} (\alpha|x,I)-V(\alpha|x,I)$.
As the variance of $\widehat{V} (\alpha|x,I)$ is difficult to compute due to the implicit definition of this estimator, an expansion for the variance of its leading term $\widetilde{V} (\alpha|x,I)$ is given instead.

\begin{theorem}	
\label{CLT}
Under Assumptions \ref{Auct},
\ref{Kernel}, \ref{Specification} and for $\overline{V} (\cdot|\cdot,I)$, $\widetilde{V} (\cdot|\cdot,I)$ as in (\ref{WidetildeV}), (\ref{BarV}) respectively, it holds for all $\alpha$ in $[0,1]$ and all $x$ in $\mathcal{X}$,
\begin{eqnarray}
\overline{V} (\alpha|x,I)
& = &
V(\alpha|x,I)
+
h^{s+1}
\mathsf{Bias}_{h}(\alpha|x,I)
+
o(h^{s+1}),
\label{VBias}\\
\mathrm{Var}
\left[
 \widetilde{V} (\alpha|x,I) 
\right]
& = &
\frac{x_{1}^{\prime} \Sigma_{h} (\alpha|I) x_{1}+O(h)}{LIh},
\label{Vvar}
\end{eqnarray}
with $\sup_{(\alpha,x) \in [0,1]\times \mathcal{X}} | \mathsf{Bias}_{h}(\alpha|x,I)|=O(1)$ and 
$\sup_{\alpha \in [0,1]}  \left\| \Sigma_{h}(\alpha|I) \right\|=O(1)$.

If $\alpha \neq 0$, $x_{1}^{\prime}\Sigma_{h}\left(
\alpha|I\right)  x_{1}$ stays bounded away from $0$ for all $x$ and 
\[
\left(  \frac{LIh}{x_{1}^{\prime}\Sigma_{h}\left(
\alpha|I\right)  x_{1}  }\right)  ^{1/2}\left(  \widehat{V}\left(
\alpha|x,I\right)  -\overline{V} (\alpha|x,I)  \right)
\]
converges in distribution to a standard normal.
\end{theorem}

These expansions give a better understanding of potential boundary effects affecting $\widehat{V} (\alpha|x,I)$. For $\mathsf{Bias} (\alpha|x,I)$ and $\Sigma (\alpha|I)$ defined before Theorem \ref{IMSE}, it holds 
\[
\mathsf{Bias}_{h}(\alpha|x,I)=\mathsf{Bias}(\alpha|x,I)\text{ and }\Sigma
_{h}\left(  \alpha|I\right)  =\Sigma\left(  \alpha|I\right)  \text{ for all
}\alpha\text{ in }\left[  h,1-h\right]
\]
since the support of the kernel $K(\cdot)$ is $[-1,1]$. Hence a pointwise optimal bandwidth for central quantile levels is 
$h_{\ast} (\alpha|x,I)=\left(  \frac{  \Sigma (\alpha|I)}{2\left(
	s+1\right)  \mathsf{Bias}^{2} (\alpha|x,I)}\frac{1}{LI}\right)  ^{\frac{1}%
	{2s+3}}
$, which is obtained by minimizing the  leading term of the Mean Squared Error obtained from (\ref{VBias}) and (\ref{Vvar}).

As the asymptotic bias and standard deviation are proportional to $\alpha$, more bias and variance are expected for higher quantile levels and smaller $I$. This follows from the fact that the leading term of $\widehat{V} (\alpha|x,I)$ is $\alpha \widehat{B}^{(1)} (\alpha|x,I) / (I-1)$, which is proportional to $\alpha/(I-1)$.
Boundary effects can only occur for quantile levels in $[0,h]$ or $[1-h,1]$,  which differ from this respect. As $\widehat{V} (\alpha|x,I) = \widehat{B} (\alpha|x,I) + \alpha \widehat{B}^{(1)} (\alpha|x,I) / (I-1)$, $\widehat{V} (\alpha|x,I)$ is close to $\widehat{B} (\alpha|x,I)$ when $\alpha$ is in $[0,h]$.  In particular, $\widehat{V} (0|x,I)=\widehat{B} (0|x,I)$ which converges to $B(0|X,I)$ with  the rate $1/\sqrt{LI}+o(h^{s+1})$ at least.

For upper quantile levels $\alpha$ in $[1-h,1]$, the consistency rate of $\widehat{V} (\alpha|x,I)$ is the slower $1/\sqrt{Lh}+O(h^{s+1})$ as for central quantile levels. Bias and variance also involve the matrix
\[
\left(  \int_{-\frac{\alpha}{h}}%
^{\frac{1-\alpha}{h}}\pi\left(  t\right)  \pi\left(  t\right)  ^{\prime
}K\left(  t\right)  dt\right)^{-1}
=
\left(  \int_{-1}
^{\frac{1-\alpha}{h}}\pi\left(  t\right)  \pi\left(  t\right)  ^{\prime
}K\left(  t\right)  dt\right)  ^{-1}
\]
for $h$ small enough. This matrix increases with $\alpha$, so that higher 
$\left|\mathsf{Bias} (\alpha|x,I)\right|$ and $\Sigma_{h} (\alpha|I)$ can take place near $\alpha=1$ compared to central quantile levels. See Fan and Gijbels (1996) and the references therein for similar discussions. Accordingly, our simulations show an increase of the bias and variance for $\alpha$ approaching 1,  see Figures \ref{figgpvaqr} and \ref{PVAQRsim} in Section \ref{PVERsim}.

\subsection{Functional estimation \label{Funcest}}

The plug-in estimators of $\theta\left(  x\right)  $ and $\theta$ in
(\ref{Thetafunc}) are
\[
\widehat{\theta}\left(  x\right)  =\int_{0}^{1}\mathcal{F}\left[
\alpha,x,\widehat{B}\left(  \alpha|x,I\right)  ,\widehat{B}^{\left(  1\right)
}\left(  \alpha|x,I\right)  ;I\in\mathcal{I}\right]  d\alpha,\quad
\widehat{\theta}=\int_{\mathcal{X}}\widehat{\theta}\left(  x\right)  dx,
\]
with AQR estimators $\widehat{B}\left(  \alpha|x,I\right)  $ and $\widehat{B}%
^{\left(  1\right)  }\left(  \alpha|x,I\right)  $. Let us now introduce the asymptotic variances of $\widehat{\theta}\left(
x\right)  $ and $\widehat{\theta}$. The variances depend upon the matrices%
\begin{align*}
\mathbf{P}\left(  I\right)   =
\mathbb{E}\left[  \mathbb{I}\left(  I_{\ell}=I\right)  X_{1\ell} X_{1\ell}^{\prime}  \right]  ,
\quad
\mathbf{P}_{0}\left(  \alpha|I\right)    =\mathbb{E}\left[  \frac
{\mathbb{I}\left(  I_{\ell}=I\right)  X_{1\ell} X_{1\ell}^{\prime}  }{B^{\left(  1\right)  }\left(  \alpha|X_{\ell},I_{\ell}\right)  }\right]  ,
\end{align*}
and of the functions, recalling $b_{0I}$ and $b_{1I}$ stand for $B\left(
\alpha|x,I\right)  $ and $B^{\left(  1\right)  }\left(  \alpha|x,I\right)  $
respectively,
\begin{align*}
\varphi_{0I}\left(  \alpha,x\right)   &  
=
\partial_{b_{0I}}
\mathcal{F}\left[
\alpha,x,
B\left(  \alpha|x,I\right)  ,
B^{\left(  1\right)  }\left(\alpha|x,I\right)  
;I\in\mathcal{I}
\right],\\
\varphi_{1I}\left(  \alpha,x\right)   & 
=
\partial_{b_{1I}}
\mathcal{F}
\left[
\alpha,x,B\left(  \alpha|x,I\right)  ,B^{\left(  1\right)  }\left(
\alpha|x,I\right)  ;I\in\mathcal{I}
\right] .
\end{align*}
Let $A$ be a $\mathcal{U}_{[0,1]}$ random variable, 
$\mathbf{1} = [1,\ldots,1]^{\prime}$ a $(D+1) \times 1$ vector, and define, recalling $x_1= [1,x^{\prime}]^{\prime}$, 
\begin{align*}
\sigma_{L}^{2}\left(  x|I\right)   
&  =
\operatorname*{Var}
\left[    
\int_{0}^{A}
\left(  \varphi_{0I}\left(
\alpha|x\right)  
-
\partial_{\alpha} \varphi_{1I}\left(  \alpha|x\right)
\right)  
\mathbf{1}^{\prime}
\mathbf{P}_{0}\left(  \alpha|I\right)^{-1}
\mathbf{P}\left(  I\right) ^{1/2}x_1
d\alpha    \right]    ,
\\
\sigma_{L}^{2}\left(  I\right)   
&  =
\operatorname*{Var}
\left[  
\int_{\mathcal{X}}
\left\{  
\int_{0}^{A}
\left(  \varphi_{0I}\left(
\alpha|x\right)  
-
\partial_{\alpha} \varphi_{1I}\left(  \alpha|x\right)
\right)  
\mathbf{1}^{\prime}
\mathbf{P}_{0}\left(  \alpha|I\right)^{-1}
\mathbf{P}\left(  I\right) ^{1/2}x_1
d\alpha 
\right\}
dx
\right]    ,
\\
\sigma_{L}^{2}\left(  x\right)   &  
=
\sum_{I\in\mathcal{I}}
\frac{\sigma_{L}^{2}\left(  x|I\right)}{I}  ,
\quad
\sigma_{L}^{2}=\sum_{I\in\mathcal{I}}\frac{\sigma_{L}^{2}\left(  I\right)}{I}  .
\end{align*}
The proof of Theorem \ref{FuncCLT} in \ref{App:Proofsmain} shows that the asymptotic
variances of $\widehat{\theta}\left(  x\right)  $ and $\widehat{\theta}$ are
$\sigma_{L}^{2}\left(  x\right)  /  L  $ and
$\sigma_{L}^{2}/L$ respectively provided they are bounded away from $0$. This holds if
\begin{equation}
\text{for all $x$ in $\mathcal{X}$, }
\varphi_{0I}\left(  \alpha|x \right)  \neq
\partial_{\alpha} \varphi_{1I}\left(\alpha|x\right)   
\text{ for some $(\alpha,I)$ of $\left[  0,1\right]  
	\times\mathcal{I}$.}
\label{Condvar}%
\end{equation}
It may be indeed  that  $\sigma_{L}^{2}\left(  x|I\right)  =0$ and $\sigma_{L}^{2}=0$, in which case $\widehat{\theta}\left(
x\right)  $ and $\widehat{\theta}$ can converge to $\theta\left(  x\right)  $
and $\theta$ with \textquotedblleft superefficient\textquotedblright\ rates,
that is faster than  $1/L^{1/2}$. Why it is possible is
better understood in our quantile context, through an example of functionals
for which (\ref{Condvar}) does not hold. Consider, for some given $I_{0}$ of $\mathcal{I}$,%
\[
\mathcal{F}\left[  \alpha,x,B\left(  \alpha|x,I\right)  ,B^{\left(
1\right)  }\left(  \alpha|x,I\right)  ;I\in\mathcal{I}\right]  =2B\left(
\alpha|x,I_{0}\right)  B^{\left(  1\right)  }\left(  \alpha|x,I_{0}\right)
\]
which gives $\left(  \varphi_{0I_{0}}\left(  \alpha|x\right)  ,\varphi
_{1I_{0}}\left(  \alpha|x\right)  \right)  =2\left(  B^{\left(  1\right)
}\left(  \alpha|x,I_{0}\right)  ,B\left(  \alpha|x,I_{0}\right)  \right)  $. Hence
(\ref{Condvar}) does not hold and $\sigma_{L}^{2}\left(  x\right)
=\sigma_{L}^{2}=0$. Why $\widehat{\theta}\left(  x\right)  $ and
$\widehat{\theta}$ can converge with superefficient rates for these
functionals is in fact not surprising observing that they estimate%
\[
\theta \left(  x\right)  =B^{2}\left(  1|x,I_{0}\right)  -B^{2}\left(
0|x,I_{0}\right)  ,\quad\theta=\int_{\mathcal{X}}\theta\left(
x\right)  dx,
\]
respectively. For these examples, the parameters of interest only
depend upon extreme quantiles, in which case superefficient estimation is
possible, see e.g. Hirano and Porter (2003) and the references therein. The next
Theorem establishes the asymptotic normality of $\widehat{\theta}\left(
x\right)  $ and $\widehat{\theta}$.

\begin{theorem}
	\label{FuncCLT}	
	Suppose Assumptions  \ref{Auct},
	\ref{Kernel} with $\frac{\log^2 L}{L h^3  }=o\left(  1\right)  $, Assumption \ref{Specification} with $s \geq 2$ and (\ref{Condvar}) hold. Then, for all $x$ in $\mathcal{X}$,
	$\sigma_{L}^{2}\left(  x\right)  $ and $\sigma_{L}^{2}$ are bounded away from
	$0$ and infinity when $L$ grows, and
	\[
			\frac{\sqrt{L}\left(
			\widehat{\theta}\left(  x\right)  -\theta\left(  x\right)  -\mathsf{bias}_{L,\theta\left(  x\right)  }\right)}{  \sigma_{L}\left(  x\right)},
		\quad
		\frac{\sqrt{L}\left(  \widehat{\theta
		}-\theta-\mathsf{bias}_{L,\theta}\right)}{  \sigma_{L}}
	\]
	both converge in distribution to a standard normal, the bias items $\mathsf{bias}_{L,\theta\left(
		x\right)  }$ and $\mathsf{bias}_{L,\theta}$ being $o(h^s)$.
\end{theorem}

Note that the conditional functional estimator $\widehat{\theta} (x)$ converges with a parametric rate, under a bandwidth condition slightly stronger than in Assumption \ref{Kernel}. This bandwidth condition corresponds to the fact that the linearization error term $\widehat{B}^{(1)} (\alpha|x,I) -\widetilde{B} (\alpha|x,I)$ is of order $\log L/(Lh^{3/2})$ as guessed from (\ref{IMSEBaha}) and must be $o(1/\sqrt{L})$.

The bias term order $o\left(h^s\right)$ is given by the estimation of $B^{\left(  1\right)
}\left(  \alpha|x,I\right)  $, and is of order $O(h^{s+1})$  when $\mathcal{F}\left(  \cdot\right)  $
depends upon $\alpha B^{\left(  1\right)  }\left(  \alpha|x,I\right)  $ as in
all the Examples. Let $\mathcal{G}_{b_{1I}}\left(  \cdot\right)  $ be the
partial derivative of $\mathcal{F}\left(  \cdot\right)  $ with respect to
$\alpha B^{\left(  1\right)  }\left(  \alpha|x,I\right)  $, and $\mathsf{Bias}_{h}\left(  \alpha|x,I\right)  $ be as in
(\ref{AQR_bias}). 
Then
\begin{align*}
\mathsf{bias}_{L,\theta\left(  x\right)  }  &  =h^{s+1}\left(  1+o\left(
1\right)  \right)   \\
&  \quad\quad\times 
\sum_{I\in\mathcal{I}}\int_{0}^{1}\mathcal{G}_{b_{1I}%
}\left[  \alpha,x,B\left(  \alpha|x,I\right)  ,\alpha B^{\left(  1\right)
}\left(  \alpha|x,I\right)  ;I\in\mathcal{I}\right]
\mathsf{Bias}_{h}\left(  \alpha|x,I\right)  d\alpha
\end{align*}
and $\mathsf{bias}_{L,\theta}=\int_{\mathcal{X}}\mathsf{bias}_{L,\theta\left(
x\right)  }dx$. The estimators $\widehat{\theta
}\left(  x\right)  $ or $\widehat{\theta}$ are therefore asymptotically
unbiased if $h^{s+1}\sqrt{Lh}=o\left(  1\right)  $ or
$h^{s+1}\sqrt{L}=o\left(  1\right)  $ respectively.

Theorem \ref{FuncCLT} applies to our functional Examples, but the resulting variance can be somehow involved, so that the use of the bootstrap can be preferred as discussed below. We first detail the variance obtained for the cdf estimator of Example 3 to illustrate the influence of the bandwidth $\eta$ on its variance.

\paragraph{Example 3 (cont'd).}

For the cdf estimator $\widehat{F}_{\eta}\left(
v|x,I\right)  =\int_{0}^{1}\mathbb{I}_{\eta}\left[  v-\widehat{V}\left(
\alpha|x,I\right)  \right]  d\alpha$,%
\begin{align*}
\varphi_{0I}\left(  \alpha|x\right)   &  =-\frac{1}{\eta}k\left(
\frac{v-V\left(  \alpha|x,I\right)  }{\eta}\right)  ,\quad\varphi_{1I}\left(
\alpha|x\right)  =\frac{\alpha}{\left(  I-1\right)  \eta}k\left(
\frac{v-V\left(  \alpha|x,I\right)  }{\eta}\right)  ,\\
\partial_{\alpha} \varphi_{1I}\left(  \alpha|x\right)   &
=\frac{1}{\left(  I-1\right)  \eta}k\left(  \frac{v-V\left(  \alpha
|x,I\right)  }{\eta}\right)  -\frac{\alpha}{\left(  I-1\right)  \eta^{2}%
}k^{\left(  1\right)  }\left(  \frac{v-V\left(  \alpha|x,I\right)  }{\eta
}\right)  V^{\left(  1\right)  }\left(  \alpha|x,I\right)  .
\end{align*}
When $\eta$ goes to $0$, the dominant part of the variance is, for inner $v$,
integrating by parts and setting $V_{x,I}=V\left(  A|x,I\right)  $%
\begin{align*}
&  \frac{I}{L}\operatorname*{Tr}\left\{
\operatorname*{Var}\left[  \left(  \int_{0}^{A}
\partial_{\alpha} \varphi_{1I}\left(  \alpha|x\right)  
\mathbf{P}_{0}\left(\alpha|I\right)  ^{-1}d\alpha\right)  \mathbf{P}\left(  I\right)
^{1/2}x_1  \right]  \right\} \\
&  \quad=\frac{\left(  1+o\left(  1\right)  \right)  I}{L}
\operatorname*{Tr}\left\{  \operatorname*{Var}\left[  \varphi_{1I}\left(
A|x\right)  
\partial_{\alpha} \left[ \mathbf{P}_{0}\left(  A|I\right)^{-1} \right]
\mathbf{P}\left(  I\right)  ^{1/2}x_1  \right]  \right\} \\
&  \quad=\frac{\left(  1+o\left(  1\right)  \right)  I}{\left(  I-1\right)
^{2}L}\\
&  \quad\quad\quad\times\operatorname*{Tr}\left\{  \operatorname*{Var}\left[
\frac{F\left(  V_{x,I}|x,I\right)  }{f\left(  V_{x,I}|x,I\right)  }%
\frac{k\left(  \frac{v-V_{x,I}}{\eta}\right)  }{\eta}
\partial_{\alpha}
\left[ \mathbf{P}_{0}\left(  F\left(  V_{x,I}|x,I\right)  |I\right)^{-1} \right]
\mathbf{P}\left(  I\right)^{1/2}x_1  \right]  \right\} \\
&  \quad=\frac{\left(  1+o\left(  1\right)  \right)  I\int k^{2}\left(
t\right)  dt}{\left(  I-1\right)  ^{2}L\eta }\left(
\frac{F\left(  v|x,I\right)  }{f\left(  v|x,I\right)  }\right)  ^{2}\\
&  \quad\quad\quad\times\operatorname*{Tr}\left\{  
\partial_{\alpha} \left[ \mathbf{P}_{0}\left(  F\left(  v|x,I\right)  |I\right)  ^{-1} \right]
\mathbf{P}\left(  I\right)  ^{1/2}x_1
x_1 ^{\prime}\mathbf{P}\left(  I\right)  ^{1/2}
\partial_{\alpha} 
\left[ \mathbf{P}_{0}\left(  F\left(  v|x,I\right)  |I\right)^{-1}\right]
\right\}  .
\end{align*}
Hence the order of the variance of $\widehat{F}_{\eta}\left(  v|x,I\right)  $
is $1/\left(  L\eta \right)  $ when $\eta$ goes to $0$. Its bias  has two components: the first is $\mathsf{bias}%
_{L,F_{\eta}\left(  v|x,I\right)  }$ due to the bias of $\widehat{V}\left(
\alpha|x,I\right)  $ and is of order $O\left(  h^{s+1}\right)  $, while the
second is $F_{\eta}\left(  v|x,I\right)  -F\left(  v|x,I\right)  =O\left(
\eta^{s+1}\right)  $ if $k\left(  \cdot\right)  $ is a kernel of order $s$. 
Further work is needed to determine at which rate $\eta$ can go to $0$ in this heuristic.

\paragraph{Bootstrap inference.}
Earlier theoretical works considering quantile-regression bootstrap inference are Rao and Zhao (1992), for the weighted bootstrap, and Hahn (1995) for the pairwise bootstrap. See also Liu and Luo (2017) for quantile-based auction testing procedures. For standard and sieve quantile-regression estimators, Belloni et al. (2019) establish consistency of several bootstrap procedures for functionals and uniformly with respect to the quantile levels. It is expected that it carries over to our AQR estimators when bias terms can be neglected, but is out of the scope of the present paper.

\setcounter{equation}{0}

\section{Simulation experiments\label{Simexp}}

The first simulation experiments compare the AQR estimation method with GPV and its homogenized-bid extension. Other simulations illustrate its performances for estimation of the private-value quantile function, expected seller revenue and optimal reserve price, or risk-aversion parameter. All experiments involve $L=100$ auctions with $I=2$ or $I=3$ bidders, so that small samples of $200$ or $300$ are considered. In most of the experiments, three auction-specific auction covariates are considered. The number of replications is $1,000$ in all experiments. AQR are computed over the estimation grid $\alpha
	=0,0.01,\ldots,0.99,1$. The AQR local polynomial order $s+1$ is set to $2$ and the AQR kernel is the Epanechnikov $K\left(
	t\right)  =\frac{3}{4}\left(  1-t^2\right)  \mathbb{I}\left(  t\in\left[  -1,1\right]
	\right)  $.

Since the asymptotic bias and variance of $\widehat{V} (\alpha|x,I)$ tend to decrease with $I$, choosing a small number $I$ of bidders  is challenging. Hickman and Hubbard (2015) used $5$ bids while $I=3$ or $5$
in Marmer and Shneyerov (2012) and Ma, Marmer and Shneyerov (2019). The number
of bids $LI$  ranges from $1,000$ for Hickman and Hubbard (2015) to $4,200$ for Marmer
and Shneyerov (2012). In a simulation experiment focused on nonparametric
estimation of the utility function of risk-averse bidders, Zincenko (2018)
considers $I=2$ with $L=300$ and $I=4$ with $L=150$. These references do not consider covariate, with the exception of Zincenko (2018) for $L=900$ auctions
with one or two covariates. Therefore our simulation setting correspond to rather demanding small sample situations.

\subsection{Comparison with GPV and  homogenized-bid  \label{Comp}}

\begin{figure}[ht]
	\centering
	\includegraphics[width=.8\linewidth]{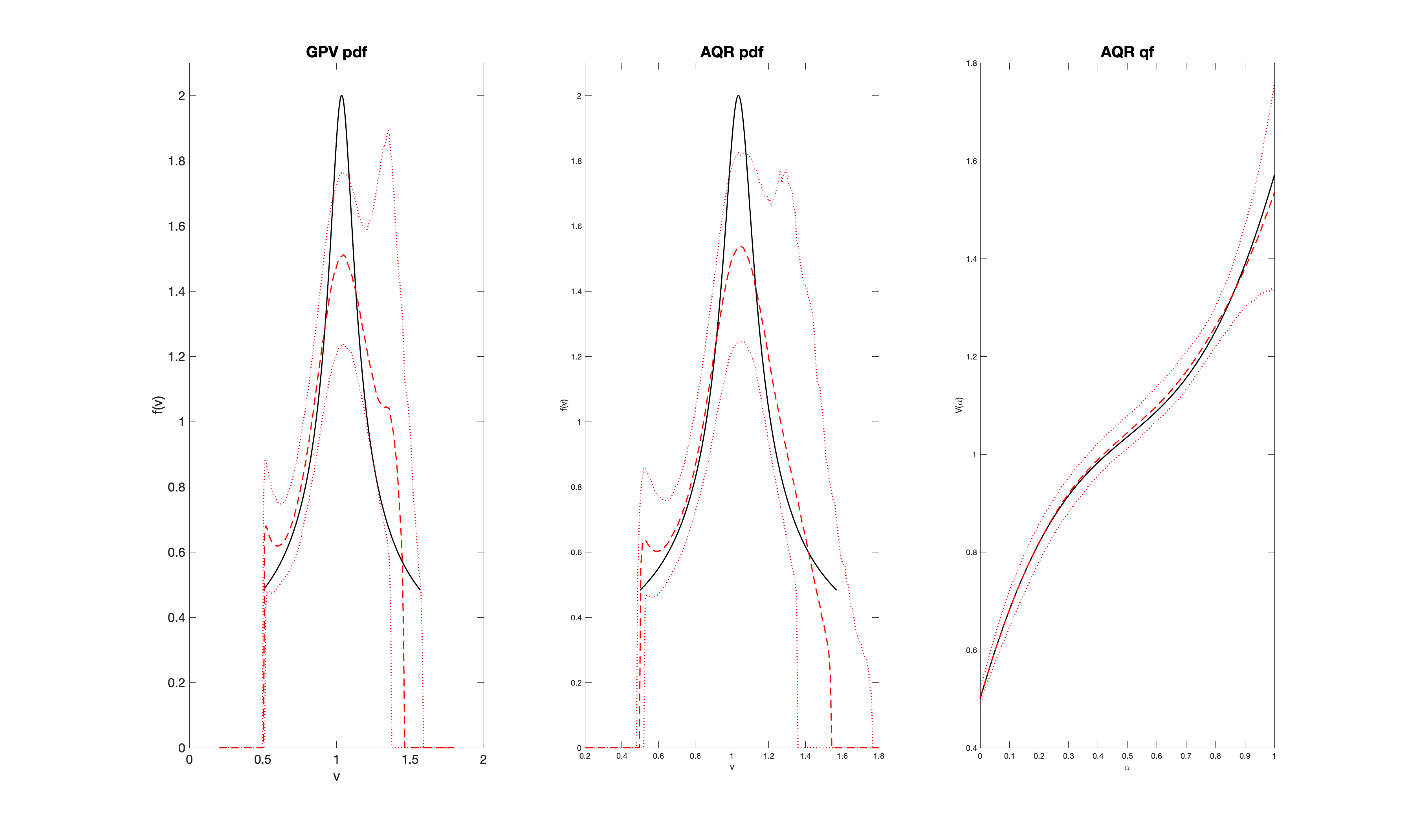}
	\caption{Private value pdf estimation using Hickman and Hubbard (2015) bias corrected version of GPV (left), Private value pdf  estimation based upon AQR and Hickman and Hubbard (2015) (center), and AQR $\widehat{V} (\cdot)$ (right) , $L=100$ and $I=2$. Black line: true functions. Dashed
		red and dotted lines: pointwise median and $2.5\%-97.5\%$
		quantiles of the estimated functions across $1,000$
		simulations.}
	\label{figgpvaqr}
\end{figure}

The simulation experiments of this Section makes use of a ``trigonometric'' quantile function
\begin{equation}
T(\alpha)
=
\frac{1}{2}
\left(
(\pi+1)\alpha+\cos(\pi \alpha)
\right)
\label{Trigo}
\end{equation}
whose probability density function has a compact support and is bounded away from $0$, with a shape similar to a peaked Gaussian one as seen from the left panel of Figure \ref{figgpvaqr}.\footnote{The pdf graph is obtained noting that $T^{(1)} (\alpha)=1/f(T(\alpha))$, so that the graph $\alpha \in [0,1] \mapsto (T(\alpha),1/T^{(1)}(\alpha)$ is the one of the pdf $f(\cdot)$. The associated bid quantile function can be obtained using (\ref{V2B}), which is used to simulate bids via a quantile transformation of uniform draws, as performed elsewhere in our simulation experiments.}

	\paragraph{Comparison with GPV.}
	This experiment considers private values drawn from $T(\cdot)$ and does not include covariate. It compares first the boundary bias corrected GPV two-step pdf estimator with triweight kernel and rule of thumb bandwidth of Hickman and Hubbard (2015) with the AQR pdf estimator
	\[
	\widehat{f} (v) 
	= 
	\int_{0}^{1} \frac{1}{h_{AQR}} 
	\widetilde{K}_{tri} \left(\frac{v-\widehat{V} (\alpha)}{h_{AQR}} \right)
	d \alpha,
	\quad
	h_{AQR} = \frac{3}{(LI)^{1/5}} 
	\left( 
	\int_{0}^{1}
	\left(
	  \widehat{V} (\alpha)
	  -
	  \int_{0}^{1} 
	  \widehat{V} (t)
	  dt
	\right)^2 d\alpha
	\right)^{1/2},
	\]
	where the AQR bandwidth is $h=.3$, $I=2$ and
	$\widetilde{K}_{tri} (\cdot)$ is the Hickman and Hubbard (2015) boundary bias corrected triweight kernel.

	The results are reported in Figure \ref{figgpvaqr}, which illustrates how much harder estimating a pdf can be compared to estimating a quantile function. The variance and the bias of the two private-value pdf estimators look much higher, especially just after the density peak. This peak causes a small AQR bias for central quantiles.

Other features are common to the two pdf estimation procedures. 
The first stage of the GPV procedure is based upon an estimation of the private values from (\ref{GPV}), which is likely to have more bias and variance for small bids than for large ones. Accordingly, the left panel of Figure \ref{figgpvaqr} reveals that the GPV pdf estimator performs quite well before the density peak, but that both its bias and variance increase for higher private values. 
The AQR pdf estimator in the center panel looks less affected by these issues.
By contrast, the quantile estimation procedure in the right panel of Figure \ref{figgpvaqr} is only affected by a variance increase for upper quantiles as expected.

\begin{figure}[ht]
	\centering
	\includegraphics[width=.8\linewidth]{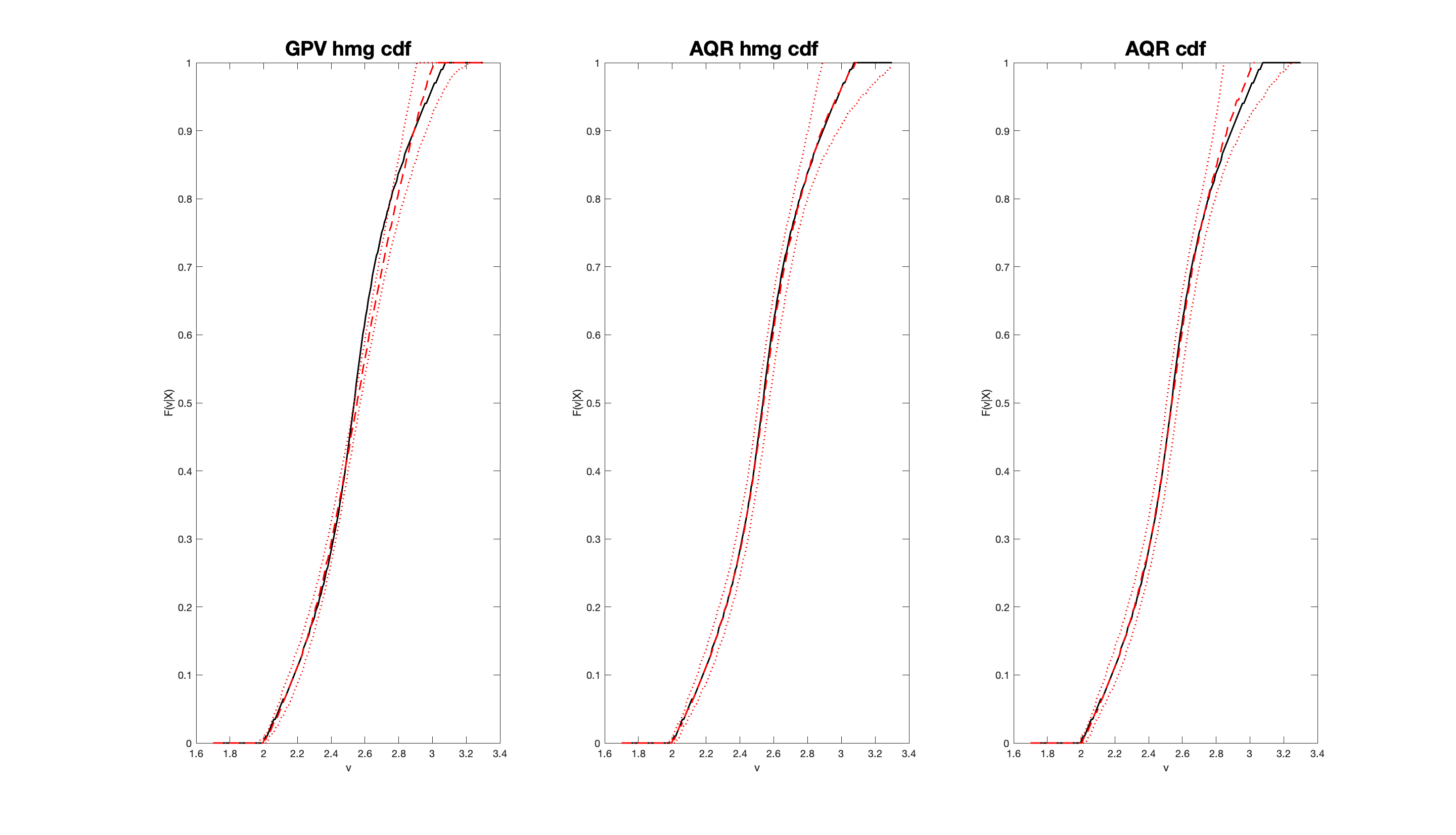}
	\caption{Private value cdf estimation using the homogenized bids and GPV version of Hickmann and Hubbard (2005) (left), AQR with homogenized bids (center) and AQR (right). Black line: true functions. Dashed
		red and dotted lines: pointwise median and $2.5\%-97.5\%$
		quantiles of the estimated functions across $1,000$
		simulations.}
	\label{fig:hmg}
\end{figure}

\paragraph{Cdf estimation with homogenized bids and AQR}
The bids considered in this experiment are associated with private values satisfying
\[
V_{i\ell}= X_{1\ell} + X_{2\ell} + X_{3i\ell} + v_{i\ell}
\]
where the $X_{j\ell}$'s are independently drawn from the uniform, independent from $v_{i\ell}$, whose quantile function is $T(\cdot)$ in (\ref{Trigo}). The bids are regressed on the covariate and the constant to obtain the homogenized bids $\widehat{b}_{i\ell}$. The latter are used as in the first-stage of Hickman and  Hubbard (2015) to obtain homogenized pseudo private values $\widehat{v}_{i\ell}$, and the estimated private-value conditional cdf at $x_1=x_{2}=x_3=1/2$ is
\[
\widehat{F} (v|x)
=
\frac{1}{LI}
\sum_{\ell}^{L}
\sum_{i=1}^I
\mathbb{I}
\left(
\frac{1}{2}(\widehat{\beta}_1+\widehat{\beta}_{2}+\widehat{\beta}_3)
+
\widehat{v}_{i \ell}
\leq v
\right),
\]
where the $\widehat{\beta}_j$ are the OLS slope estimators computed in the homogenized-bid regression.
Two AQR conditional estimators are computed. The first uses the homogenized bids and (\ref{Cdf}) while the second is based on the standard AQR $\widehat{V}(\alpha|x,I)$, both using the bandwidth $h=.3$. The performances of these three cdf estimators are reported in Figure \ref{fig:hmg}.

As expected, considering estimation of the private-value cdf gives smaller bias and variance than estimating pdf. All procedures have a similar variability. However the homogenized-bid GPV procedure has a larger bias, dominating its variability in the right centre  part of the private-value distribution, than its AQR counterparts.  Applying the AQR to the homogenized bids seems to slightly dominate the other AQR procedure.

\subsection{Private value and expected revenue \label{PVERsim}}

\paragraph{Quantile-regression model and estimation details.}
The private-value quantile function is given by a quantile-regression model
with an intercept and three independent covariates with the uniform
distribution over $\left[  0,1\right]  $,%
\[
V\left(  \alpha|X\right)  =\gamma_{0}\left(  \alpha\right)  +\gamma_{1}\left(
\alpha\right)  X_{1}+\gamma_{2}\left(  \alpha\right)  X_{2}+\gamma_{3}\left(
\alpha\right)  X_{3}%
\]
with%
\begin{align*}
\gamma_{0}\left(  \alpha\right)   &  =1+0.5\exp(5(\alpha-1)),\quad\gamma
_{1}\left(  \alpha\right)  =1,\\
\gamma_{2}\left(  \alpha\right)   &  =0.5(1-\exp(-5\alpha)),\quad
\gamma_{3}\left(  \alpha\right)  =0.8+0.15((2\pi+1)\alpha+\cos(2\pi\alpha)).
\end{align*}
The coefficient $\gamma_{0}\left(  \cdot\right)  $ is flat near $0$ and strongly
increases near $1$ while $\gamma_{2}\left(  \cdot\right)  $ strongly increases near $0$
and is flat after. The slope $\gamma_{3}\left(  \cdot\right)  $ is as the 
trigonometric quantile function (\ref{Trigo}), but with stronger oscillations which makes it harder to estimate.

The performances of the private-value quantile estimation procedure are evaluated through the individual estimation of each slope function or estimation of $V(\alpha|x)$ when the $x_j$ are set to their median $1/2$. The curvature of the expected revenue is mostly due to $\gamma_{2} (\cdot)$, the other coefficients having a rather flat contribution. The performances of the expected revenue estimation procedure are therefore evaluated removing the intercept, setting $x_{1}$ and $x_{3}$ to $0$ and taking $x_{2}=0.8$. This choice gives a unique optimal reserve price achieved for $\alpha_{\ast}=.3$, which is not too close to the boundaries so that the expected revenue function has a substantial concave shape which is suppose to make estimation more difficult. This is also used for evaluating estimation of the optimal reserve price $R_{\ast}=.8 \gamma_{2} (\alpha_{\ast})$.

\paragraph{Simulation results.}
Table \ref{PVER} summarizes the simulation results for the estimation of the
private-value quantile function, the expected revenue and the optimal reserve
price. The Bias and square Root Integrated Mean Squared Error (RIMSE) lines
for $\widehat{V}\left(  \cdot|\cdot\right)  $ gives the simulation
counterparts of, respectively%
\[
\left(  \frac{1}{4}\sum_{j=0}^{3}\int_{0}^{1}\left(  \mathbb{E}\left[
\widehat{\gamma}_{j}\left(  \alpha\right)  \right]  -\gamma_{j}\left(
\alpha\right)  \right)  ^{2}d\alpha\right)  ^{1/2}\text{and }\left(  \frac
{1}{4}\sum_{j=0}^{3}\int_{0}^{1}\mathbb{E}\left[  \left(  \widehat{\gamma}%
_{j}\left(  \alpha\right)  -\gamma_{j}\left(  \alpha\right)  \right)
^{2}\right]  d\alpha\right)  ^{1/2}.
\]
The Bias and RIMSE for the expected revenue are computed similarly. Table
\ref{PVER} also gives the Bias and square Root Mean Squared Error (RMSE) of the optimal
reserve price estimator. All these quantities are computed for bandwidths
$.2,.3,\ldots,.9$. \begin{table}[h]
\centering%
\begin{tabular}
[c]{cl|cccccccc|}\cline{2-10}
& \multicolumn{1}{|l|}{$h$} & $.2$ & $.3$ & $.4$ & $.5$ & $.6$ & $.7$ & $.8$ &
$.9$\\\hline\hline
\multicolumn{1}{|c|}{$\widehat{V}\left(  \cdot|\cdot\right)  $} & Bias &
\multicolumn{1}{|l}{.131} & \multicolumn{1}{l}{.141} &
\multicolumn{1}{l}{.143} & \multicolumn{1}{l}{.145} & \multicolumn{1}{l}{.150}
& \multicolumn{1}{l}{.159} & \multicolumn{1}{l}{.166} &
\multicolumn{1}{l|}{.176}\\
\multicolumn{1}{|c|}{} & RIMSE & \multicolumn{1}{|l}{.433} &
\multicolumn{1}{l}{.386} & \multicolumn{1}{l}{.355} & \multicolumn{1}{l}{.332}
& \multicolumn{1}{l}{.322} & \multicolumn{1}{l}{.309} &
\multicolumn{1}{l}{.303} & \multicolumn{1}{l|}{.305}\\\hline
\multicolumn{1}{|c|}{$\widehat{ER}\left(  \cdot\right)  $} & Bias &
\multicolumn{1}{|l}{.036} & \multicolumn{1}{l}{.044} &
\multicolumn{1}{l}{.049} & \multicolumn{1}{l}{.050} & \multicolumn{1}{l}{.051}
& \multicolumn{1}{l}{.049} & \multicolumn{1}{l}{.047} &
\multicolumn{1}{l|}{.045}\\
\multicolumn{1}{|c|}{} & RIMSE & \multicolumn{1}{|l}{.109} &
\multicolumn{1}{l}{.104} & \multicolumn{1}{l}{.102} & \multicolumn{1}{l}{.100}
& \multicolumn{1}{l}{.099} & \multicolumn{1}{l}{.098} &
\multicolumn{1}{l}{.097} & \multicolumn{1}{l|}{.096}\\\hline
\multicolumn{1}{|c|}{$\widehat{R}_{\ast}$} & Bias & \multicolumn{1}{|l}{-.036}
& \multicolumn{1}{l}{-.031} & \multicolumn{1}{l}{-.014} &
\multicolumn{1}{l}{-.002} & \multicolumn{1}{l}{.009} &
\multicolumn{1}{l}{.022} & \multicolumn{1}{l}{.037} &
\multicolumn{1}{l|}{.043}\\
\multicolumn{1}{|c|}{} & RMSE & \multicolumn{1}{|l}{.129} &
\multicolumn{1}{l}{.099} & \multicolumn{1}{l}{.075} & \multicolumn{1}{l}{.067}
& \multicolumn{1}{l}{.062} & \multicolumn{1}{l}{.064} &
\multicolumn{1}{l}{.066} & \multicolumn{1}{l|}{.066}\\\hline\hline
\end{tabular}
\caption{Private value quantile function, expected revenue, and optimal
reserve price}%
\label{PVER}%
\end{table}
\begin{figure}[h]
	\centering
	\includegraphics[width=.8\linewidth]{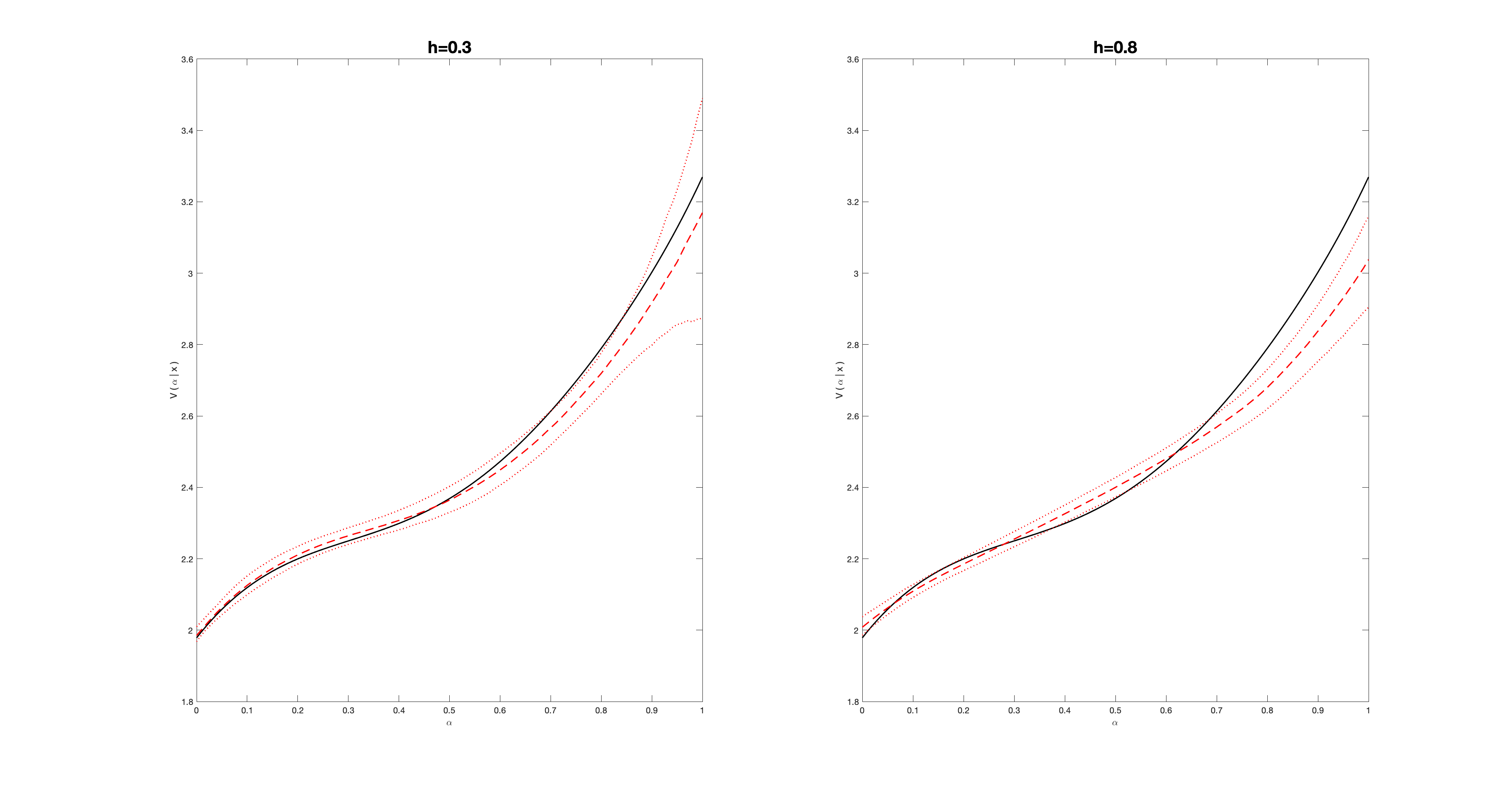}
	\caption{Private value quantile estimation for $h=0.3$ (left) and $h=0.8$
		(right) for average covariate. True $V(\alpha|x)=\gamma_{0}\left(
		\alpha\right)  +\left(  \gamma_{1}\left(  \alpha\right)  +\gamma_{2}\left(
		\alpha\right)  +\gamma_{3}\left(  \alpha\right)  \right)  /2$ in black. Dashed
		red and dotted lines: pointwise median and $2.5\%-97.5\%$
		quantiles of $\protect\widehat{V}\left(  \alpha|x\right)  $ across $1,000$
		simulations.}
	\label{PVAQRsim}
\end{figure}

Estimation of the private-value slope coefficients seems  more
sensitive to the bandwidth parameter than the expected revenue or optimal
reserve price. It has also a much higher RIMSE. The bandwidth behavior of
$\widehat{V}\left(  \alpha|x\right)  $ is also illustrated in Figure \ref{PVAQRsim}%
, which considers the small bandwidth $h=0.3$ and the larger $h=0.8$. 
As
expected from Theorem \ref{CLT}, the dispersion of $\widehat{V}\left(
\alpha|x\right)  $ increases with $\alpha$ and decreases with $h$, while the
bias increases with $\alpha$ and $h$.
%TCIMACRO{\FRAME{fhFU}{5.7873in}{3.2197in}{0pt}{\Qcb{Private value quantile
%estimation for $h=0.3$ (left) and $h=0.8$ (right) for average covariate. True
%$V(\alpha|X)=\gamma_{0}\left(  \alpha\right)  +\left(  \gamma_{1}\left(
%\alpha\right)  +\gamma_{2}\left(  \alpha\right)  +\gamma_{3}\left(
%\alpha\right)  \right)  /2$ in black. Dashed red line: average estimation.
%Dotted red line: pointwise $2.5\%-97.5\%$ quantiles of $\protect\widehat{V}%
%\left(  \alpha|X\right)  $ across $1,000$ simulations.}}{\Qlb{PVAQRsim}%
%}{pvqi2h38.bmp}{\special{ language "Scientific Word";  type "GRAPHIC";
%width 5.7873in;  height 3.2197in;  depth 0pt;  original-width 19.1997in;
%original-height 10.6398in;  cropleft "0";  croptop "1";  cropright "1";
%cropbottom "0";  filename 'PVQI2H38.bmp';file-properties "XNPEU";}} }%
%BeginExpansion
Figure \ref{PVAQRsim} also suggests that choosing a large bandwidth as
recommended by Table \ref{PVER} may lead to important bias issues, including
underestimating the private-value quantile function for high $\alpha$. This is mostly due to the slope $\gamma_3(\cdot)$ which is an important source of bias.

%TCIMACRO{\FRAME{ftbpFU}{5.7873in}{3.4238in}{0pt}{\Qcb{Expected revenue
%estimation for $h=0.3$ (left) and $h=0.8$ (right). True $ER\left(
%\alpha|X\right)  $ in black. Dashed red line: average estimation. Dotted red
%line: pointwise $2.5\%-97.5\%$ quantiles of $\protect\widehat{ER}\left(
%\alpha|X\right)  $ across $1,000$ simulations.}}{\Qlb{ERsim}}{erh38.bmp}%
%width 5.7873in;  height 3.4238in;  depth 0pt;  original-width 19.1997in;
%original-height 11.3204in;  cropleft "0";  croptop "1";  cropright "1";
%cropbottom "0";  filename 'ERH38.bmp';file-properties "XNPEU";}} }%
%BeginExpansion
\begin{figure}[tbp]
	\centering
	\includegraphics[width=.8\linewidth]{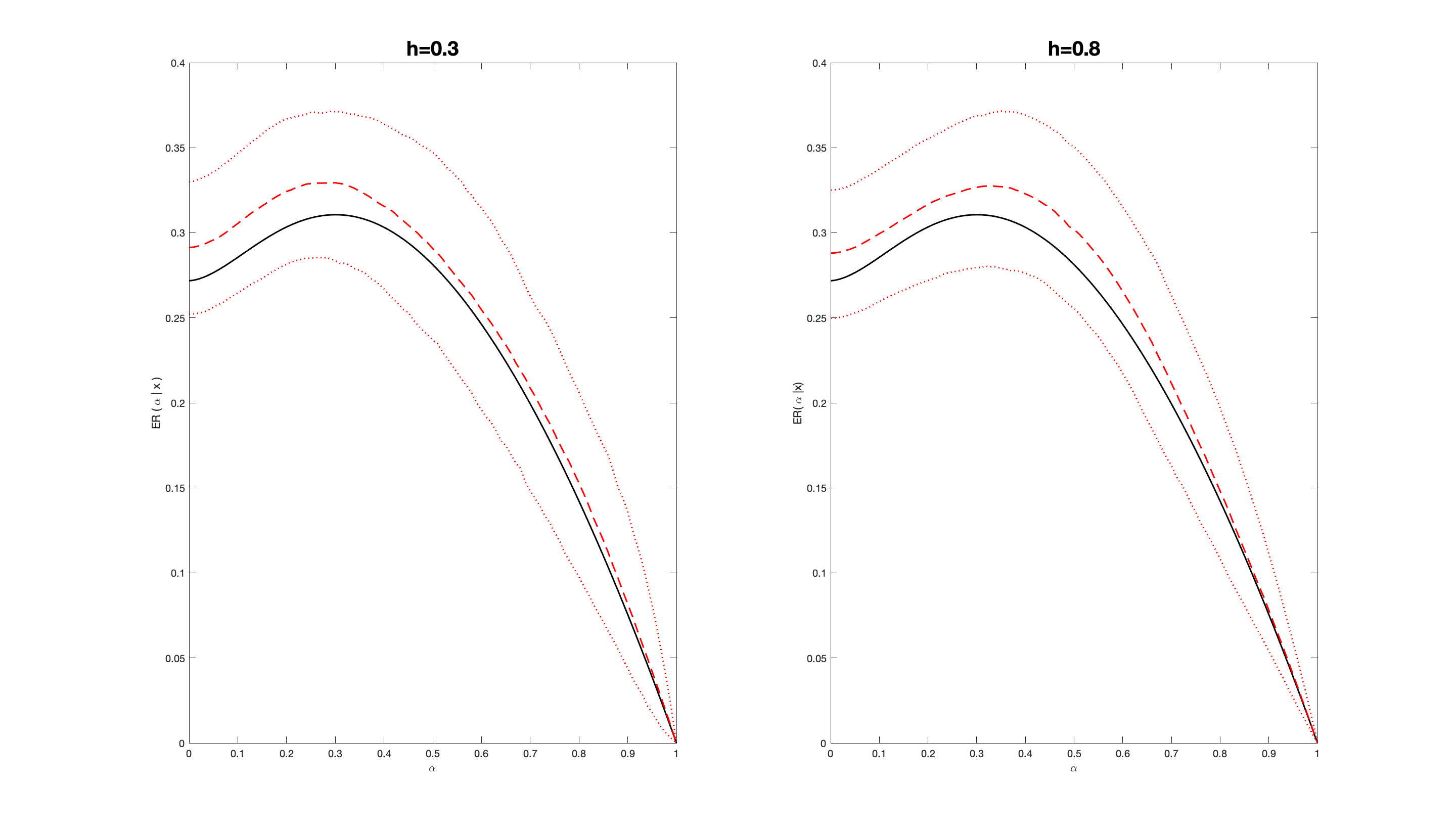}
	\caption{Expected revenue estimation for $h=0.3$ (left) and $h=0.8$ (right).
		True $ER\left(  \alpha|x\right)  $ in black. Dashed
		red and dotted lines: pointwise median and $2.5\%-97.5\%$
		quantiles of $\protect\widehat{ER}\left(  \alpha|x\right)  $ across $1,000$
		simulations.
		}
	\label{ERsim}
\end{figure}
This contrasts with estimation of the expected revenue and optimal reserve
price, which seems mostly unaffected by the bandwidth. This is partly because the
expected revenue depends upon $\left(  1-\alpha\right)  V\left(
\alpha|x\right)  $: multiplying the private-value quantile function by
$\left(  1-\alpha\right)  $ mitigates larger bias and variance near the
boundary $\alpha=1$, see also Figure \ref{ERsim}. For the considered
experiment, the true expected revenue is always in the $95\%$ band of Figure
\ref{ERsim} while the true private-value quantile function is out for large $\alpha$
when $h=0.8$.

\subsection{Risk-aversion parameter}

Two risk-aversion estimators are considered. The first estimator
$\widehat{\nu}_{fp}$ is based upon (\ref{Crrafp}) and uses two independent
samples of size $L=100$ with 2 and 3 bidders from the model above, which
corresponds to a CRRA utility function $t^{\nu}$ with $\nu
=1$.\footnote{The optimal bid functions can be computed explicitly under the
risk-neutral case $\nu=1$. Considering other values of $\nu$ would
request to use numerical computations of the bid functions.} Integrals with
respect to $\alpha$ are computed using Riemann sums whereas integrals with
respect to $x$ are replaced with sample means over the two auction samples.
The second estimator $\widehat{\nu}_{asc}$ is based upon (\ref{Crraasc})
and uses an additional sample of size $L=100$ of ascending auctions with two
bidders. In this case, it is possible to consider various values of $\nu$
and the simulation experiment considers the values $0.2$, $0.6$ and $1$.
Indeed, if $B\left(  \alpha|X\right)  $ is the first-price auction quantile
bid function with $I=2$, the observed bids drawn from $B\left(  \alpha
|X\right)  $ are rationalized by a CRRA utility function $t^{\nu}$ if the
private-value quantile function is set to
\[
V_{\nu}\left(  \alpha|X\right)  =B\left(  \alpha|X,2\right)  +\nu\alpha
B^{\left(  1\right)  }\left(  \alpha|X,2\right)
\]
provided $V_{\nu}^{\left(  1\right)  }\left(  \cdot|X\right)  >0$ for all
$X$. As $V_{\nu
}^{\left(  1\right)  }\left(  \cdot|\cdot\right)  >0$ holds in our case, we
use $V_{\nu}\left(  \alpha|X\right)  $ to generate two ascending bids for
each auction. Following Gimenes (2017), $V_{\nu}\left(  \alpha|X\right)  $
can be estimated from winning bids in these ascending auctions using AQR for
quantile level $2\alpha-\alpha^{2}$.

%The performance of the two estimators are summarized in the next Table.
\begin{table}[h]
\centering%
\begin{tabular}
[c]{cll|cccccccc|}\cline{2-11}
& \multicolumn{1}{|l|}{$\nu$} & \multicolumn{1}{|l|}{$h$} & $.2$ & $.3$ &
$.4$ & $.5$ & $.6$ & $.7$ & $.8$ & $.9$\\\hline\hline
\multicolumn{1}{|c}{$\widehat{\nu}_{fp}$} & \multicolumn{1}{|l|}{$1$} &
Bias & \multicolumn{1}{|l}{-.795} & \multicolumn{1}{l}{-.564} &
\multicolumn{1}{l}{-.412} & \multicolumn{1}{l}{-.288} &
\multicolumn{1}{l}{-.178} & \multicolumn{1}{l}{.-.080} &
\multicolumn{1}{l}{.003} & \multicolumn{1}{l|}{.053}\\
\multicolumn{1}{|c}{} & \multicolumn{1}{|l|}{} & RMSE &
\multicolumn{1}{|l}{.891} & \multicolumn{1}{l}{.681} &
\multicolumn{1}{l}{.545} & \multicolumn{1}{l}{.471} & \multicolumn{1}{l}{.404}
& \multicolumn{1}{l}{.380} & \multicolumn{1}{l}{.393} &
\multicolumn{1}{l|}{.436}\\\hline
\multicolumn{1}{|c}{$\widehat{\nu}_{asc}$} & \multicolumn{1}{|l|}{$1$} &
Bias & \multicolumn{1}{|l}{-.016} & \multicolumn{1}{l}{-.019} &
\multicolumn{1}{l}{-.037} & \multicolumn{1}{l}{-.061} &
\multicolumn{1}{l}{-.085} & \multicolumn{1}{l}{-.100} &
\multicolumn{1}{l}{-.109} & \multicolumn{1}{l|}{-.111}\\
\multicolumn{1}{|c}{} & \multicolumn{1}{|l|}{} & RMSE &
\multicolumn{1}{|l}{.240} & \multicolumn{1}{l}{.247} &
\multicolumn{1}{l}{.248} & \multicolumn{1}{l}{.254} & \multicolumn{1}{l}{.260}
& \multicolumn{1}{l}{.267} & \multicolumn{1}{l}{.276} &
\multicolumn{1}{l|}{.282}\\
\multicolumn{1}{|c}{} & \multicolumn{1}{|l|}{$.6$} & Bias &
\multicolumn{1}{|l}{.028} & \multicolumn{1}{l}{.023} &
\multicolumn{1}{l}{.009} & \multicolumn{1}{l}{-.008} &
\multicolumn{1}{l}{-.025} & \multicolumn{1}{l}{-.035} &
\multicolumn{1}{l}{-.040} & \multicolumn{1}{l|}{-.042}\\
\multicolumn{1}{|c}{} & \multicolumn{1}{|l|}{} & RMSE &
\multicolumn{1}{|l}{.172} & \multicolumn{1}{l}{.176} &
\multicolumn{1}{l}{.174} & \multicolumn{1}{l}{.175} & \multicolumn{1}{l}{.175}
& \multicolumn{1}{l}{.179} & \multicolumn{1}{l}{.184} &
\multicolumn{1}{l|}{.188}\\
\multicolumn{1}{|c}{} & \multicolumn{1}{|l|}{$.2$} & Bias & .088 & .083 &
.075 & .066 & .058 & .053 & .052 & .053\\
\multicolumn{1}{|c}{} & \multicolumn{1}{|l|}{} & RMSE & .135 & .133 & .126 &
.122 & .117 & .116 & .116 & .118\\\hline\hline
\end{tabular}
\caption{Risk-aversion estimation}%
\label{RAest}%
\end{table}Table \ref{RAest} shows that $\widehat{\nu}_{asc}$
dominates $\widehat{\nu}_{fp}$ in this experiment. While the RMSE\ and bias
of $\widehat{\nu}_{asc}$ do not seem sensitive to $h$, this is not the case
for $\widehat{\nu}_{fp}$ which has a high downward bias, and then RMSE, for
small $h$. Further investigations suggest this is due to an unbalanced
variable issue, the difference $\widehat{B}\left(  \alpha|X,3\right)
-\widehat{B}\left(  \alpha|X,2\right)  $ being very smooth while
$\alpha\left(  \widehat{B}^{\left(  1\right)  }\left(  \alpha|X,3\right)
/2-\widehat{B}^{\left(  1\right)  }\left(  \alpha|X,2\right)  \right)  $ is
more erratic, especially when $\alpha$ is close to $1$. This issue is
addressed in the application by restricting $\alpha$ to $\left[  0,.8\right]
$ for risk-aversion estimation.

\begin{figure}[t]
	\centering
	\includegraphics[width=.8\linewidth]{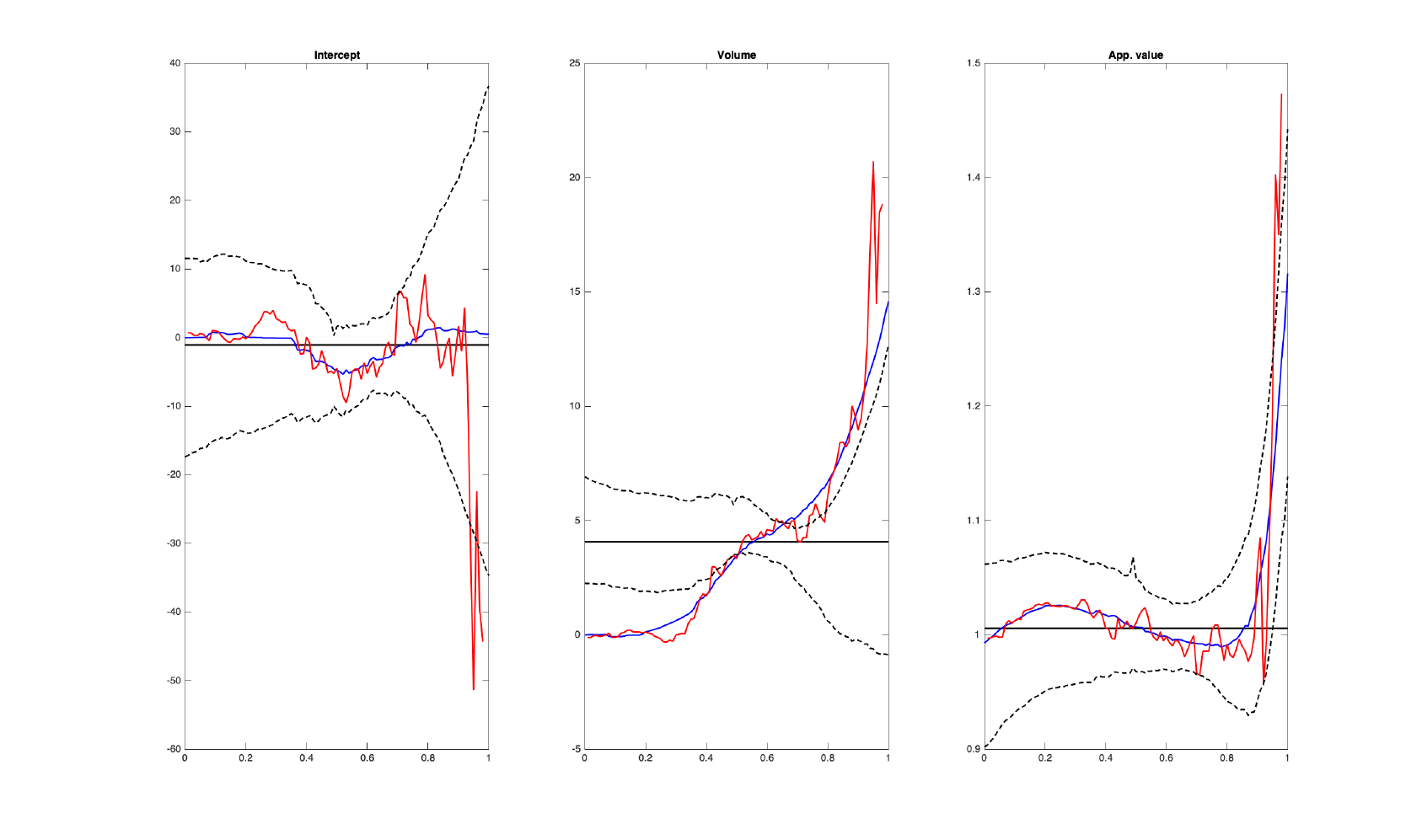}
	\caption{Two bidders first-price auction bid quantile\ slope coefficients:
		Intercept (left), volume (center) and appraisal value (right). AQR with $h=.3$
		(blue), standard QR (red) and OLS regression (black), and pointwise $90\%$
		confidence intervals for the AQR-regression difference (black dashed line)
		centered at the regression coefficients. A regression or AQR estimated slope
		coefficients outside the confidence band indicates a potential
		misspecification of the homogenized-bid regression model.}
	\label{BidI2}
\end{figure}

\section{Timber data application\label{Appli}}

Timber auctions data have
been used in several empirical studies (see Athey and Levin (2001), Athey,
Levin and Seira (2011) Li and Zheng (2012), Aradillas-Lopez, Gandhi and Quint
(2013) among others). Some other works have investigated risk-aversion in
timber auctions (e.g., Lu and Perrigne (2008), Athey and Levin (2001), Campo et
al. (2011)). This section uses data from 
timber auctions run by the US Forest Service (USFS) from Lu and Perrigne (2008) and Campo
et al. (2011), which aggregates auctions of 1979 from the states covering the western
half of the United States (regions 1--6 as labeled by the USFS). It contains bids and a set of variables characterizing each timber
tract, including the estimated volume of the timber measured in thousands of
board feet (mbf) and its estimated appraisal value given in dollars per unit
of volume. We consider the 107 first-price auctions with two bidders, the
first-price auctions with three bidders ($L=108$) and ascending auctions with two
bidders ($L=241$). The considered covariates are the appraisal value and the
timber volume taken in log. AQR is implemented with bandwidth $h=.3$ and the Epanechnikov kernel. Pointwise confidence intervals are computed using 10,000 pairwise bootstrap replications.
As in the simulation experiments, the CRRA parameter estimator has a high variance, and 
	risk-neutrality cannot be rejected, see Gimenes and Guerre (2019a). The rest of the application therefore assumes risk-neutral bidders.

\paragraph{Specification testing.}
Table \ref{Spectest} reports first the results of Rothe and Wied (2013)   test for the four following null hypotheses: correct specification of the quantile-regression (QR), of the homogenized-bid (HHS) model, exogeneity of the auction format (Format),  participation exogeneity (Entry). The three last null hypotheses are also tested using quantile-regression coefficients comparison tests.
Quantile-regression coefficient test statistics are based upon a discretized version of Liu and Luo (2017) integral statistic (\ref{Liuluo}), using Riemann sum over a grid $\alpha=0,1/100,\ldots,1$.
For HHS, the intercept of $\widehat{\beta}_{H_0} (\cdot )$ is from the AQR estimator while the slope are OLS. For the Format null hypothesis, $\widehat{\beta}_{H_0} (\cdot )= \alpha^{-1} \int_0^{\alpha}  \widehat{\gamma}_{asc} (a|2) da$, $\widehat{\gamma}_{asc} (\cdot|2)$ being an AQR version of Gimenes (2017) using the ascending auction sample with two bidders. For Entry, $\widehat{\beta}_{H_0} (\cdot )$ is an AQR version of (\ref{Betai1i2}) using  first-price auction with three bidders data.
The Rothe and Wied (2013) statistic uses the unconstrained and constrained cdf estimators computed from the two bidder sample
\begin{align*}
\widehat{G}(b,x) 
& =
\frac{1}{2L}
\sum_{\ell=1}^{L}
\sum_{i=1}^{2}
\mathbb{I}
\left( B_{i\ell} \leq b \text{ and } X_{\ell} \leq x \right)
,
\\
\widehat{G}_{H_0}(b,x) 
& =
\frac{1}{2L}
\sum_{\ell=1}^{L}
\sum_{i=1}^{2}
\mathbb{I}
\left( X_{\ell} \leq x \right) \widehat{G} \left(B_{i\ell}\left| X_{\ell}, \widehat{\beta}_{H_0} (\cdot)\right.\right)
\text{ with }
\\
\widehat{G} \left(b\left| x, \widehat{\beta}_{H_0} (\cdot)\right.\right)
& =
\int_{0}^{1}
\mathbb{I}
\left[
x_1^{\prime} \widehat{\beta}_{H_0} (\alpha)
\leq b
\right]
d \alpha
\end{align*}
which are compared using the Cramer-von Mises statistic
\[
\frac{1}{2L}
\sum_{\ell=1}^{L}
\sum_{i=1}^{2}
\left(
\widehat{G}_{H_0}(B_{i\ell},X_{\ell})
-
\widehat{G}(B_{i\ell},X_{\ell}) 
\right)^2
.
\]
\begin{table}[ht]
	\centering
	\begin{tabular}
		[c]{ll|ccccc}
		Tests & & QR & HHS & Format & Entry \\
		\hline \hline
		Rothe-Wied (2013) & Stat. value & .022 & .084 & .031 & .031 \\
		                  & $p$-value   & .07 & .00 & .22 & .58\\
		Test stat. (\ref{Liuluo})      & Stat. value &   x  & 2.95 & 0.70  & 0.42\\
		                  & $p$-value   &   x  & .03 & .01 & .07\\
	\end{tabular}
\caption{Specification tests}%
\label{Spectest}%
\end{table}
The $p$-values of the tests based upon Rothe and Wied (2013) use 10,000 replications of the bootstrap procedure proposed by these authors, while the other $p$-values are from  10,000 pairwise bootstrap replications. The Rothe and Wied (2013) procedure does not reject the quantile-regression specification at the $5\%$ level. This test also gives very high $p$-values for the Format and Entry null hypotheses, which both correspond to quantile-regression models,  estimated in a different way than from the null hypothesis QR.\footnote{Rothe and Wied (2013) testing procedure seems very sensitive to the estimation variance of the considered quantile model. Attempts not reported here show that it also holds for the Escanciano and Goh (2014) bootstrap procedure, which gives smaller $p$-values. However, it does not include a re-estimation of the quantile-regression specifications, which may underestimate $p$-values in small samples. While Rothe and Wied (2013) bootstrap combines pairwise bootstrap, which draws auctions with replacement, with a semiparametric one, which draws bids from the considered model, only the semiparametric bootstrap is implemented here.
} Both tests reject the homogenized-bid specification at  $5\%$ level. The coefficient-based test also rejects at this level  exogeneity of the auction format, disagreeing with Rothe and Wied (2013).

\begin{figure}[h]
	\centering
	\includegraphics[width=.9\linewidth]{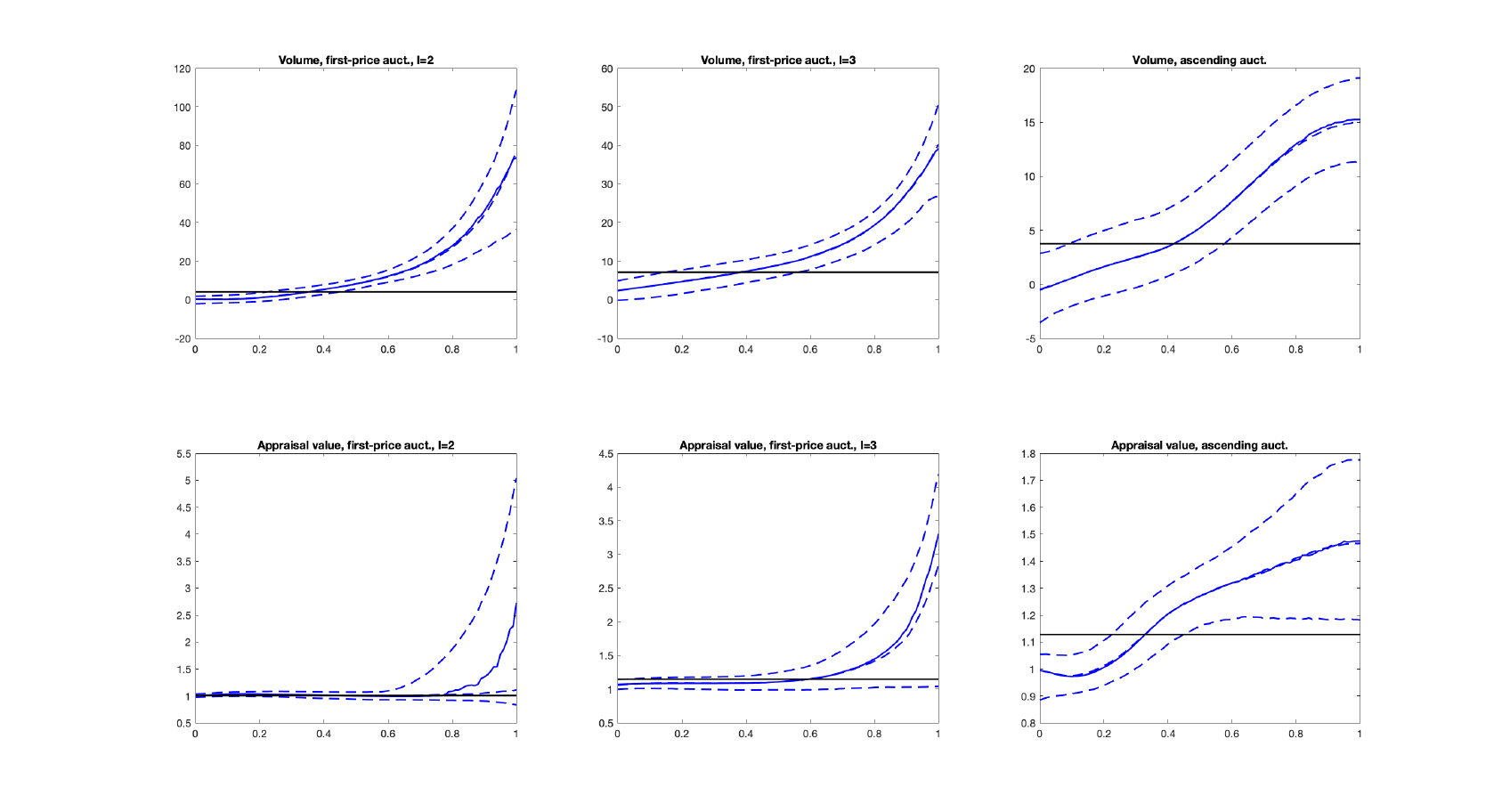}
	\caption{Volume (top) and appraisal value (bottom) estimated private-value
		slope function for first-price auctions with two bidders (left), three bidders
		(center) and ascending auctions (right), for $h=.3$. AQR estimation (full
		line), regression (full straight line) and $5\%,50\%,95\%$ bootstrapped
		quantile (dashed line).}
	\label{RNslope}
\end{figure}

\paragraph{Bid quantile functions.}

\begin{figure}[t]
	\centering
	\includegraphics[width=.7\linewidth]{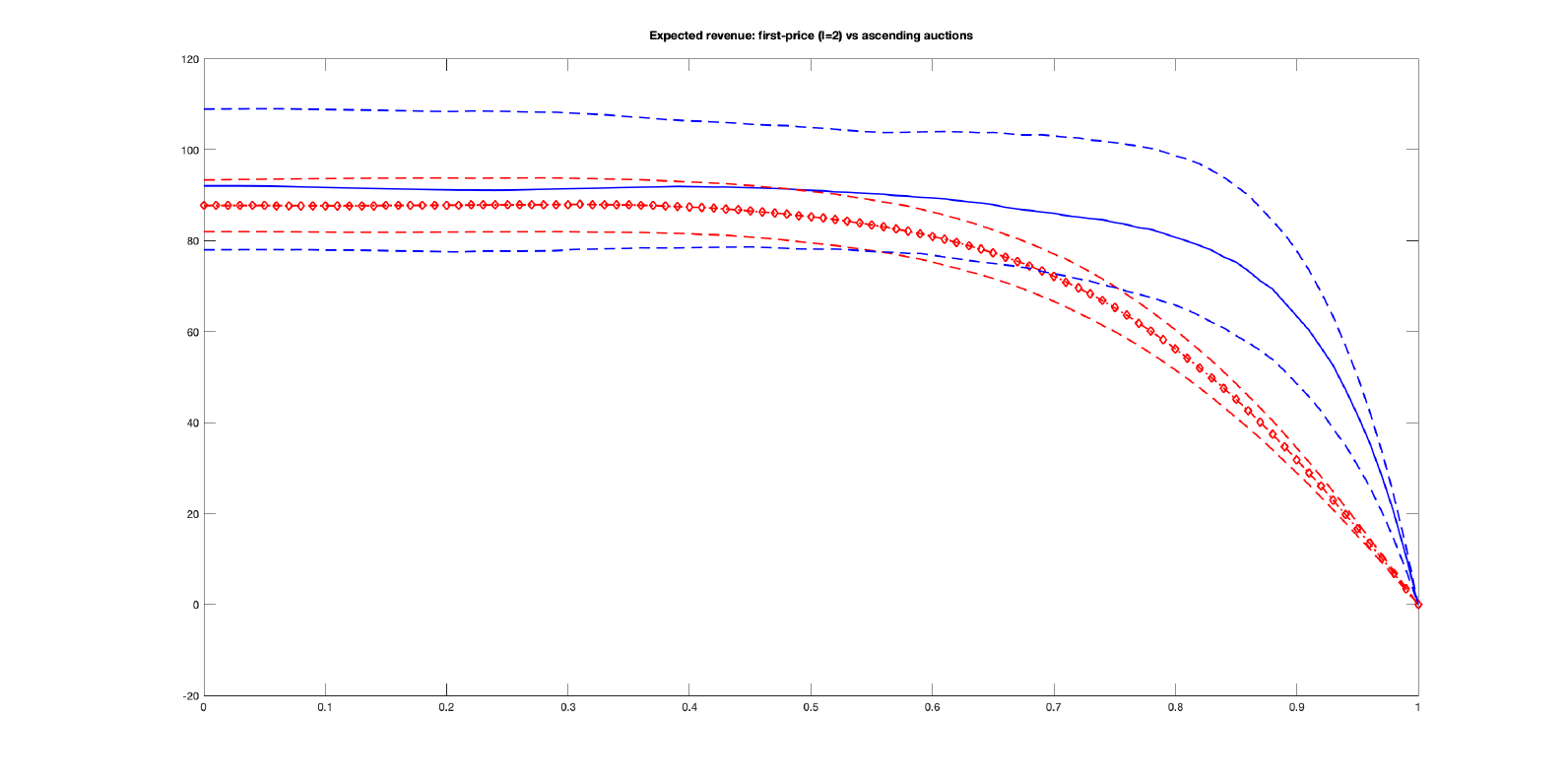}
	\caption{Estimated expected revenue as a function of the screening level for first-price (full line) and ascending
		(diamond) auctions with two bidders ($h=.4$). Volume and appraisal value set
		to median of the first-price auctions. $5\%-95\%$ bootstrap quantiles in
		dashed lines.}%
	\label{Exprev}
\end{figure}

Table \ref{Hmg} gives the results of  bid OLS regressions. The
dependent variables are the bids for  first-price auctions and the
winning bid  for the ascending auction. 
\begin{table}[h]
\centering
\begin{tabular}
[c]{lc|cccc}%
Auctions &  & Intercept & Volume & Appraisal value & $R^{2}$\\\hline\hline
First-price & $I=2$ & $\underset{\left(  6.67\right)  }{-1.06}$ &
$\underset{(1.12)}{4.07}$ & $\underset{(0.04)}{1.01}$ & $0.77$\\
& $I=3$ & $\underset{\left(  9.55 \right)  }{-20.79}$ & $\underset{\left(
1.34 \right)  }{7.10}$ & $\underset{\left(  0.06 \right)  }{1.15}$ & $0.70$\\
Ascending & $I=2$ & $\underset{\left(  15.05 \right)  }{2.76}$ &
$\underset{\left(  1.85 \right)  }{3.76}$ & $\underset{\left(  0.06 \right)
}{1.12}$ & $0.67$%
\end{tabular}
\caption{Auction bid regressions}%
\label{Hmg}%
\end{table}The appraisal value coefficient is close to $1$ in all auctions,
but is found significantly distinct at the $5\%$ level when comparing the
first-price auction with $I=2$ with the one with $I=3$ and the ascending
auction. Similarly the volume coefficient of the first-price auction with
$I=2$ differs from the one with $I=3$ at the $10\%$ level.  The appraisal value and volume coefficients of the first-price auctions with
$I=2$ and $I=3$ are  statistically distinct at the $5\%$ level. This is not
compatible with a homogenized-bid regression model assuming entry exogeneity.

Figure \ref{BidI2} gives the estimated slope for the first-price
auction bids with $I=2$. The volume OLS coefficient 
 is consistently outside the pointwise $90\%$ bootstrap confidence
interval of its AQR counterpart.  The appraisal value OLS estimate
lies outside the AQR confidence intervals for high quantile level in $[.9,1]$.  Figure
\ref{BidI2} also reports standard quantile-regression estimators, which exhibit a
similar pattern.  The intercept function does not look significant. Therefore, the intercept will be
kept constant and estimated using OLS in the rest
of the application. Comparison of the augmented and standard quantile-regression estimation also shows that the former produces smoother
slope coefficients.

\paragraph{Private value quantile function and expected revenue.}

Figure
\ref{RNslope} gives the private-value slope function of the volume and
appraisal variables. The  shape of the volume slope  varies across the type of auctions: while convex and in the $[20,100]$ range for
high $\alpha$ in the first-price case, it is in the $\left[  8,15\right]  $
range and more oscillating for ascending auctions. This suggests that the
private-value distribution and the auction mechanism are not independent, as also reported in Table \ref{Spectest} for the test based upon (\ref{Liuluo}).

The appraisal value slope seems statistically different from its OLS
counterpart for ascending auctions. For all auctions, the estimated appraisal value slopes start at
$1$ for $\alpha$ near $0$, suggesting that low type bidders do not get added value from
the appraisal value. This contrasts with high
type bidders with higher $\alpha$, which markup can be very high, in a
significant way for the case of ascending auction. This illustrates  the
important difference between low type and high type bidders.

A possible discrepancy between first-price and ascending auctions with two
bidders also appears in the expected revenue computed for median values of the
two explanatory variables, see Figure \ref{Exprev}. The ascending auction expected revenue is always below the first-price one. This seems statistically significant for high screening levels.  However, this may not be
relevant for the seller as the optimal revenue is achieved for a wide range
$\left[  0,.5\right]  $ of screening levels over which the two expected revenue
curves seem flat.

%EndExpansion

\section{Final remarks\label{conclusion}}

This paper proposes a quantile-regression modeling strategy for
first-price auction  under the independent private value
paradigm, which applies quantile-level local-polynomial to estimate the private-value quantile regression. This new framework can also be used to estimate some private-value random-coefficient models and to test some specification and exogeneity hypotheses of economic interest.
This approach is found to work well both in simulations, and in a timber
auction application where a strong low type/high type bidder heterogeneity is
detected. Another empirical
finding is that the seller expected revenue in a median auction is higher in
first-price than in ascending auctions, but
flat in a large zone around the optimal reserve prices. This suggests that the choice of a reserve price and of an auction mechanism may not be so important, at least for the median auction considered in the application.

Many aspects of the paper deserve further investigations. 
The estimated constant relative risk-aversion
exhibits a quite large variance, suggesting that a better understanding of
efficiency issues is needed.  
Various extensions can also be considered, such as endogenous entry as in Marmer, Shneyerov and Xu (2013a) or
Gentry and Li (2014). Our
quantile approach can be extended to exchangeable affiliated values as
considered in Hubbard, Li and Paarsch (2012), see also Gimenes and Guerre (2019b) for the more involved case of interdependent values.
The approach of Wei and Carroll (2009)  can be used
to tackle unobserved heterogeneity as in Krasnokutskaya (2011). 

%TCIMACRO{\TeXButton{TeX field}{\setcounter{equation}{0}}}%
%BeginExpansion
\setcounter{equation}{0}%
%EndExpansion

\pagebreak

\renewcommand{\thesection}{Appendix \Alph{section}}\setcounter{section}{0}

\section{\!\!\!\!\!\! - Sieve extension and assumptions\label{App:Sieve}}

\renewcommand{\thesection}{A.\arabic{section}}
\renewcommand{\thetheorem}{A.\arabic{theorem}}\setcounter{section}{0}
\renewcommand{\thehp}{\Alph{hp}.A}
\renewcommand{\thefootnote}{A.\arabic{footnote}}
\setcounter{footnote}{0}
\setcounter{equation}{0} \setcounter{theorem}{0} 

\section{Quantile interactive specification and localized sieve}

We introduce here a sieve counterpart of AQR, the \emph{Augmented Sieve Quantile Regression}, or in short ASQR, which gives primary conditions for (\ref{Sieve}), see Proposition \ref{SeriesB2} below.

\subsection{Quantile interactive specification}

The private-value quantile-regression model (\ref{Vqr}) assumes linearity  with respect to the covariate $X$. This
may be too strong and can be relaxed using a quantile nonparametric additive
specification, which was considered in Horowitz and Lee (2005). The latter writes, recalling
$X=\left(  X_{1},\ldots,X_{D}\right)  $,
\begin{equation}
V\left(  \alpha|X,I\right)  =\sum_{j=1}^{D}V_{j}\left(  \alpha;X_{j},I\right)
\label{AQ}%
\end{equation}
where each function $V_{j}\left(  \alpha;X_{j},I\right)  $ is specific to the
entry $X_{j}$. The effective dimension\ involved in the nonparametric
estimation of this model is $1$ because it can be estimated with the  rate
applying for a nonparametric model with a unique covariate as shown in Horowitz and
Lee (2005). This parsimonious model can be generalized following Andrews and
Whang (1990) to allow for more covariate interactions. This leads to the
additive interactive quantile specification with $D_{\mathcal{M}}$
interactions%
\begin{eqnarray*}
V\left(  \alpha|X,I\right) 
&=& 
\sum_{\mathbf{j}:1\leq
	j_{1}<\cdots<j_{D_{\mathcal{M}}}}
V_{\mathbf{j}}\left(  \alpha
;X,I\right)
,
\quad
D_{\mathcal{M}}\leq D, \quad
\mathbf{j} = \left(j_1,\ldots,j_{D_{\mathcal{M}}}\right)^{\prime}
\label{AIQ}%
\end{eqnarray*}
where each function $V_{\mathbf{j}}\left(  \alpha
;X,I\right)=V_{\mathbf{j}}\left(  \alpha;X_{j_{1}%
},\ldots X_{j_{D_{\mathcal{M} }}},I\right)  $ depends upon only $ D_{\mathcal{M}}$ entries of
$X$. Setting $D_{\mathcal{M}}=D$ gives the general, or \emph{saturated}, quantile
specification. As seen from Andrews and Whang (1990) for the regression case,
such specification can be estimated with the nonparametric rate applying for a  function of
$D_{\mathcal{M}}$ variables, so that $D_{\mathcal{M}}$ can be viewed as the
effective dimension of this model.

The stability property in Proposition \ref{Ident}-(i) ensures that a private-value quantile specification with $D_{\mathcal{M}}$ interactions will generate
a bid one with the same interactions: if
(\ref{AIQ}) holds, then the bid quantile function satisfies%
\[
B\left(  \alpha|X,I\right)  =
\sum_{\mathbf{j}:1\leq j_{1}<\cdots<j_{d}\leq D_{\mathcal{M}}} 
B_{\mathbf{j}}
	\left(  \alpha
;X,I\right)
\]
and the private-value components of the specification can be recovered using
Proposition \ref{Ident}-(ii).

\subsection{Localized sieve \label{Intersieve}}

The proposed estimation of the private-value quantile specification (\ref{AIQ}) is based upon a localized sieve which depends upon a smoothing parameter analogous to a bandwidth, as in regressogram methods. It is tailored for the approximation of function $\mu (\cdot)$ with $D_{\mathcal{M}}$ interactions,  for
$\mathbf{j}=(j_1,\ldots,j_{D_{\mathcal{M}}})^{\prime}$ and $\mu_{\mathbf{j}} \left( 
x \right)=\mu_{\mathbf{j}} \left( 
x_{j_{1}},\ldots x_{j_{D_{\mathcal{M}}}} \right)$
\begin{eqnarray}
\mu \left(  x \right)  
& = &
\sum_{\mathbf{j}:1\leq j_{1}<\cdots<j_{D_{\mathcal{M}}}\leq D} 
\mu_{\mathbf{j}} \left( 
x \right).
\label{Interactive}
\end{eqnarray}

For the sake of brevity, the localized sieve smoothing parameter will be taken identical to the AQR bandwidth $h$.  Assume that the support of $X$ is $\mathcal{X}=[0,1]^D$. The localized sieves considered here for additive interactive quantile function of order
$D_{\mathcal{M}}$ as in (\ref{AIQ}) depend upon a real function $p(\cdot)$ with compact support.
The considered  localized sieve is a collection of product functions
\begin{equation}
P_{\mathbf{i},\mathbf{j},h}  (x)
=
\frac{1}{h^{D_{\mathcal{M}}/2}}
p
\left(
\frac{x_{j_1}-hi_1}{h}
\right)
\times \cdots \times p\left(\frac{x_{j_{D_{\mathcal{M}}}}-hi_{D_{\mathcal{M}}}}{h} 
\right)
\label{AIS}%
\end{equation}
where the entries of the $D_{\mathcal{M}}\times 1$ $\mathbf{i}=\left( i_1, \ldots, i_{D_{\mathcal{M}}}\right)^{\prime}$ consist in all positive and negative integer numbers such that the support of $P_{\mathbf{i},\mathbf{j},h}  (\cdot)$ has a non empty intersection with $[0,1]^{D}$, and $\mathbf{j}$ is as in (\ref{Interactive}).

The functions  $P_{\mathbf{i},\mathbf{j},h}  (\cdot)$ can be reordered into a collection $\{P_{n} (\cdot),n=1,\ldots,N\}$, with $N \asymp h^{-D_{\mathcal{M}}}$, using the lexicographic order for $\mathbf{j}$ and, for a given $\mathbf{j}$, defining the successor of $\mathbf{i}$ as 
$\arg \max_{\mathbf{k}} \int_{[0,1]^D} \left|P_{\mathbf{k},\mathbf{j},h} (x) P_{\mathbf{i},\mathbf{j},h} (x)  \right| dx$. Note that this ordering does not depend upon the bandwidth $h$ and that the resulting $N \times N$ cross product matrix
\[
\int_{[0,1]^D} P (x) P(x)^{\prime} dx
,
\quad
P (x)
=
\left[
P_1 (x), \ldots, P(x)
\right]^{\prime}
\]
is a band matrix, ie its entries satisfy 
$\int_{[0,1]^D} P_{n_1} (x) P_{n_{2}} (x) dx = 0$ provided $|n_{2}-n_1| \geq c$, where the band size $c$ only depends upon the support of $p (\cdot)$,  but not upon $h$.
The standardization by $h^{-D_{\mathcal{M}}/2}$ in (\ref{AIS}) ensures that its diagonal entries satisfy
$1/C \leq \int_{[0,1]^D} P_{n}^2 (x) dx <C$ for all $h>0$ and $n=1,\ldots,N$. The Euclidean norm of the sieve vector satisfies $\max_{x \in \mathcal{X}} \| P(x) \| = O(h^{-{D_{\mathcal{M}}/2}})$  because there are only $C$ non $0$ entries $P_n (x)$, in which case the bound $|P_n (x)| \leq C h^{-{D_{\mathcal{M}}/2}}$ applies by (\ref{AIS}).

A simple choice of function $p(\cdot)$ is $p(t) = \mathbb{I} \left( t \in [0,1]\right)$, in which case the $\int_{[0,1]^D} P (x) P(x)^{\prime} dx$ is the $N \times N$ identity matrix $\mathrm{Id}_N$ provided $1/h$ is an integer number. This  sieve corresponds to regressogram methods, which has good approximation properties for Lipshitz functions $\mu (\cdot)$ satisfying (\ref{Interactive}).
In particular, for $\mu = \mu_h= \int_{\mathcal{X}} \mu (x) P (x) dx$, it holds 
$\max_{x \in \mathcal{X}} \left| P(x)^{\prime} \mu - \mu (x)\right|=O(h)$.
This rate can be improved for functions $\mu (\cdot)$ $(s+1)$-th differentiable using a proper choice of $p(\cdot)$, see the two examples below. Consider an extension of $\mu(\cdot)$, also denoted $\mu(\cdot)$ for the sake of brevity, defined over an enlargement $\mathcal{X}_{\epsilon}=[-\epsilon,1+\epsilon]^{D}$ of $\mathcal{X}$, $\epsilon>0$ . Define the modulus of continuity of the partial derivatives $\partial_{x_d}^r \mu(x) = \frac{\partial^r}{\partial x_d^{r}} \mu(x)$ as
\[
\mathrm{mc}_{r} (\mu;h)
=
\sum_{d=1}^{D}
\sup_{x \in \mathcal{X}_{\epsilon}}
\sup_{t \in [-h,h]\cap[-\epsilon-x_{d},1+\epsilon-x_{d}]}
\left| 
\partial^{r}_{x_d}
\mu
\left(
x_1,\ldots,x_{d-1},x_{d}+t,x_{d+1},\ldots,x_{D}
\right)
-
\partial^{r}_{x_d}
\mu (x)
\right|
.
\]
We shall assume later that the choice of the function $p (\cdot)$ ensures that:
\begin{quote}
	\textbf{Approximation property S.}
	\emph{There exists a constant $C=C_{p(\cdot),s,D}$ such that for any  interactive function $\mu(\cdot)$ satisfying (\ref{Interactive}) and, for $r \leq s+1$, $r$-th differentiable over $\mathcal{X}_{\epsilon}$, there exists  a $N \times 1$ vector $\mu = \mu_{\mu(\cdot),h,p(\cdot)}$ such that
	\[
	\sup_{x \in \mathcal{X}}
	\left|
	  P(x)^{\prime} \mu - \mu (x)
	\right|
	\leq 
	C h^{r} \sum_{\mathbf{j}: 1 \leq j_1 < \cdots < j_{D_{\mathcal{M}}} \leq D}\mathrm{mc}_{r} (\mu_{\mathbf{j}};h) .
	\]}
\end{quote}
In other words, the sieve should allow to approximate interactive functions with $r$ bounded derivatives up to an $O(h^{r})$ error, which can be improved to $o(h^{r})$ for continuous derivatives of order $r$, for all $r \leq s+1$. This is sufficient to ensure a negligible bias contribution from the sieve component of the estimation compared to the one of quantile level smoothing as shown in Theorem \ref{Bias} below.

\paragraph{Sieve Example 1:  wavelets with a compact support.}
Wavelet methods are a natural extension of the regressogram. They are based upon a scaling function $p(\cdot)$ which generates a multiresolution analysis, i.e. a nested sequence of linear spaces $\mathcal{P}_k$, $k$ in $\mathbb{N}$, generated by the functions $2^{-k/2} p \left( 2^{k} (x-2^{-k}i)\right)$, $i \in \mathbb{Z}$, such that $\mathcal{P}_k$ is asymptotically dense in the space $L_{2} (\mathbb{R})$ of squared integrable functions.\footnote{This exposition differs from the literature, which  builds on expansions
\[
\sum_{i} a_i 2^{-k_0/2} p \left( 2^{k_0} (x-2^{-k_0}i)\right)
+
\sum_{k\geq k_0}
\sum_{i} b_{ik} 2^{-k/2} q \left( 2^{k} (x-2^{-k}i)\right)
\]
where the so called ``father wavelets'' $\left\{2^{-k/2} p \left( 2^{k} (x-2^{-k}i)\right), i \in \mathbb{Z} \right\}$ is a basis of $\mathcal{P}_k$, and the bigger space $\mathcal{P}_{k+1}$ is generated by $\left\{2^{-k/2} p \left( 2^{k} (x-2^{-k}i)\right), i \in \mathbb{Z} \right\}$ and the  ``mother wavelets'' $\left\{2^{-k/2} q \left( 2^{k} (x-2^{-k}i)\right), i \in \mathbb{Z} \right\}$. Most of the statistical applications build on the expansion above, with a fixed $k_0$ and a truncation of the mother wavelet expansion, see  H\"{a}rdle et al. (1998), Chen (2007) and the references therein. We consider instead a father wavelet expansion $\sum_{i} a_i 2^{-k/2} p \left( 2^{k} (x-2^{-k}i)\right)$ where $k$ grows with the sample size and replaces $2^{-k}$ by a bandwidth $h$ as permitted for approximation purposes by the moment condition satisfied by $\mathcal{K} (\cdot,\cdot)$, see also  H\"{a}rdle et al. (1998, Chap. 8). As these two kind of expansions are equivalent, our framework also applies to father/mother expansions. However the mother wavelets with different resolution have an overlapping support, which do not fit our disjoint support assumption \ref{Riesz}-(ii) below, so that using a father wavelet expansion is better suited here. Although not detailed here, thresholding can be applied to the mother/father wavelet expansion associated to our estimated father one, see H\"{a}rdle et al. (1998) and the references therein.
} As developed in  H\"{a}rdle, Kerkyacharian, Picard and Tsybakov (1998), wavelet methods share many common features with kernel nonparametric estimation and, as explained here, can be applied for approximation of functions with compact support using a bandwidth $h$ instead of a dyadic power.
H\"{a}rdle et al.(1998, Chap. 8) consider a scaling function $p (\cdot)$ which satisfies
\begin{enumerate}
	\item For all integer number $i \neq 0$, $\int p(t) p(t-i) dt =0$ and $\int p^2(t) dt=1$; \label{Orthp} 
	\item \label{Momp} The associated kernel $\mathcal{K}(t_1,t_{2})= \sum_{i=-\infty}^{\infty} p(t_1-i) p(t_{2}-i)$ satisfies the moment conditions
	\[\int_{-\infty}^{\infty} \mathcal{K}(t_1,t_{2}) dt_{2} =1
	\text{ and }
	\int_{-\infty}^{\infty} \left(t_{2}-t_1\right)^{r}\mathcal{K}(t_1,t_{2}) dt_{2} =0,
	\quad
	r=0,\ldots, s+1.
	\]
\end{enumerate}
We assume in addition that $p(\cdot)$ has a compact support, which ensures existence of all the integrals above as $\left|\mathcal{K}(t_1,t_{2})\right|\leq C \mathbb{I} \left(|t_{2}-t_1| \leq C \right)$.  H\"{a}rdle et al. (1998, Theorem 8.3-(i)) show that (\ref{Orthp}) and (\ref{Momp}) are satisfied by the Daubechies scaling function of order $s+2$, see Daubechies (1992, Chaps. 6-8) and H\"{a}rdle et al. (1998, Remark 8.1 and Chap. 7). Daubechies wavelet scaling functions do not have an explicit expression but can be implemented using standard scientific softwares such as Matlab. Part (\ref{Orthp}) implies that the translated functions 
$p(\cdot - i)$, $i$ in $\mathbb{Z}$, form an orthonormal system while (\ref{Momp}) is the analogous of the vanishing moment condition used for the bias in kernel nonparametric estimation.

Wavelet sieve for additive interactive functions can be based upon (\ref{AIS}). To check that it satisfies the Approximation Property S, it is sufficient to consider $r=s+1$ and the saturated case $D=D_{\mathcal{M}}$ for any values of $D$ and to apply the approximation procedure detailed now to each function $\mu_{\mathbf{j}} (\cdot)$ in the decomposition (\ref{Interactive}). Observe that, as $\mathbf{j}$ is set to $(1,\ldots,D)^{\prime}$, $P_{\mathbf{i},\mathbf{j},h}  (\cdot)$ can be abbreviated into $P_{\mathbf{i},h}  (\cdot)$. The multivariate counterpart of the kernel function $\mathcal{K}(t_1,t_{2})$ is $\mathcal{K}(x,y)=\prod_{d=1}^D \mathcal{K} \left(x_d,y_d\right)$. Observe that $\mathcal{K}_h (x,y)=h^{-1} \mathcal{K} (x/h,y/h)$ is such that
\begin{equation}
\mathcal{K}_h (x,y)
=
\sum_{\mathbf{i} \in \mathbb{Z}^D}
P_{\mathbf{i},h} (x) P_{\mathbf{i},h} (y)
,
\label{Kh}
\end{equation}
and that $\mathcal{K}_h (x,y) = \sum_{n=1}^N P_n (x) P_n (y)$ when $X$ belongs to $[0,1]^D$.
In view of (\ref{Orthp}), a natural approximation $P(x)^{\prime} \mu$ of $\mu(x)$ is its orthogonal projection on the sieve, given by  
\begin{equation}
\mu_n = \int \mu (x) P_n (x) dx,
\label{Mun}
\end{equation}
assuming that $h$ is small enough, so that the support of $P_n(\cdot)$ is in $\mathcal{X}_{\epsilon}$ for all $n$.
This gives
\[
\mu'P(x) = \int \mu (y) \mathcal{K}_h (x,y) dy.
\]
A Taylor expansion of order $s+1$ with integral remainder gives, for any $d$
\begin{eqnarray*}
\lefteqn{
\mu (y) - \mu \left(y_1,\ldots,y_{d-1},x_d,y_{d+1},\ldots,y_D \right)
=
\sum_{r=1}^{s+1}
\left(y_d-x_d\right)^r
\partial^{r}_{x_d}
\mu \left(y_1,\ldots,y_{d-1},x_d,y_{d+1},\ldots,y_D \right)
}
\\
& + &
\frac{\left(y_d-x_d\right)^{s+1}}{s!}
\int_0^1
(1-t)^s
\left(
\partial^{s+1}_{x_d}
\mu \left(y_1,\ldots,y_{d-1},x_d+t(y_d-x_d),y_{d+1},\ldots,y_D \right)
\right.
\\
&&
\quad\quad\quad\quad\quad\quad\quad\quad\quad\quad\quad\quad\quad\quad\quad\quad\quad\quad
\left.
-
\partial^{s+1}_{x_d}
\mu \left(y_1,\ldots,y_{d-1},x_d,y_{d+1},\ldots,y_D \right)
\right)
dt.
\end{eqnarray*}
Hence (\ref{Momp}) gives, starting with $d=1$,
\begin{eqnarray*}
\lefteqn{
\sup_{x \in \mathcal{X}}
\left|
\int \mu (y) \mathcal{K}_h (x,y) dy
-
\int
\mu (x_1,y_{2},\ldots,y_D)
\frac{1}{h^{D-1}}
\prod_{d=2}^D
\mathcal{K} \left(\frac{x_d}{h},\frac{y_d}{h}\right)
dy_d
\right|
}
\\
& \leq &
C
\textrm{mc}_{s+1} (\mu;h)
\int
\frac{\left|X_1-x_d\right|^{s+1}}{h^{D}}
\mathbb{I}
\left(
\max_{d=1,\ldots,D}
\left|
y_d-x_d
\right|
\leq C \cdot h
\right)
dy
\leq
C h^{s+1}
\textrm{mc}_{s+1} (\mu;h).
\end{eqnarray*}
Iterating over index $d$ then shows that 
$
\sup_{x \in \mathcal{X}}
\left|
P(x)^{\prime} \mu
-
\mu (x)
\right|
\leq
C h^{s+1}
\textrm{mc}_{s+1} (\mu;h)
$.
Hence the Approximation Property S holds.

\paragraph{Sieve Example 2: Cardinal B-spline.} For $m\geq s+2$, set $\left(
t\right)  _{+}^{m-1}=t^{m-1}$ if $t>0$ and $\left(  t\right)  _{+}^{m-1}=0$
otherwise. The cardinal B-spline sieve is based upon the uniformly spaced
simple knots $B-$spline function of order $m$ (Schumaker (2007), p.135)%
\[
p\left(  t\right)  =\sum_{i=0}^{m}\frac{\left(  -1\right)  ^{i}\binom{m}%
	{i}\left(  t-i\right)  _{+}^{m-1}}{(m-1)!}%
\]
which has $m-2$ continuous derivatives over the straight line and whose
support is $\left[  0,m\right]  $. Theorem 12.8 of Schumaker (2007) shows that the Approximation property S holds for this choice of $p(\cdot)$,\footnote{Following Schumaker (2007) shows that it is possible to modify $p(\cdot)$ for indices $\mathbf{i},\mathbf{j}$ such that the support of $P_{\mathbf{i},\mathbf{j},h}  (\cdot)$ is not a subset of $\mathcal{X}$. However considering an extension of $\mu(\cdot)$ over the enlarged $\mathcal{X}_{\epsilon}$ shows it is not necessary.
Note also that Schumaker(2007) Theorem 12.8 uses a different modulus of continuity. Equivalence with the one used here follows from Schumaker (2007), Theorem 13.24 and (13.62). 
} constructing a spline approximation of each $\mu_{\mathbf{j}} (\cdot)$ in (\ref{Interactive}) as possible with the interactive sieve (\ref{AIS}). Note that the coefficients $\mu$ used for the spline approximation are not given explicitly here.  Schumaker (2007, Theorem 12.8) gives integral expression of coefficients that can be used for splines. In both examples, it is however clear that these coefficients are not unique. For instance, in the Wavelet Example 1, using $\widetilde{\mu}_n$ close enough to $\mu_n$ in (\ref{Mun}) could also work.

\setcounter{equation}{0}
\section{Main sieve assumptions \label{Sieveestassump}}

	The set of Assumptions below extends the one used for the AQR to the ASQR estimation method. These extended Assumptions are identified by adding ``.A'' to the labels in the main section. For instance, Assumption \ref{Spec.A} below corresponds to Assumption \ref{Specification} in the main body of paper and extends it to cover both the AQR and ASQR cases. As a consequence, proofs refer to the Assumptions stated here. Assumption \ref{Funct} remains unchanged. Assumption \ref{Riesz} is new and specific to the ASQR case. Hence the labels of these assumptions do not have an additional ``.A''. 
Instead of 	$\max\left(
a,b\right)  $ and $\min\left(  a,b\right)  $, the notations	
$a\vee b$ and $a\wedge b$ are used, respectively. Recall that $\| \cdot \|$ is the Euclidean norm.

\setcounter{hp}{0}

\begin{hp}
	\label{Auct.A} 
	Assumption \ref{Auct} holds in the AQR case. For the ASQR case, it is assumed in addition in (i) that the support of $X$ given $I$ is the set $\mathcal{X}=[0,1]^D$, and that the pdf $f(x,i)$ of $(X,I)$ are all bounded away from $0$ and infinity on $\mathcal{X}$.
\end{hp}

\setcounter{hp}{18}

	\begin{hp}
		\label{Spec.A}
		(i) In the AQR case, $V(\alpha|X,I)$ is as in Assumption \ref{Specification} and $D_{\mathcal{M}}$ is set to $0$. (ii) In the ASQR case,
		$V\left(  \alpha|X,I\right)  $ has $D_{\mathcal{M}} \geq 1$
		interactions as in (\ref{AIQ}) and its components
		$V_{\mathbf{j}} \left(\alpha|X,I\right)  $ are $\left(  s+1\right)  -$times continuously
		differentiable over $\left[  0,1\right]  \times\mathcal{X}$.
	\end{hp}

\setcounter{hp}{7}

	\begin{hp}
		\label{Kernel.A} The kernel function $K\left(  \cdot\right)  $ is as in Assumption \ref{Kernel} and the bandwidth satisfies
		\[
		\lim_{L\rightarrow\infty}\frac{\log^2 L}{Lh^{3D_{\mathcal{M}}+2}}=0.
		\]
		For the ASQR estimator, $P\left(  x\right)  =\left[  P_{1}\left(  x\right)
		,\ldots,P_{N}\left(  x\right)  \right]  ^{\prime}$ where $P_{n}\left(
		x\right)  =P_{hn}\left(  x\right)  $ and $N\asymp h^{-D_{\mathcal{M}}}$. 
	\end{hp}

\renewcommand{\thehp}{\Alph{hp}}

\setcounter{hp}{17}

\begin{hp}
	\label{Riesz} 
	The sieve $P(x)$  satisfies the Approximation property S, and: (i) The eigenvalues of the Gram matrix $\int_{\mathcal{X}%
		}P\left(  x\right)  P^{\prime}\left(  x\right)  dx$ stay bounded away from $0$
		and infinity, in an interval $[1/C,C]$ for some $C>0$, when $h$ goes to $0$ and
		$
		\max_{x\in\mathcal{X}}\left\Vert P\left(  x\right)  \right\Vert 
		=O\left(h^{-D_{\mathcal{M}}/2}\right) 
		$, $
		\min_{x\in\mathcal{X}}\left\Vert P\left(  x\right)  \right\Vert 
		\geq C h^{-D_{\mathcal{M}}/2}  
		$ for some $C>0$, $
		\max_{n\leq N}\left\{  \int_{\mathcal{X}}\left\vert P_{n}\left(  x\right)
		\right\vert dx\right\}  =O\left(  h^{D_{\mathcal{M}}/2}\right) 
		$.

		(ii) The sieve is 
		localized, i.e. it satisfies the following ``Disjoint Support Property'': there is a $c>0$ such that $P_{n_{1}}\left(
		\cdot\right)  P_{n_{2}}\left(  \cdot\right)  =0$ as soon as $\left\vert
		n_{2}-n_{1}\right\vert \geq c/2$.

	(iii) For some $\eta\in\left(  0,1\right]  $ and $\overline{N}_{1L}$ with
		$\log\overline{N}_{1L}=O\left(  \log L\right)  $, it holds that%
		\[
		\left\Vert P\left(  x\right)  -P\left(  X^{\prime}\right)  \right\Vert
		\leq\overline{N}_{1L}\left\Vert x-x^{\prime}\right\Vert ^{\eta}\text{ for all
		}x\text{,}x^{\prime}\text{ of }\mathcal{X}.
		\]
\end{hp}

The bandwidth rate in Assumption \ref{Kernel.A} is driven by the linearization of $\widehat{B}^{(1)} (\cdot|\cdot,I)$, see the discussion of Theorem \ref{Baha}.

Assumption \ref{Riesz} covers the sieve (\ref{AIS}) as seen from (i), but is slightly more general. In particular the product structure of the $P_{\mathbf{i},\mathbf{j},h} (\cdot)$ in (\ref{AIS}) is not needed. It includes the disjoint support property (ii) which is an important characteristic of localized sieves. 	
It is useful to obtain bounds for scalar products of the form $\mathbf{1}_N^{\prime} P(x)$, where $\mathbf{1}_N$ is a $N\times 1$ vector with unit entries, so that $\| \mathbf{1}_N\| = N^{1/2}=O(h^{-{D_{\mathcal{M}}/2}})$. Hence, while the Cauchy-Schwarz inequality $|  P(x)^{\prime}\mathbf{1}_N| \leq \|P(x)\| \|\mathbf{1}_N\|  $ gives $\max_{x \in \mathcal{X}} | P(x)^{\prime}\mathbf{1}_N| = O(h^{-{D_{\mathcal{M}}}})$, the fact that $P_n (x) \neq 0$ for only $c$ entries of $P(x)$ implies the better bound $\max_{x \in \mathcal{X}} | P(x)^{\prime}\mathbf{1}_N| = O(h^{-{D_{\mathcal{M}}}/2})$ as 
$
\max_{n \leq N} \max_{x \in \mathcal{X}} \left| P_n (x )\right|
\leq \max_{x \in \mathcal{X}} \left\| P(x) \right\| 
= O(h^{-{D_{\mathcal{M}}}/2})
$.\footnote{This also follows from the  bound
	$\max_{x \in \mathcal{X}} | P(x)^{\prime}\beta| \leq \max_{x \in \mathcal{X}} \sum_n \| \beta \|_{\infty}|P_n(x)| \leq Ch^{-D_{\mathcal{M}}/2} \| \beta \|_{\infty}$, where $\| \beta \|_{\infty}$ is the largest entry of $\beta$ in absolute value. }
	For sieve as in (\ref{AIS}), Assumption
	\ref{Riesz}-(iii) holds for bandwidth $h$ as in Assumption \ref{Kernel.A}, provided $p\left(  \cdot\right)  $ is H\"{o}lder
	with exponent $\eta$. This allows for cardinal B-splines for which $\eta=1$,
	but also for wavelets which are not always differentiable but H\"{o}lder with
	$\eta<1$, see Daubechies (1992). 

While $
\max_{x\in\mathcal{X}}\left\Vert P\left(  x\right)  \right\Vert 
=O\left(h^{-D_{\mathcal{M}}/2}\right) 
$ and  
$
\max_{n\leq N}\left\{  \int_{\mathcal{X}}\left\vert P_{n}\left(  x\right)
\right\vert dx\right\}  =O\left(  h^{D_{\mathcal{M}}/2}\right) 
$ follows from (\ref{AIS}) for a compactly supported $p(\cdot)$, 
a less intuitive condition of Assumption \ref{Riesz} is $
\min_{x\in\mathcal{X}}\left\Vert P\left(  x\right)  \right\Vert 
\geq C h^{-D_{\mathcal{M}}/2}  
$. This is may be easier to understand considering the Wavelet examples, for which  $\mu$ in the Approximation property S can be set to
$\int_{\mathcal{X}} \mu (x) P(x) dx$, which is such that $\sup_{n \leq N} |\mu_n| =O(h^{D_{\mathcal{M}}/2})$. Having 
$
\min_{x\in\mathcal{X}}\left\Vert P\left(  x\right)  \right\Vert 
= o( h^{-D_{\mathcal{M}}/2})  
$
which gives for a localized sieve $\inf_{x \in \mathcal{X}} | P(x)^{\prime} \mu | \leq  c h^{D_{\mathcal{M}}/2} o( h^{-D_{\mathcal{M}}/2}) = o(1)$, which would contradict the Approximation Property S taking for instance $\mu (\cdot)$ bounded away from $0$, such as $\mu(\cdot)=1$. This extends to general localized sieve using the least square choice of $\mu$ (\ref{LS}) used in Proposition \ref{SeriesB2} which satisfies $\sup_{n \leq N} |\mu_n| =O(h^{D_{\mathcal{M}}/2})$.\footnote{See the proof of Proposition \ref{SeriesB2}, which does not use   $
	\min_{x\in\mathcal{X}}\left\Vert P\left(  x\right)  \right\Vert 
	\geq C h^{-D_{\mathcal{M}}/2}  
	$ for some $C>0$.}

	\subparagraph{Localized sieve and unknown covariate support.}
	Assumption \ref{Auct.A} imposes a known $[0,1]^{D}$ covariate support. As its purpose is to impose a finite number of localized sieve coefficients, it can be assumed that the covariate support is the closure of a bounded open set. A conjecture is that assuming a known support can be relaxed using a data-driven selection of the index $\mathbf{i}$ used in the localized sieve inspired by Cuevas and Fraiman (1997). These authors proposed to estimate the support $\mathcal{X}$ using $\{x; \widehat{f}_X (x) \geq \tau_L \}$
	where $\widehat{f}_X (\cdot)$ is a covariate pdf kernel estimator and $\tau_L$ a threshold sequence that goes with $0$ with the sample size. In our framework, the sieve entry $P_{\mathbf{i},\mathbf{j},h} (\cdot)$ is useful if there are enough observations in the support of neighboring $P_{\mathbf{i}^{v},\mathbf{j},h} (\cdot)$, such that $P_{\mathbf{i},\mathbf{j},h} (\cdot)P_{\mathbf{i}^{v},\mathbf{j},h} (\cdot) \neq 0$. Following Cuevas and Fraiman (1997), this suggests to select, for each given $\mathbf{j}$, all $\mathbf{i}$ and $\mathbf{i}^{v}$ such that 
	all $\frac{1}{L} \sum_{\ell=1}^{L} \left| P_{\mathbf{i}^{v},\mathbf{j},h} (X_{\ell}) \right|
	\mathbb{I} (I_{\ell}=I)$ are large enough compared to
	$ h^{D_{\mathcal{M}/2}}$.
	Note that, for $\mathbf{j}=(j_1,\ldots,j_{D_{\mathcal{M}}})$, the union of the support of the selected $P_{\mathbf{i},\mathbf{j},h} (\cdot)$ is an estimator for the support of $(x_{j_1},\ldots,x_{j_{D_{\mathcal{M}}}})$ given $I$. The sieve estimator should deliver its best performance  over this estimated support.

\section{Sieve quantile regression and first-price auctions}
The Approximation Property S ensures that there exists a sieve coefficient vector $\gamma (\alpha|I)=\gamma_{h} (\alpha|I)$ such that, for each $\alpha$, $I$ and $X$
\begin{equation*}
V\left(  \alpha|X,I\right)  =\lim_{h\rightarrow0} P(X)^{\prime} \gamma (\alpha|I) 
\end{equation*}
for an interactive private-value quantile function $V(\alpha|X,I)$  satisfying some smoothness conditions. The LHS of the equation above is an asymptotic sieve quantile regression which can be estimated using an augmented quantile regression. Making this rigorous necessitates first to elicit the smoothness properties of the coefficients $\gamma (\cdot|I)$, which, as suggested by their expression (\ref{Mun}) for the Wavelet Example 1, should inherit the smoothness of the parent $V(\alpha|X,I)$. 
As the sieve coefficients used in the Approximation Property S are not unique, it is necessary to be more specific in their choice and, from now on, we use a Least Square choice of the sieve coefficient $\gamma (\cdot|I)$ of $V(\cdot|\cdot,I)$\footnote{In the AQR case, the slope coefficient in (\ref{LS}) is identical to the quantile-regression slope as $P(X)=X_1$ and $V(\alpha|X,I)=X_1^{\prime} \gamma (\alpha|I)$.}
\begin{equation}
\gamma (\alpha|I)
=
\gamma_h (\alpha|I)
=
\left(
\mathbb{E}
\left[
\left.
P(X )
P(X)^{\prime}
\right|
I
\right]
\right)^{-1}
\mathbb{E}
\left[
\left.
P(X )
V(\alpha|X,I)
\right|
I
\right]
\label{LS}
\end{equation}
assuming that 
$
\mathbb{E}
\left[
\left.
P(X )
P(X)^{\prime}
\right|
I
\right]
$ is full rank.
Second, a similar task should be carried for the asymptotic sieve bid quantile-regression model generated by (\ref{Vsqr}). These issues are addressed in the next Proposition, which proof is given in \ref{App:Proofsprelim}. Note that it is also useful in the AQR case, in which case (\ref{LS}) is identical to the slope in $V(\alpha|X,I)=X_{1}^{\prime} \gamma(\alpha|I)$ setting $P(X)=X_1$ and $B(\alpha|X,I)=X_1^{\prime} \beta(\alpha|I)$ in (ii) and (iii).
\begin{proposition}
	\label{SeriesB2}
	Suppose Assumptions \ref{Auct.A}, \ref{Spec.A} and \ref{Riesz} holds.
	Then  $\gamma (\alpha|I)$ in (\ref{LS}) is $(s+1)$-th continuously differentiable and it holds:
	\begin{equation}
	\max_{(\alpha,x) \in [0,1] \times \mathcal{X}}
	\left|
	P(x)^{\prime} \gamma^{(p)} (\alpha|I) -  V^{(p)}(\alpha|x,I)
	\right|
	=
	o \left( h^{s+1-p} \right),
	\ldots
	p=0,\ldots, s+1.
	\label{Vsqr}
	\end{equation}

	Moreover for the bid quantile function $B\left(  \alpha|X,I\right)  $ as in (\ref{V2B}) and 
	\begin{equation}
	\beta (\alpha|I) = \frac{I-1}{\alpha^{I-1}}\int_0^{\alpha}a^{I-2} \gamma(a|I)da,
	\text{ $\gamma (\cdot|I)$ from (\ref{LS}),}
	%\label{Bs2Vs}
	\label{Bsqr}
	\end{equation}
	it holds:
	\begin{enumerate}
		\item $\min_{\left(  \alpha,x\right)  \in\left[  0,1\right]  \times
			\mathcal{X}}B^{\left(  1\right)  }\left(  \alpha|x,I\right)  >0$,
		$\max_{\left(  \alpha,x\right)  \in\left[  0,1\right]  \times\mathcal{X}%
		}B^{\left(  1\right)  }\left(  \alpha|x,I\right)  <\infty$ and, for each $(x,I)$ of $\mathcal{X} \times \mathcal{I}$, $\alpha \mapsto B\left(
		\alpha|x,I\right)  $ is $(s+1)$th continuously differentiable over $\left[
		0,1\right]  $, and  $(s+2)$th continuously differentiable over $\left(
		0,1\right]$ with
		$
		\lim_{\alpha\downarrow0}\sup_{  x  \in\mathcal{X}
		}\left\vert \alpha B^{\left(  s+2\right)  }\left(  \alpha|X,I\right)
		\right\vert =0
		$.  Moreover $\alpha \mapsto \alpha B(\alpha|x,I)$ is $(s+2)$th continuously differentiable over $[0,1]$. 
		\item $\beta (\alpha|I)$ is $(s+1)$-th continuously differentiable over $[0,1]$ with
		\begin{equation}
		\max_{(\alpha,x) \in [0,1] \times \mathcal{X}}
		\left|
		P(x)^{\prime} \beta^{(p)} (\alpha|I)
		-
		B^{(p)} (\alpha|x,I)
		\right|
		=
		o \left(h^{s+1-p}\right),
		\quad p=0,\ldots,s+1.
		\label{Bsqr2}
		\end{equation}
		\item $\alpha \beta^{(s+2) }(\alpha|I)$ is continuous over $(0,1]$ with $\lim_{\alpha \downarrow 0} \alpha \beta^{(s+2) }(\alpha|I)=0$. Moreover 
		\begin{eqnarray*}
		\max_{(\alpha,x) \in [0,1] \times \mathcal{X}}
		\left|
		P(x)^{\prime} \alpha \beta^{(p)} (\alpha|I)
		-
		\alpha
	    B^{(p)}(\alpha|x,I)
		\right|
		&=&
		o \left(h^{s+2-p}\right),
		\quad p=1,\ldots,s+2,
		\\
		\max_{(\alpha,x) \in [0,1] \times \mathcal{X}}
		\left|
		P(x)^{\prime} \alpha \beta^{(s+2)} (\alpha|I)
		-
		\alpha B^{(s+2)}(\alpha|x,I)
		\right|
		&=&
		o(1).
		\end{eqnarray*}
	\end{enumerate}
\end{proposition}

Note that the sieve bid quantile slope in (\ref{Bsqr}) can be defined through a least square formula similar to (\ref{LS}), that is
\begin{equation}
\beta^{(p)} (\alpha|I)
=
\left(
\mathbb{E}
\left[
\left.
P(X )
P(X)^{\prime}
\right|
I
\right]
\right)^{-1}
\mathbb{E}
\left[
\left.
P(X )
B^{(p)}(\alpha|X,I)
\right|
I
\right],
\quad
p=0,1,\ldots,s+2.
\label{BetaLS}
\end{equation}

Proposition \ref{SeriesB2} gathers results on various rate of approximation which allows to extend the AQR approach to a sieve setup, starting from the asymptotic sieve quantile regression (\ref{Vsqr}) with slope $\gamma (\cdot)$ for the private value. Defining the sieve bid coefficients as in (\ref{Bsqr}) gives an asymptotic quantile regression (\ref{Bsqr2}) 
\[
B(\alpha|x,I) = P(x)^{\prime} \beta (\alpha|I) + o \left(h^{s+1}\right)
\]
with a differentiable slope $\beta(\cdot|\cdot)$  which identifies the private-value slope using 
\begin{equation}
\gamma (\alpha|I)
=
\beta (\alpha|I) + \frac{\alpha \beta^{(1)} (\alpha|I)}{I-1}
\label{Bs2Vs}
\end{equation}
as in  the quantile-regression case, see (\ref{Sieve}). In addition, Proposition \ref{SeriesB2} establishes rates for the approximation of $B^{(p)} (\cdot|\cdot,I)$ using the $p$-th slope derivatives $\beta^{(p)} (\cdot|I)$ as necessary for the augmented approach presented in the next section.
Note however that Proposition \ref{SeriesB2}-(iii) does not give a rate for 
$
\max_{(\alpha,x) \in [0,1] \times \mathcal{X}}
\left|
P(x)^{\prime} \alpha \beta (\alpha|I)
-
\alpha
B (\alpha|x,I)
\right|
$
which is at least $o(h^{s+1})$ by (ii). This is specific to the the ASQR case, as arguing with Taylor expansions as for the proof of Proposition \ref{SeriesB2} gives a better rate $o(h^{s+2})$ for the AQR.

\paragraph{Augmented sieve quantile regression and functional estimation.}
	The AQR can be modified in a straightforward way to account for our sieve extension, leading to the Augmented sieve quantile regression (ASQR), which proceeds by using a vector $b$ of dimension $(s+2)N$ instead of $(s+2)(D+1)$ and to redefine $P(x,t)$ in (\ref{Pxt}) as
	\begin{equation}
	P\left(  x,t\right)  
	=
	\pi\left(  t\right)  
	\otimes
	P(x)
	\in \mathbb{R}^{(s+2)N},
	\quad
	P(x)^{\prime}
	=
	\left[P_1 (x),\ldots,P_N(x)\right].
	\label{Pxtsieve}
	\end{equation}
	The expression of the objective function $\widehat{\mathcal{R}}\left(  b;\alpha,I\right)$ is unchanged and as in (\ref{CalR}).
	The estimation of $b\left(
	\alpha|I\right) = [\beta(\alpha|I)^{\prime}, \ldots, \beta^{(s+1)}(\alpha|I)^{\prime}]  $ is 
	$\widehat{b}\left(  \alpha|I\right)  
	=
	\arg\min_{b\in \mathbf{R}^{(s+2)N}} \widehat{\mathcal{R}}\left(  b;\alpha,I\right)  $ and the ASQR estimators are
	\begin{eqnarray*}
		\widehat{V}\left(  \alpha|X,I\right)  
		& = &
		P\left(  x\right)  ^{\prime
		}\widehat{\gamma}\left(  \alpha|I\right)
		=  \widehat{B}\left(  \alpha|X,I\right)+\frac{\alpha\widehat{B}^{\left(
				1\right)  }\left(  \alpha|X,I\right) }{I-1},		
		\text{ where by (\ref{Bs2Vs})}
		\\
		\widehat{\gamma}\left(
		\alpha|I\right)  &=& \widehat{\beta}_{0}\left(  \alpha|I\right)  +\frac
		{\alpha\widehat{\beta}_{1}\left(  \alpha|I\right)  }{I-1},
		\text{ }
		\widehat{B}\left(  \alpha|X,I\right)   =  P\left(  x\right)  ^{\prime
		}\widehat{\beta}_{0}\left(  \alpha|I\right),
		\text{ }
		\widehat{B}^{\left(
			1\right)  }\left(  \alpha|X,I\right)  =P\left(  x\right)  ^{\prime
		}\widehat{\beta}_{1}\left(  \alpha|I\right).
	\end{eqnarray*}
	The functional estimators use a plug-in construction as the AQR case.

\setcounter{equation}{0}
\section{Results for the sieve extension}

We give here the sieve version of Theorems \ref{IMSE} and \ref{CLT}. The pseudo true $\overline{V}(\cdot|\cdot)$ and the Bahadur leading term $\widetilde{V}(\cdot|\cdot)$ of $\widehat{V}(\cdot|\cdot)$ are also defined, respectively, as in (\ref{BarV}) and (\ref{WidetildeV}) for the sieve case. While, in the ASQR case, the bias items $\mathsf{Bias}_{I}$ and $\mathsf{Bias}_h (\alpha|X,I)$ are defined as for the AQR, the variance items $\Sigma_{IL}$ and $\Sigma_{h}(\alpha|I)$ below are defined using $P(X)$ instead of the covariates in the variance items $\Sigma_{I}$ and $\Sigma_{h}(\alpha|I)$ introduced for Theorems \ref{IMSE} and \ref{CLT}.
All these ASQR results are proved in \ref{App:Proofsmain} with their AQR counterparts.

\begin{theorem}
	\label{A.IMSE} 
	Suppose Assumptions \ref{Auct.A},
	\ref{Kernel.A}, \ref{Riesz}, \ref{Spec.A} and for $\widetilde{V} (\cdot|\cdot,I)$ as in (\ref{WidetildeV}), it holds for all $I$ in $\mathcal{I}$
	\begin{eqnarray}
	\mathbb{E}
	\left[  \int_{\mathcal{X}}\int_{0}^{1}\left(  \widetilde{V} \left(
	\alpha|x,I\right)  -V\left(  \alpha|x,I\right)  \right)  ^{2}d\alpha
	dx\right]   
	&  = &
	h^{2\left(  s+1\right)  }\mathsf{Bias}_{IL}^{2}+\frac
	{\Sigma_{IL}}{LIh^{D_{\mathcal{M}}+1}}
	\label{A.IMSEexp}
	\\
	&  &+o\left(  h^{2\left(  s+1\right)  }+\frac{1}{Lh^{D_{\mathcal{M}}+1}%
	}\right)
	\nonumber
\end{eqnarray}
with $\Sigma_{IL} = O(1)$ and
\begin{eqnarray}
	\sup_{(\alpha,x) \in [0,1] \times \mathcal{X}}
	\sqrt{LIh^{D_{\mathcal{M}}+1}}
	\left|
	\widehat{V}\left(  \alpha|x,I\right)
	-\widetilde{V}\left(  \alpha|x,I\right)
	\right|
	%\nonumber \\
	%& &
	%\quad\quad\quad\quad\quad\quad\quad\quad\quad\quad\quad\quad 
	= 
	O_{\mathbb{P}}
	\left(
		\left(  \frac{\log^2 L}{
		Lh^{3D_{\mathcal{M}}+2  }
	}\right)^{1/2}  
	\right)
	=
	o_{\mathbb{P}}(1).
	\label{A.IMSEBaha}
	\end{eqnarray}
	It also holds, for each $I$ of $\mathcal{I}$,
	\begin{eqnarray}
	\sup_{(\alpha,x) \in [0,1] \times \mathcal{X}}
	\left|
	\widehat{V} (\alpha|x,I) - V (\alpha|x,I)
	\right|
	& = &
	O_{\mathbb{P}}
	\left(\sqrt{\frac{\log L}{L I h^{D_{\mathcal{M}}+1}}}\right)
	+
	O(h^{s+1}),
	\label{A.CVU.V}
	\\
	\sup_{(\alpha,x) \in [0,1] \times \mathcal{X}}
	\left|
	\widehat{B} (\alpha|x,I) - B (\alpha|x,I)
	\right|
	& = &
	O_{\mathbb{P}}
	\left(\sqrt{\frac{\log L}{L I h^{D_{\mathcal{M}}}}}\right)
	+
	o(h^{s+1}).
	\label{A.CVU.B}
	\end{eqnarray}
\end{theorem}

\begin{theorem}
	\label{A.CLT} 	
	Under Assumptions \ref{Auct.A},
	\ref{Kernel.A}, \ref{Riesz}, \ref{Spec.A} 
	and for $\overline{V} (\cdot|\cdot)$, $\widetilde{V} (\cdot|\cdot)$ as in (\ref{WidetildeV}), (\ref{BarV}) respectively, it holds for all $\alpha$ in $[0,1]$ and all $x$ in $\mathcal{X}$,
	\begin{eqnarray}
	\overline{V} (\alpha|x,I)
	& = &
	V(\alpha|x,I)
	+
	h^{s+1}
	\mathsf{Bias}_{h}(\alpha|x,I)
	+
	o(h^{s+1}),
	\label{A.VBias}\\
	\mathrm{Var}
	\left[
	\widetilde{V} (\alpha|x,I)
	\right]
	& = &
	\frac{h^{D_{\mathcal{M}}}P(x)^{\prime} \Sigma_{h} (\alpha|I) P(x)+o(1)}{LIh^{D_{\mathcal{M}}+1}},
	\label{A.Vvar}
	\end{eqnarray}
	with $\sup_{(\alpha,x) \in [0,1] \times \mathcal{X}} | \mathsf{Bias}_{h}(\alpha|x,I)|=O(1)$ and, for any $\alpha>0$,
	$h^{D_{\mathcal{M}}}P\left(  x\right)^{\prime}\Sigma_{h}\left(  \alpha|I\right)  P\left(  x\right)  \asymp\alpha^2
	$ uniformly in $x$ belonging to $\mathcal{X}$.
	
	If $\alpha \neq 0$,   
	$
	\left(  \frac{LIh^{D_{\mathcal{M}}+1}}{h^{D_{\mathcal{M}}}P\left(  x\right)  ^{\prime}\Sigma_{h}\left(
		\alpha|I\right)  P\left(  x\right)  }\right)  ^{1/2}\left(  \widehat{V}\left(
	\alpha|x,I\right)  -\overline{V} (\alpha|x,I)  \right)
	$
	converges in distribution to a standard normal.
\end{theorem}

	For the sieve counterpart of Theorem \ref{FuncCLT}, change the covariate $x_1$ into the sieve $P(x)$ in the definition of $\mathbf{P}_{0}\left(  \alpha|I\right)$ and $\mathbf{P}\left(  I\right)$. 
	Recall 
	$
	\mathbf{1}_{N}
	=
	\left[1,\ldots,1 \right]^{\prime}
	$
	is a $N\times 1$ column vector and
	redefine $\sigma_{L}^{2}\left(  x|I\right)$ and $\sigma_{L}^{2}\left(  I\right)$ as
	\begin{align*}
	\sigma_{L}^{2}\left(  x|I\right) 
	& =
	\operatorname*{Var}
	\left[
	\int_{0}^{A}\left(
	\varphi_{0}\left(  \alpha|x\right)  -\partial_{\alpha} \varphi_{1}\left(
	\alpha|x\right)  \right)  
	\mathbf{1}_N^{\prime}
	\mathbf{P}_{0}\left(
	\alpha|I\right)  ^{-1}\mathbf{P}\left(  I\right)  ^{1/2}
	h^{D_{\mathcal{M}}/2}P\left(  x\right)
	d\alpha
	\right],
	\\
	\sigma_{L}^{2}\left(  I\right)  
	&
	=
	\operatorname*{Var}
	\left[
	\int_{\mathcal{X}}
	\left\{
	\int_{0}^{A}\left(
	\varphi_{0}\left(  \alpha|x\right)  -\partial_{\alpha} \varphi_{1}\left(
	\alpha|x\right)  \right)  
	\mathbf{1}_N^{\prime}
	\mathbf{P}_{0}\left(
	\alpha|I\right)  ^{-1}\mathbf{P}\left(  I\right)  ^{1/2}
	P\left(  x\right)
	d\alpha
	\right\}
	dx
	\right] 
	\end{align*} 
with
$\sigma_{L}^{2}\left(  x\right)   
=
\sum_{I\in\mathcal{I}}
\frac{\sigma_{L}^{2}\left(  x|I\right)}{I}$ 
and
$
\sigma_{L}^{2}=\sum_{I\in\mathcal{I}}\frac{\sigma_{L}^{2}\left(  I\right)}{I}
$.

\begin{theorem}
	\label{A.FuncCLT}
	Suppose Assumptions \ref{Auct.A}, \ref{Funct}, \ref{Kernel.A},
	\ref{Spec.A}, \ref{Riesz} with $s\geq 2$ and (\ref{Condvar}) hold. Then
	$\sigma_{L}^{2}\left(  x\right)  $ and $\sigma_{L}^{2}$ are bounded away from
	$0$ and infinity and
	\begin{enumerate}
		\item If $\frac{\log^2 L}{Lh^{3(D_{\mathcal{M}}+1)  }}=o\left(  1\right)  $ , $\sqrt{Lh^{D_{\mathcal{M}}}}\left(
		\widehat{\theta}\left(  x\right)  -\theta\left( x\right)  -\mathsf{bias}%
		_{L,\theta\left(  x\right)  }\right)  /\sigma_{L}\left(  x\right)  $ converges
		in distribution to a standard normal for all $x$ in $\mathcal{X}$, where $\mathsf{bias}_{L,\theta\left(
			x\right)  }$ is a $o\left(h^s\right)$ bias term.
		\item If $\frac{\log^2 L}{Lh^{4D_{\mathcal{M}}+3 }}=o\left(  1\right)  $, $\sqrt{L}\left(  \widehat{\theta
		}-\theta-\mathsf{bias}_{L,\theta}\right)  /\sigma_{L}$ converges in
		distribution to a standard normal, where $\mathsf{bias}_{L,\theta}$ is a
		$o\left(h^s\right)$ bias term.
	\end{enumerate}
\end{theorem}

As expected, the number of interactions $D_{\mathcal{M}}$ plays the role of a covariate dimension in these sieve results, showing its ability to circumvent the curse of dimensionality.
It is worth mentioning that, as in GPV, our results do not request a smoothness $s$ increasing with $D_{\mathcal{M}}$. This is due to the use of a localized sieve and specific proof techniques.
When applied their results established for general sieve to cardinal B-spline,,	
Belloni, Chernozhukov, Chetverikov and Fern\'{a}ndez-Val (2019, Corollaries 1 and 2, Comment 4)
have a $s \geq D$ and $s \geq 3D/2$ restrictions for  quantile estimator, where their covariate dimension $D$ plays a role similar to $D_{\mathcal{M}}$. 
Aryal, Gabrielli and Vuong (2019, Assumption A2-(ii)) use a condition $s>D+1$ for a semiparametric version of GPV
based on a local polynomial estimation of the private values. Such restrictions limit applications with many covariates or interactions, as $s$ is usually set to $2$, or even to 1 as in our applications.

Theorem \ref{A.FuncCLT} holds under specific bandwidth conditions, which are driven by the order $\log L/(Lh^{4 D_{\mathcal{M}}+3} )$ of the linearization error term
$\widehat{V} (\alpha|x,I) - \widetilde{V} (\alpha|x,I)$
in (\ref{A.IMSEBaha}), which must be $o(1/\sqrt{Lh^{D_{\mathcal{M}}}})$ for $\widehat{\theta} (x)$ and $o(1/\sqrt{L})$ for $\widehat{\theta}$.

\setcounter{equation}{0}

\pagebreak

\renewcommand{\thesection}{Appendix \Alph{section}}\setcounter{section}{1}

\section{\!\!\!\!\!\!\! - Structure, notations and first results\label{App:Prelims}}

\renewcommand{\thesection}{B.\arabic{section}}
\renewcommand{\thetheorem}{B.\arabic{theorem}}
\renewcommand{\theequation}{B.\arabic{equation}}\setcounter{section}{0}
\renewcommand{\thefootnote}{B.\arabic{footnote}}
\setcounter{footnote}{0}
\setcounter{equation}{0} \setcounter{theorem}{0} 

\section{Appendices structure, notations, norm equivalences \label{Structure}}

\subparagraph{Proof main arguments.}
The proofs use some  notations detailed below which allows to study the AQR and ASQR using the same framework. The intuition behind the various proofs are rather simple, although execution is long. Starting with Lemma \ref{Phipsi}-(i), it consists in finding a vicinity of the renormalized true parameter value
$\mathsf{b}(\alpha|I)$ such that $t \mapsto P(x,t)\mathsf{b}$ is strictly increasing in a neighborhood of $0$, for all $x$. This ensures
that the sample and population objective functions are both twice differentiable and strictly convex asymptotically over this vicinity, with a probability tending to 1 for the sample one.  This vicinity is shown to be large enough to contain a local minimum of the objective functions, which is also the global minimum as they are convex over the full parameter space and strictly convex in the considered vicinity.\footnote{\label{Convex}To see this, suppose there is another local minimum outside the considered vicinity, which has an inner local maximum. Consider the segment joining their locations. Strict convexity over the vicinity implies the objective function must increase on this segment near its inner local minimum. It cannot increase all along the segment. That it decreases or is constant later would violate global convexity, so that the vicinity local minimum should also be a global one.} The fact that the sample and population Hessians are Lipshitz then allows to obtain a bias approximation, see Theorem \ref{Bias}, and a Bahadur representation for $\widehat{b}(\cdot|\cdot,I)$, (i.e. a linear expansion given in Theorem \ref{Baha}) which rate is important to cope with the fact that quantile and derivative slope estimators converge with different rates. Our main results follow from this two building blocks and from Lemma \ref{Leadterm} which studies the stochastic component of $\widehat{b}(\cdot|\cdot,I)$. For the sieve case, the proof arguments crucially rely on the   vector supremum norm $\| \cdot \|_{\infty}$ and on various important properties induced by the disjoint support condition of Assumption \ref{Riesz}-(ii) which implies that the matrix $P(x)P(x)^{\prime}$ is a band matrix. This starts with Lemma \ref{Invband} below which studies the supremum norm of the inverse of band matrices depending upon the quantile level, as the AQSR Hessian matrix which is a band one up to a basis permutation.

\subparagraph{Structure of appendices.}	
The preliminary lemmas of these appendices deals with the monotonicity of $t \mapsto P(x,t)^{\prime} \mathsf{b}$ and $\mathsf{b} \mapsto P(x,t)^{\prime} \mathsf{b}$, see Lemma \ref{Phipsi}, which are key to get invertible  population and sample Hessian, a distinctive feature of the augmented procedure. See Lemmas \ref{R2} and \ref{HatR2}, which are used to study the bias in \ref{App:Bias} and derive a Bahadur expansion in \ref{App:Bahadur}. Lemma \ref{HatR1}, which is used for the latter, deals with the score function. Lemma \ref{Leadterm} studies the leading term (\ref{WidetildeV}) of $\widehat{V}(\cdot|\cdot)$. Proposition \ref{SeriesB2} and Lemmas \ref{Invband}-\ref{Leadterm} are proven in \ref{App:Proofsprelim}, which also gathers the proof of Lemmas \ref{Cmin} and \ref{VarI}, which are more specific to Theorems \ref{FuncCLT} and \ref{A.FuncCLT}. These results are derived under the assumptions listed in Section \ref{Sieveestassump}, which extends to the sieve case the ones of the main part of the paper. 

The most important intermediary results are Theorem \ref{Bias} in  \ref{App:Bias} for the bias, and the Bahadur expansion of Theorem \ref{Baha} in \ref{App:Bahadur}. Our main results, Theorems \ref{IMSE} and \ref{A.IMSE}, \ref{CLT} and \ref{A.CLT}, \ref{FuncCLT} and \ref{A.FuncCLT} are established in corresponding Sections of \ref{App:Proofsmain}.

\subparagraph{Notations.}	
As mentioned earlier, $\wedge$ and $\vee$ stand for $\min$ and $\max$ respectively. The usual order for symmetric matrices is denoted $\preceq$, i.e. $A \preceq B$ if $B-A$ is a nonnegative symmetric matrix. Our default norms, $\|\cdot\|$, are the Euclidean norm of a vector or the associated operator norm of a matrix that we also denote $\|\cdot\|_{2}$, and the $\|\cdot\|_{\infty}$ norm for vectors, 
$\|\mathsf{b}\|_{\infty}=\max_{j} |\mathsf{b}_j|$. Note that for $\mathsf{b}$ of dimension $d_h=N$, $N(s+2)$ or $D$,
\begin{equation}
\| \mathsf{b} \|_{\infty} 
\leq
\| \mathsf{b} \|_{2} = \| \mathsf{b} \| \leq d_h^{1/2} \| \mathsf{b}\|_{\infty}. 
\label{Vecnormeq}
\end{equation}
For the associated balls $\mathcal{B}_j (\mathsf{b},\epsilon) = \{\mathsf{a};\|\mathsf{a}-\mathsf{b}\|_j \leq \epsilon\}$, $j=2,\infty$, it therefore holds
$\mathcal{B} (\mathsf{b},\epsilon)= \mathcal{B}_{2}  (\mathsf{b},\epsilon)\subset \mathcal{B}_{\infty} (\mathsf{b},\epsilon)$.
Note that for two conformable vectors $\mathsf{b}_1$, $\mathsf{b}_{2}$, it holds $| \mathsf{b}_1^{\prime} \mathsf{b}_{2} | \leq \| \mathsf{b}_1 \| \| \mathsf{b}_{2} \|$ which is the standard Cauchy-Schwarz inequality, but also $| \mathsf{b}_1^{\prime} \mathsf{b}_{2} | \leq \| \mathsf{b}_1 \|_1 \| \mathsf{b}_{2} \|_{\infty}$ with $\| \mathsf{b}_1 \|_1=\sum_{j} |\mathsf{b}_{1j}|$. In particular,
$| P(x)^{\prime} \beta | \leq \| P(x) \|_1 \|\beta\|_{\infty}$, noting that the disjoint support property in Assumption \ref{Riesz}-(ii) ensures $\| P(x) \|_1 \leq c^{1/2} \| P(x) \|_{2} \leq c \| P(x) \|_{\infty}$, so that
$
\max_{x \in \mathcal{X} }
| P(x)^{\prime} \beta |
\leq 
C h^{-D_{\mathcal{M}}/2}\|\beta\|_{\infty}
$
by Assumption \ref{Riesz}-(i).

The operator norms associated to $\|\cdot\|_{2}$ and $\|\cdot\|_{\infty}$ are, for a matrix $A$,  $\|A \|_{j}=\sup_{\mathsf{b}:\|\mathsf{b}\|_{j}=1} \|A \mathsf{b} \|_{j}$, so that $\|A \mathsf{b} \|_{j} \leq \|A \|_{j} \|\mathsf{b}\|_{j}$, $j=2,\infty$. We also use $|A|_{\infty}=\max_{i,j}|A_{ij}|$.
As well-known, $\|A\|_{2}$ is the largest eigenvalue in absolute value of a symmetric $A$.

\subparagraph{Matrix norm equivalence for band matrix.}	
 When restricted to $c$-band matrices with $A_{ij}=0$ for $|i-j|\geq c/2$, the considered matrix norms satisfy\footnote{
The inequalities for  $\| A \|_{\infty}$ follow
$
\| A \|_{\infty}
= 
\max_{\|\mathsf{b}\|_{\infty}=1} 
\max_{i} 
\left|
\sum_{j:|j-i|< c/2}
A_{ij} \mathsf{b}_j
\right|
=
\max_{i} 
\sum_{j:|j-i|< c/2}
\left|
A_{ij} 
\right|
$.
For $\| A \|_{2}$, if $\{ s_i \}$ is the canonical basis, $| A_{ij} | = |s_i^{\prime} A s_j|\leq \|s_i \|_{2} \| A \|_{2} \| s_j \|_{2} =  \| A \|_{2}$, so that
$| A |_{\infty} \leq \| A \|_{2}$. The inequality $\| A \|_{2} \leq c | A |_{\infty}$ follows from  $\|A\|_{2} = \sup_{\mathsf{b}:\|\mathsf{b}\|_{2}=1} \| A\mathsf{b} \|_{2}$ and
\begin{align*}
\| A \mathsf{b} \|_{2} 
& 
=
\sum_i
\left(
\sum_{j:|i-j|<c/2}
A_{ij} \mathsf{b}_j
\right)^2 
\leq
| A |_{\infty}^2
\sum_i
\left(
\sum_{j:|i-j|<c/2}
| \mathsf{b}_j |
\right)^2 
\leq
| A |_{\infty}^2
\sum_i
c
\sum_{j:|i-j|<c/2}
\mathsf{b}_j^2
\\
& 
\leq
| A |_{\infty}^2
c
\sum_j
\sum_i
\mathbb{I}
\left( |i-j|<c/2 \right)
\mathsf{b}_j^2
\leq
| A |_{\infty}^2
c^2 \| \mathsf{b} \|_{2}^2.
\end{align*}
}
\begin{equation}
\begin{array}{l}
|A|_{\infty} \leq \| A \|_{\infty} \leq c |A|_{\infty},
\\
|A|_{\infty} \leq \| A \|_{2} = \|A \| \leq c |A|_{\infty}
\end{array}
\label{Matnormeq}
\end{equation}
which establishes equivalence of the three matrix norms over the linear space of $c$-band matrices. 
Note that this holds independently of the matrix dimension, and this also holds for permutations $A_{\sigma} = \Big[A_{\sigma (i),\sigma (j)} \Big]$ of $c$-band matrices, where $\sigma (\cdot)$ is an index permutation.
This in particular covers $\widehat{\mathsf{R}}^{(2)} (\mathsf{b};\alpha,I)$, $\overline{\mathsf{R}}^{(2)} (\mathsf{b};\alpha,I)$ in addition to $\mathbb{E} \left[ \mathbb{I} \left(I_{\ell}=I\right)P(X_{\ell})P(X_{\ell})^{\prime}\right]$ as $P(x,t)P(x,t)^{\prime}=\pi(t) \pi (t)^{\prime} \otimes P(x) P(x)^{\prime}$ is derived through an index permutation from the $c(s+2)/2$-band matrix $P(x) P(x)^{\prime} \otimes \pi(t) \pi (t)^{\prime}$.
We shall also make use of the following Lemma, which deals with the $\|\cdot\|_{\infty}$ norm of the inverse of band matrices. The proof of Lemma \ref{Invband} can be found in \ref{App:Proofsprelim}.

\begin{lemma}
	\label{Invband}
	Let $\{M(\alpha); \alpha \in [0,1]\}$ be such that, for a a permutation matrix $S$, 
	$S^{-1} M(\alpha) S^{-1}$ is a $c$-band matrix for all $\alpha$. Then if $\sup_{\alpha \in [0,1]} \left\| M(\alpha) \right\|_{2}$ and $\sup_{\alpha \in [0,1]} \left\| M(\alpha)^{-1} \right\|_{2}$ are less than $C_0 <\infty$, there exists a constant $C_1$, which depends upon $c$ and $C_0$ but not on the dimension of $M(\cdot)$, such that
	\[
	   \sup_{\alpha \in [0,1]}
	   \left\|
	   	M(\alpha)^{-1}
	   \right\|_{\infty}
	   \leq C_1.
	\]
\end{lemma}

\section{Additional notations, score function and Hessian.}

We start with additional notations used all along the proof section and some
preliminary lemmas which are established in \ref{App:Proofsprelim}. 
Set $N=D+1$ and $D_{\mathcal{M}}=0$ in the AQR case, $N$ being the dimension of the sieve $P(\cdot)$ in the ASQR case.
Recall that the AQR and ASQR estimators write
\[
\widehat{b} (\alpha|I)
=
\left[
\widehat{\beta}_0 (\alpha|I)^{\prime},\widehat{\beta}_1 (\alpha|I)^{\prime}, \ldots, \widehat{\beta}_{s+1} (\alpha|I)^{\prime}
\right]^{\prime}
\]
where the $\widehat{\beta}_{j} (\alpha|I)^{\prime}$ are $1\times (D+1)$ in the AQR case and $1 \times N$ in the ASQR case. Recall that $\widehat{\beta}_{j} (\alpha|I)$ estimate $\beta^{(j)} (\alpha|I)$ in the AQR, see (\ref{Bsqr}) in Proposition \ref{SeriesB2} for the ASQR case.\footnote{The notation $\widehat{\beta}^{(j)} (\alpha|I)^{\prime}$ is avoided because $\widehat{\beta}_{j} (\alpha|I)^{\prime}$ is not the $j$-th derivative of $\widehat{\beta}_{0} (\alpha|I)$.}
Set
\[
P\left(  x\right)  =\left\{
\begin{array}
[c]{ll}%
\left[  1,x^{\prime}\right]  ^{\prime}=x_1 & \text{in the AQR case,  $N=D+1$ and $D_{\mathcal{M}}=0$}%
\\
\left[  P_{1}\left(  x\right)  ,\ldots,P_{N}\left(  x\right)  \right]
^{\prime} & \text{in the ASQR case}
\end{array}
\right.
\]
and recall $P(x,t) =\pi (t) \otimes P(x)$ in both case, 
allowing an unified treatment of the two estimators, although the proof focus
is on the more difficult ASQR case. Recall that $\left\Vert P\left(  x\right)
\right\Vert =\left(  P\left(  x\right)  ^{\prime}P\left(  x\right)  \right)
^{1/2}$ is the standard Euclidean norm and that, under Assumptions \ref{Auct.A} and \ref{Riesz}-(i),
$
\max_{\left(
x,t\right)  \in\mathcal{X\times}\left[  -1,1\right]  }\left\Vert P\left(
x,t\right)  \right\Vert =O\left(  h^{-D_{\mathcal{M}}/2}\right)$.

\subparagraph{Parameter renormalization.}
Since
\[
P\left(  x,ht\right)  =\pi\left(  ht\right)  \otimes P\left(  x\right)
,\quad\pi\left(  ht\right)  ^{\prime}=\left[  1,ht,\ldots,\frac{\left(
ht\right)  ^{s+1}}{\left(  s+1\right)  !}\right]
\text{ with $\lim_{h \downarrow 0} \pi\left(  ht\right) =[1,0,\ldots,0]^{\prime}$,}
\]
the \textquotedblleft design\textquotedblright\ matrix $\mathbb{E}%
\left[  P\left(  X,ht\right)  P\left(  X,ht\right)  ^{\prime
}\right]  $ degenerates asymptotically. To avoid this, consider the change of
parameters $\mathsf{b}=Hb$ with $H=\operatorname*{Diag}\left(  1,\ldots
,h^{s+1}\right)  \otimes\operatorname*{Id}_{N}$,%
\begin{equation}
\mathsf{b}
=
\left[\mathsf{b}_{0}^{\prime},\ldots,\mathsf{b}_{s+1}^{\prime}\right]^{\prime}
=
\left[  \underset{\mathsf{b}_{0}^{\prime}=\beta_{0}^{\prime
}}{\underbrace{\beta_{0,1},\ldots,\beta_{0,N}}},\underset{\mathsf{b}_{1}^{\prime}=h\beta_{1}^{\prime}}{\underbrace{h\beta_{1,1},\ldots
,h\beta_{1,N}}},\ldots,\underset{\mathsf{b}_{s+1}^{\prime}=h^{s+1}\beta
_{s+1}^{\prime}}{\underbrace{h^{s+1}\beta_{s+1,1},\ldots,h^{s+1}\beta_{s+1,N}}}\right]^{\prime}
\label{Ssb}%
\end{equation}
so that $P\left(  x,ht\right)  ^{\prime}\beta=P\left(  x,t\right)  ^{\prime}\mathsf{b}$. 
Let $S_0 = [1,0,\ldots,0]$ and $S_1 = [0,1,0,\ldots,0]$ be $1 \times (s+2)$ row vectors
and define $\mathsf{S}_0 = S_0 \otimes \mathrm{Id}_N$, $\mathsf{S}_1 = S_1 \otimes \mathrm{Id}_N$, with $N=D+1$ in the AQR case. These selection matrices are such that
\begin{equation}
\mathsf{S}_0 \mathsf{b} = \mathsf{b}_0 =\beta_0,
\quad
P(x)^{\prime} \beta_0 = P(x)^{\prime}\mathsf{S}_0 \mathsf{b}, 
\quad
\mathsf{S}_1 \mathsf{b} = \mathsf{b}_1 = h \beta_1,
\quad
P(x)^{\prime} \beta_1= \frac{P(x)^{\prime}\mathsf{S}_1 \mathsf{b}}{h}.
\label{Selmat}
\end{equation} 
Define accordingly
\begin{align*}
\widehat{\mathsf{R}}\left(  \mathsf{b};\alpha,I\right)   &  =\frac{1}{LIh}%
\sum_{\ell=1}^{L}\mathbb{I}\left(  I_{\ell}=I\right)  \sum_{i=1}^{I_{\ell}%
}\int_{0}^{1}\rho_{a}\left(  B_{i\ell}-P\left(  X_{\ell},\frac{a-\alpha}%
{h}\right)  ^{\prime}\mathsf{b}\right)  K\left(  \frac{a-\alpha}{h}\right)
da\\
&  =\frac{1}{LI}\sum_{\ell=1}^{L}\mathbb{I}\left(  I_{\ell}=I\right)
\sum_{i=1}^{I_{\ell}}\mathsf{\int_{-\frac{\alpha}{h}}^{\frac{1-\alpha}{h}}%
}\rho_{a+ht}\left(  B_{i\ell}-P\left(  X_{\ell},t\right)  ^{\prime}%
\mathsf{b}\right)  K\left(  t\right)  dt,\\
\overline{\mathsf{R}}\left(  \mathsf{b};\alpha,I\right)   &  =\mathbb{E}%
\left[  \widehat{\mathsf{R}}\left(  \mathsf{b};\alpha,I\right)  \right]  ,
\end{align*}
which are such that
\[
\widehat{\mathcal{R}}\left(  b;\alpha,I\right)=\widehat{\mathsf{R}}\left(  Hb;\alpha,I\right),
\quad
\mathbb{E}
\left[
\widehat{\mathcal{R}}\left(  b;\alpha,I\right)
\right]
=
\overline{\mathsf{R}}\left(  Hb;\alpha,I\right).
\]
Note that $\mathsf{b}\mapsto\mathsf{\int_{-\frac{\alpha}{h}}%
^{\frac{1-\alpha}{h}}}\rho_{a+ht}\left(  B_{i\ell}-P\left(  X_{\ell},t\right)
^{\prime}\mathsf{b}\right)  K\left(  t\right)  dt$ is convex as an integral of
convex functions. It follows that $\widehat{\mathsf{R}}\left(  \mathsf{b}%
;\alpha,I\right)  $ and $\overline{\mathsf{R}}\left(  \mathsf{b}%
;\alpha,I\right)  $ have minimizers,%
\begin{align*}
\widehat{\mathsf{b}}\left(  \alpha|I\right)   &  =\arg\min_{\mathsf{b}%
}\widehat{\mathsf{R}}\left(  \mathsf{b};\alpha,I\right)  =H\widehat{b
}\left(  \alpha|I\right)  ,\\
\overline{\mathsf{b}}\left(  \alpha|I\right)   &  =\arg\min_{\mathsf{b}%
}\overline{\mathsf{R}}\left(  \mathsf{b};\alpha,I\right)  ,
\end{align*}
which uniqueness will be established in the next section. Set $\overline
{b}\left(  \alpha|I\right)  =H^{-1}\overline{\mathsf{b}}\left(  \alpha
|I\right)  $ recalling $\overline{b}\left(  \alpha|I\right)  =\left[
\overline{\beta}_{0}\left(  \alpha|I\right)  ^{\prime},\ldots,\overline{\beta
}_{s+1}^{\prime}\left(  \alpha|I\right)  \right]  ^{\prime}$ and define
$\overline{B}\left(  \alpha|x,I\right)  =P\left(  x\right)  ^{\prime}%
\overline{\beta}_{0}\left(  \alpha|I\right)  ,$%
\[
\overline{\gamma}_{0}\left(  \alpha|I\right)  =\overline{\beta}_{0}\left(
\alpha|I\right)  +\frac{\alpha\overline{\beta}_{1}\left(  \alpha|I\right)
}{I-1},\quad\overline{V}\left(  \alpha|x,I\right)  =P\left(  x\right)
^{\prime}\overline{\gamma}_{0}\left(  \alpha|I\right)  .
\]

\subparagraph{Pseudo-true value and value of interest.}
To sum up, we interpret $\overline{\mathsf{b}}\left(  \alpha|I\right)$ as a pseudo-true value, while the aim is to estimate a slope 
$\mathsf{b}\left(
\cdot|\cdot\right)$ derived from (\ref{Bqr}) in the AQR case or (\ref{Bsqr}) for the ASQR.
Observe there exists some $b \left(  \cdot|\cdot\right)  $ in $\mathbb{R}^{N(s+2)}$,
\[
b\left(  \alpha|I\right)  ^{\prime}=\left[  \beta\left(
\alpha|I\right)  ^{\prime},\beta^{(1)} \left(  \alpha|I\right)  ^{\prime
},\ldots,\beta^{(s+1)} \left(  \alpha|I\right)  ^{\prime}\right]  ,
\]
such that
\[
\sup_{\left(  \alpha,x\right)  \in\left[  0,1\right]  \times\mathcal{X}%
}\left\vert P\left(  x\right)  b\left(  \alpha|I\right)  -B\left(
\alpha|x,I\right)  \right\vert   =o\left(  h^{s+1}\right)  .
\]
by Proposition \ref{SeriesB2} with $\beta(\alpha|I)$ as in (\ref{Bsqr}) for the ASQR case and, for the AQR one, with $\beta(\alpha|I)$ as in (\ref{Bqr}) and a remainder term 
$o\left(  h^{s+1}\right)$  set to $0$. An intermediate aim of the proof section is to estimate $\mathsf{b}\left(
\cdot|\cdot\right)  =Hb\left(  \cdot|\cdot\right)  $. 

\subparagraph{The score function.}
The next notations deal with the two first derivatives of the objective functions
$\widehat{\mathsf{R}}\left(  \mathsf{\cdot};\alpha,I\right)  $. Since%
\[
\partial_{\mathsf{b}}
\left[
\rho_{\alpha+ht}\left(  B-P\left(  X_{\ell},t\right)  ^{\prime
}\mathsf{b}\right)
\right]  
=
\left\{
\mathbb{I}\left(  B_{i\ell}\leq P\left(  X_{\ell},t\right)  ^{\prime
}\mathsf{b}\right)  -\left(  \alpha+ht\right)  \right\}  P\left(  X_{\ell
},t\right)  ,
\]
almost everywhere, it follows that $\widehat{\mathsf{R}}\left(  \mathsf{\cdot
};\alpha,I\right)  \ $is differentiable with%
\begin{align}
\widehat{\mathsf{R}}^{\left(  1\right)  }\left(  \mathsf{b};\alpha,I\right)
&
=
\frac{1}{LI}\sum_{\ell=1}^{L}\mathbb{I}\left(  I_{\ell}=I\right) 
\sum_{i=1}^{I_{\ell}}
\int_{-\frac{\alpha}{h}}^{\frac{1-\alpha}{h}}
\left\{  \mathbb{I}\left(  B_{i\ell}\leq P\left(  X_{\ell},t\right)
^{\prime}\mathsf{b}\right)
\right. 
\nonumber
\\
&
\quad\quad\quad\quad\quad\quad
\quad\quad\quad\quad\quad\quad
\quad\quad\quad
\left.
  -\left(  \alpha+ht\right)  \right\}  P\left(
X_{\ell},t\right)  K\left(  t\right)  dt,
\label{Der1R} \\
\overline{\mathsf{R}}^{\left(  1\right)  }
\left(  \mathsf{b};\alpha,I\right)  &
=\mathbb{E}\left[  \widehat{\mathsf{R}}^{\left(  1\right)
}\left(  \mathsf{b};\alpha,I\right)  \right],
\nonumber  
\end{align}
by the Dominated Convergence Theorem. Note also that $\widehat{\mathsf{R}}^{\left(  1\right)  }\left(  \mathsf{b};\alpha,I\right)$ is equal to
\[
\frac{1}{LI}\sum_{\ell=1}^{L}\mathbb{I}\left(  I_{\ell}=I\right)  
\sum_{i=1}^{I_{\ell}}
\int_{0}^{1}
\left\{  \mathbb{I}\left(  B_{i\ell}\leq P\left(  X_{\ell},\frac{a-\alpha}{h}\right)
^{\prime}\mathsf{b}\right)  -a  \right\}  
P\left(  X_{\ell},\frac{a-\alpha}{h}\right)  
K\left(  \frac{a-\alpha}{h}\right)  da.
\]

\subparagraph{The Hessian.}
To compute the Hessian $\widehat{\mathsf{R}}^{\left(  2\right)  }\left(  \mathsf{b};\alpha,I\right)$, it is convenient to remove the
 indicator 
$\mathbb{I}\left(  B_{i\ell}\leq P\left(  X_{\ell},t\right)^{\prime}\mathsf{b}\right)$
in (\ref{Der1R}) by showing that $t \mapsto P\left(  X_{\ell},t\right)^{\prime}\mathsf{b}$ is strictly increasing for suitable $\mathsf{b}$.
The relevant interval for $t$ in the integral expression of $\widehat{\mathsf{R}}^{\left(  1\right)  }\left(  \mathsf{b};\alpha,I\right)$  
is
\[
\mathcal{T}_{\alpha,h}=\left[  \underline{t}_{\alpha,h},\overline{t}_{\alpha,h}\right]  
=\left[  -\min\left(  1,\frac{\alpha}{h}\right)
,\min\left(  1,\frac{1-\alpha}{h}\right)  \right]  =\left[  -1,1\right]
\cap\left[  -\frac{\alpha}{h},\frac{1-\alpha}{h}\right],
\]
as the support of $K(\cdot)$ is $[-1,1]$.
It is convenient to redefine $P(x,t)^{\prime} \mathsf{b}$ as a constant function outside $\mathcal{T}_{\alpha,h}$, that is\footnote{In principle $\Psi\left(
	\cdot|\cdot\right)  $ should be denoted $\Psi_{\alpha,h}\left(  \cdot
	|\cdot\right)  $ to acknowledge that its definition depends upon $\alpha$ and
	$h$. Instead, $t$ is restricted to lie in $\mathcal{T}_{\alpha,h}$ in the
	sequel. The same comment applies for the functions $\Psi\left(  \cdot
	|\cdot\right)  $ and $\Delta\left(  \cdot|\cdot\right)  $ introduced below.}
\[
\Psi\left(  t|x,\mathsf{b}\right)  =\left\{
\begin{array}
[c]{ll}%
P\left(  x,\overline{t}_{\alpha,h}\right)  ^{\prime}\mathsf{b} &
t>\overline{t}_{\alpha,h}\\
P\left(  x,t\right)  ^{\prime}\mathsf{b} & t\in\mathcal{T}_{\alpha,h}\\
P\left(  x,\underline{t}_{\alpha,h}\right)  ^{\prime}\mathsf{b} &
t<\underline{t}_{\alpha,h}%
\end{array}
\right.  .
\]
When $\mathsf{b}=\mathsf{b}\left(  \alpha|I\right)  $, $\Psi\left(  t|x,\mathsf{b}\left(  \alpha|I\right)\right)  =P\left(
x,ht\right)  ^{\prime}b\left(  \alpha|I\right)  $ is close to
$B\left(  \alpha+ht|x,I\right)  $, which inverse as a function of $t$ 
is
\[
\frac{G\left(  u|x,I\right)  -\alpha}{h}\text{,\quad}u\in\left[  B\left(
\alpha+h\underline{t}_{\alpha,h}|x,I\right)  ,B\left(  \alpha+h
\overline{t}_{\alpha,h}|x,I\right)  \right]  .
\]
When $h$ is small enough,  Lemma \ref{Phipsi}
below shows that 
$\Psi\left(  \cdot|x,\mathsf{b}\left(  \alpha|I\right)\right)$ is strictly increasing provided 
$\mathsf{b}$ is in a suitable vicinity of $\mathsf{b}\left(  \alpha|I\right)$.
In such case, define
\begin{align*}
\Phi\left(  u|x,\mathsf{b}\right)   &  =\left\{
\begin{array}
[c]{ll}%
\alpha+h\overline{t}_{\alpha,h} & u>\Psi\left(  \overline{t}_{\alpha
,h}|x,\mathsf{b}\right) \\
\alpha+h\Psi^{-1}\left(  u|x,\mathsf{b}\right)  & u\in\Psi\left(
\mathcal{T}_{\alpha,h}|x,\mathsf{b}\right) \\
\alpha+h\underline{t}_{\alpha,h} & u<\Psi\left(  \underline{t}_{\alpha
,h}|x,\mathsf{b}\right)
\end{array}
\right.  ,\\
\Delta\left(  u|x,\mathsf{b}\right)   &  =\frac{\Phi\left(  u|x,\mathsf{b}%
\right)  -\alpha}{h}=\left\{
\begin{array}
[c]{ll}%
\overline{t}_{\alpha,h} & u>\Psi\left(  \overline{t}_{\alpha,h}|x,\mathsf{b}%
\right) \\
\Psi^{-1}\left(  u|x,\mathsf{b}\right)  & u\in\Psi\left(  \mathcal{I}%
_{\alpha,h}|x,\mathsf{b}\right) \\
\underline{t}_{\alpha,h} & u<\Psi\left(  \underline{t}_{\alpha,h}%
|x,\mathsf{b}\right)
\end{array}
\right.  ,
\end{align*}
which is such that, as seen above, the central part of $\Phi\left(
u|x,\mathsf{b}\left(  \alpha|I\right)  \right)  $ is close to $G\left(
u|x,I\right)  $ when $u$ is in $\Psi\left(  \mathcal{T}_{\alpha,h}%
|x,\mathsf{b}\right)  $. Observe now that, provided $\Psi\left(
\cdot|x,\mathsf{b}\right)  $ is increasing and since the support of $K\left(
\cdot\right)  $ is $\left[  -1,1\right]  $%
\begin{align*}
&  \int_{\underline{t}_{\alpha,h}}^{\overline{t}_{\alpha,h}}\left\{
\mathbb{I}\left(  B_{i\ell}\leq\Psi\left(  t|X_{\ell},\mathsf{b}\right)
\right)  -\left(  \alpha+ht\right)  \right\}  P\left(  X_{\ell},t\right)
K\left(  t\right)  dt\\
&  \quad=\int_{\underline{t}_{\alpha,h}}^{\overline{t}_{\alpha,h}}\left\{
\mathbb{I}\left(  \frac{\Phi\left(  B_{i\ell}|X_{\ell},\mathsf{b}\right)
-\alpha}{h}\leq t\right)  -\left(  \alpha+ht\right)  \right\}  P\left(
X_{\ell},t\right)  K\left(  t\right)  dt\\
&  \quad=\int_{\frac{\Phi\left(  B_{i\ell}|X_{\ell},\mathsf{b}\right)
-\alpha}{h}}^{\overline{t}_{\alpha,h}}P\left(  X_{\ell},t\right)  K\left(
t\right)  dt-\int_{\underline{t}_{\alpha,h}}^{\overline{t}_{\alpha,h}}\left(
\alpha+ht\right)  P\left(  X_{\ell},t\right)  K\left(  t\right)  dt
\end{align*}
which is differentiable with respect to $\mathsf{b}$, with by the Implicit Function Theorem and for $B_{i\ell}$\ in
$\Psi\left(  \mathcal{T}_{\alpha,h}|x,\mathsf{b}\right)  $
\[
\partial_{\mathsf{b}}
\Phi\left(  B_{i\ell}|X_{\ell},\mathsf{b}\right)  
=-\frac{P\left(  x,\Delta\left(  B_{i\ell
}|X_{\ell},\mathsf{b}\right)  \right)  }{\Psi^{\left(  1\right)  }\left(
\Delta\left(  B_{i\ell}|X_{\ell},\mathsf{b}\right)  |X_{\ell},\mathsf{b}%
\right)  /h}\mathbb{I}\left[  B_{i\ell}\in\Psi\left(  \mathcal{T}_{\alpha
,h}|X_{\ell},\mathsf{b}\right)  \right]  .
\]
Since $K(\cdot)$ must vanish at its frontier boundaries and is continuous,  $\widehat{\mathsf{R}}\left(
\mathsf{b};\alpha,I\right)  $ and $\overline{\mathsf{R}}\left(  \mathsf{b}%
;\alpha,I\right)  $ are twice continuously differentiable over a vicinity of $\mathsf{b}\left(  \alpha|I\right)  $ for all $h$ small enough with,%
\begin{align}
&  \widehat{\mathsf{R}}^{\left(  2\right)  }\left(  \mathsf{b};\alpha
,I\right)  =\frac{1}{LIh}\sum_{\ell=1}^{L}\sum_{i=1}^{I_{\ell}}\mathbb{I}%
\left[  B_{i\ell}\in\Psi\left(  \mathcal{T}_{\alpha,h}|X_{\ell},\mathsf{b}%
\right)  ,I_{\ell}=I\right] 
\nonumber
\\
&  
\quad\quad\quad\quad\quad\quad\quad\quad\quad\frac{P\left(
X_{\ell},\Delta\left(  B_{i\ell}|X_{\ell},\mathsf{b}\right)  \right)  P\left(
X_{\ell},\Delta\left(  B_{i\ell}|X_{\ell},\mathsf{b}\right)  \right)
^{\prime}}{\Psi^{\left(  1\right)  }\left(  \Delta\left(  B_{i\ell}|X_{\ell
},\mathsf{b}\right)  |X_{\ell},\mathsf{b}\right)  /h}K\left(  \Delta\left(
B_{i\ell}|X_{\ell},\mathsf{b}\right)  \right)  ,
\label{Der2R}
\\
&  \overline{\mathsf{R}}^{\left(  2\right)  }\left(  \mathsf{b};\alpha
,I\right)  =\mathbb{E}\left[  \widehat{\mathsf{R}}^{\left(  2\right)  }\left(
\mathsf{b};\alpha,I\right)  \right]  .
\nonumber
\end{align}
This existence of a sample second derivative contrasts with standard quantile-regression methods.
\section{Main intermediary lemmas}

\subparagraph{Properties of $\Psi\left(
	\cdot|x,\mathsf{b}\right)  $ and $\Phi\left(  \cdot|x,\mathsf{b}\right)  $.}
Lemma \ref{Phipsi}-(i) below gives necessary conditions 
ensuring $\Psi\left(
\cdot|x,\mathsf{b}\right)$ is strictly increasing for all $\mathsf{b}$ in a suitable vicinity, and then existence of $\widehat{\mathsf{R}}^{\left(  2\right)  }\left(  \cdot;\alpha
,I\right)$.  Part (ii) recalls derivatives of
$\Phi\left(  \cdot|x,\mathsf{b}\right)$ derived from the Implicit Function Theorem, as used above to sketch existence of the Hessian. Lemma \ref{Phipsi}-(iii) completes Proposition \ref{SeriesB2}, and gives an expansion for $\alpha \left(P(x,t)^{\prime} \mathsf{b} (\alpha|I)-B(\alpha+ht|x,I)\right)$ which is used for the bias of the AQR and ASQR estimators. Lemma \ref{Phipsi}-(iv) establishes that $\mathsf{b} \mapsto  \Psi\left(
t|x,\mathsf{b}\right), \Phi\left(  t|x,\mathsf{b}\right)$ are Lipshitz, with a Lipshitz factor diverging with the sample size. 

Define
\begin{align*}
\mathcal{BI}_{\alpha,h}  &  =\left\{  \mathsf{b};\min_{\left(  t,x\right)
\in\mathcal{T}_{\alpha,h}\times\mathcal{X}}
 \Psi^{(1)}\left(t|x,\mathsf{b}\right)  >0\right\}  ,\\
\underline{\mathcal{BI}}_{\alpha,h}  &  =\left\{  \mathsf{b};\min_{\left(
t,x\right)  \in\mathcal{T}_{\alpha,h}\times\mathcal{X}}
\Psi^{(1)}\left(  t|x,\mathsf{b}\right)  >h/\underline{f}%
,\max_{p=1,\ldots,s+1}\left(  \frac{\max_{x\in\mathcal{X}}\left\vert P\left(
x\right)  ^{\prime}\mathsf{b}_{p}\right\vert }{h}\right)  <\overline
{f}\right\}  
\end{align*}
 where $\underline{f}$ and
$\overline{f}$ will be taken large enough later one. While $\mathcal{BI}_{\alpha,h}$ is
used to bound the first derivative of $\Psi\left(  \cdot|x,\mathsf{b}\right)
$ away from $0$ to get strict monotonicity, $\underline{\mathcal{BI}}_{\alpha,h}$ is used to bound the
successive derivatives $\Psi^{\left(  p\right)  }\left(  \cdot|x,\mathsf{b}%
\right)  $, $p=1,\ldots,s+1$, away from infinity. As made possible by Lemma
\ref{Phipsi}-(i), below, a ball $\mathcal{B}_{\infty} \left(  \mathsf{b}
\left(  \alpha|I\right)  ,Ch^{D_{\mathcal{M}}/2+1}\right)  $ with a
small enough constant $C>0$ can be considered instead of the sets
$\mathcal{BI}_{\alpha,h}$ and $\underline{\mathcal{BI}}_{\alpha,h}$.

\begin{lemma}
\label{Phipsi} Suppose Assumptions \ref{Auct.A}, \ref{Riesz}-(i), \ref{Spec.A} hold, that $\underline{f}$ and $\overline{f}$ are large
enough and $h$ is small enough. Then,  for all $I$ in $\mathcal{I}$,
\begin{enumerate}
\item $\mathsf{b}\left(  \alpha|I\right)  $ belongs to
$\underline{\mathcal{BI}}_{\alpha,h}\subset\mathcal{BI}_{\alpha,h}$ and for
$C_0$ small enough, $\mathcal{B}_{\infty} \left(  \mathsf{b}\left(  \alpha|I\right)
,C_0 h^{D_{\mathcal{M}}/2+1}\right)  $ is a subset of $\underline{\mathcal{BI}%
}_{\alpha,h}$, for all $\alpha$ in $\left[  0,1\right]  $. 
Hence $\widehat{\mathsf{R}} \left(\cdot;\alpha,I\right)$ and $\overline{\mathsf{R}} \left(\cdot;\alpha,I\right)$ are twice continuously differentiable over $\mathcal{B}_{\infty} \left(  \mathsf{b}\left(  \alpha|I\right)
	,C_0 h^{D_{\mathcal{M}}/2+1}\right)  $, with first and second derivatives given by (\ref{Der1R}) and (\ref{Der2R}) respectively, for all $\alpha$ in $[0,1]$.
\item For all $\mathsf{b}$ in $\mathcal{BI}_{\alpha,h}$ and all $u$ in
$\Psi\left(  \mathcal{T}_{\alpha,h}|x,\mathsf{b}\right)  $%
\begin{align*}
\partial_{\mathsf{b}} \Phi\left(  u|x,\mathsf{b}\right)    &  =-\frac{P\left(  x,\Delta\left(  u|x,\mathsf{b}\right)  \right)  }%
{\Psi^{(1)}\left(  \Delta\left(  u|x,\mathsf{b}\right)  |x,\mathsf{b}\right)  /h},\\
\partial_{u}\Phi\left(  u|x,\mathsf{b}\right)   &  =\frac
{1}{\Psi^{(1)}\left(  \Delta\left(  u|x,\mathsf{b}\right)  |x,\mathsf{b}\right)
/h}.
\end{align*}
\item It holds, setting $\left. \alpha B^{\left(
	s+2\right)  }\left(  \alpha|x,I\right) \right|_{\alpha=0}
=
\lim_{\alpha \downarrow 0}
\alpha B^{\left(
	s+2\right)  }\left(  \alpha|x,I\right)
=
0
$,
\begin{align*}
&  \max_{\left(  \alpha,x\right)  \in\left[  0,1\right]  \times\mathcal{X}%
}\max_{t\in\mathcal{T}_{\alpha,h}}\left\vert \Psi\left(  t|x,\mathsf{b}\left(  \alpha|I\right)  \right)  -B\left(  \alpha+ht|x,I\right)  \right\vert
=o\left(  h^{s+1}\right)  ,\\
&  \max_{\left(  \alpha,x\right)  \in\left[  0,1\right]  \times\mathcal{X}%
}\max_{t\in\mathcal{T}_{\alpha,h}}\left\vert \alpha  B\left(
\alpha+ht|x,I\right)  
- 
\alpha B(\alpha|x,I)
  \right. \\
&  \quad\quad\quad\quad\quad\quad
\left.
-
\sum_{p=1}^{s+1} \frac{t^p}{p!}
P(x)^{\prime} \alpha \mathsf{b}_p (\alpha|I)
-\frac{\left(  ht\right)  ^{s+2}}{\left(  s+2\right)  !}\alpha B^{\left(
s+2\right)  }\left(  \alpha|x,I\right)  \right\vert =o\left(  h^{s+2}\right)  ,
\end{align*}
and
\begin{align*}
\max_{\left(  \alpha,x\right)  \in\left[  0,1\right]  \times\mathcal{X}%
}\left\vert P\left(  x\right)  ^{\prime}\alpha\beta^{(1)}\left(
\alpha|I\right)  -\alpha B^{\left(  1\right)  }\left(  \alpha|x,I\right)
\right\vert  &  =o\left(  h^{s+1}\right)  ,\\
\max_{(\alpha,x) \in [0,1] \times \mathcal{X}}%
\max_{u\in\Psi\left[  \mathcal{T}_{\alpha,h}|x,\mathsf{b}\left(
\alpha|I\right)  \right]  }\left\vert \Phi\left(  u|x,\mathsf{b}\left(
\alpha|I\right)  \right)  -G\left(  u|x,I\right)  \right\vert  &  =o\left(
h^{s+1}\right)  .
\end{align*}
\item There is a $C>0$ such that for any $\mathsf{b}^{0}$ and $\mathsf{b}^{1}$
in $\underline{\mathcal{BI}}_{\alpha,h}$ and all $(\alpha,x)$ in $\left[
0,1\right] \times \mathcal{X} $%
\begin{align*}
&  \max_{t\in\mathcal{T}_{\alpha,h}}\left\vert \Psi\left(  t|x,
\mathsf{b}^{1}\right)  -\Psi\left(  t|x,\mathsf{b}^{0}\right)  \right\vert ,\\
&  \max_{u\in\Psi\left[  \mathcal{T}_{\alpha,h}|x,\mathsf{b}^{0}\right]
\cap\Psi\left[  \mathcal{T}_{\alpha,h}|x,\mathsf{b}^{1}\right]  }\left\vert
\Phi\left(  u|x,\mathsf{b}^{1}\right)  -\Phi\left(  u|x,\mathsf{b}^{0}\right)
\right\vert ,\\
&  
 \max_{u\in\Psi\left[  \mathcal{T}_{\alpha,h}|x,\mathsf{b}^{0}\right]
\cap\Psi\left[  \mathcal{T}_{\alpha,h}|x,\mathsf{b}^{1}\right]  }h \left\vert
\partial_{u}\Phi\left(  u|x,\mathsf{b}^{1}\right)
-\partial_{u}\Phi\left(  u|x,\mathsf{b}^{0}\right)  \right\vert
,\\
&  \max_{u\in\Psi\left[  \mathcal{T}_{\alpha,h}|x,\mathsf{b}^{0}\right]
\cap\Psi\left[  \mathcal{T}_{\alpha,h}|x,\mathsf{b}^{1}\right]  }\left\vert
\Psi^{\left(  1\right)  }\left(  \Delta\left(  u|x,\mathsf{b}^{1}\right)
|x,\mathsf{b}^{1}\right)  -\Psi^{\left(  1\right)  }\left(  \Delta\left(
u|x,\mathsf{b}^{0}\right)  |x,\mathsf{b}^{0}\right)  \right\vert
,
\end{align*}
are all smaller or equal to $Ch^{-D_{\mathcal{M}}/2}\left\Vert 
\mathsf{b}^{1}-\mathsf{b}^{0}\right\Vert_{\infty} $.
\end{enumerate}
\end{lemma}

Lemma \ref{Phipsi}-(i) implies that the sample and population objective function are twice continuously differentiable in the vicinity of the true value $\mathsf{b} (\alpha|I)$. This will allow to use standard first-order linearization technique to study the AQR and ASQR estimators. The order $h^{D_{\mathcal{M}}/2+1}$ for the ball radius in (i) can be understood from (\ref{BetaLS}), assuming that
$\mathbb{E} [ P(X)P(X)^{\prime}|I]$ is the identity matrix to simplify the discussion.\footnote{If not, Lemma \ref{Invband} can also be used to show that $\max_{\alpha \in [0,1]}\| \beta^{(p)} (\alpha) \|_{\infty} = O (h^{D_{\mathcal{M}}/2})$ for all $p\leq s+1$.} If so $\max_{1 \leq n \leq N} \int_{\mathcal{X}} |P_n (x)| dx = O (h^{D_{\mathcal{M}}/2})$ implies that $\max_{\alpha \in [0,1]}\| \beta^{(p)} (\alpha) \|_{\infty} = O (h^{D_{\mathcal{M}}/2})$. This gives $\Psi^{(1)}(t|x,\mathbf{b}(\alpha|I)) = h P(x)^{\prime} \beta^{(1)} (\alpha|I)+O(h^2)$, with a leading term $h P(x)^{\prime} \beta^{(1)} (\alpha|I)$ which is positive for all $\alpha,x$ provided $h$ is small enough by Proposition \ref{SeriesB2}-(ii). Therefore, $\Psi^{(1)}(t|x,\mathsf{b})$ will be positive and $\Psi(\cdot|x,\mathsf{b})$ increasing if $\|\mathsf{b} - \mathbf{b}(\alpha|I) \|_{\infty}$ is small compared to $h\beta^{(1)} (\alpha|I)$, which is of order $h^{D_{\mathcal{M}}/2+1}$.    

\subparagraph{Population Hessian.}
Let $\Omega_{h}\left(  \alpha\right)  $, $\Omega\left(  0\right)  $,
$\Omega\left(  1\right)  $, $\Omega=\Omega\left(  0\right)  +\Omega\left(
1\right)  $ and $\Omega_{1h}\left(  \alpha\right)  $ be the $\left(
s+2\right)  \times\left(  s+2\right)  $ matrices%
\begin{align*}
\Omega_{h}\left(  \alpha\right)   &  =\int_{\underline{t}_{\alpha,h}%
}^{\overline{t}_{\alpha,h}}\pi\left(  t\right)  \pi\left(  t\right)  ^{\prime
}K\left(  t\right)  dt=\left[  \int_{-\frac{\alpha}{h}}^{\frac{1-\alpha}{h}%
}t^{p_{1}+p_{2}}K\left(  t\right)  dt,0\leq p_{1},p_{2}\leq s+1\right]  ,\\
\Omega\left(  0\right)   &  =\int_{-1}^{0}\pi\left(  t\right)  \pi\left(
t\right)  ^{\prime}K\left(  t\right)  dt,\quad\Omega\left(  1\right)
=\int_{0}^{1}\pi\left(  t\right)  \pi\left(  t\right)  ^{\prime}K\left(
t\right)  dt,\\
\Omega_{1h}\left(  \alpha\right)   &  =\int_{\underline{t}_{\alpha,h}%
}^{\overline{t}_{\alpha,h}}t\pi\left(  t\right)  \pi\left(  t\right)
^{\prime}K\left(  t\right)  dt,
\end{align*}
While $\Omega_{h}\left(  \alpha\right)  \preceq\Omega$ for all $\alpha$ and
$h$, it holds that for $h$ small enough $\Omega_{h}\left(  \alpha\right)
\succeq\Omega\left(  0\right)  $ for all $\alpha$ in $\left[  0,1/2\right]  $
and $\Omega_{h}\left(  \alpha\right)  \succeq\Omega\left(  1\right)  $ for all
$\alpha$ in $\left[  1/2,1\right]  $, ensuring that the eigenvalues of these matrices stay bounded away from $0$ and infinity when $h$ goes to $0$. Define also
\begin{align*}
\mathbf{P}\left(  I\right)  & =\mathbb{E}\left[  P\left(  X_{\ell}\right)  P\left(  X_{\ell
}\right)^{\prime} \mathbb{I}\left(  I_{\ell}=I\right) 
	\right]  ,\\
\mathbf{P}_{0}\left(  \alpha
|I\right)  & =\mathbb{E}\left[  P\left(  X_{\ell}\right)  P\left(  X_{\ell}\right)^{\prime}  \frac{\mathbb{I}\left(  I_{\ell}=I\right)  }{B^{\left(  1\right)
	}\left(  \alpha|X_{\ell},I_{\ell}\right)  }\right]  ,\\
\mathbf{P}_{1}\left(  \alpha
|I\right)  & =-\mathbb{E}\left[  P\left(  X_{\ell}\right)  P\left(  X_{\ell
}\right)^{\prime}  \frac{\mathbb{I}\left(  I_{\ell}=I\right)  B^{\left(  2\right)
	}\left(  \alpha|X_{\ell},I_{\ell}\right)  }{\left(  B^{\left(  1\right)
	}\left(  \alpha|X_{\ell},I_{\ell}\right)  \right)  ^{2}}\right]  .
\end{align*} 
Note that Assumptions \ref{Riesz}-(i), \ref{Auct.A} and Proposition \ref{SeriesB2}-(i) ensures that $\mathbf{P}_{0}\left(  \alpha
|I\right)$ is well-conditioned for all $\alpha$ in $[0,1]$, as $\int_{\mathcal{X}} P(x) P(x)^{\prime} dx$ is  and $f(x,I)/B^{(1)} (\alpha|x,I)$ is bounded away from $0$ and infinity so that
\begin{align*}
\min_{(\alpha,x,I) \in [0,1] \times \mathcal{X} \times \mathcal{I}}
\left\{
\frac{f(x,I)}{B^{(1)}(\alpha|x,I)}
\right\}
\int_{\mathcal{X}} P(x) P(x)^{\prime} dx
 \preceq
\mathbf{P}_{0}\left(  \alpha
|I\right)
=
\int_{\mathcal{X}} P(x) P(x)^{\prime} \frac{f(x,I)}{B^{(1)}(\alpha|x,I)} dx
&
\\ 
\preceq
\max_{(\alpha,x,I) \in [0,1] \times \mathcal{X} \times \mathcal{I}}
\left\{
\frac{f(x,I)}{B^{(1)}(\alpha|x,I)}
\right\}
\int_{\mathcal{X}} P(x) P(x)^{\prime} dx.
&
\end{align*}
These two inequalities yield that for any $N \times 1$ vector $S$,
$
S^{\prime} \mathbf{P}_{0}\left(  \alpha
|I\right) S
\asymp
S^{\prime} \int_{\mathcal{X}} P(x) P(x)^{\prime} dx S
$
uniformly in $\alpha$ and $S$. Taking the infimum and supremum over those $S$ with $\left\| S \right\|=1$  gives that the smallest and largest eigenvalues of $\mathbf{P}_{0}\left(  \alpha
|I\right)$ are, up to some constants, between the smallest and largest ones of
$\int_{\mathcal{X}} P(x) P(x)^{\prime} dx$ for all $\alpha$, so that the eigenvalues of  $\mathbf{P}_{0}\left(  \alpha
|I\right)$ are bounded away from $0$ and infinity for all $\alpha$ and all $h$ small enough. Similarly, $\mathbf{P}\left(  I\right)$ is well-conditioned provided $h$ is small enough. 

\begin{lemma}
\label{R2}Suppose Assumptions \ref{Auct.A}, \ref{Riesz}-(i) and \ref{Spec.A} hold and that $h$ and $C_0$ below are small enough. Then (i) $\mathsf{b} \mapsto \overline{\mathsf{R}}^{\left(  2\right)  }\left(
\mathsf{\cdot};\alpha,I\right)  $ is continuously differentiable over
$\mathcal{B}_{\infty}
\left(  \mathsf{b}\left(  \alpha|I\right)
,C_0h^{D_{\mathcal{M}}/2+1}\right)  $ with
\begin{align*}
%\max_{\alpha\in\left[  0,1\right]  }\max_{\mathsf{b}^{1},\mathsf{b}%
%^{0}\mathsf{\in}\mathcal{B}\left(  \mathsf{b}\left(  \alpha|I\right)
%,C_0h^{D_{\mathcal{M}}/2+1}\right)  }\frac{\left\Vert \overline{\mathsf{R}%
%}^{\left(  2\right)  }\left(  \mathsf{b}^{1};\alpha,I\right)  \mathsf{-}%
%\overline{\mathsf{R}}^{\left(  2\right)  }\left(  \mathsf{b}^{0}
%;\alpha,I\right)  \right\Vert }{\left\Vert \mathsf{b}^{1}-\mathsf{b}^{0}\right\Vert 
%/\left(  \alpha\left(  1-\alpha\right)  +h\right)  } & =O\left(
%h^{-D_{\mathcal{M}}/2}\right),
%\\
\max_{\alpha\in\left[  0,1\right]  }
\max_{
\mathsf{b}^{1},\mathsf{b}^{0}
\mathsf{\in}
\mathcal{B}_{\infty}\left(  \mathsf{b}\left(  \alpha|I\right),C_0h^{D_{\mathcal{M}}/2+1}\right)  }
\frac{
\left\Vert 
\overline{\mathsf{R}}^{\left(  2\right)  }
\left(  \mathsf{b}^{1};\alpha,I\right)  
-
\overline{\mathsf{R}}^{\left(  2\right)  }
\left(  \mathsf{b}^{0};\alpha,I\right)  
\right\Vert_{\infty} 
}{
\left\Vert 
\mathsf{b}^{1}
-
\mathsf{b}^{0}
\right\Vert_{\infty} 
	/\left(  \alpha\left(  1-\alpha\right)  +h\right)  } & 
=O\left(
h^{-D_{\mathcal{M}}/2}\right).
\end{align*}

(ii) The eigenvalues of $\overline{\mathsf{R}}^{\left(  2\right)  }\left[
\mathsf{b}\left(  \alpha|I\right)  ;\alpha,I\right]  $ belongs to
$\left[  1/C,C\right]  $ for a large enough $C$, for all $\alpha$ in $\left[
0,1\right]  $, and
\begin{align*}
&  \max_{\alpha\in\left[  0,1\right]  }\left\Vert \overline{\mathsf{R}%
}^{\left(  2\right)  }\left[  \mathsf{b}\left(  \alpha|I\right)
;\alpha,I\right]  -\Omega_{h}\left(  \alpha\right)  \otimes
\mathbf{P}_{0}\left(  \alpha
|I\right)  
%\right. \\
%&  \quad\quad\quad\quad\quad\quad\quad\quad\quad\left.  
-h\Omega_{1h}\left(
\alpha\right)  \otimes\mathbf{P}_{1}\left(  \alpha
|I\right)  \right\Vert =o\left(  h\right)  .
\end{align*}
\end{lemma}

In (i) the normalization item $\alpha(1-\alpha)+h$ indicates a rate dependence with respect to the quantile level $\alpha$, and more specifically some boundary effects. For instance, if $\alpha = O(h)$ or  $1-\alpha = O(h)$, then for all $\mathsf{b}^{0}$, $\mathsf{b}^{1}$ as in the Lemma,
\[
\left\Vert \overline{\mathsf{R}%
}^{\left(  2\right)  }\left(  \mathsf{b}^{1};\alpha,I\right)  \mathsf{-}%
\overline{\mathsf{R}}^{\left(  2\right)  }\left(  \mathsf{b}^{0}%
;\alpha,I\right)  \right\Vert_{\infty} 
\leq
\frac{O\left(
	h^{-D_{\mathcal{M}}/2}\right)}{h}
\left\Vert \mathsf{b}^{1}-\mathsf{b}^{0}\right\Vert _{\infty}
\]
while for $\alpha$ in the vicinity of $1/2$
\[
\left\Vert \overline{\mathsf{R}%
}^{\left(  2\right)  }\left(  \mathsf{b}^{1};\alpha,I\right)  \mathsf{-}%
\overline{\mathsf{R}}^{\left(  2\right)  }\left(  \mathsf{b}^{0}%
;\alpha,I\right)  \right\Vert_{\infty} 
\leq
O\left(
h^{-D_{\mathcal{M}}/2}\right)
\left\Vert \mathsf{b}^{1}-\mathsf{b}^{0}\right\Vert_{\infty} 
\]
indicating a Lipschitz constant inflated by a $1/h$ factor for extreme quantile levels compared to central ones.
The use of the factor $\alpha(1-\alpha)+h$ is repeatedly used below to capture some quantile level boundary effects.

Lemma \ref{R2}-(i) yields, for any $C>0$ and by the matrix norm equivalence (\ref{Matnormeq}),
\begin{align*}
\max_{\alpha\in\left[  0,1\right]  }\max_{\mathsf{b\in}\mathcal{B}_{\infty}\left(
\mathsf{b}\left(  \alpha|I\right)  ,Ch^{s+1+D_{\mathcal{M}}/2}\right)  }\left\Vert
\overline{\mathsf{R}}^{\left(  2\right)  }\left(  \mathsf{b};\alpha,I\right)
\mathsf{-}\overline{\mathsf{R}}^{\left(  2\right)  }\left(  \mathsf{b}\left(  \alpha|I\right)  ;\alpha,I\right)  \right\Vert    =O\left(
h^{s}\right) =o(1), &\\
\max_{\alpha\in\left[  0,1\right]  }\max_{\mathsf{b\in}\mathcal{B}_{\infty}\left(
\mathsf{b}\left(  \alpha|I\right)  ,C\left(  \frac{\log L}{L}\left(
\alpha\left(  1-\alpha\right)  +h\right)  \right)  ^{1/2}\right)  }%
\frac{\left\Vert \overline{\mathsf{R}}^{\left(  2\right)  }\left(
\mathsf{b};\alpha,I\right)  \mathsf{-}\overline{\mathsf{R}}^{\left(  2\right)
}\left(  \mathsf{b}\left(  \alpha|I\right)  ;\alpha,I\right)
\right\Vert }{\left(  \frac{\log L}{L\left(  \alpha\left(  1-\alpha\right)
+h\right)  }\right)  ^{1/2}}    =O\left(  h^{-D_{\mathcal{M}}/2}\right) & \\
\text{\quad if }\left(  \frac{\log L}{L}\right)  ^{1/2}    =o\left(
h^{D_{\mathcal{M}}/2+1}\right), &
\end{align*}
noting that the bandwidth condition above holds under Assumption \ref{Kernel.A}.

It then follows that the eigenvalues of $\overline{\mathsf{R}}^{\left(
2\right)  }\left(  \mathsf{b};\alpha,I\right)  $ stays bounded away from $0$
and infinity uniformly in $\alpha$ and in $\mathsf{b}$ in the two
neighborhoods considered above.
The fact that this holds for all $\alpha$, including $0$ and $1$, is useful to establish existence and uniqueness of the pseudo-true value $\mathsf{b} (\alpha|I)$. Note however that the matrix expansion leading term in Lemma \ref{R2}-(ii) involves $\Omega_h (\alpha)$, which is such that $\Omega_h (\alpha)$, which is constant over $[h,1-h]$ for $h$ small enough, but depends upon $\alpha$ and $h$ otherwise on  $[0,h]$ and $[1-h,1]$, another example of boundary effects.

\subparagraph{Sample Hessian and score function.}

The two next Lemmas study the first and second derivatives of
$\widehat{\mathsf{R}}\left(  \mathsf{\cdot};\alpha,I\right)  $ in a shrinking
vicinity of $\mathsf{b}\left(  \alpha|I\right)  $. In particular, Lemma
\ref{HatR2} implies that $\widehat{\mathsf{R}}\left(  \mathsf{\cdot}%
;\alpha,I\right)  $ is strictly convex over such a vicinity with a probability
tending to $1$.

\begin{lemma}
\label{HatR2}Suppose Assumptions \ref{Auct.A}, \ref{Riesz}-(i,ii) and S hold, and $\log L/\left(
Lh^{D_{\mathcal{M}}+1}\right)  =o\left(  1\right)  $. Then, for any $C>0$
small enough,%
\[
\max_{\alpha\in\left[  0,1\right]  }\max_{\mathsf{b\in}\mathcal{B}_{\infty}\left(
\mathsf{b}\left(  \alpha|I\right)  ,Ch^{D_{\mathcal{M}}/2+1}\right)
}\left\Vert \widehat{\mathsf{R}}^{\left(  2\right)  }\left(  \mathsf{b}%
;\alpha,I\right)  \mathsf{-}\overline{\mathsf{R}}^{\left(  2\right)  }\left(
\mathsf{b};\alpha,I\right)  \right\Vert_{\infty} =O_{\mathbb{P}}\left(  \left(
\frac{\log L}{Lh^{D_{\mathcal{M}}+1}}\right)  ^{1/2}\right)
\]

\end{lemma}

\begin{lemma}
\label{HatR1}Suppose Assumptions \ref{Auct.A}, \ref{Riesz}-(i,ii) and S hold, and $\log L/\left(
Lh^{D_{\mathcal{M}}+1}\right)  =o\left(  1\right)  $. Then $\max_{\alpha \in [0,1]}\left\| \mathrm{Var} \left[ \widehat{\mathsf{R}}^{(1)} \left(\overline{\mathsf{b}} (\alpha|I);\alpha,I\right) \right] \right\|_{j}=O\left(\frac{1}{LI}\right)$, $j=2,\infty$, and  for any $C>0$,%
\begin{align*}
\max_{\alpha\in\left[  0,1\right]  }\max_{\mathsf{b\in}\mathcal{B}_{\infty} \left(
\mathsf{b}\left(  \alpha|I\right)  ,Ch^{D_{\mathcal{M}}/2+1}\right)
}\left\Vert \frac{\widehat{\mathsf{R}}^{\left(  1\right)  }\left(
\mathsf{b};\alpha,I\right)  \mathsf{-}\overline{\mathsf{R}}^{\left(  1\right)
}\left(  \mathsf{b};\alpha,I\right)  }{\left(  h+\alpha\left(  1-\alpha
\right)  \right)  ^{1/2}}\right\Vert_{\infty} 
&
=
O_{\mathbb{P}}\left(  \left(  \frac{\log
L}{L}\right)  ^{1/2}\right) ,
\\
\max_{\alpha\in\left[  0,1\right]  }\max_{\mathsf{b\in}\mathcal{B}\left(
	\mathsf{b}\left(  \alpha|I\right)  ,Ch^{D_{\mathcal{M}}/2+1}\right)
}\left\Vert \frac{\widehat{\mathsf{R}}^{\left(  1\right)  }\left(
	\mathsf{b};\alpha,I\right)  \mathsf{-}\overline{\mathsf{R}}^{\left(  1\right)
	}\left(  \mathsf{b};\alpha,I\right)  }{\left(  h+\alpha\left(  1-\alpha
	\right)  \right)  ^{1/2}}\right\Vert
&
=
O_{\mathbb{P}}\left(  \left(  \frac{\log
	L}{Lh^{D_{\mathcal{M}}}}\right)  ^{1/2}\right).
\end{align*}
\end{lemma}

Since $\overline{\mathsf{R}}^{\left(  1\right)  }\left(  \overline{\mathsf{b}%
}\left(  \alpha|I\right)  ;\alpha,I\right)  =0$ and assuming  $\sup_{\alpha\in\left[  0,1\right]
}\left\Vert \overline{\mathsf{b}}\left(  \alpha|I\right)  -\mathsf{b}\left(  \alpha|I\right)  \right\Vert =o\left(  ^{D_{\mathcal{M}}/2+1}\right)  $ as
established in (\ref{Bar2*b}), it holds that%
\[
\max_{\alpha\in\left[  0,1\right]  }\left\Vert \frac{\widehat{\mathsf{R}%
}^{\left(  1\right)  }\left(  \overline{\mathsf{b}}\left(  \alpha|I\right)
;\alpha,I\right)  }{\left(  h+\alpha\left(  1-\alpha\right)  \right)  ^{1/2}%
}\right\Vert_{\infty} =O_{\mathbb{P}}\left(  \left(  \frac{\log L}{L}\right)  ^{1/2}\right)  .
\]
The normalization factor $h+\alpha\left(  1-\alpha\right)$ gives that 
$
\widehat{\mathsf{R}}^{\left(  1\right)  }
\left(  \overline{\mathsf{b}}\left(  \alpha|I\right)
	;\alpha,I\right)
$
has the fast rate $\left(\frac{h \log L}{L} \right)^{1/2}$ for boundary $\alpha$ in $[0,h]$ and $[1-h,1]$, while it is of order $\left( \frac{\log L}{L} \right)^{1/2}$ for central quantile levels.

Lemmas \ref{HatR2} and \ref{R2} implies that the second derivative
$\widehat{\mathsf{R}}^{\left(  2\right)  }\left(  \mathsf{b};\alpha,I\right)$ is well-conditioned for $\mathsf{b}$ in a vicinity of $\mathsf{b} (\alpha|I)$, implying local strict convexity of the sample objective function for all $\alpha$ in $[0,1]$ with a probability tending to 1. This is used later on to show existence and uniqueness of the AQR and ASQR for all $\alpha$ including extreme quantile levels, with a probability tending to $1$.

\subparagraph{Bahadur leading term.}
The next Lemma studies the leading term $\widehat{\mathsf{e}}\left(
\alpha|I\right)  $ of $\widehat{\mathsf{b}}\left(  \alpha|I\right)
-\overline{\mathsf{b}}\left(  \alpha|I\right)  $,%
\[
\widehat{\mathsf{e}}\left(  \alpha|I\right)  =-\left[  \overline{\mathsf{R}%
}^{\left(  2\right)  }\left(  \overline{\mathsf{b}}\left(  \alpha|I\right)
;\alpha,I\right)  \right]  ^{-1}\widehat{\mathsf{R}}^{\left(  1\right)
}\left(  \overline{\mathsf{b}}\left(  \alpha|I\right)  ;\alpha,I\right)
\]
see Theorem \ref{Baha} below. Note that $\overline{\mathsf{R}}^{\left(
2\right)  }\left(  \overline{\mathsf{b}}\left(  \alpha|I\right)
;\alpha,I\right)  $ has, by Lemma \ref{R2}, an inverse provided $\sup_{\alpha
\in\left[  0,1\right]  }\left\Vert \overline{\mathsf{b}}\left(  \alpha
|I\right)  -\mathsf{b}\left(  \alpha|I\right)  \right\Vert =o\left(
h^{s+1+D_{\mathcal{M}}/2}\right)=o\left(h^{D_{\mathcal{M}}/2+1}\right)  $ as therefore assumed and established in the proof of Theorem
\ref{Bias} below, see (\ref{Bar2*b}). Let $\mathbf{P}\left(  I\right)$ and $\mathbf{P}_0\left(  \alpha|I\right)$ be as for Lemma \ref{R2}
and recall $\Sigma_h (\alpha|I)=\alpha^2 v_{h}^{2}\left(  \alpha\right) 
\mathbf{P}_0 (\alpha|I)^{-1}
\mathbf{P}\left(  I\right)
\mathbf{P}_0 (\alpha|I)^{-1}/(I-1)$.

\begin{lemma}
\label{Leadterm} Suppose Assumptions \ref{Auct.A}, \ref{Kernel.A}, \ref{Riesz} and \ref{Spec.A} hold, with $1/\left(
Lh^{D_{\mathcal{M}}+1}\right)  =o\left(  1\right)  $, $s\geq 1$ and $\sup_{\alpha\in\left[  0,1\right]  }\left\Vert \overline{\mathsf{b}%
}\left(  \alpha|I\right)  -\mathsf{b}\left(  \alpha|I\right)
\right\Vert =o\left(  h^{s+1+D_{\mathcal{M}}/2}\right)  $. Then, for $h$ small enough: (i) uniformly in $\left(
\alpha,x\right)  $ in $\left[  0,1\right]  \times\mathcal{X}$, 
\begin{align*}
&\left\| \operatorname*{Var}\left[  \widehat{\mathsf{e}
}_{0}\left(  \alpha|I\right)  \right] \right\| =O\left(  \frac{1}{L}\right)
\text{ and }
\operatorname*{Var}\left[  P\left(  x\right)  ^{\prime}\widehat{\mathsf{e}
}_{0}\left(  \alpha|I\right)  \right]  =O\left(  \frac{1}{Lh^{D_{\mathcal{M}}%
}}\right),
\\
&
\operatorname*{Var}\left[
\frac{\widehat{\mathsf{e}}_{1}\left(  \alpha|I\right)}{h}\right] 
=
\frac{
v_{h}^{2}\left(  \alpha\right) 
\mathbf{P}_0 (\alpha|I)^{-1}
\mathbf{P}\left(  I\right)
\mathbf{P}_0 (\alpha|I)^{-1}
 +o(1)}{Lh},
\end{align*}
where $1/C \leq v_{h}^{2}\left(  \alpha\right) <C$ for all $\alpha$ in $[0,1]$ and $h$ small enough, 
so that, for all $\alpha>0$,  
\[
Lh^{D_{\mathcal{M}}+1} \operatorname*{Var}\left[  \frac{\alpha P\left(  x\right)  ^{\prime}
\widehat{\mathsf{e}}_{1}\left(  \alpha|I\right)}{h (I-1)}  \right]
=
h^{D_{\mathcal{M}}}
P(x)^{\prime}
\Sigma_h (\alpha|I)
P(x)
+o(1)
\] stays bounded away from $0$ and infinity when $h$ goes to $0$, uniformly in $x$ in $ \mathcal{X}$.

(ii) It also holds%
\begin{align*}
\sup_{\left(  \alpha,x\right)  \in\left[  0,1\right]  \times\mathcal{X}%
}\left\vert P\left(  x\right)  ^{\prime}\widehat{\mathsf{e}}_{0}\left(
\alpha|I\right)  \right\vert  &  =O_{\mathbb{P}}\left(  \left(  \frac{\log
L}{Lh^{D_{\mathcal{M}}}}\right)  ^{1/2}\right)  ,\\
\sup_{\left(  \alpha,x\right)  \in\left[  0,1\right]  \times\mathcal{X}%
}\left\vert P\left(  x\right)  ^{\prime}\frac{\widehat{\mathsf{e}}_{1}\left(
\alpha|I\right)  }{h}\right\vert  &  =O_{\mathbb{P}}\left(  \left(  \frac{\log
L}{Lh^{D_{\mathcal{M}}+1}}\right)  ^{1/2}\right)  .
\end{align*}
\end{lemma}

As explained now, Lemma \ref{Leadterm} improves Lemma \ref{HatR1}, which gives taking $\mathsf{b} = \overline{\mathsf{b}} (\alpha|I)$ 
\begin{equation}
\label{Maxe}
\max_{\alpha \in [0,1]} 
\frac{
	\left\|
	  \widehat{e} (\alpha|I)
	\right\|_{\infty}
}{
\left(
h+\alpha(1-\alpha)
\right)^{1/2}
}
=
O_{\mathbb{P}}
\left(
\left(
\frac{\log L}{L}
\right)^{1/2}
\right)
\end{equation}
by Lemmas \ref{R2} and \ref{Invband}, $\| A b \|_{\infty} \leq \|A \|_{\infty} \| b \|_{\infty}$ for a matrix $A$ and a conformable $b$. This implies that 
$\widehat{e}_1 (\cdot|I) = \mathsf{S}_{1}\widehat{e} (\cdot|I)/h$ is of order $\frac{1}{h} (\log L /L)^{1/2}$ in $\| \cdot \|_{\infty}$ norm.  Lemma \ref{Leadterm} improves this order to $\frac{1}{h^{1/2}} (\log L /L)^{1/2}$. This could be seen here noting that the variance of $\widehat{e}_1 (\alpha|I)$ is of order $1/(Lh)$ instead of $1/(Lh^2)$.

A precise evaluation of this variance is the key tool to obtain this improvement, which is important to relate the estimation rate of $\beta^{(1)} (\cdot)$ with the nonparametric density estimation rate. To see this, consider the AQR case, ie $D_{\mathcal{M}}=0$ and $P(x)=[1,x]^{\prime}$.
The leading term of the estimation error for $\beta(\alpha|I)$ is, by (\ref{Selmat})
$\mathsf{e}_{0}\left(
\alpha|I\right)$, which has a parametric rate $1/\sqrt{L}$ by Lemma \ref{Leadterm}-(i). The leading term of the estimation error of  $\beta^{(1)}(\alpha|I)$
is $\mathsf{e}_{1}\left(
\alpha|I\right)/h$
is of order $1/\sqrt{Lh}$, as it would hold for a kernel density estimator with bandwidth $h$.

\pagebreak

\renewcommand{\thesection}{Appendix \Alph{section}}\setcounter{section}{2}

\section{\!\!\!\!\!\! - Asymptotic bias and ASQR properties}
\label{App:Bias}

\renewcommand{\thesection}{C.\arabic{section}}
\renewcommand{\thetheorem}{C.\arabic{theorem}}
\renewcommand{\theequation}{C.\arabic{equation}}\setcounter{section}{0}
\renewcommand{\thefootnote}{C.\arabic{footnote}}
\setcounter{footnote}{0}
\setcounter{equation}{0} \setcounter{theorem}{0}

The study of the bias $\overline{V}\left(  \alpha|X,I\right)  -V\left(
\alpha|X,I\right)  $ and $\overline{B}\left(  \alpha|X,I\right)  -B\left(
\alpha|X,I\right)  $ is based on the following Lemma which is a consequence of
the Kantorovitch-Newton Theorem, see e.g. Gragg and Tapia (1974).

\begin{lemma}
\label{Kantorovich} Let $\|\cdot\|_{A}$ be an arbitrary norm and $\mathcal{F}\left(  \cdot\right)  :\mathbb{R}%
^{D}\rightarrow\mathbb{R}$ be a function. Suppose that there is a
$\mathbf{x}^{\ast}\mathbf{\in}\mathbb{R}^{D}$ and some real numbers
$\epsilon>0$ and $C_{0}>0$ such that $\mathcal{F}\left(  \cdot\right)  $ is
twice differentiable on $\mathcal{B}_A\left(  \mathbf{x}^{\ast}\mathbf{,}%
2C_{0}\epsilon\right)  =\left\{  x\mathbf{\in}\mathbb{R}^{D};\left\Vert
x-\mathbf{x}^{\ast}\right\Vert_A <2C_{0}\epsilon\right\}  $. If, in addition,

\begin{enumerate}
\item $\left\Vert \mathcal{F}^{\left(  1\right)  }\left(  \mathbf{x}^{\ast
}\right)  \right\Vert_A \leq\epsilon$ and $\left\Vert \left[  \mathcal{F}%
^{\left(  2\right)  }\left(  \mathbf{x}^{\ast}\right)  \right]  ^{-1}%
\right\Vert_A \leq C_{0}$, where the norm for the inverse of the second derivative is the operator norms associated with $\|\cdot\|_{A}$;

\item There is a $C_{1}>0$ such that $\left\Vert \mathcal{F}^{\left(
2\right)  }\left(  x\right)  -\mathcal{F}^{\left(  2\right)  }\left(
x^{\prime}\right)  \right\Vert_A \leq C_{1}\left\Vert x-x^{\prime}\right\Vert_A $
for all $x,x^{\prime}\in\mathcal{B}_A\left(  \mathbf{x}^{\ast}\mathbf{,}%
2C_{0}\epsilon\right)  $;

\item $C_{0}^{2}C_{1}\epsilon\leq1/2$.
\end{enumerate}

Then there is a unique $\overline{\mathbf{x}}\ $such that $\left\Vert
\overline{\mathbf{x}}\mathbf{-x}^{\ast}\right\Vert_A <2C_{0}\epsilon$ and
$\mathcal{F}^{\left(  1\right)  }\left(  \overline{\mathbf{x}}\right)  =0$.
\end{lemma}

\bigskip

The next lemma gathers results for intermediary bias terms. Define, 
$\mathbf{P}_0 (\alpha|I)$ being as for Lemma \ref{R2},
	\begin{align*}
	\overline{g}\left(  \alpha|t,x,I\right)  
	& =
	\int_{0}^{1}
	g\left(  \Psi\left(
	t|x,\overline{\mathsf{b}}\left(  \alpha|I\right)  \right)  +u\left(  B\left(
	\alpha+ht|x,I\right)  -\Psi\left(  t|x,\mathsf{b}\left(  \alpha
	|I\right)  \right)  \right)  |x,I\right)  du,
	\\
	\check{\mathsf{R}}^{(2)}
	\left(\alpha|I\right)
	& =
	\int\left(  \int%
	_{\underline{t}_{\alpha,h}}^{\overline{t}_{\alpha,h}}\overline{g}\left(
	\alpha|t,x,I\right)  P\left(  x,t\right)  P\left(  x,t\right)  ^{\prime
	}K\left(  t\right)  dt\right)  f\left(  x,I\right)  dx,
	\\
	\overline{\mathfrak{bias}}_{h} (\alpha|I)
	& =
	\left[\check{\mathsf{R}}^{(2)}
	\left(\alpha|I\right)\right]^{-1}
	\int 
	\left\{
	\int_{\underline{t}_{\alpha,h}}^{\overline{t}_{\alpha,h}} 
	\overline{g} (\alpha|t,x,I)
	\left(
	P(x)^{\prime} \beta (\alpha|I) - B(\alpha|x,I)
	\right)
	\right.\\
	&
	\quad\quad\quad\quad\quad\quad\quad\quad\quad
	\quad\quad\quad\quad\quad\quad\quad\quad\quad
	\quad\quad\quad
	%\left.
	\times
	P(x,t) K(t) dt
	\Bigg\}
	f(x,I) dx,
	\\
	\check{b}_{s+2} (\alpha|I)
	& =
	\left[\check{\mathsf{R}}^{(2)}
	\left(\alpha|I\right)\right]^{-1}
	\int 
	\left\{
	\int_{\underline{t}_{\alpha,h}}^{\overline{t}_{\alpha,h}} 
	\overline{g} (\alpha|t,x,I)
	\frac{t^{s+2}}{(s+2)!}
	\alpha B^{(s+2)} (\alpha|x,I)
	\right.\\
	&
	\quad\quad\quad\quad\quad\quad\quad\quad\quad
	\quad\quad\quad\quad\quad\quad\quad\quad\quad
	\quad\quad\quad
	%\left.
	\times
	P(x,t) K(t) dt
	\Bigg\}
	f(x,I) dx.
\end{align*}
	
Recall $\mathsf{S}_0 = S_0 \otimes \mathrm{Id}_{N}$ and $\mathsf{S}_1 = S_1 \otimes \mathrm{Id}_{N}$, where the row vectors
$S_0 = [1,0,\ldots,0]$ and $S_1 = [0,1,0,\ldots,0]$ have dimension $s+2$, see (\ref{Selmat}).
\begin{lemma}
	\label{Intbiasterm}
	Suppose that Assumptions \ref{Auct.A}, \ref{Kernel.A} and \ref{Riesz}-(i,ii)
	hold, implying $s\geq 1$. Suppose also that (\ref{Bar2*b}) in the proof of Theorem \ref{Bias} is true. Then it holds, for $\mathsf{Bias}_h (\cdot|\cdot,I)$ as in (\ref{AQR_bias}),
	\begin{enumerate}
		\item 
		$\sup_{(\alpha,x,I) \in [0,1] \times \mathcal{X} \times \mathcal{I}}
		\left\| \left[\check{\mathsf{R}}^{(2)}
		\left(\alpha|I\right)\right]^{-1} \right\|_{\infty}
		=O(1)
		$,
		$\sup_{(\alpha,x,I) \in [0,1] \times \mathcal{X} \times \mathcal{I}}
		\left|
		P(x)^{\prime} \mathsf{S}_0  \overline{\mathfrak{bias}}_{h} (\alpha|I)
		\right|  =  o(h^{s+1})$ and 
		$\sup_{(\alpha,x,I) \in [0,1] \times \mathcal{X} \times \mathcal{I}}
		\left|
		P(x)^{\prime} \mathsf{S}_1  \overline{\mathfrak{bias}}_{h} (\alpha|I)
		\right|  =  o(h^{s+2})$.
		\item 
		$\sup_{(\alpha,x,I) \in [0,1] \times  \mathcal{X} \times \mathcal{I}}
		\left|
		P(x)^{\prime} \mathsf{S}_1 \check{b}_{s+2} (\alpha|I) - (I-1) \mathsf{Bias}_h (\alpha|x,I)
		\right|=o(1)$.
	\end{enumerate}
\end{lemma}

Lemma \ref{Intbiasterm} will be used in the proof of Theorem \ref{Bias} after having established (\ref{Bar2*b}). 
Lemma \ref{Intbiasterm}-(i) shows that the approximation error $P(x)^{\prime} \beta (\alpha|I) - B(\alpha|x,I)$, which is $o(h^{s+1})$ by Proposition \ref{SeriesB2}-(ii), does not contribute to the bias for the estimation of $\alpha B^{(1)} (\alpha|x,I)$ as
$P(x)^{\prime} \mathsf{S}_1  \overline{\mathfrak{bias}}_{h} (\alpha|I)$ has the better order $o(h^{s+2})$.
The proof of Lemma  \ref{Intbiasterm} is given at the end of this Appendix, after the proof of Theorem \ref{Bias}.

\section{Main bias result and ASQR properties}

Theorem \ref{Bias} gives a bias expansion for $\widehat{V} (\alpha|x,I)$ and $\alpha \widehat{B}^{(1)} (\alpha|x,I)$, and the order of the bias of $\widehat{B} (\alpha|x,I)$. It shows that the leading term of the bias of $\widehat{V} (\alpha|x,I)$ is the one of $\alpha \widehat{B}^{(1)} (\alpha|x,I)$ for central quantile levels, as the leading term of the bias of $\alpha \widehat{B}^{(1)} (\alpha|x,I)/(I-1)$ vanishes for $\alpha =0$ by (\ref{AQR_bias}) and $\lim_{\alpha \downarrow 0}  \alpha B^{(s+2)} (\alpha|x,I)=0$ by Proposition \ref{SeriesB2}-(i). It follows that the bias of $\widehat{V} (\alpha|x,I)$ is expected to be smaller for quantile levels close to $0$ than for central or upper ones.

\begin{theorem}
\label{Bias} Suppose that Assumptions \ref{Auct.A}, \ref{Kernel.A} and \ref{Riesz}-(i,ii)
hold. Then, for $h$ small enough and all $\alpha$
in $\left[  0,1\right]  $, $\overline
{\mathsf{b}}\left(  \alpha|I\right)  =\arg\min_{\mathsf{b}}\overline
{\mathsf{R}}\left(  \mathsf{b};\alpha,I\right)  $ is unique. It also holds
\begin{align*}
\sup_{\left(  \alpha,x,I\right)  \in\left[  0,1\right]  \times\mathcal{X\times
I}}\left\vert \overline{V}\left(  \alpha|x,I\right)  -V\left(  \alpha
|x,I\right)  -h^{s+1}\mathsf{Bias}
_{h}\left(  \alpha|x,I\right)\right\vert & =o\left(  h^{s+1}\right),
\\
\sup_{\left(  \alpha,x,I\right)  \in\left[  0,1\right]  \times\mathcal{X\times
		I}}\left\vert \alpha \overline{B}^{(1)}\left(  \alpha|x,I\right)  -\alpha B^{(1)}\left(  \alpha
|x,I\right)  -h^{s+1}(I-1)\mathsf{Bias}
_{h}\left(  \alpha|x,I\right) \right\vert & =o\left(  h^{s+1}\right),
\end{align*}
where $\mathsf{Bias}
_{h}\left(  \alpha|x,I\right)$ is as in (\ref{AQR_bias}) and satisfies $\sup_{\left(  \alpha,x,I\right)  \in\left[  0,1\right]  \times
\mathcal{X\times I}}\left| \alpha \mathsf{Bias}
_{h}\left(  \alpha|x,I\right)\right| =O\left(  1\right)  $.
Moreover
\begin{align*}
\sup_{\left(  \alpha,x,I\right)  \in\left[  0,1\right]  \times\mathcal{X\times
I}}\left\vert \overline{B}\left(  \alpha|x,I\right)  -B\left(  \alpha
|X,I\right)  \right\vert  &  =o\left(  h^{s+1}\right)  ,\\
\sup_{\left(  \alpha,x,I\right)  \in\left[  0,1\right]  \times\mathcal{X\times
I}}\left\vert \overline{B}^{\left(  1\right)  }\left(  \alpha|x,I\right)
-B^{\left(  1\right)  }\left(  \alpha|x,I\right)  \right\vert  &  =o\left(
h^{s}\right)  .
\end{align*}

\end{theorem}

The proof of Theorem \ref{Bias} is given after discussing some properties of the ASQR estimator.

\subparagraph{Local polynomial and sieve bias.}
The bias $\overline{V} (\cdot|\cdot) - V(\cdot|\cdot)$ is given by the bias of the bid quantile derivative, which has two components: one which is due to the sieve procedure and a second induced by local polynomials. That the AQR and ASQR have the same asymptotic bias suggests that the bias is due to the local polynomial method. This can be seen noticing that the sieve component of the bias is
due to the item $P(x)^{\prime} \alpha \beta^{(1)} (\alpha|I) - \alpha B^{(1)} (\alpha|I)$, which is $o(h^{s+1})$ by Proposition \ref{SeriesB2}-(iii) and is negligible compared to the Bias leading term
$h^{s+1} \mathsf{Bias}_{h}\left(  \alpha|x,I\right)$. The expression of $\mathsf{Bias}
_{h}\left(  \alpha|x,I\right)$ given in (\ref{AQR_bias}) depends upon the local polynomial vector $\pi(t)$ and upon the kernel, which are associated to the local polynomial component of the ASQR estimator.

\subparagraph{Asymptotic uniqueness of the estimator.}
The proof of Theorem \ref{Bias} establishes that $\sup_{\alpha\in\left[
0,1\right]  }\left\Vert \overline{\mathsf{b}}\left(  \alpha|I\right)
-\mathsf{b}\left(  \alpha|I\right)  \right\Vert_{\infty} =o\left(
h^{s+1+D_{\mathcal{M}}/2}\right)  $, see (\ref{Bar2*b}) below. Hence by a second order Taylor expansion, Lemmas \ref{HatR2} and \ref{R2}, it holds
\begin{align*}
&  \sup_{\alpha\in\left[  0,1\right]  }
\sup_{
	\mathsf{b\in}
	\mathcal{B}_{\infty} \left(\overline{\mathsf{b}}\left(  \alpha|I\right)  ,C_0h^{D_{\mathcal{M}}/2+1}\right)  
}
h^{-2\left(D_{\mathcal{M}}/2+1\right)  }
\left\vert
\widehat{\mathsf{R}}\left(  \mathsf{b}%
;\alpha,I\right)  -\widehat{\mathsf{R}}\left(  \overline{\mathsf{b}}\left(
\alpha|I\right)  ;\alpha,I\right)  
    \right. \\
& \left. 
-
\left(  \mathsf{b-}\overline{\mathsf{b}}\left(  \alpha|I\right)  \right)^{\prime}
\widehat{\mathsf{R}}^{\left(1\right)  }
\left(  \overline{\mathsf{b}}\left(  \alpha|I\right)
;\alpha,I\right)
-
\frac{1}{2}
\left(
\mathsf{b-}\overline{\mathsf{b}}\left(  \alpha|I\right)  \right)  ^{\prime
}\overline{\mathsf{R}}^{\left(  2\right)  }\left(  \mathsf{b}
\left(  \alpha|I\right)  ;\alpha,I\right)  \left(  \mathsf{b-}\overline
{\mathsf{b}}\left(  \alpha|I\right)  \right)  \right\vert =o_{\mathbb{P}%
}\left(  1\right)  .
\end{align*}
Then,
as the eigenvalues of
$\overline{\mathsf{R}}^{\left(  2\right)  }\left(  \mathsf{b}
\left(  \alpha|I\right)  ;\alpha,I\right)$ stay bounded away from $0$ and infinity 
by Lemma \ref{R2}-(ii), the 
linear approximation of the sample objective function $\widehat{\mathsf{R}}\left(  \mathsf{b}%
;\alpha,I\right)  -\widehat{\mathsf{R}}\left(  \overline{\mathsf{b}}\left(
\alpha|I\right)  ;\alpha,I\right)$,
\[
\left(  \mathsf{b-}\overline{\mathsf{b}}\left(  \alpha|I\right)  \right)^{\prime}
\widehat{\mathsf{R}}^{\left(1\right)  }
\left(  \overline{\mathsf{b}}\left(  \alpha|I\right)
;\alpha,I\right)
+
\frac{1}{2}
\left(
\mathsf{b-}\overline{\mathsf{b}}\left(  \alpha|I\right)  \right)  ^{\prime
}\overline{\mathsf{R}}^{\left(  2\right)  }\left(  \mathsf{b}
\left(  \alpha|I\right)  ;\alpha,I\right)  \left(  \mathsf{b-}\overline
{\mathsf{b}}\left(  \alpha|I\right)  \right)
\]
has a unique minimizer $\overline{\mathsf{b}}^{\ast} (\alpha|x,I)$,
\[
\overline{\mathsf{b}}^{\ast} (\alpha|x,I)
=
\overline{\mathsf{b}} (\alpha|x,I)
-
\left[
\overline{\mathsf{R}}^{\left(  2\right)  }\left(  \mathsf{b}
\left(  \alpha|I\right)  ;\alpha,I\right)
\right]^{-1}
\widehat{\mathsf{R}}^{\left(1\right)  }
\left(  \overline{\mathsf{b}}\left(  \alpha|I\right)
;\alpha,I\right),
\]
which belongs to $\mathcal{B}_{\infty} \left(\overline{\mathsf{b}}\left(  \alpha|I\right)  ,C_0h^{D_{\mathcal{M}}/2+1}\right)$ for all $\alpha$ in $[0,1]$ with a probability tending to 1 by Lemmas \ref{HatR1} and \ref{R2} as $(\log L/L)^{1/2} = o \left(h^{D_{\mathcal{M}}/2+1}\right)$ under Assumption \ref{Kernel.A}.
Then, since  $\widehat{\mathsf{R}}\left(
\mathsf{\cdot};\alpha,I\right)  $ is also strictly convex over $\mathcal{B}_{\infty} \left(\overline{\mathsf{b}}\left(  \alpha|I\right)  ,C_0h^{D_{\mathcal{M}}/2+1}\right)$ for all $\alpha$ in $[0,1]$ with a probability tending to 1, its minimizer $\widehat{\mathsf{b}}^{\ast} (\alpha|I)=\arg\min_{\mathsf{b} \in \mathcal{B}_{\infty} \left(\overline{\mathsf{b}}\left(  \alpha|I\right)  ,C_0h^{D_{\mathcal{M}}/2+1}\right)} \widehat{\mathsf{R}}\left(
\mathsf{b};\alpha,I\right)$ and by the Argmax Theorem 2.1 in Newey and McFadden (1994)\footnote{It is easy to extend this result to the case where the objective functions depend upon a parameter $\alpha$.}  is unique and satisfies
\[
\sup_{  \alpha \in [0,1] }
\left\|
\widehat{\mathsf{b}}^{\ast} (\alpha|I)
-
\overline{\mathsf{b}} (\alpha|I)
\right\|_{\infty}
=
o_{\mathbb{P}}
\left(h^{D_{\mathcal{M}}/2+1}\right).
\] 
Hence $\widehat{\mathsf{b}}^{\ast} (\alpha|I)$ is an interior point of the considered ball, over which the sample objective function is strictly convex. Arguing as in Footnote \ref{Convex} implies
\[
\lim_{L\uparrow\infty} \mathbb{P}
\left( \widehat{\mathsf{b}} (\alpha|I) = \widehat{\mathsf{b}}^{\ast} (\alpha|I) \text{ for all $\alpha \in [0,1]$}\right)=1.
\]
It follows that $\widehat{\mathsf{b}} (\alpha|I)$ is unique for all $\alpha$ of $[0,1]$ with a probability tending to 1.

\subparagraph{Asymptotic smoothness of $\widehat{\mathsf{b}}(\cdot|I)$.}
As
$\widehat{\mathsf{b}}(\alpha|I)$ belongs to $\mathcal{B}_{\infty} \left(\overline{\mathsf{b}}\left(  \alpha|I\right)  ,C_0h^{D_{\mathcal{M}}/2+1}\right)$ for all $\alpha$ in $[0,1]$ with a probability tending to 1,
Lemmas \ref{Phipsi}-(i), \ref{R2} and \ref{HatR2} allow to apply the Implicit Function Theorem to the FOC
$\widehat{\mathsf{R}}^{\left(1\right)  }
\left(  \widehat{\mathsf{b}}\left(  \alpha|I\right)
;\alpha,I\right)=0$. As noted in Fernandes, Guerre and Horta (2019), it then follows that  $\widehat{\mathsf{b}}(\cdot|I)$ is differentiable with a probability tending to $1$ with derivative
\[
\widehat{\mathsf{b}}^{(1)}(\alpha|I)
=
-
\left[
\widehat{\mathsf{R}}^{\left(2\right)  }
\left(  \widehat{\mathsf{b}}\left(  \alpha|I\right)
;\alpha,I\right)
\right]^{-1}
\partial_{\alpha}
\widehat{\mathsf{R}}^{\left(1\right)  }
\left(  \mathsf{b}
;\alpha,I\right)
\bigg|_{\mathsf{b}=\widehat{\mathsf{b}}\left(  \alpha|I\right)}
\]
where, by (\ref{Der1R})
\[
\partial_{\alpha}
\widehat{\mathsf{R}}^{\left(1\right)  }
\left(  \mathsf{b}
;\alpha,I\right)
=
-
\frac{1}{L}
\sum_{\ell=1}^{L}
\mathbb{I}
\left(
I_{\ell} = I
\right)
\int_{-\frac{\alpha}{h}}^{\frac{1-\alpha}{h}}
P
\left(
X_{\ell},t
\right)
K(t)
dt.
\]
As a consequence, $\widehat{\beta}_0 (\cdot|I)$ and $B(\cdot|x,I)$  are differentiable with a probability tending to $1$, contrasting with the standard quantile-regression estimator, as also illustrated in Figure \ref{BidI2} in the application Section. 
If the derivative of $B(\cdot|x,I)$ can be used as an alternative estimator of $B^{(1)} (\cdot|x,I)$ as in Fernandes et al. (2019) deserves further work.

\section{Proof of Theorem \ref{Bias}} 

That $\sup_{\left(  \alpha,x,I\right)  \in\left[  0,1\right]  \times
	\mathcal{X\times I}}\left| \alpha \mathsf{Bias}
_{h}\left(  \alpha|x,I\right)\right| =O\left(  1\right)  $ follows from (\ref{AQR_bias}) and Proposition \ref{SeriesB2}-(i).

\subparagraph{Step 1.}
The first step of the proof works by establishing that there is a solution of the first-order
condition in a open ball where $\overline{\mathsf{R}}\left(  \mathsf{b}%
;\alpha,I\right)  $ is strictly convex by checking the conditions of Lemma
\ref{Kantorovich}, which will also gives the rate stated in the Theorem and
the uniqueness of $\overline{\mathsf{b}}\left(  \alpha|I\right)  $. Observe first that the definition of 
$
\overline{\mathsf{R}}^{\left(  1\right)  }\left(  \mathsf{b}\left(  \alpha|I\right)  ;\alpha,I\right)
$
gives, by the Taylor inequality as $\alpha+ht = G \left(\left. B(\alpha+ht|x,I) \right| x,I\right)$,
\begin{align}
&  \max_{\left(  \alpha,I\right)  \in\left[  0,1\right]  \times\mathcal{I}%
}\left\Vert \overline{\mathsf{R}}^{\left(  1\right)  }\left(  \mathsf{b}\left(  \alpha|I\right)  ;\alpha,I\right)  \right\Vert_{\infty} =\epsilon
_{L}\text{ with }\label{FOC}\\
&  
\quad\quad\quad
\epsilon_{L}=O\left(  \max_{\left(  \alpha,x\right)  \in\left[  0,1\right]
\times\mathcal{X}}\max_{t\in\mathcal{T}_{\alpha,h}}\left\vert \Psi\left(
t|x,\mathsf{b}\left(  \alpha|I\right)  \right)  -B\left(
\alpha+ht|x,I\right)  \right\vert 
\max_{1\leq n \leq N}
\mathbb{E}
\left[
\left|
P_n (X)
\right|
\right]
\right)  
\nonumber
\\
&
\quad\quad\quad
\quad\quad\quad
=o\left(  h^{s+1+D_{\mathcal{M}}/2}\right)
,\nonumber
\end{align}
where $\epsilon_{L}=o\left(  h^{s+1+D_{\mathcal{M}}/2}\right)  $ follows from Lemma
\ref{Phipsi}-(iii) and Assumptions \ref{Riesz}-(i) and \ref{Auct.A} which gives
$\max_{1\leq n \leq N}
\mathbb{E}
\left[
\left|
P_n (X)
\right|
\right]
\leq 
C
\max_{1\leq n \leq N}
\int_{\mathcal{X}}
|P_n(x)| dx
=
O (h^{D_{\mathcal{M}}/2})
$.  
The second part of Condition (i) in Lemma
\ref{Kantorovich}  follows from Lemma
\ref{R2}-(ii) and the norm equivalence
(\ref{Matnormeq}), which ensures that there is a $C_{0}>0$ such that, for all $h$ small enough, 
\[
\sup_{\left(  \alpha,I\right)  \in\left[  0,1\right]  \times\mathcal{I}%
}\left\Vert \left[  \overline{\mathsf{R}}^{\left(  2\right)  }\left(
\mathsf{b}\left(  \alpha|I\right)  ;\alpha,I\right)  \right]
^{-1}\right\Vert_{\infty} \leq C_{0}.
\]
Condition (ii) in Lemma \ref{Kantorovich} follows from Lemma \ref{R2}-(i),
which ensures that for $C_{1L}=O\left(  h^{-D_{\mathcal{M}}/2-1}\right)  $,%
\[
\left\Vert \overline{\mathsf{R}}^{\left(  2\right)  }\left(  
\mathsf{b}^{1};\alpha,I\right)  
-
\overline{\mathsf{R}}^{\left(  2\right)  }\left(
\mathsf{b}^{0};\alpha,I\right)  \right\Vert_{\infty} \leq C_{1L}\left\Vert
\mathsf{b}^{1}-\mathsf{b}^{0}\right\Vert_{\infty}
\]
for all $\mathsf{b}^{1}$, $\mathsf{b}^{0}$ in $\mathcal{B}_{\infty} \left(
\mathsf{b}\left(  \alpha|I\right)  ,2C_{0}\epsilon_{L}\right)  $ and
all $\alpha$, $I$. For condition (iii) in Lemma \ref{Kantorovich},
$\epsilon_{L}=o\left(  h^{s+1+D_{\mathcal{M}}/2}\right)  $  implies
$C_{0}^{2}C_{1L}\epsilon_{L}=o\left(  h^{s}\right)
=o\left(  1\right)  <1/2$ for $h$ small enough. Hence Lemma \ref{Kantorovich}
ensures that, for $h$ small enough, all $\alpha$ and all $I$, there is a
unique $\overline{\mathsf{b}}\left(  \alpha|I\right)  $ in $\mathcal{B}_{\infty} \left(
\mathsf{b}\left(  \alpha|I\right)  ,2C_{0}\epsilon_{L}\right)  $ such
that%
\[
\overline{\mathsf{R}}^{\left(  1\right)  }\left(  \overline{\mathsf{b}}\left(
\alpha|I\right)  ;\alpha,I\right)  =0
\]
and is therefore the unique minimizer of $\overline{\mathsf{R}}\left(
\cdot;\alpha,I\right)  $ over $\mathcal{B}_{\infty} \left(  \mathsf{b}\left(
\alpha|I\right)  ,2C_{0}\epsilon_{L}\right)  $. Since the convex function
$\overline{\mathsf{R}}\left(  \cdot;\alpha,I\right)  $ cannot have distinct
separated minimizers as discussed in Footnote \ref{Convex}, $\overline{\mathsf{b}}\left(  \alpha|I\right)  $ is also the
unique global minimizer of $\overline{\mathsf{R}}\left(  \cdot;\alpha
,I\right)  $. Since $\epsilon_{L}=o\left(  h^{s+1+D_{\mathcal{M}}/2}\right)  $, it follows that%
\begin{equation}
\sup_{\left(  \alpha,I\right)  \in\left[  0,1\right]  \times\mathcal{I}%
}\left\Vert \overline{\mathsf{b}}\left(  \alpha|I\right)  -\mathsf{b}\left(  \alpha|I\right)  \right\Vert_{\infty} =o\left(  h^{s+1+D_{\mathcal{M}}/2}\right)  .
\label{Bar2*b}%
\end{equation}
Note that this together with Proposition \ref{SeriesB2}-(ii) gives, as $\mathsf{S}_0 \mathsf{b}\left(  \alpha|I\right) = \beta (\alpha|I)$ and
$\mathsf{S}_1 \mathsf{b}\left(  \alpha|I\right) = h \beta^{(1)} (\alpha|I)$, $\overline{B} (\alpha|x,I) = P(x)^{\prime} \mathsf{S}_0 \mathsf{b}\left(  \alpha|I\right)$ and $\overline{B}^{(1)} (\alpha|x,I) = P(x)^{\prime} \mathsf{S}_1 \mathsf{b}\left(  \alpha|I\right)/h$,
\begin{align*}
\sup_{(\alpha,x,I) \in [0,1] \times \mathcal{X} \times \mathcal{I}}
\left|
\overline{B} (\alpha|x,I)
-
B(\alpha|x,I)
\right|
\leq
\sup_{(\alpha,x,I) \in [0,1] \times \mathcal{X} \times \mathcal{I}}
\left|
P(x)^{\prime}
\mathsf{S}_0
\left(
\overline{\mathsf{b}}\left(  \alpha|I\right)  -\mathsf{b}\left(  \alpha|I\right)
\right)
\right|
&
\\
+
\sup_{(\alpha,x,I) \in [0,1] \times \mathcal{X} \times \mathcal{I}}
\left|
P(x)^{\prime}
\beta (\alpha|I)
-
B(\alpha|x,I)
\right|
&
\\
=
O\left(  h^{-D_{\mathcal{M}}/2}\right)o\left(  h^{s+1+D_{\mathcal{M}}/2}\right)
+
o\left(  h^{s+1}\right)=o\left(  h^{s+1}\right),
&
\\
\sup_{(\alpha,x,I) \in [0,1] \times \mathcal{X} \times \mathcal{I}}
\left|
\overline{B}^{(1)} (\alpha|x,I)
-
B^{(1)}(\alpha|x,I)
\right|
\leq
\sup_{(\alpha,x,I) \in [0,1] \times \mathcal{X} \times \mathcal{I}}
\left|
P(x)^{\prime}
\frac{
\mathsf{S}_1
\left(
\overline{\mathsf{b}}\left(  \alpha|I\right)  -\mathsf{b}\left(  \alpha|I\right)
\right)}{h}
\right|
&
\\
+
\sup_{(\alpha,x,I) \in [0,1] \times \mathcal{X} \times \mathcal{I}}
\left|
P(x)^{\prime}
\beta^{(1)} (\alpha|I)
-
B^{(1)}(\alpha|x,I)
\right|
 =
o\left(  h^{s}\right),
&
\end{align*} 
so that the results for $B(\cdot|\cdot)$ and $B^{(1)} (\cdot|\cdot)$ are now established.

\subparagraph{Step 2.}
For $V(\alpha|x,I)$ and $\alpha B^{(1)} (\alpha|x,I)$, we shall obtain a more accurate
expansion of $\alpha\overline{\mathsf{b}}\left(  \alpha|I\right)
-\alpha\mathsf{b}\left(  \alpha|I\right)  $. Observe that, for  $\overline{g} (\alpha|t,x,I)$ as in Lemma \ref{Intbiasterm},
\begin{align*}
& 
G
\left[
\left.  
\Psi
\left(  
t|x,
\overline{\mathsf{b}}\left(  \alpha|I\right)
\right)
\right|
x,I
\right]  
-
\left(  \alpha+ht\right)
=
G
\left[
\left.  
\Psi
\left(  
t|x,
\overline{\mathsf{b}}\left(  \alpha|I\right)
\right)
\right|
x,I
\right]    
-
G
\left(\left. 
B(  \alpha+ht|x,I) 
\right| 
x,I
\right)
\\
& 
\quad\quad\quad
=
\int_{\Psi\left(  t|x,\overline{\mathsf{b}}\left(  \alpha|I\right)\right)}
^{B\left(  \alpha+ht|x,I \right)} 
g(y|x,I) dy
=
\left(
\Psi\left(  t|x,\overline{\mathsf{b}}\left(  \alpha|I\right) \right)
-
B(\alpha+ht|x,I)
\right)
\overline{g} (\alpha|t,x,I)
\end{align*}
The first-order condition (\ref{FOC}) then gives
\begin{align*}
0  &  
=\int\left(  \int_{\underline{t}_{\alpha,h}}^{\overline{t}_{\alpha,h}%
}\left\{  G\left[  \Psi\left(  t|x,\overline{\mathsf{b}}\left(  \alpha
|I\right)  \right)  |x,I\right]  -\left(  \alpha+ht\right)  \right\}  P\left(
x,t\right)  K\left(  t\right)  dt\right)  f\left(  x,I\right)  dx\\
&  =\int\left(  \int_{\underline{t}_{\alpha,h}}^{\overline{t}_{\alpha,h}%
}\overline{g}\left(  \alpha|t,x,I\right)  \left\{  \Psi\left(  t|x,\overline
{\mathsf{b}}\left(  \alpha|I\right)  \right)  -B\left(  \alpha+ht|x,I\right)
\right\}  P\left(  x,t\right)  K\left(  t\right)  dt\right)  f\left(
x,I\right)  dx\\
&  =
\int\left(  \int_{\underline{t}_{\alpha,h}}^{\overline{t}_{\alpha,h}%
}\overline{g}\left(  \alpha|t,x,I\right)  \left\{  \Psi\left(  t|x,\overline
{\mathsf{b}}\left(  \alpha|I\right)  \right)  -\Psi\left(  t|x,\mathsf{b}\left(  \alpha|I\right)  \right)  \right\}  P\left(  x,t\right)
K\left(  t\right)  dt\right)  f\left(  x,I\right)  dx\\
&  +\int\left(  \int_{\underline{t}_{\alpha,h}}^{\overline{t}_{\alpha,h}%
}\overline{g}\left(  \alpha|t,x,I\right)  
\left\{  
\Psi\left(  t|x,\mathsf{b}\left(  \alpha|I\right)  \right)  
-
B\left(  \alpha+ht|x,I\right)
\right\}  
P\left(  x,t\right)  K\left(  t\right)  dt\right)  f\left(
x,I\right)  dx
\end{align*}
where, for $\check{\mathsf{R}}^{(2)}
\left(\alpha|I\right)$ as in Lemma \ref{Intbiasterm},
\begin{align*}
&
\alpha
\int
\left(  
\int_{\underline{t}_{\alpha,h}}^{\overline{t}_{\alpha,h}}
\overline{g}\left(  \alpha|t,x,I\right)  
\left\{  
\Psi
\left(  
t|x,
\overline{\mathsf{b}}\left(\alpha|I\right)\right)  
-
\Psi
\left(  
t|x,
\mathsf{b}\left(\alpha|I\right)  
\right)  
\right\}  
P\left(  x,t\right)
K\left(  t\right)  dt\right)  f\left(  x,I\right)  dx
\\
& \quad
= 
\left[
\int
\left\{ 
\int_{\underline{t}_{\alpha,h}}^{\overline{t}_{\alpha,h}}
\overline{g}\left(  \alpha|t,x,I\right)
P\left(x,t\right)
P\left(x,t\right)^{\prime}
K(t) dt
\right\}
f(x,I) dx
\right]
\alpha
\left(
\overline{\mathsf{b}}\left(\alpha|I\right)
-
\mathsf{b}\left(\alpha|I\right)
\right)
\\
& 
\quad
= 
\check{\mathsf{R}}^{(2)}
\left(\alpha|I\right)
\left(
\alpha
\overline{\mathsf{b}}\left(\alpha|I\right)
-
\alpha
\mathsf{b}\left(\alpha|I\right)
\right).
\end{align*}
For the other integral, the definition of $\Psi (t|x,\mathsf{b})$ gives, as
$\mathsf{b}_0 (\alpha|I) = \beta(\alpha|I)$,
\begin{align*}
&
\alpha
B\left(  \alpha+ht|x,I\right)
-
\alpha
\Psi\left(  t|x,\mathsf{b}\left(  \alpha|I\right)  \right)  
=
\alpha
B\left(  \alpha+ht|x,I\right)
-
\sum_{p=0}^{s+1}
\frac{t^p}{p!}
P(x)^{\prime}
\mathsf{b}_p (\alpha|I)
-
\alpha
B\left(  \alpha+ht|x,I\right)
\\
&
\quad\quad\quad
=
\alpha
B\left(  \alpha+ht|x,I\right)
-
\alpha
B\left(  \alpha|x,I\right)
-
\sum_{p=1}^{s+1}
\frac{t^p}{p!}
P(x)^{\prime}
\alpha
\mathsf{b}_p (\alpha|I)
+
\alpha
\left(
B\left(  \alpha|x,I\right)
-
P(x)^{\prime}
\beta(\alpha|I) 
\right)
\\
&
\quad\quad\quad
=
\frac{\left(ht\right)^{s+2}}{(s+2)!}
\alpha B^{(s+2)} (\alpha|x,I)
+
o\left(h^{s+2}\right)
+
\alpha
\left(
B\left(  \alpha|x,I\right)
-
P(x)^{\prime}
\beta(\alpha|I) 
\right)
\end{align*}
by Lemma \ref{Phipsi}-(iii) with a $o\left(h^{s+2}\right)$ which is uniform in $\alpha$ and $x$.
This gives, for $\check{b}_{s+2} (\alpha|I)$ and 
$\overline{\mathfrak{bias}}_{h} (\alpha|I)$  as in Lemma \ref{Intbiasterm},
\begin{align*}
&
\alpha 
\int\left(  \int_{\underline{t}_{\alpha,h}}^{\overline{t}_{\alpha,h}%
}\overline{g}\left(  \alpha|t,x,I\right)  
\left\{  
\Psi\left(  t|x,\mathsf{b}\left(  \alpha|I\right)  \right)  
-
B\left(  \alpha+ht|x,I\right)
\right\}  
P\left(  x,t\right)  K\left(  t\right)  dt\right)  f\left(
x,I\right)  dx
\\
&
\quad\quad\quad
=
-
\check{\mathsf{R}}^{(2)}
\left(\alpha|I\right)
\left(
h^{s+2} \check{b}_{s+2} (\alpha|I)
+
\alpha
\overline{\mathfrak{bias}}_{h} (\alpha|I)
\right)
\\
&
\quad\quad\quad
-
\int
\left(  
\int_{\underline{t}_{\alpha,h}}^{\overline{t}_{\alpha,h}%
}\overline{g}\left(  \alpha|t,x,I\right)   
o\left(h^{s+2}\right) 
P\left(  x,t\right)  K\left(  t\right)  dt
\right)  f\left(
x,I\right)  dx
.
\end{align*}
Hence the first-order condition (\ref{FOC}) gives 
\begin{align*}
&
\sup_{\alpha \in [0,1]}
\left\|
\alpha
\overline{\mathsf{b}}\left(\alpha|I\right)
-
\alpha
\mathsf{b}\left(\alpha|I\right)
-
h^{s+2}
\check{b}_{s+2} (\alpha|I)
-
\alpha
\overline{\mathfrak{bias}}_{h} (\alpha|I)
\right\|_{\infty}
\\
& \leq 
\sup_{\alpha \in [0,1]}
\left\|
\left[
\check{\mathsf{R}}^{(2)}
\left(\alpha|I\right)
\right]^{-1}
\int
\left(  
\int_{\underline{t}_{\alpha,h}}^{\overline{t}_{\alpha,h}}
\overline{g}\left(  \alpha|t,x,I\right)    
o\left(h^{s+2}\right)  
P\left(  x,t\right)  K\left(  t\right)  dt
\right)  
f\left(
x,I\right)  dx
\right\|_{\infty}
\\
&
\leq
\sup_{\alpha \in [0,1]}
\left\|
\left[
\check{\mathsf{R}}^{(2)}
\left(\alpha|I\right)
\right]^{-1}
\right\|_{\infty}
o\left(h^{s+2}\right)
\max_{1 \leq n \leq N}
\int_{\mathcal{X}} |P_n (x) | dx
=
o\left(h^{s+2+D_{\mathcal{M}}/2}\right)
\end{align*}
by Assumption \ref{Riesz}-(i) and Lemma \ref{Intbiasterm}-(i).

As $\overline{B}^{(1)} (\alpha|x,I) = P(x)^{\prime} \mathsf{S}_1 \mathsf{b}\left(  \alpha|I\right)/h$ with
$\sup_{(\alpha,x,I) \in [0,1] \times \mathcal{X} \times \mathcal{I}}
\left|
P(x)^{\prime} \mathsf{S}_1  \overline{\mathfrak{bias}}_{h} (\alpha|I)/h
\right|  =  o(h^{s+1})
$ and 
$\sup_{(\alpha,x,I) \in [0,1] \times  \mathcal{X} \times \mathcal{I}}
\left|
P(x)^{\prime} \mathsf{S}_1 \check{b}_{s+2} (\alpha|I) - (I-1) \mathsf{Bias}_h (\alpha|x,I)
\right|=o(1)$
by Lemma \ref{Intbiasterm}, Proposition \ref{SeriesB2}-(iii) gives
\begin{align*}
&
\sup_{  (\alpha,x,I) \in [0,1] \times \mathcal{X} \times \mathcal{I}}
\left|
\alpha
\overline{B}^{(1)}
(\alpha,x,I)
-
\alpha
B^{(1)}
(\alpha,x,I)
-
(I-1)h^{s+1}
\mathsf{Bias}_h (\alpha|x,I)
\right|
\\
& 
\quad \quad \quad
\leq
\sup_{  (\alpha,x,I) \in [0,1] \times \mathcal{X} \times \mathcal{I}}
\left|
P(x)^{\prime}
\mathsf{S}_1
\frac{\alpha
	\overline{\mathsf{b}}\left(\alpha|I\right)
	-
	\alpha
	\mathsf{b}\left(\alpha|I\right)
	-
	h^{s+2}
	\check{b}_{s+2} (\alpha|I)
	-
	\alpha
	\overline{\mathfrak{bias}}_{h} (\alpha|I)
	}{h}
\right|
\\
&
\quad \quad \quad
+ 
h^{s+1}
\sup_{(\alpha,x,I) \in [0,1] \times  \mathcal{X} \times \mathcal{I}}
\left|
P(x)^{\prime} \mathsf{S}_1 \check{b}_{s+2} (\alpha|I) - (I-1) \mathsf{Bias}_h (\alpha|x,I)
\right|
\\
&
\quad \quad \quad 
+
\frac{\sup_{(\alpha,x,I) \in [0,1] \times \mathcal{X} \times \mathcal{I}}
	\left|
	P(x)^{\prime} \mathsf{S}_1  \alpha \overline{\mathfrak{bias}}_{h} (\alpha|I)/h
	\right|}{h}
\\
&
\quad \quad \quad 
+
\sup_{(\alpha,x,I) \in [0,1] \times \mathcal{X} \times \mathcal{I}}
\left|
P(x)^{\prime} \beta^{(1)} (\alpha|I) - B^{(1)} (\alpha|x,I)
\right|
\\
&
\quad \quad \quad 
=
O\left(h^{-D_{\mathcal{M}}/2}\right)
o\left(h^{s+1+D_{\mathcal{M}}/2}\right)
+
o\left(h^{s+1}\right)
=
o\left(h^{s+1}\right).
\end{align*}
As $\overline{V} (\alpha|x,I) = \overline{B} (\alpha|x,I) + \alpha \overline{B}^{(1)} (\alpha|x,I)/(I-1)$ with
$
\sup_{(\alpha,x,I) \in [0,1] \times \mathcal{X} \times \mathcal{I}}
\left|
\overline{B} (\alpha|x,I)
-
B(\alpha|x,I)
\right|
=
o(h^{s+1})$, this also establishes the claimed result for $V(\cdot|\cdot)$.
\hfill$\Box$

\section{Proof of Lemma \ref{Intbiasterm}}
\subparagraph{Proof of (i).}
Set
\begin{align*}
\mathfrak{bias}_{h} (\alpha|I)
& =
\int 
\Bigg\{ 
\frac{1}{B^{(1)}(\alpha|x,I)}
\Bigg[
P(x)^{\prime} \beta (\alpha|I) - B(\alpha|x,I),
0, \ldots 0
\Bigg]^{\prime}
\\
&
\quad\quad\quad\quad\quad\quad\quad\quad\quad
\quad\quad\quad\quad\quad\quad\quad\quad\quad
\otimes \left[\mathbf{P}_0 (\alpha|I)^{-1}P(x)\right]  
\Bigg\}
f\left(
x,I\right)  dx
.
\end{align*}
Define, for $\Omega_h (\alpha)$ as in Lemma \ref{R2} and $S_{0} =[1,0,\ldots,0]^{\prime}$,
so that $\Omega_h (\alpha) S_{0} = \int_{\underline{t}_{\alpha,h}}^{\overline{t}_{\alpha,h}} \pi (t) K(t) dt$,
\begin{align*}
\overline{\mathsf{r}} (\alpha|I)
& =
\int 
\left\{
\int_{\underline{t}_{\alpha,h}}^{\overline{t}_{\alpha,h}} 
\overline{g} (\alpha|t,x,I)
\left(
P(x)^{\prime} \beta (\alpha|I) - B(\alpha|x,I)
\right)
\times
P(x,t) K(t) dt
\right\}
f(x,I) dx
\\
\mathsf{r} (\alpha|I)
&
=
\left[\Omega_h (\alpha) S_{0} \right]
\otimes
\int 
\left(
P(x)^{\prime} \beta (\alpha|I) - B(\alpha|x,I)
\right)
\frac{
	P(x) 
	f(x,I)}{B^{(1)}(\alpha|x,I)} dx
\\
& =
\int 
\left\{
\Omega_h (\alpha)
\left[
P(x)^{\prime} \beta (\alpha|I) - B(\alpha|x,I),
0, \ldots, 0
\right]^{\prime}
\right\}
\otimes
P(x)
\frac{ 
	f(x,I)}{B^{(1)}(\alpha|x,I)} dx
\\
& = 
\int 
\left\{
\int_{\underline{t}_{\alpha,h}}^{\overline{t}_{\alpha,h}} 
\frac{
	P(x)^{\prime} \beta (\alpha|I) - B(\alpha|x,I)
}{B^{(1)} (\alpha|x,I)}
P(x,t)
dt 
\right\}
f(x,I)
dx
\end{align*}
so that 
$
\overline{\mathfrak{bias}}_{h} (\alpha|I)
=
\left[\check{\mathsf{R}}^{(2)}
\left(\alpha|I\right)\right]^{-1}
\overline{\mathsf{r}} (\alpha|I)
$ and
$
\mathfrak{bias}_{h} (\alpha|I)
=
\left[
\Omega_h (\alpha) \otimes \mathbf{P}_0 (\alpha|I)
\right]^{-1}
\overline{\mathsf{r}} (\alpha|I)
$.

Under (\ref{Bar2*b})
$
\max_{(\alpha,x,I) \in [0,1] \times \mathcal{I} \times \mathcal{I}}
\max_{t \in \mathcal{T}_{\alpha,h}} 
\left| 
\Psi\left(  t|x,\overline{\mathsf{b}}\left(  \alpha|I\right)  \right)
-
B\left(\alpha+ht|x,I\right)  
\right|
= o \left(h^{s+1}\right)
=
o(h^2)
$
by Lemma \ref{Phipsi}-(iii,iv) and $s\geq 1$ by Assumption \ref{Spec.A}. Set
\begin{align*}
\epsilon_{BL} 
=
\max_{\alpha,u,x,I \in [0,1]^2 \times \mathcal{X} \times \mathcal{I}}
\max_{t \in \mathcal{T}_{\alpha,h}}
\Big|
\Psi\left(
t|x,\overline{\mathsf{b}}\left(  \alpha|I\right)  \right)  +u\left(  B\left(
\alpha+ht|x,I\right)  -\Psi\left(  t|x,\mathsf{b}\left(  \alpha
|I\right)  \right)  \right) 
\quad\quad\quad
&
\\ 
-
B (\alpha+ht|x,I)
\Big|
&
\end{align*}
which therefore satisfies $\epsilon_{BL}=o(h^2)$. It follows that, for $C$ large enough, all $u \in [0,1]$ and all $t$ in
$
\left[
\underline{t}_{\alpha,h}
+
C \epsilon_{BL}/h
,
\overline{t}_{\alpha,h}
-
C \epsilon_{BL}/h
\right]
$,
\[
\Psi\left(
t|x,\overline{\mathsf{b}}\left(  \alpha|I\right)  \right)  +u\left(  B\left(
\alpha+ht|x,I\right)  -\Psi\left(  t|x,\mathsf{b}\left(  \alpha
|I\right)  \right)  \right)
\in
\left[
B(0|x,I),B(1|x,I)
\right].
\]
Hence since $g(\cdot|x,I)$ is differentiable and $g\left( \left. B(\alpha|x,I) \right|x,I\right)=1/B^{(1)}(\alpha|x,I)$
\begin{equation}
\label{Barg}
\max_{\alpha,u,x,I \in [0,1]^2 \times \mathcal{X} \times \mathcal{I}}
\max_{t \in \left[
	\underline{t}_{\alpha,h}
	+
	C \epsilon_{BL}/h
	,
	\overline{t}_{\alpha,h}
	-
	C \epsilon_{BL}/h
	\right]}
\Big|
\overline{g}\left(  \alpha|t,x,I\right)
-
\frac{1}{B^{(1)}(\alpha|x,I)}
\Big|
=
O
\left( h \right).
\end{equation}
It follows that, as $P(x,t)=\pi (t) \otimes P(x)$ and for $\Omega_h (\alpha)$ as for Lemma \ref{R2},
\begin{align*}
\max_{\alpha \in [0,1]}
\left|
\check{\mathsf{R}}^{(2)}
\left(\alpha|I\right)
-
\Omega_h (\alpha|I)
\otimes
\mathbf{P}_0 (\alpha|I)
\right|_{\infty}
& =
O(h),
\\
\max_{\alpha \in [0,1]}
\left\|
\overline{\mathsf{r}}(\alpha|I)
-
\mathsf{r}(\alpha|I)
\right\|_{\infty}
& =
o
\left(
h^{s+2+D_{\mathcal{M}}/2}
\right)
\end{align*}
where the latter follows from (\ref{Barg}),
$\max_{(\alpha,x) \in [0,1] \times \mathcal{X}}
|P(x)^{\prime}\beta(\alpha)-B(\alpha|x,I)|=O(h^{s+1})$ by Proposition \ref{SeriesB2}-(ii), and
$\max_{1\leq n \leq N} \mathbb{E} \left[ |P_n(X)| \right]=O(h^{D_{\mathcal{M}}/2})$ by Assumptions \ref{Riesz}-(i) and \ref{Auct.A}.
As 
$
\check{\mathsf{R}}^{(2)}
\left(\alpha|I\right)
-
\Omega_h (\alpha|I)
\otimes
\mathbf{P}_0 (\alpha|I)$ is an index-permuted $c(s+2)/2$-band matrix, and since $\Omega_h (\alpha|I)
\otimes
\mathbf{P}_0 (\alpha|I)$ which has an inverse with a $\|\cdot\|_{2}$ norm bounded away from infinity by Assumption \ref{Riesz}-(i)  and definition of $\Omega_h (\alpha|I)$, it holds\footnote{The details, which are standard, are as follows. Set
	$
	\mathcal{O}_h
	=
	\check{\mathsf{R}}^{(2)}
	\left(\alpha|I\right)
	-
	\Omega_h (\alpha|I)
	\otimes
	\mathbf{P}_0 (\alpha|I)
	$ which is such that $\|\mathcal{O}_h \|_{2}$ and $\|\mathcal{O}_h \|_{\infty}$ are $O(h)$ by (\ref{Matnormeq}). Since $\left\|\left[\Omega_h (\alpha|I)
	\otimes
	\mathbf{P}_0 (\alpha|I) \right]^{-1} \right\|_{2}$ stays bounded away from infinity,
	\begin{align*}
	\left[ 
	\check{\mathsf{R}}^{(2)}
	\left(\alpha|I\right) 
	\right]^{-1}
	& =
	\left[ 
	\Omega_h (\alpha|I)
	\otimes
	\mathbf{P}_0 (\alpha|I)
	+
	\mathcal{O}_h
	\right]^{-1}
	=
	\left[ 
	\Omega_h (\alpha|I)
	\otimes
	\mathbf{P}_0 (\alpha|I)
	\right]^{-1}
	\left[ 
	\mathrm{Id}
	+
	\mathcal{O}_h
	\left[ 
	\Omega_h (\alpha|I)
	\otimes
	\mathbf{P}_0 (\alpha|I)
	\right]^{-1}
	\right]^{-1}
	\\
	& =
	\left[ 
	\Omega_h (\alpha|I)
	\otimes
	\mathbf{P}_0 (\alpha|I)
	\right]^{-1}
	\left\{
	\mathrm{Id}
	+
	\sum_{k=1}^{\infty}
	\left(-\mathcal{O}_h
	\left[ 
	\Omega_h (\alpha|I)
	\otimes
	\mathbf{P}_0 (\alpha|I)
	\right]^{-1}
	\right)^{k}
	\right\}
	\end{align*}
	where the series converges in the $\|\cdot\|_{2}$ sense for all $h$ small enough. Now Lemma \ref{Invband} gives, for a constant $C$ independent of $h$,
	since the eigenvalues of $\int_{\mathcal{X}}P(x)P(x)^{\prime} dx$ and $f(\cdot,I)$ are bounded away from $0$ and infinity,
	\begin{align*}
	&
	\left\|
	\left[ 
	\check{\mathsf{R}}^{(2)}
	\left(\alpha|I\right) 
	\right]^{-1}
	-
	\left[ 
	\Omega_h (\alpha|I)
	\otimes
	\mathbf{P}_0 (\alpha|I)
	\right]^{-1}
	\right\|_{\infty}
	\\
	&
	\quad\quad\quad
	\leq 
	C
	\left\|
	\left[ 
	\Omega_h (\alpha|I)
	\otimes
	\mathbf{P}_0 (\alpha|I)
	\right]^{-1}
	\right\|_{\infty} 
	\left\{
	\sum_{k=1}^{\infty}
	\left\|
	\mathcal{O}_h
	\right\|_{\infty}^k
	\left\|
	\left[ 
	\Omega_h (\alpha|I)
	\otimes
	\mathbf{P}_0 (\alpha|I)
	\right]^{-1}
	\right\|_{\infty}^{k}
	\right\}
	\\
	&
	\quad \quad \quad
	\leq
	C
	\sum_{k=1}^{\infty}
	\left(Ch\right)^k
	=
	\frac{Ch}{1-Ch}=O(h).
	\end{align*} 
}
\[
\max_{\alpha \in [0,1]}
\left\|
\left[
\check{\mathsf{R}}^{(2)}
\left(\alpha|I\right)
\right]^{-1}
-
\Omega_h (\alpha|I)^{-1}
\otimes
\mathbf{P}_0 (\alpha|I)^{-1}
\right\|_{\infty}
=
O(h).
\]
This implies $\sup_{(\alpha,x,I) \in [0,1] \times \mathcal{X} \times \mathcal{I}}
\left\| \left[\check{\mathsf{R}}^{(2)}
\left(\alpha|I\right)\right]^{-1} \right\|_{\infty}
=O(1)
$ by Lemma \ref{Invband}, Assumptions \ref{Riesz}-(i) and \ref{Auct.A}.
As 
$
\max_{\alpha \in [0,1]}
\left\|
\overline{\mathsf{r}}(\alpha|I)
\right\|_{\infty}
=
o
\left(
h^{s+1+D_{\mathcal{M}}/2}
\right)
$,
$
\overline{\mathfrak{bias}}_{h} (\alpha|I)
=
\left[\check{\mathsf{R}}^{(2)}
\left(\alpha|I\right)\right]^{-1}
\overline{\mathsf{r}} (\alpha|I)
$
and
$
\mathfrak{bias}_{h} (\alpha|I)
=
\left[
\Omega_h (\alpha) \otimes \mathbf{P}_0 (\alpha|I)
\right]^{-1}
\overline{\mathsf{r}} (\alpha|I)
$,
it follows
\begin{align*}
\lefteqn{
	\max_{\alpha \in [0,1] }
	\left\|
	\overline{\mathfrak{bias}}_{h} (\alpha|I)
	-
	\mathfrak{bias}_{h} (\alpha|I)
	\right\|_{\infty}
}
\\
& 
\leq
\max_{\alpha \in [0,1]}
\left\|
\left[
\check{\mathsf{R}}^{(2)}
\left(\alpha|I\right)
\right]^{-1}
-
\Omega_h (\alpha|I)^{-1}
\otimes
\mathbf{P}_0 (\alpha|I)^{-1}
\right\|_{\infty}
\max_{\alpha \in [0,1]}
\left\|
\overline{\mathsf{r}}(\alpha|I)
\right\|_{\infty}
\\
& +
\max_{\alpha \in [0,1]}
\left\|
\Omega_h (\alpha|I)^{-1}
\otimes
\mathbf{P}_0 (\alpha|I)^{-1}
\right\|_{\infty}
\max_{\alpha \in [0,1]}
\left\|
\overline{\mathsf{r}}(\alpha|I)
-
\mathsf{r}(\alpha|I)
\right\|_{\infty}
\\
& =
o
\left(
h^{s+2+D_{\mathcal{M}}/2}
\right).
\end{align*}
Hence, for $\mathsf{S}=\mathsf{S}_0$ or $\mathsf{S}_1$,
\begin{align*}
&
\sup_{(\alpha,x,I) \in [0,1] \times \mathcal{X} \times \mathcal{I}}
\left|
P(x)^{\prime} \mathsf{S}  \left(\overline{\mathfrak{bias}}_{h} (\alpha|I)-\mathfrak{bias}_{h} (\alpha|I)\right)
\right|
\\
&
\quad\quad\quad
\leq
\sup_{x \in \mathcal{X}}
\| P(x) \|
\sup_{(\alpha,I) \in [0,1] \times  \mathcal{I}}
\max_{\alpha \in [0,1] }
\left\|
\overline{\mathfrak{bias}}_{h} (\alpha|I)
-
\mathfrak{bias}_{h} (\alpha|I)
\right\|_{\infty}
\\
& 
\quad\quad\quad
=
O(h^{-D_{\mathcal{M}}/2})
o
\left(
h^{s+2+D_{\mathcal{M}}/2}
\right)
=
o
\left(
h^{s+2}
\right).
\end{align*}
This gives the  results of the Lemma  since
$\sup_{(\alpha,x,I) \in [0,1] \times \mathcal{X} \times \mathcal{I}}
\left|
P(x)^{\prime} \mathsf{S}_0  \mathfrak{bias}_{h} (\alpha|I)
\right|  =  o(h^{s+1})$
and
$\mathsf{S}_1  \mathfrak{bias}_{h} (\alpha|I)=0$. 

\subparagraph{Proof of (ii).}
By the Approximation Property S and since $(\alpha,x) \in [0,1] \times \mathcal{X} \mapsto \alpha B^{(s+2)} (\alpha|x,I)$ is continuous by Proposition \ref{SeriesB2}-(i), there exists for each $\alpha$ of $[0,1]$ a $\beta^{\ast}(\alpha|I)$ satisfying
\[
\sup_{(\alpha,x) \in [0,1] \times \mathcal{X}}
\left|
P(x)^{\prime} \beta^{\ast} (\alpha|I) - \alpha B^{(s+2)} \left(\alpha|x,I\right)
\right|
=
o(1).
\] 
Define 
\[
\beta^{\ast}_h (\alpha|I) 
=
\beta^{\ast} (\alpha|I) 
S_1 
\Omega_h (\alpha)^{-1} 
\int_{-\frac{\alpha}{h}}^{\frac{1-\alpha}{h}}
\frac{t^{s+2}}{(s+2)!}
K(t) dt  
\]
so that
$
\sup_{(\alpha,x) \in [0,1] \times \mathcal{X}}
\left|
P(x)^{\prime} \beta^{\ast}_h (\alpha|I) - (I-1) \mathsf{Bias}_h \left(\alpha|x,I\right)
\right|
=
o(1)
$. Define also
\begin{align*}
\check{b}^{\ast} (\alpha|I)
& =
\left[\check{\mathsf{R}}^{(2)}
\left(\alpha|I\right)\right]^{-1}
\int 
\left\{
\int_{\underline{t}_{\alpha,h}}^{\overline{t}_{\alpha,h}} 
\overline{g} (\alpha|t,x,I)
\frac{t^{s+2}}{(s+2)!}
P(x)^{\prime} \beta^{\ast} (\alpha|I)
\right.\\
&
\quad\quad\quad\quad\quad\quad\quad\quad\quad
\quad\quad\quad\quad\quad\quad\quad\quad\quad
\quad\quad\quad
%\left.
\times
P(x,t) K(t) dt
\Bigg\}
f(x,I) dx,
\\
b^{\ast}_h (\alpha|I)
& = 
\left[\Omega_h (\alpha) \otimes \mathbf{P}_0 (\alpha|I)\right]^{-1}
\int 
\left\{
\int_{\underline{t}_{\alpha,h}}^{\overline{t}_{\alpha,h}} 
\frac{t^{s+2}}{(s+2)!}
\frac{P(x)^{\prime} \beta^{\ast} (\alpha|I)}{B^{(1)}(\alpha|x,I)}
P(x,t)
K(t) dt
\right\}
f(x,I) dx
\\
& = 
\left[
\Omega_h (\alpha)^{-1}
\int_{-\frac{\alpha}{h}}^{\frac{1-\alpha}{h}}
\frac{t^{s+2}}{(s+2)!} \pi (t) dt
\right]
\otimes \beta^{\ast} (\alpha|I),
\end{align*}
which is such that $\mathsf{S}_1 b^{\ast}_h (\alpha|I) = \beta^{\ast}_h (\alpha|I)$. Now, arguing as above gives
\begin{align*}
\sup_{\alpha \in [0,1]}
\left\| 
\check{b}_{s+2} (\alpha|I)
-
\check{b}^{\ast} (\alpha|I)
\right\|_{\infty}
& \leq
\left(
\sup_{\alpha \in [0,1]}
\left\|
\left[\check{\mathsf{R}}^{(2)}
\left(\alpha|I\right)\right]^{-1}
\right\|_{\infty}
\right)
o(1) \max_{1\leq n \leq N}
\int_{\mathcal{X}} |P_n (x)| dx 
\\
&
=
o
\left(
h^{D_{\mathcal{M}}/2}
\right),
\\
\sup_{\alpha \in [0,1]}
\left\| 
\check{b}^{\ast} (\alpha|I)
-
b^{\ast}_h (\alpha|I)
\right\|_{\infty}
& \leq
Ch^{D_{\mathcal{M}}/2}
\sup_{\alpha \in [0,1]}
\left\|
\left[\check{\mathsf{R}}^{(2)}
\left(\alpha|I\right)\right]^{-1}
-
\left[\Omega_h (\alpha) \otimes \mathbf{P}_0 (\alpha|I)\right]^{-1}
\right\|_{\infty}
\\
=
o
\left(
h^{D_{\mathcal{M}}/2}
\right).
\end{align*}
Hence 
$
\sup_{\alpha \in [0,1]}
\left\| 
\check{b}_{s+2} (\alpha|I)
-
b^{\ast}_h (\alpha|I)
\right\|_{\infty}
=
o
\left(
h^{D_{\mathcal{M}}/2}
\right)
$
which gives uniformly in $\alpha$ and $x$
\begin{align*}
P(x)^{\prime} \mathsf{S}_1 \check{b}_{s+2} (\alpha|I)
& =
P(x)^{\prime} \mathsf{S}_1 b^{\ast}_h (\alpha|I)
+
o(1)
=
P(x)^{\prime} \beta^{\ast}_h (\alpha|I)
+
o(1)
\\
& = (I-1) \mathsf{Bias}_h \left(\alpha|x,I\right) + o(1).
\end{align*}
This ends the proof of the Lemma.
\hfill $\Box$

\pagebreak

\renewcommand{\thesection}{Appendix \Alph{section}}\setcounter{section}{3}

\section{\!\!\!\!\!\! -  Bahadur representation \label{App:Bahadur}}

\renewcommand{\thesection}{D.\arabic{section}}
\renewcommand{\thetheorem}{D.\arabic{theorem}}
\renewcommand{\theequation}{D.\arabic{equation}}\setcounter{section}{0}
\renewcommand{\thefootnote}{D.\arabic{footnote}}
\setcounter{footnote}{0}
\setcounter{equation}{0} \setcounter{theorem}{0} 

Let $\widehat{\mathsf{e}}\left(  \alpha|I\right)  $ be a candidate
linearization leading term for $\widehat{\mathsf{b}}\left(  \alpha|I\right)
-\overline{\mathsf{b}}\left(  \alpha|I\right)  $ and $\widehat{\mathsf{d}%
}\left(  \alpha|I\right)  $ the associate linearization error term, or Bahadur
remainder term,%
\begin{align}
\widehat{\mathsf{e}}\left(  \alpha|I\right)   &  =-\left(  \overline
{\mathsf{R}}^{\left(  2\right)  }\left(  \overline{\mathsf{b}}\left(
\alpha|I\right)  ;\alpha,I\right)  \right)  ^{-1}\widehat{\mathsf{R}}^{\left(
1\right)  }\left(  \overline{\mathsf{b}}\left(  \alpha|I\right)
;\alpha,I\right)  ,\label{Llt}\\
\widehat{\mathsf{d}}\left(  \alpha|I\right)   &  =\widehat{\mathsf{b}}\left(
\alpha|I\right)  -\overline{\mathsf{b}}\left(  \alpha|I\right)
-\widehat{\mathsf{e}}\left(  \alpha|I\right)  . \label{Let}%
\end{align}
This section goal is to study the order of $\widehat{\mathsf{d}}\left(
\alpha|I\right)  $ and of the remainder term $P^{\prime}\left(
x\right)  \widehat{\mathsf{d}}_{0}\left(  \alpha|I\right)  $ and $P^{\prime
}\left(  x\right)  \widehat{\mathsf{d}}_{1}\left(  \alpha|I\right)  /h$, which are the differences of $\widehat{B} (\alpha|x,I)$ and $\widehat{B}^{(1)} (\alpha|x,I)$ to their respective linear expansions $P(x)^{\prime}\widehat{\mathsf{e}}_0 (\alpha|I)$ and $P(x)^{\prime}\widehat{\mathsf{e}}_1 (\alpha|I)/h$, with $\widehat{\mathsf{e}}_0 (\alpha|I) = \mathsf{S}_0 \widehat{\mathsf{e}} (\alpha|I)$
and $\widehat{\mathsf{e}}_1 (\alpha|I) = \mathsf{S}_1 \widehat{\mathsf{e}} (\alpha|I)$, $\mathsf{S}_0$ and $\mathsf{S}_1$ as in (\ref{Selmat}).

\begin{theorem}
\label{Baha}Suppose Assumptions \ref{Auct.A}, \ref{Kernel.A}, \ref{Riesz} and \ref{Spec.A} hold.
Then for $\widehat{\mathsf{e}} (\alpha|I)$ as in (\ref{Llt}),
\begin{align*}
&  \sup_{\alpha\in\left[  0,1\right]  }
\left\Vert 
\frac{Lh^{(3D_{\mathcal{M}}+1)/2}}{\left(  h+\alpha\left(
1-\alpha\right)  \right)  ^{1/2}\log L}\left(  \widehat{\mathsf{b}}\left(
\alpha|I\right)  -\overline{\mathsf{b}}\left(  \alpha|I\right)  
%\right.\right. 
%\\
%&  \quad\quad\quad\quad\quad\quad\left.   
 -\widehat{\mathsf{e}} (\alpha|I)  
\right)
 \right\Vert_{\infty} =O_{\mathbb{P}}\left(  1\right).
\end{align*}
Moreover
\begin{align*}
&
\left(L h^{D_{\mathcal{M}}} \right)^{1/2}
\sup_{(\alpha,x) \in [0,1] \times \mathcal{X}}
\left|
\widehat{B} (\alpha|x,I) - P(x)^{\prime} \widehat{\mathsf{e}}_{0} (\alpha|I)
\right|
= 
O_{\mathbb{P}}
\left(
\left(
\frac{\log^2 L}{L h^{3 D_{\mathcal{M}}+1}}
\right)^{1/2}
\right)
=
o_{\mathbb{P}}
\left(h^{1/2}\right),
\\
&
\left(L h^{D_{\mathcal{M}}+1} \right)^{1/2}
\sup_{(\alpha,x) \in [0,1] \times \mathcal{X}}
\left|
\widehat{B}^{(1)} (\alpha|x,I) 
- 
\frac{P(x)^{\prime} \widehat{\mathsf{e}}_{1} (\alpha|I)}{h}
\right|
= 
O_{\mathbb{P}}
\left(
\left(
\frac{\log^2 L}{L h^{3 D_{\mathcal{M}}+2}}
\right)^{1/2}
\right)
=
o_{\mathbb{P}}
(1).
\end{align*}
\end{theorem}

The order  
$\left(
\frac{\log^2 L}{L h^{3 D_{\mathcal{M}}+2}}
\right)^{1/2}$
obtained for 
$\left(L h^{D_{\mathcal{M}}+1} \right)^{1/2}
\sup_{(\alpha,x) \in [0,1] \times \mathcal{X}}
\left|
\widehat{B}^{(1)} (\alpha|x,I) 
- 
\frac{P(x)^{\prime} \widehat{\mathsf{e}}_{1} (\alpha|I)}{h}
\right|
$
is the reason of the bandwidth condition in Assumption \ref{Kernel.A}. It holds when
$\frac{\log^2 L}{Lh^2} = o(1)$ for the AQR case ($D_{\mathcal{M}}=0$), see also Assumption \ref{Kernel}.
For a purely local-polynomial version of the ASQR conditional quantile estimator, Guerre and Sabbah (2017) allow for a better condition $\log L/(Lh^{D+1}) = o(1)$, where $D$ plays the role of $D_{\mathcal{M}}$.

As Theorem \ref{Baha} and Lemma \ref{Leadterm} gives, for $\widehat{\mathsf{d}} (\alpha|I)$ as in (\ref{Let}),
\begin{align*}
\sup_{  \alpha \in [0,1] }
\left\|
\widehat{\mathsf{d}} (\alpha|I)
\right\|_{\infty}
=
O_{\mathbb{P}}
\left(
\frac{\log L}{L h^{3 D_{\mathcal{M}}/2+1/2}}
\right),
\quad
\sup_{  \alpha \in [0,1] }
\left\|
\widehat{\mathsf{e}} (\alpha|I)
\right\|_{\infty}
=
O_{\mathbb{P}}
\left(
\left(
\frac{\log L}{L}
\right)^{1/2}
\right)
\\
\text{ with }
\frac{\log L}{L h^{3 D_{\mathcal{M}}/2+1/2}}
=
\left(
\frac{\log L}{L}
\right)^{1/2}
\left(
\frac{\log L}{Lh^{3 D_{\mathcal{M}}+1}}
\right)^{1/2}
=
\left(
\frac{\log L}{L}
\right)^{1/2}
o(1),
\end{align*}
and the order for $\widehat{\mathsf{e}} (\alpha|I)$ is, up to a $\log^{1/2} L$ term, sharp as its variance is proportional to $1/L$, $\widehat{\mathsf{e}} (\alpha|I)$ is the leading term of $\widehat{\mathsf{b}} (\alpha|I)-\overline{\mathsf{b}} (\alpha|I)$.
Note that all these rates above are driven by central quantiles and can be improved for extreme $\alpha$ with
$\alpha = O(h)$ or $\alpha=1-O(h)$.

As
$\frac{\log L}{Lh^{ D_{\mathcal{M}}+2} }
=o(1)$ by Assumption \ref{Kernel.A}, it follows that 
$\left(
\frac{\log L}{L}
\right)^{1/2}
=
o
\left(
h^{D_{\mathcal{M}}/2+1}
\right)
$, Lemma \ref{Leadterm}, Theorem \ref{Baha} and (\ref{Bar2*b}) show that
\[
\sup_{  \alpha \in [0,1] }
\left\|
\widehat{\mathsf{b}} (\alpha|I)-\mathsf{b} (\alpha|I)
\right\|_{\infty}
=
o
\left(
h^{s+1+D_{\mathcal{M}}/2}
\right)
+
O_{\mathbb{P}}
\left(
\left(
\frac{\log L}{L}
\right)^{1/2}
\right)
=
o_{\mathbb{P}}
\left(
h^{D_{\mathcal{M}}/2+1}
\right),
\]
so that Lemma \ref{Phipsi}-(i) shows that, uniformly in $\alpha$, the ASQR objective function $\widehat{\mathsf{R}}(\cdot;\alpha,I)$ is twice continuously differentiable in a vicinity $\mathcal{B}_{\infty}\left(\widehat{\mathsf{b}} (\alpha|I),Ch^{D_{\mathcal{M}}/2+1} \right)$, with a probability tending to $1$. In addition of simplifying the proof of Theorem \ref{Baha} compared to Guerre and Sabbah (2012, 2014), it can have practical implications for the numerical computation of the estimator as well as for inference.

\paragraph{Proof of Theorem \ref{Baha}. }
Let $\widehat{\mathsf{d}} (\alpha|I)$ be as in (\ref{Let}).	
We first introduce some normalizations. Let, for $\widehat{\mathsf{e}}\left(  \alpha|I\right)  $ as in (\ref{Llt}),
	\begin{align*}
	\varrho_{\alpha L}  &  
	=
	\frac{\left(  h+\alpha\left(  1-\alpha\right)\right)^{1/2}\log L}{Lh^{3D_{\mathcal{M}}/2+1/2}},\\
	\widehat{\boldsymbol{R}}\left(  \mathsf{d};\alpha,I\right)   &  =\widehat{\mathsf{R}%
	}\left(  \overline{\mathsf{b}}\left(  \alpha|I\right)  +\widehat{\mathsf{e}%
	}\left(  \alpha|I\right)  +\varrho_{\alpha L}\mathsf{d};\alpha,I\right)
	-\widehat{\mathsf{R}}\left(  \overline{\mathsf{b}}\left(  \alpha|I\right)
	+\widehat{\mathsf{e}}\left(  \alpha|I\right)  ;\alpha,I\right)  .
	\end{align*}

The definition of $\widehat{\boldsymbol{R}}\left(  \mathsf{d};\alpha,I\right)$ ensures that
\[
	\frac{\widehat{\mathsf{d}}\left(  \alpha|I\right)  }{\varrho_{\alpha L}}%
	=\arg\min_{\mathsf{d}}\widehat{\boldsymbol{R}}\left(  \mathsf{d};\alpha,I\right)  .
\]
It follows that
\begin{align*}
&  \left\{  \sup_{\alpha\in\left[  0,1\right]  }\left\Vert \frac
{\widehat{\mathsf{d}}\left(  \alpha|I\right)  }{\varrho_{\alpha L}}\right\Vert_{\infty}
\geq t\right\}  =
{\displaystyle\bigcup\limits_{\alpha\in\left[  0,1\right]  }}
\left\{  \left\Vert \frac{\widehat{\mathsf{d}}\left(  \alpha|I\right)
}{\varrho_{\alpha L}}\right\Vert_{\infty} \geq t\right\} \\
&  \quad\subset
{\displaystyle\bigcup\limits_{\alpha\in\left[  0,1\right]  }}
\left\{  \inf_{\left\Vert \mathsf{d}\right\Vert_{\infty} \geq t}\widehat{\boldsymbol{R}}\left(
\mathsf{d};\alpha,I\right)  \leq\inf_{\left\Vert \mathsf{d}\right\Vert_{\infty} \leq t}
\widehat{\boldsymbol{R}}\left(  \mathsf{d};\alpha,I\right)  \right\}  \subset
{\displaystyle\bigcup\limits_{\alpha\in\left[  0,1\right]  }}
\left\{  \inf_{\left\Vert \mathsf{d}\right\Vert_{\infty} \geq t}\widehat{\boldsymbol{R}}\left(
\mathsf{d};\alpha,I\right)  \leq0\right\}
\end{align*}
since $\inf_{\left\Vert \mathsf{d}\right\Vert_{\infty} \leq t}\widehat{\boldsymbol{R}}\left(
\mathsf{d};\alpha,I\right)  \leq\widehat{\boldsymbol{R}}\left(  0;\alpha,I\right)  =0$.
The next step uses a convexity argument that can be found in Pollard (1991).
For any $\mathsf{d}$ with $\left\Vert \mathsf{d}\right\Vert_{\infty} \geq t$, convexity yields%
\begin{align*}
\widehat{\boldsymbol{R}}\left(  \mathsf{d};\alpha,I\right)   &  =\frac{\left\Vert
	\mathsf{d}\right\Vert_{\infty} }{t}\left\{  \frac{t}{\left\Vert \mathsf{d}\right\Vert_{\infty} }
\widehat{\boldsymbol{R}}\left(  \left\Vert \mathsf{d}\right\Vert_{\infty} \frac{\mathsf{d}}{\left\Vert
	\mathsf{d}\right\Vert_{\infty} };\alpha,I\right)  +\left(  1-\frac{t}{\left\Vert \mathsf{d}\right\Vert_{\infty}
}\right)  \widehat{\boldsymbol{R}}\left(  0;\alpha,I\right)  \right\} \\
&  \geq\frac{\left\Vert \mathsf{d}\right\Vert_{\infty} }{t}\widehat{\boldsymbol{R}}\left(
t\frac{\mathsf{d}}{\left\Vert \mathsf{d}\right\Vert };\alpha,I\right)
\end{align*}
so that $\inf_{\left\Vert \mathsf{d}\right\Vert_{\infty} \geq t}\widehat{\boldsymbol{R}}\left(
\mathsf{d};\alpha,I\right)  \leq0$ implies $\inf_{\left\Vert \mathsf{d}\right\Vert_{\infty}
	=t}\widehat{\boldsymbol{R}}\left(  \mathsf{d};\alpha,I\right)  \leq0$ and then
\begin{equation}
\left\{  \sup_{\alpha\in\left[  0,1\right]  }\left\Vert \frac
{\widehat{\mathsf{d}}\left(  \alpha|I\right)  }{\varrho_{\alpha L}}\right\Vert_{\infty}
\geq t\right\}  \subset\left\{  \inf_{\alpha\in\left[  0,1\right]  }%
\inf_{\mathsf{d} : \left\Vert \mathsf{d}\right\Vert_{\infty} =t}\widehat{\boldsymbol{R}}\left(
\mathsf{d};\alpha,I\right)  \leq0\right\}  . \label{Largehatd}%
\end{equation}
Thus it is sufficient to consider those $\mathsf{d}$ with $\left\Vert \mathsf{d}\right\Vert_{\infty} =t$, as done from now on.

Note that, for any $t>0$,
$\left(\frac{\log L}{L}\right)^{1/2} + t \varrho_{\alpha L} \leq 2 \left(\frac{\log L}{L}\right)^{1/2} = o\left(h^{D_{\mathcal{M}}/2+1}\right)$. It follows that 
$
\lim_{L\uparrow \infty}
\mathbb{P}
\left(
\max_{\alpha \in [0,1]}
\max_{\mathsf{d}: \|\mathsf{d}\|_{\infty}=t}
\left\|
\widehat{\mathsf{e}} (\alpha|I)
+
\varrho_{\alpha L} \mathsf{d}
\right\|_{\infty}
\leq
C_0 h^{D_{\mathcal{M}}/2+1}
\right)
=1
$
for $C_0$ as in Lemma \ref{Phipsi}-(i), ensuring with (\ref{Bar2*b}) that $\mathsf{d} \mapsto \widehat{\boldsymbol{R}}\left(  \mathsf{d};\alpha,I\right)$ is twice continuously differentiable for $L$ large enough with a large probability.

Under this condition and by (\ref{Maxe}),
the expression of $\widehat{\boldsymbol{R}}\left(  \mathsf{d};\alpha,I\right)  $ gives for all $\mathsf{d}$ with $\|\mathsf{d}\|_{\infty}=t$,
with a probability tending to $1$ for any $t$,
\begin{align*}
\widehat{\boldsymbol{R}}\left(  \mathsf{d};\alpha,I\right)   &  =\varrho_{\alpha
	L}\mathsf{d}^{\prime}\widehat{\mathsf{R}}^{\left(  1\right)  }\left(  \overline
{\mathsf{b}}\left(  \alpha|I\right)  +\widehat{\mathsf{e}}\left(
\alpha|I\right)  ;\alpha,I\right) \\
&  +\varrho_{\alpha L}^{2}\mathsf{d}^{\prime}\left[  \int_{0}^{1}\widehat{\mathsf{R}%
}^{\left(  2\right)  }\left(  \overline{\mathsf{b}}\left(  \alpha|I\right)
+\widehat{\mathsf{e}}\left(  \alpha|I\right)  +u\varrho_{\alpha L}%
\mathsf{d};\alpha,I\right)  \left(  1-u\right)  du\right]  \mathsf{d}\\
&  =\varrho_{\alpha L}\mathsf{d}^{\prime}\widehat{\mathsf{R}}^{\left(  1\right)
}\left(  \overline{\mathsf{b}}\left(  \alpha|I\right)  ;\alpha,I\right) \\
&  +\varrho_{\alpha L}\mathsf{d}^{\prime}\left[  \int_{0}^{1}\widehat{\mathsf{R}%
}^{\left(  2\right)  }\left(  \overline{\mathsf{b}}\left(  \alpha|I\right)
+u\widehat{\mathsf{e}}\left(  \alpha|I\right)  ;\alpha,I\right)  du\right]
\widehat{\mathsf{e}}\left(  \alpha|I\right) \\
&  +\varrho_{\alpha L}^{2}\mathsf{d}^{\prime}\left[  \int_{0}^{1}\widehat{\mathsf{R}%
}^{\left(  2\right)  }\left(  \overline{\mathsf{b}}\left(  \alpha|I\right)
+\widehat{\mathsf{e}}\left(  \alpha|I\right)  +u\varrho_{\alpha L}%
\mathsf{d};\alpha,I\right)  \left(  1-u\right)  du\right]  \mathsf{d}.
\end{align*}
Since $\widehat{\mathsf{R}}^{\left(  1\right)  }\left(  \overline{\mathsf{b}%
}\left(  \alpha|I\right)  ;\alpha,I\right)  +\widehat{\mathsf{R}}^{\left(
	2\right)  }\left(  \overline{\mathsf{b}}\left(  \alpha|I\right)
;\alpha,I\right)  \widehat{\mathsf{e}}\left(  \alpha|I\right)  =0$ by
(\ref{Llt}), it follows that%
\begin{align*}
\widehat{\boldsymbol{R}}\left(  \mathsf{d};\alpha,I\right)   &  =\varrho_{\alpha
	L}\mathsf{d}^{\prime}
\left[  
  \int_{0}^{1}\left\{\widehat{\mathsf{R}}^{\left(
	2\right)  }\left(  \overline{\mathsf{b}}\left(  \alpha|I\right)
+u\widehat{\mathsf{e}}\left(  \alpha|I\right)  ;\alpha,I\right)
-\widehat{\mathsf{R}}^{\left(  2\right)  }\left(  \overline{\mathsf{b}}\left(
\alpha|I\right)  ;\alpha,I\right)  \right\}  du
\right]  \widehat{\mathsf{e}%
}\left(  \alpha|I\right) \\
&  +
\varrho_{\alpha L}^{2}\mathsf{d}^{\prime}\left[  \int_{0}^{1}\widehat{\mathsf{R}%
}^{\left(  2\right)  }\left(  \overline{\mathsf{b}}\left(  \alpha|I\right)
+\widehat{\mathsf{e}}\left(  \alpha|I\right)  +u\varrho_{\alpha L}%
\mathsf{d};\alpha,I\right)  \left(  1-u\right)  du\right]  \mathsf{d}
\\
& = 
\widehat{\boldsymbol{R}}_1 \left(  \mathsf{d};\alpha,I\right) + \widehat{\boldsymbol{R}}_{2}\left(  \mathsf{d};\alpha,I\right).
\end{align*}
We now give an upper bound for $\widehat{\boldsymbol{R}}_1 \left(  \mathsf{d};\alpha,I\right)$ and a lower one for 
$\widehat{\boldsymbol{R}}_{2} \left(  \mathsf{d};\alpha,I\right)$.

\subparagraph{Upper bound for $\left| \widehat{\boldsymbol{R}}_1 \left(  \mathsf{d};\alpha,I\right) \right|$.}
The Cauchy-Schwarz inequality, the norm equivalences (\ref{Vecnormeq}) and (\ref{Matnormeq}) under the disjoint support property of Assumption \ref{Riesz}-(ii) give, for all $\alpha$ of $[0,1]$ and $\mathsf{d}$ with $\| \mathsf{d} \|_{\infty} = t$,
\begin{align*}
\left|
\widehat{\boldsymbol{R}}_1 \left(  \mathsf{d};\alpha,I\right)
\right|
& 
\leq
\varrho_{\alpha L}
\| \mathsf{d} \|_{2} 
\left\|
\widehat{\mathsf{e}} (\alpha|I)
\right\|_{2} 
\\
&
\times
\left\|
\int_{0}^{1}
\left\{
\widehat{\mathsf{R}}^{\left(
	2\right)  }\left(  \overline{\mathsf{b}}\left(  \alpha|I\right)
+u\widehat{\mathsf{e}}\left(  \alpha|I\right)  ;\alpha,I\right)
-\widehat{\mathsf{R}}^{\left(  2\right)  }\left(  \overline{\mathsf{b}}\left(
\alpha|I\right)  ;\alpha,I\right)  
\right\}  du
\right\|_{2}
\\
& \leq 
\varrho_{\alpha L}
(s+2)N
\| \mathsf{d} \|_{\infty} 
\left\|
\widehat{\mathsf{e}} (\alpha|I)
\right\|_{\infty}
\\
&
\times
c
\left\|
\int_{0}^{1}
\left\{
\widehat{\mathsf{R}}^{\left(
	2\right)  }\left(  \overline{\mathsf{b}}\left(  \alpha|I\right)
+u\widehat{\mathsf{e}}\left(  \alpha|I\right)  ;\alpha,I\right)
-\widehat{\mathsf{R}}^{\left(  2\right)  }\left(  \overline{\mathsf{b}}\left(
\alpha|I\right)  ;\alpha,I\right)  
\right\}  du
\right\|_{\infty}
.
\end{align*}
Observe that, by Assumption \ref{Riesz}-(i) and (\ref{Maxe}),
\[
\max_{\alpha \in [0,1]}
\frac{
	(s+2)N
	\| \mathsf{d} \|_{\infty} 
	\left\|
	\widehat{\mathsf{e}} (\alpha|I)
	\right\|_{\infty}}{
	\left(h+\alpha (1-\alpha) \right)^{1/2}
}
=
t
\cdot
h^{-D_{\mathcal{M}}}
O_{\mathbb{P}}
\left(
\left(
\frac{\log L}{L}
\right)^{1/2}
\right).
\]
For the matrix norm, it holds by (\ref{Bar2*b}),  Lemmas \ref{HatR2} and \ref{R2}-(i) with (\ref{Maxe}),
\begin{align*}
& 
\max_{\alpha\in\left[  0,1\right]  }
\left\|
\int_{0}^{1}
\left\{
\widehat{\mathsf{R}}^{\left(
	2\right)  }\left(  \overline{\mathsf{b}}\left(  \alpha|I\right)
+u\widehat{\mathsf{e}}\left(  \alpha|I\right)  ;\alpha,I\right)
-\widehat{\mathsf{R}}^{\left(  2\right)  }\left(  \overline{\mathsf{b}}\left(
\alpha|I\right)  ;\alpha,I\right)  
\right\}  du
\right\|_{\infty}
\\
& 
\quad\quad\quad
\leq
O_{\mathbb{P}} (1)
\max_{\alpha\in\left[  0,1\right]  }\max_{\mathsf{b\in}\mathcal{B}_{\infty}\left(
	\mathsf{b}\left(  \alpha|I\right)  ,Ch^{D_{\mathcal{M}}/2+1}\right)
}\left\Vert \widehat{\mathsf{R}}^{\left(  2\right)  }\left(  \mathsf{b}%
;\alpha,I\right)  \mathsf{-}\overline{\mathsf{R}}^{\left(  2\right)  }\left(
\mathsf{b};\alpha,I\right)  \right\Vert_{\infty}
\\
& 
\quad\quad\quad
+
\max_{\alpha\in\left[  0,1\right]  }
\left\|
\int_{0}^{1}
\left\{
\overline{\mathsf{R}}^{\left(
	2\right)  }\left(  \overline{\mathsf{b}}\left(  \alpha|I\right)
+u\widehat{\mathsf{e}}\left(  \alpha|I\right)  ;\alpha,I\right)
-\overline{\mathsf{R}}^{\left(  2\right)  }\left(  \overline{\mathsf{b}}\left(
\alpha|I\right)  ;\alpha,I\right)  
\right\}  du
\right\|_{\infty}
\\
& 
\quad\quad\quad
=
O_{\mathbb{P}}
\left(
\left(
\frac{\log L}{Lh^{D_{\mathcal{M}}+1}}
\right)^{1/2}
\right)
+
h^{-D_{\mathcal{M}}/2}
O_{\mathbb{P}}
\left(
\left(
\frac{\log L}{L}
\right)^{1/2}
\right)
=
O_{\mathbb{P}}
\left(
\left(
\frac{\log L}{Lh^{D_{\mathcal{M}}+1}}
\right)^{1/2}
\right).
\end{align*}
Combining this bound then gives, uniformly in $\alpha$ in $[0,1]$, $t>0$ and $\mathsf{d}$ with $\|\mathsf{d}\|_{\infty}=t$,
\begin{align*}
\left| \widehat{\boldsymbol{R}}_1 \left(  \mathsf{d};\alpha,I\right) \right|
&
\leq
t 
\cdot
\varrho_{\alpha L}
\cdot
\left(h+\alpha(1-\alpha)\right)^{1/2}
\left|
O_{\mathbb{P}}
\left(
\frac{\log L}{Lh^{3D_{\mathcal{M}}/2+1}}
\right)
\right|
=
t 
\cdot
\varrho_{\alpha L}^2
\cdot
\widehat{R}_1
\\
& \text{ with $\widehat{R}_1 \geq 0$ and }
\sup_{\alpha \in [0,1], t>0, \mathsf{d} \text{ \small with } \|\mathsf{d}\|_{\infty}=t}
\widehat{R}_1
=
O_{\mathbb{P}} (\frac{\log L}{Lh^{3D_{\mathcal{M}}/2+1}})
=
o_{\mathbb{P}} (1).
\end{align*}

\subparagraph{Lower bound for 
	$\widehat{\boldsymbol{R}}_{2} \left(  \mathsf{d};\alpha,I\right)$.}
Note that there is a diverging $t_{L}>0$ such that
\[
t_{L} \max_{\alpha \in [0,1]} \varrho_{\alpha L}
\asymp
t_{L} \frac{\log L}{L h^{3 D_{\mathcal{M}}/2+1/2}}
=
o 
\left(
h^{D_{\mathcal{M}}/2+1}
\right).
\]
Lemma \ref{HatR2},
the matrix norm equivalence (\ref{Matnormeq}), Lemma \ref{R2}-(i,ii) and (\ref{Bar2*b}) then give, uniformly in $\alpha \in [0,1]$, $t \in [0,t_{L}]$ and $\mathsf{d} \in \mathcal{B}_{\infty} (0,t)$
\begin{align*} 
&
\widehat{\boldsymbol{R}}_{2} \left(  \mathsf{d};\alpha,I\right)
\geq
\varrho_{\alpha L}^{2}
\mathsf{d}^{\prime}
\left[  
\overline{\mathsf{R}}^{\left(  2\right)  }
\left(  \mathsf{b}\left(  \alpha|I\right)
;\alpha,I\right) \right]  \mathsf{d}
\\
&
\quad
-
C
\varrho_{\alpha L}^{2}
t^2
\sup_{(\alpha,\mathsf{d}) \in [0,1] \times \mathcal{B}_{\infty} \left(0,\varrho_{\alpha L}t\right)}
\left\|
\widehat{\mathsf{R}}^{\left(  2\right)  }
\left(  \overline{\mathsf{b}}\left(  \alpha|I\right)
+\widehat{\mathsf{e}}\left(  \alpha|I\right)  +
\mathsf{d};\alpha,I\right)
%\right.
%&
%\\
%\left.
-
\overline{\mathsf{R}}^{\left(  2\right)  }
\left(  \mathsf{b}\left(  \alpha|I\right)
;\alpha,I\right)
\right\|_{\infty}
\\
&
\quad
\geq 
C
\varrho_{\alpha L}^{2}
t^2
\left[
1-
\widehat{R}_2
\right]
\text{ with $\widehat{R}_2>0$ and }
\\
& 
\sup_{(\alpha,t) \in [0,1] \times [0,t_{L}]}
\sup_{\mathsf{d} \in \mathcal{B}_{\infty} (0,t)}
\widehat{R}_2
=
O_{\mathbb{P}}
\left(h+\left(\frac{\log L}{L h^{D_{\mathcal{M}}/2+1}}\right)^{1/2}\right)
=
o_{\mathbb{P}} (1).
\end{align*}

\subparagraph{Order of $\widehat{\mathsf{d}}\left(  \alpha|I\right)$.}
The bounds for 
$\widehat{\boldsymbol{R}}_{1} \left(  \mathsf{d};\alpha,I\right)$ and $\widehat{\boldsymbol{R}}_{2} \left(  \mathsf{d};\alpha,I\right)$
then give for $L$ large enough, which allows to take $t\leq t_{L}$ large enough
\begin{align*}
\inf_{\alpha \in [0,1]}
\inf_{\mathsf{d} : \left\Vert \mathsf{d}\right\Vert_{\infty} =t}\widehat{\boldsymbol{R}}\left(
\mathsf{d};\alpha,I\right)
&
\geq
\inf_{\alpha \in [0,1]}
\inf_{\mathsf{d} : \left\Vert \mathsf{d}\right\Vert_{\infty} =t}
\left\{
C
\varrho_{\alpha L}^{2}
t
\left(
t
\left[
1-
\widehat{R}_2
\right]
-
\widehat{R}_1
\right)
\right\}  
\\
&
\geq
\varrho_{\alpha L}^{2}
t
\left(
t
-
o_{\mathbb{P}} (1)
\right).
\end{align*}
Hence (\ref{Largehatd}) shows
that
\begin{equation}
\label{Orderd}
\sup_{\alpha\in\left[  0,1\right]  }\left\Vert \frac
{\widehat{\mathsf{d}}\left(  \alpha|I\right)  }{\varrho_{\alpha L}}\right\Vert_{\infty}
=
O_{\mathbb{P}}
\left(1 \right).
\end{equation}

\subparagraph{Estimation of bid quantile function and its first derivatives.}
As
$\max_{\alpha \in [0,1]} \varrho_{\alpha L}
\asymp
\frac{\log L}{L h^{3 D_{\mathcal{M}}/2+1/2}}$ and
$
\max_{x \in \mathcal{X}}
\left|
P(x)^{\prime}
\beta
\right|
\leq 
C 
h^{-D_{\mathcal{M}}/2}
\| \beta \|_{\infty}
$
under Assumption \ref{Riesz}-(i,ii), (\ref{Orderd}) implies, using also Assumption \ref{Kernel.A},
\begin{align*}
\left(L h^{D_{\mathcal{M}}} \right)^{1/2}
\sup_{(\alpha,x) \in [0,1] \times \mathcal{X}}
\left|
\widehat{B} (\alpha|x,I) - P(x)^{\prime} \widehat{\mathsf{e}}_{0} (\alpha|I)
\right|
= 
\left(L h^{D_{\mathcal{M}}} \right)^{1/2}
\sup_{(\alpha,x) \in [0,1] \times \mathcal{X}}
\left|
P(x)^{\prime} \mathsf{S}_0 \widehat{\mathsf{d}} (\alpha|I)
\right|
&
\\
\leq 
C L^{1/2} 
\sup_{\alpha \in [0,1]}
\left\|
\widehat{\mathsf{d}} (\alpha|I)
\right\|_{\infty}
=
O_{\mathbb{P}}
\left(
\left(
\frac{\log^2 L}{L h^{3 D_{\mathcal{M}}+1}}
\right)^{1/2}
\right)
=
o_{\mathbb{P}}
\left(h^{1/2}\right)
&
\\
\left(L h^{D_{\mathcal{M}}+1} \right)^{1/2}
\sup_{(\alpha,x) \in [0,1] \times \mathcal{X}}
\left|
\widehat{B}^{(1)} (\alpha|x,I) 
- 
\frac{P(x)^{\prime} \widehat{\mathsf{e}}_{1} (\alpha|I)}{h}
\right|
= 
\left(L h^{D_{\mathcal{M}}+1} \right)^{1/2}
\sup_{(\alpha,x) \in [0,1] \times \mathcal{X}}
\left|
P(x)^{\prime} \mathsf{S}_0 \widehat{\mathsf{d}} (\alpha|I)
\right|
&
\\
\leq 
C \left(\frac{L}{h}\right)^{1/2} 
\sup_{\alpha \in [0,1]}
\left\|
\widehat{\mathsf{d}} (\alpha|I)
\right\|_{\infty}
=
O_{\mathbb{P}}
\left(
\left(
\frac{\log^2 L}{L h^{3 D_{\mathcal{M}}+2}}
\right)^{1/2}
\right)
=
o_{\mathbb{P}}
(1).
&
\end{align*}
This ends the proof of the Theorem.
\hfill$\Box$

\pagebreak

\renewcommand{\thesection}{Appendix \Alph{section}}\setcounter{section}{4}

\section{\!\!\!\!\!\! - Proofs of main results \label{App:Proofsmain}}

\renewcommand{\thesection}{E.\arabic{section}}
\renewcommand{\thetheorem}{E.\arabic{theorem}}
\renewcommand{\theequation}{E.\arabic{equation}}\setcounter{section}{0}
\renewcommand{\thefootnote}{E.\arabic{footnote}}
\setcounter{footnote}{0}
\setcounter{equation}{0} \setcounter{theorem}{0}
\setcounter{footnote}{0}\label{Mainproof}

\section{Proof of Theorems \ref{IMSE} and \ref{A.IMSE}}

Recall that $S_{1}$ is the row vector $\left[  0,1,0,\ldots,0\right]
$ of dimension $s+2$ and that $S_{0}=\left[  1,0,\ldots,0\right]  $,
$\mathsf{S}_{0}=S_{0}\otimes\operatorname*{Id}_{N}$, $\mathsf{S}_{1}=S_{1}\otimes\operatorname*{Id}_{N}$ so that $\widehat{\beta}%
_{j}\left(  \alpha|I\right)  =\mathsf{S}_{j}\widehat{\beta}\left(
\alpha|I\right)  $, $j=0,1$ and
\begin{align*}
\widehat{V}\left(  \alpha|x,I\right)   &  =P\left(  x\right)  ^{\prime}\left[
\mathsf{S}_{0}+\frac{\alpha\mathsf{S}_{1}}{h\left(  I-1\right)  }\right]
\widehat{\mathsf{b}}\left(  \alpha|I\right)  \text{,}\\
\overline{V}\left(  \alpha|x,I\right)   &  =P\left(  x\right)  ^{\prime
}\left[  \mathsf{S}_{0}+\frac{\alpha\mathsf{S}_{1}}{h\left(  I-1\right)
}\right]  \overline{\mathsf{b}}\left(  \alpha|I\right),
\end{align*}
see (\ref{Selmat}). Recall that (\ref{WidetildeV}) gives, for $\widehat{\mathsf{e}}\left(  \alpha|I\right)  $ as in (\ref{Llt})
\begin{equation}
\widetilde{V}\left(  \alpha|x,I\right)  =\overline{V}\left(  \alpha|x,I\right)
+P\left(  x\right)  ^{\prime}\left[  \mathsf{S}_{0}+\frac{\alpha\mathsf{S}%
_{1}}{h\left(  I-1\right)  }\right]  \widehat{\mathsf{e}}\left(
\alpha|I\right)  
\label{TildeV}
\end{equation}
which is such, for $\widehat{\mathsf{d}}\left(  \alpha|I\right)  $ as in
(\ref{Let}),%
\[
\widehat{V}\left(  \alpha|x,I\right)  -\widetilde{V}\left(  \alpha|x,I\right)
=P\left(  x\right)  ^{\prime}\left[  \mathsf{S}_{0}+\frac{\alpha\mathsf{S}%
_{1}}{h\left(  I-1\right)  }\right]  \widehat{\mathsf{d}}\left(
\alpha|I\right)  .
\]
Then Theorem \ref{Baha} implies, as
$\widehat{V} (\alpha|x,I) = \widehat{B} (\alpha|x,I) + \alpha \widehat{B}^{(1)} (\alpha|x,I)/(I-1)$,
\[
\sup_{(\alpha,x) \in [0,1] \times \mathcal{X}}
\left(L h^{D_{\mathcal{M}}+1}\right)^{1/2}
\left| \widehat{V}\left(  \alpha|x,I\right)  -\widetilde{V}\left(  \alpha|x,I\right) \right|
=
O_{\mathbb{P}}
\left(
\left(
\frac{\log^2 L}{Lh^{3D_{\mathcal{M}}+2}}
\right)^{1/2}
\right)
\]
which gives (\ref{IMSEBaha}) and (\ref{A.IMSEBaha}). (\ref{CVU.V}) and (\ref{A.CVU.V}), (\ref{CVU.B}) and (\ref{A.CVU.B})  follow from Lemma \ref{Leadterm}-(ii) together with Theorem \ref{Baha} and Theorem \ref{Bias}.

Consider now (\ref{IMSEexp}) and (\ref{A.IMSEexp}). It holds since $\mathbb{E}\left[  \widehat{\mathsf{e}}\left(  \alpha|I\right)
\right]  =\overline{\mathsf{R}}^{\left(  2\right)  }\left(  \overline
{\mathsf{b}}\left(  \alpha|I\right)  ;\alpha,I\right)  ^{-1}\overline
{\mathsf{R}}^{\left(  1\right)  }\left(  \overline{\mathsf{b}}\left(
\alpha|I\right)  ;\alpha,I\right)  =0$ for all $\alpha$ in $\left[
0,1\right]  $
\begin{align*}
&  \mathbb{E}\left[  \int_{\mathcal{X}}\int_{0}^{1}\left(  \widetilde{V}\left(
\alpha|x,I\right)  -V\left(  \alpha|x,I\right)  \right)  ^{2}d\alpha
dx\right]  =\int_{\mathcal{X}}\int_{0}^{1}\left(  \overline{V}\left(
\alpha|x,I\right)  -V\left(  \alpha|x,I\right)  \right)  ^{2}d\alpha dx\\
&  \quad\quad\quad+\int_{\mathcal{X}}\int_{0}^{1}\mathbb{E}\left[  \left(
P\left(  x\right)  ^{\prime}\left[  \mathsf{S}_{0}+\frac{\alpha\mathsf{S}_{1}%
}{h\left(  I-1\right)  }\right]  \widehat{\mathsf{e}}\left(  \alpha|I\right)
\right)  ^{2}\right]  d\alpha dx.
\end{align*}
For the bias part, Theorem \ref{Bias} gives%
\begin{align*}
\int_{\mathcal{X}}\int_{0}^{1}\left(  \overline{V}\left(  \alpha
|x,I\right)  -V\left(  \alpha|x,I\right)  \right)  ^{2}d\alpha dx
& =
\int
_{\mathcal{X}}\int_{0}^{1}\left(  h^{s+1}\mathsf{Bias}_{h}\left(  \alpha|x,I\right)+o\left(
h^{s+1}\right)  \right)  ^{2}d\alpha dx\\
& 
=
h^{2\left(  s+1\right)  }
\int_{\mathcal{X}}\int_{0}^{1}
\mathsf{Bias}_{h}^2 \left(  \alpha|x,I\right)d\alpha dx+o\left(  h^{2\left(s+1\right)  }\right)
\\
& 
=
h^{2\left(  s+1\right)  }
\left(
\mathsf{Bias}_I^2 +o(h)
\right) 
+o\left(  h^{2\left(s+1\right)  }\right).
\end{align*}
For the stochastic part, Lemma \ref{Leadterm}-(i) yields, recalling $\mathsf{S}_{1} \widehat{\mathsf{e}}\left(  \alpha|I\right) = \widehat{\mathsf{e}}_1 \left(  \alpha|I\right)$,
\begin{align*}
&  \int_{\mathcal{X}}\int_{0}^{1}\mathbb{E}\left[  \left(  P\left(  x\right)
^{\prime}\left[  \mathsf{S}_{0}+\frac{\alpha\mathsf{S}_{1}}{h\left(
	I-1\right)  }\right]  \widehat{\mathsf{e}}\left(  \alpha|I\right)  \right)
^{2}\right]  d\alpha dx
\\
&  
\quad
=
\int_{\mathcal{X}}
\left\{
P(x)^{\prime}
\left[
\int_{0}^{1}
\frac{\alpha^2}{(I-1)^2}
\mathrm{Var} \left( \frac{\widehat{\mathsf{e}}_1 \left(  \alpha|I\right)}{h}\right)
d \alpha
\right]
P(x)
\right\}
dx+O\left(  \frac{1}{Lh^{D_{\mathcal{M}}}}\right) \\
& 
\quad
=
\frac{1}{LIh}
\int_{x \in \mathcal{X}}
\left\{
P(x)^{\prime}
\left[
\int_0^1
\Sigma_h (\alpha|I)
d \alpha
\right] 
P(x)
\right\}
dx
+
O\left(  \frac{1}{Lh^{D_{\mathcal{M}}}}\right)
\\
&
\quad=\frac{\Sigma_{LI}^{2}}{LIh^{D_{\mathcal{M}}+1}}+o\left(  \frac
{1}{Lh^{D_{\mathcal{M}}+1}}\right)  .
\end{align*}
Note that
$\mathrm{Var} \left( P(x)^{\prime} \widehat{\mathsf{e}}_1 (\alpha|I) /h\right) \asymp 1/(L h^{D_{\mathcal{M}}+1})$ uniformly in $\alpha$ and $x$ gives $\Sigma_{LI}^{2}=O(1)$.
Substituting in the bias-variance decomposition of the integrated mean squared
error ends the proof of the Theorem. $\hspace*{\fill}\square$

\section{\textbf{Proof of Theorems \ref{CLT}} and \ref{A.CLT}}

Note that (\ref{VBias}) and (\ref{A.VBias}) follow from Theorem \ref{Bias}, while (\ref{Vvar}) and (\ref{A.Vvar}) follow from Lemma \ref{Leadterm}-(i). 

Lemma \ref{Leadterm}-(i) implies that $h^{D_{\mathcal{M}}} P(x)^{\prime} \Sigma_h (\alpha|I) P(x) \asymp \alpha^2$ uniformly in $x$, for all $\alpha>0$. To prove the CLT part, Lemma \ref{Leadterm}-(i) and Theorem \ref{Baha} give
\[
\sqrt{LIh^{D_{\mathcal{M}}+1}}
\left(
\widehat{V} (\alpha|x,I) - \overline{V} (\alpha|x,I)
\right)
=
\sqrt{LIh^{D_{\mathcal{M}}+1}}
\frac{\alpha P(x)^{\prime} \widehat{\mathsf{e}}_1 (\alpha|I)}{h(I-1)}
+
o_{\mathbb{P}}
(1).
\]
Hence it remains to show that
\[
\left(  \frac{LIh}{P\left(  x\right)  ^{\prime}\Sigma_{h}\left(
\alpha|I\right)  P\left(  x\right)  }\right)  ^{1/2}\frac{\alpha P\left(
x\right)  ^{\prime}\mathsf{S}_{1}\widehat{\mathsf{e}}\left(  \alpha|I\right)
}{h\left(  I-1\right)  }\overset{d}{\rightarrow}\mathcal{N}\left(  0,1\right)
.
\]
Write%
\[
\left(  \frac{LIh}{P\left(  x\right)  ^{\prime}\Sigma_{h}\left(
\alpha|I\right)  P\left(  x\right)  }\right)  ^{1/2}\frac{\alpha P\left(
x\right)  ^{\prime}\mathsf{S}_{1}\widehat{\mathsf{e}}\left(  \alpha|I\right)
}{h\left(  I-1\right)  }=\sum_{\ell=1}^{L}r_{\ell}\left(  \alpha|x,I\right)
\]
with $r_{\ell}\left(  \alpha|x,I\right)  =\mathbb{I}\left(  I_{\ell}=I\right)
\sum_{i=1}^{I_{\ell}}r_{i\ell}\left(  \alpha|x,I\right)  $ and%
\begin{align*}
r_{i\ell}\left(  \alpha|x,I\right)   &  =\left(  \frac{\alpha^{2}}{LIh\left(
I-1\right)  ^{2}}\right)  ^{1/2}\frac{P\left(  x\right)  ^{\prime}}{\left(
P\left(  x\right)  ^{\prime}\Sigma_{h}\left(  \alpha|I\right)  P\left(
x\right)  \right)  ^{1/2}}\mathsf{S}_{1}\left[  \overline{\mathsf{R}}^{\left(
2\right)  }\left(  \overline{\mathsf{b}}\left(  \alpha|I\right)
;\alpha,I\right)  \right]  ^{-1}\\
&  \times\int_{-\frac{\alpha}{h}}^{\frac{1-\alpha}{h}}\left\{  \mathbb{I}%
\left(  B_{i\ell}\leq P\left(  X_{\ell},t\right)  \overline{\mathsf{b}}\left(
\alpha|I\right)  \right)  -\left(  \alpha+ht\right)  \right\}  P\left(
X_{\ell},t\right)  K\left(  t\right)  dt.
\end{align*}

Since the first-order condition gives
$\mathbb{E}\left[  r_{\ell}\left(  \alpha|x,I\right)  \right]  
=
\overline{\mathsf{R}}^{(1)}\left(\overline{\mathsf{b}}(\alpha|I);\alpha,I\right)=
0$  
and\\
$\left\vert \operatorname*{Var}\left(  r_{\ell}\left(
\alpha|x,I\right)  \right)  -1\right\vert =o\left(  1\right)  $, it is
sufficient to show that $\left\vert \mathbb{E}\left[
r_{\ell}^{3}\left(  \alpha|x,I\right)  \right]  \right\vert =o\left(
1\right)  $ holds, see e.g. Theorem
%TCIMACRO{\TEXTsymbol{<}}%
%BeginExpansion
$<$%
%EndExpansion
19%
%TCIMACRO{\TEXTsymbol{>} }%
%BeginExpansion
$>$
%EndExpansion
p.179 in Pollard (2002). But Assumption \ref{Riesz}-(i) and Proposition
\ref{SeriesB2}-(i), Lemma \ref{R2} and (\ref{Bar2*b}) give
\[
\left\vert r_{i\ell}\left(  \alpha|x,I\right)  \right\vert \leq\frac
{C}{\left(  Lh\right)  ^{1/2}}\frac{\left\Vert P\left(  x\right)  \right\Vert
}{\left\Vert P\left(  x\right)  \right\Vert }
\times\max_{x\in\mathcal{X}%
}\left\Vert P\left(  x\right)  \right\Vert =O\left(  \frac{1}{\left(
Lh^{D_{\mathcal{M}}+1}\right)  ^{1/2}}\right)  .
\]
It follows that by Assumption \ref{Kernel.A}%
\begin{align*}
\left\vert \mathbb{E}\left[  r_{\ell}^{3}\left(
\alpha|x,I\right)  \right]  \right\vert  &  \leq I
\max_{1\leq i \leq I}
\left\vert r_{i\ell}\left(  \alpha|x,I\right)  \right\vert
\left\vert \mathbb{E}\left[  r_{\ell}^{2}\left(
\alpha|x,I\right)  \right]  \right\vert 
%\\
%&  
=O\left(  \frac{1}{\left(  Lh^{D_{\mathcal{M}}+1}\right)  ^{1/2}}\right)
=o\left(  1\right)  .
\end{align*}
This ends the proof of the Theorem.$\hfill\square$

\section{Proof of Theorems \ref{FuncCLT} and \ref{A.FuncCLT}}

The proof of these theorems requests some specific additional results. The next Lemma
gives an expansion for, $f(\cdot)$ and $g(\cdot)$ being two $(s+2)\times 1$ functions and $A$ a $\mathcal{U}_{[0,1]}$ random variable,
\begin{align*}
\mathcal{C}_{h}  (f,g)&  =\int_{0}^{1}\int_{0}^{1}g\left(  \alpha
_{2}\right)^{\prime}    \left\{  \int_{0}^{1}\int_{0}%
^{1}\frac{1}{h}\pi\left(  \frac{a_{2}-\alpha_{2}}{h}\right)  K\left(
\frac{a_{2}-\alpha_{2}}{h}\right)  \right. \\
&  \quad\times\left.  \frac{1}{h}\pi^{\prime}\left(  \frac{a_{1}-\alpha_{1}%
}{h}\right)  K\left(  \frac{a_{1}-\alpha_{1}}{h}\right)  \left[  a_{1}\wedge
a_{2}-a_{1}a_{2}\right]  f\left(  \alpha_{1}\right)da_{1}da_{2}\right\}  d\alpha_{1}d\alpha_{2}
\\
& =
\mathrm{Cov}
\left[
\int_{0}^{1}
f(\alpha)^{\prime}
\left\{
\int_{0}^{1}
\mathbb{I}
\left( A \leq a \right)
\frac{1}{h}\pi\left(  \frac{a-\alpha}{h}\right)  
K\left(\frac{a-\alpha}{h}\right)
da
\right\}
d\alpha,
\right.
\\
&
\quad\quad\quad
\left.
\int_{0}^{1}
g(\alpha)^{\prime}
\left\{
\int_{0}^{1}
\mathbb{I}
\left( A \leq a \right)
\frac{1}{h}\pi\left(  \frac{a-\alpha}{h}\right)  
K\left(\frac{a-\alpha}{h}\right)
da
\right\}
d\alpha
\right].
\end{align*}
A similar item appears when computing the covariance
\[
LI
\mathrm{Cov}
\left[
\int_{0}^{1}
\left[f(\alpha) \otimes P(x) \right]^{\prime}
\widehat{\mathsf{R}}^{(1)} (\mathsf{b}(\alpha|I);\alpha,I) d\alpha,
\int_{0}^{1}
\left[g(\alpha) \otimes P(x) \right]^{\prime}
\widehat{\mathsf{R}}^{(1)} (\mathsf{b}(\alpha|I);\alpha,I) d\alpha
\right],
\]
see the score expression below (\ref{Der1R}), noting that
$
\mathbb{I} 
\left(
B_{i\ell} 
\leq
P
\left(
X_{\ell},
\frac{a-\alpha}{h}
\right)^{\prime}
\mathsf{b} (\alpha|I_{\ell})
\right)
$
is close to
\[
\mathbb{I} 
\left(
B_{i\ell} 
\leq
B(a|X_{\ell},I_{\ell})
\right)
=
\mathbb{I} 
\left(
A_{i\ell} 
\leq
a
\right).
\]
Recall that $S_{0}=\left[  1,0,\ldots,0\right]  $, $S_{1}=\left[  0,1,0,\ldots,0\right]  $ and $S_{2}=\left[  0,0,1,0,\ldots
,0\right]  $ are row vectors of dimension $s+2$. The next lemma describes the limit of 
$\mathcal{C}_{h}$ when $h$ goes to $0$, in relation with the direction $S_0$ and $S_1$ as in the variance study of Lemma \ref{Leadterm}.
\begin{lemma}
\label{Cmin} Suppose that Assumption \ref{Kernel.A} holds. Assume that $f\left(
\cdot\right)  =f_{h}\left(  \cdot\right)  $ and $g\left(  \cdot\right)
=g_{h}\left(  \cdot\right)  $ are continuously differentiable $(s+2)\times 1$ vector functions, with,
when $h$ goes to $0$,%
\begin{align*}
\sup_{\alpha\in\left[  0,1\right]  }\left\Vert f\left(  \alpha\right)
\right\Vert  &  =O\left(  1\right)  \text{ and }\sup_{\alpha\in\left[
0,1\right]  }\left\Vert g\left(  \alpha\right)  \right\Vert =O\left(
1\right)  \text{,}\\
\sup_{\alpha\in\left[  h,1-h\right]  }\left\Vert f^{\left(  1\right)  }\left(
\alpha\right)  \right\Vert  &  =O\left(  1\right)  \text{ and }\sup_{\alpha
\in\left[  h,1-h\right]  }\left\Vert g^{\left(  1\right)  }\left(
\alpha\right)  \right\Vert =O\left(  1\right)  \text{,}\\
\sup_{\alpha\in\left[  0,h\right]  \cup\left[  1-h,1\right]  }\left\Vert
f^{\left(  1\right)  }\left(  \alpha\right)  \right\Vert  &  =O\left(
\frac{1}{h}\right)  \text{ and }\sup_{\alpha\in\left[  0,h\right]  \cup\left[
1-h,1\right]  }\left\Vert g^{\left(  1\right)  }\left(  \alpha\right)
\right\Vert =O\left(  \frac{1}{h}\right)  \text{.}%
\end{align*}
Then, if $A$ is a random variable with a uniform distribution over $\left[
0,1\right]  $%
\begin{align*}
\mathcal{C}_{h} (f,g) &  \mathcal{=}\operatorname*{Cov}\left(    \int_{A}%
^{1}g\left(  a\right)^{\prime}  \Omega_{h}\left(  a\right)  da  S_{0}^{\prime},S_{0}
\int_{A}^{1}  \Omega_{h}\left(  a\right) f\left(  a\right) da
\right) \\
&  +h\left\{  \operatorname*{Cov}\left(  g\left(  A\right)^{\prime}  \Omega_{h}\left(
A\right)  S_{1}^{\prime},S_{0}  \int_{A}^{1}  \Omega_{h}\left(
a\right) f\left(  a\right) da  \right)  \right. \\
&  \quad\quad\quad+\left.  \operatorname*{Cov}\left(  \  \int_{A}%
^{1}g\left(  a\right)^{\prime}  \Omega_{h}\left(  a\right)  da  S_{0}^{\prime},  S_{1} \Omega_{h}\left(  A\right)  f\left(
A\right) \right)  \right\} \\
&  +h^{2}\operatorname*{Cov}\left(  g\left(  A\right)^{\prime}  \Omega_{h}\left(
A\right)  S_{1}^{\prime},S_{1}  \Omega_{h}\left(  A\right) f\left(  A\right) \right)  \\
&  -\frac{h^{2}}{2}\mathbb{E}\left[ g\left(
A\right)^{\prime}   \Omega_{h}\left(
A\right)  \left[  S_{0}^{\prime}S_{2}+S_{2}^{\prime}S_{0}\right]    \Omega_{h}\left(  A\right)  f\left(  A\right)\right]  +o\left(  h^{2}\right)  .
\end{align*}
\end{lemma}

%Let $\mathsf{f} (\alpha)$ and $\mathsf{g} (\alpha)$ continuously differentiable $(s+2) \times 1$ vector functions. As $\Omega_h (\alpha)^{-1}$ is differentiable with a derivative of order $1/h$ over $[0,h] $ and $[1-h,1]$, Lemma \ref{Cmin} gives
%\begin{align*}
%\mathcal{C}_h 
%\left(
%S_0 \mathsf{f} (\alpha)
%\Omega_h (\alpha)^{-1}
%,
%S_0 \mathsf{f} (\alpha)
%\Omega_h (\alpha)^{-1}
%\right)
%& =
%\mathrm{Cov}
%\left(
%S_0^{\prime} 
%\int_{A}^1
%S_0
%\mathsf{f} (\alpha)
%d\alpha,
%S_0^{\prime} 
%\int_{A}^1
%S_0
%\mathsf{f} (\alpha)
%d\alpha
%\right)
%+
%o(h^2)
%\\
%& =
%\mathrm{Var}
%\left( 
%\int_{A}^1
%S_0
%\mathsf{f} (\alpha)
%d\alpha
%\right)
%\cdot
%S_0^{\prime} S_0
%+o(h^2),
%\\
%\mathcal{C}_h 
%\left(
%S_0 \mathsf{f} (\alpha)
%\Omega_h (\alpha)^{-1}
%,
%\frac{S_1 \mathsf{g} (\alpha)
%\Omega_h (\alpha)^{-1}}{h}
%\right)
%& = o(h),
%\\
%\mathcal{C}_h 
%\left(
%\frac{S_1 \mathsf{g} (\alpha)
%	\Omega_h (\alpha)^{-1}}{h}
%,
%\frac{S_1 \mathsf{g} (\alpha)
%	\Omega_h (\alpha)^{-1}}{h}
%\right)
%& =
%\mathrm{Var}
%\left(
%S_1 \mathsf{g} (A)
%\right)
%\cdot
%S_1^{\prime} S_1
%\\
%&
%\quad
%-
%\frac{1}{2}
%\mathbb{E}
%\left[
%\left(
%S_1 \mathsf{g} (A)
%\right)^2
%\right]
%\cdot
%\left(
%S_0^{\prime} S_2+S_2^{\prime}S_0
%\right)
%+
%o(1).
%\end{align*} 

The smoothness assumption on $f(\cdot)$ and $g(\cdot)$ accounts for vector functions proportional to $\Omega_h (\alpha)$, which is constant over $[h,1-h]$ with $\Omega_h^{(1)} (\alpha) = O(1/h)$ over $[0,h]$ and $[1-h,1]$.

\paragraph{Proof of Lemma \ref{Cmin}:}
See \ref{App:Proofsprelim}.

Consider two functions $\varphi_{0}\left(  \alpha,x\right)  $ and $\varphi
_{1}\left(  \alpha|x\right)  $ and consider the real random variable
\begin{align*}
\widehat{\boldsymbol{I}}_{\varphi}\left(  x|I\right)   &  =\int_{0}^{1}\left[
\varphi_{0}\left(  \alpha|x\right)  S_{0}+\varphi_{1}\left(
\alpha|x\right)  \frac{S_{1}}{h}\right]  \otimes P^{\prime}\left(
x\right) \\
&  \quad\quad\quad\times\left[  \overline{\mathsf{R}}^{\left(  2\right)
}\left(  \overline{\mathsf{b}}\left(  \alpha|I\right)  ;\alpha,I\right)
\right]  ^{-1}\widehat{\mathsf{R}}^{\left(  1\right)  }\left(  \overline
{\mathsf{b}}\left(  \alpha|I\right)  ;\alpha,I\right)  d\alpha.
\end{align*}
The purpose of the next Lemma is to compute the variance of this integral.
Recall
\begin{align*}
\mathbf{P}  &  \mathbf{=P}\left(  I\right)  =\mathbb{E}\left[  \mathbb{I}%
\left(  I_{\ell}=I\right)  P\left(  X_{\ell}\right)  P\left(  X_{\ell}\right)
\right]  ,\\
\mathbf{P}_{0}\left(  \alpha\right)   &  =\mathbf{P}_{0}\left(  \alpha
|I\right)  =\mathbb{E}\left[  P\left(  X_{\ell}\right)  P\left(  X_{\ell
}\right)  \frac{\mathbb{I}\left(  I_{\ell}=I\right)  }{B^{\left(  1\right)
}\left(  \alpha|X_{\ell},I_{\ell}\right)  }\right]  ,\\
\mathbf{P}_{1}\left(  \alpha\right)   &  =\mathbf{P}_{1}\left(  \alpha
|I\right)  =-\mathbb{E}\left[  P\left(  X_{\ell}\right)  P\left(  X_{\ell
}\right)  \frac{\mathbb{I}\left(  I_{\ell}=I\right)  B^{\left(  2\right)
}\left(  \alpha|X_{\ell},I_{\ell}\right)  }{\left(  B^{\left(  1\right)
}\left(  \alpha|X_{\ell},I_{\ell}\right)  \right)  ^{2}}\right]  ,
\end{align*}
and set%
\[
\mathsf{M}_{0}\left(  \alpha\right)  =\Omega_{h}\left(  \alpha\right)
\otimes\mathbf{P}_{0}\left(  \alpha\right)  ,\text{ }\mathsf{M}_{1}\left(
\alpha\right)  =\Omega_{1h}\left(  \alpha\right)  \otimes\mathbf{P}_{1}\left(
\alpha\right)  .
\]
Recall $\mathbf{1}_N = [1,\ldots,1]$ is a $N \times 1$ column vector.

\begin{lemma}
\label{VarI} Suppose that Assumptions
\ref{Auct.A}, \ref{Kernel.A}, \ref{Spec.A} and \ref{Riesz} hold, with $s\geq 2$. Assume
that $\varphi_{0}\left(  \alpha|x\right)  $, $\varphi_{1}\left(
\alpha|x\right)  $ and 
$\partial_{\alpha} \varphi_{1}\left(  \alpha|x\right)
$ are continuous functions with respect to $\left(  \alpha,x\right)
\in\left[  0,1\right]  \times\mathcal{X}$. Let $A$ be a random variable with a
uniform distribution over $\left[  0,1\right]  $. Then $\operatorname*{Var}%
\left(  \sqrt{LIh^{D_{\mathcal{M}}}}\widehat{\boldsymbol{I}}_{\varphi}\left(
x|I\right)  \right)  =\sigma_{L}^{2}\left(  x|I\right)  +o\left(  1\right)  $
with%
\[
\sigma_{L}^{2}\left(  x|I\right) 
 =
\operatorname*{Var}
\left[
  \int_{0}^{A}\left(
\varphi_{0}\left(  \alpha|x\right)  -\partial_{\alpha} \varphi_{1}\left(
\alpha|x\right)  \right)  
\mathbf{1}_N^{\prime}
\mathbf{P}_{0}\left(
\alpha|I\right)  ^{-1}\mathbf{P}\left(  I\right)  ^{1/2}
h^{D_{\mathcal{M}}/2}P\left(  x\right)
d\alpha
\right]
\]
and $\operatorname*{Var}\left(  \sqrt{LI}\int_{\mathcal{X}}%
\widehat{\boldsymbol{I}}_{\varphi}\left(  x|I\right)  dx\right)  =\sigma
_{L}^{2}\left(  I\right)  +o\left(  1\right)  $ with%
\[
\sigma_{L}^{2}\left(  I\right)  
=
\operatorname*{Var}
\left[
\int_{\mathcal{X}}
\left\{
\int_{0}^{A}\left(
\varphi_{0}\left(  \alpha|x\right)  -\partial_{\alpha} \varphi_{1}\left(
\alpha|x\right)  \right)  
\mathbf{1}_N^{\prime}
\mathbf{P}_{0}\left(
\alpha|I\right)  ^{-1}\mathbf{P}\left(  I\right)  ^{1/2}
P\left(  x\right)
d\alpha
\right\}
dx
\right] .
\]

\end{lemma}

\paragraph{Proof of Lemma \ref{VarI}.}

Abbreviate $\overline{\mathsf{R}}^{\left(  2\right)  }\left(  \overline
{\mathsf{b}}\left(  \alpha|I\right)  ;\alpha,I\right)  $, $\widehat{\mathsf{R}%
}^{\left(  1\right)  }\left(  \overline{\mathsf{b}}\left(  \alpha|I\right)
;\alpha,I\right)  $ into $\overline{\mathsf{R}}^{\left(  2\right)  }\left(
\alpha\right)  $ and $\widehat{\mathsf{R}}^{\left(  1\right)  }\left(
\alpha\right)  $ respectively. We now give a suitable expansion for
$\overline{\mathsf{R}}^{\left(  2\right)  }\left(  \alpha\right)  ^{-1}$. 
Lemma \ref{R2} and (\ref{Bar2*b}) with $s\geq 1$ give, uniformly over $[0,1]$
\begin{align*}
\overline{\mathsf{R}}^{\left(  2\right)  }\left(  \alpha\right)
& = 
\overline{\mathsf{R}}^{\left(  2\right)  }\left(  \mathsf{b}\left(  \alpha|I\right);\alpha,I\right)
+
o\left(\frac{h^{s+1+D_{\mathcal{M}}/2}}{h^{1+D_{\mathcal{M}}/2}}\right)
\\
& 
=
\Omega_{h} (\alpha)
\otimes
\mathbf{P}_0 (\alpha)
+
h
\Omega_{1h} (\alpha)
\otimes
\mathbf{P}_1 (\alpha)
+
o\left(h^{s}+h\right)
%\\
%& 
=
\mathsf{M}_{0}\left(
\alpha\right)  +h\mathsf{M}_{1}\left(  \alpha\right)  +o\left(  h\right)
\end{align*}
with respect to $\|\cdot\|_2$, $\|\cdot\|_{\infty}$ or $|\cdot|_{\infty}$ by (\ref{Matnormeq}) and since the matrices above are band ones up to a basis permutation.
It then follows, uniformly over $\left[  0,1\right]  $
\begin{align}
&  \left[  \overline{\mathsf{R}}^{\left(  2\right)  }\left(  \alpha\right)  \right]
^{-1}=\left[  \operatorname*{Id}+h\mathsf{M}_{0}\left(  \alpha\right)
^{-1}\mathsf{M}_{1}\left(  \alpha\right)  +o\left(  h\right)
\right]  ^{-1}\mathsf{M}_{0}\left(  \alpha\right)  ^{-1}
\nonumber\\
&  \quad=\mathsf{M}_{0}\left(  \alpha\right)  ^{-1}-h\mathsf{M}_{0}\left(
\alpha\right)  ^{-1}\mathsf{M}_{1}\left(  \alpha\right)  \mathsf{M}_{0}\left(
\alpha\right)  ^{-1}+o\left(  h\right)  .
\label{InvR2}
\end{align}
Now $\mathsf{M}_{0}\left(  \alpha\right)  ^{-1}=\Omega_{h}\left(
\alpha\right)  ^{-1}\otimes\mathbf{P}_{0}\left(  \alpha\right)  ^{-1}$ and
\[
\mathsf{M}_{0}\left(  \alpha\right)  ^{-1}\mathsf{M}_{1}\left(  \alpha\right)
\mathsf{M}_{0}\left(  \alpha\right)  ^{-1}=\left[  \Omega_{h}\left(
\alpha\right)  ^{-1}\Omega_{1h}\left(  \alpha\right)  \Omega_{h}\left(
\alpha\right)  ^{-1}\right]  \otimes\left[  \mathbf{P}_{0}\left(
\alpha\right)  ^{-1}\mathbf{P}_{1}\left(  \alpha\right)  \mathbf{P}_{0}\left(
\alpha\right)  ^{-1}\right]
\]
with%
\begin{equation}
S_{1}\Omega_{h}\left(  \alpha\right)  ^{-1}\Omega_{1h}\left(
\alpha\right)  =S_{1}\left[
\begin{array}
[c]{ccccc}%
0 & 0 & \cdots & 0 & \times\\
1 & 0 &  & \vdots & c\left(  \alpha\right) \\
0 & 1 &  & \vdots & \times\\
\vdots &  & \ddots & 0 & \vdots\\
0 & \cdots & 0 & 1 & \times
\end{array}
\right]  =S_{0}+c\left(  \alpha\right)  S_{s+1}
\label{S1omegah1h}
\end{equation}
where 
$S_{s+1}=[0,\ldots,0,1]$,
$c\left(  \alpha\right)  =c_{h}\left(  \alpha\right)  $  satisfying the smoothness conditions
of Lemma \ref{Cmin}.

Define, for any $(s+2) \times 1$ vector functions $f\left(  \cdot\right)  $ and $g\left(
\cdot\right)  $ satisfying the conditions of Lemma \ref{Cmin}, and for two $x_f$ and $x_g$ of $\mathcal{X}$,
\begin{align*}
&  C_{h}\left(  f,g\right) 
\left[P(x_f),P(x_g)\right]
=
LI \mathrm{Cov}
\Bigg[
\int_{0}^{1}
\left[
f(\alpha)\otimes P(x_f)\right]^{\prime}
\left[\overline{\mathsf{R}}^{(2)} (\alpha) \right]^{-1}
\widehat{\mathsf{R}}^{(1)} (\alpha)
d\alpha
\\
&
\quad\quad\quad\quad\quad\quad\quad\quad\quad
\quad\quad\quad\quad\quad\quad
,
\int_{0}^{1}
\left[
g(\alpha)\otimes P(x_g)\right]^{\prime}
\left[\overline{\mathsf{R}}^{(2)} (\alpha) \right]^{-1}
\widehat{\mathsf{R}}^{(1)} (\alpha)
d\alpha
\Bigg]
\end{align*}
so that
\begin{align*}
&  C_{h}\left(  f,g\right) 
\left[P(x_f),P(x_g)\right]
\\
& \quad
=\mathbb{E}\left[  \mathbb{I}\left(  I_{\ell
}=I\right)  \int_{0}^{1}\int_{0}^{1}\int_{0}^{1}\int_{0}^{1}  
\left[ g\left(\alpha_{2}\right) \otimes P(x_g) \right]^{\prime}  \right. \\
&  \quad\times\left\{  G\left[  \left.  \min\left(  P\left(  X_{\ell}%
,\frac{a_{1}-\alpha_{1}}{h}\right)  \overline{\mathsf{b}}\left(  \alpha
_{1}|I\right)  ,P\left(  X_{\ell},\frac{a_{2}-\alpha_{2}}{h}\right)
\overline{\mathsf{b}}\left(  \alpha_{2}|I\right)  \right)  \right\vert
X_{\ell},I\right]  \right. \\
&  \quad\quad\quad\left.  -G\left[  \left.  P\left(  X_{\ell},\frac
{a_{1}-\alpha_{1}}{h}\right)  \overline{\mathsf{b}}\left(  \alpha
_{1}|I\right)  \right\vert X_{\ell},I\right]  G\left[  \left.  P\left(
X_{\ell},\frac{a_{2}-\alpha_{2}}{h}\right)  \overline{\mathsf{b}}\left(
a_{2}|I\right)  \right\vert X_{\ell},I\right]  \right\} \\
&  \quad\times\overline{\mathsf{R}}^{\left(  2\right)  }\left(  \alpha_{1}\right)
^{-1}\left\{  \left[  \pi\left(  \frac{a_{1}-\alpha_{1}}{h}\right)  \pi\left(
\frac{a_{2}-\alpha_{2}}{h}\right)  ^{\prime}\right]  \otimes\left[  P\left(
X_{\ell}\right)  P\left(  X_{\ell}\right)  ^{\prime}\right]  \right\}
\overline{\mathsf{R}}^{\left(  2\right)  }\left(  \alpha_{2}\right)  ^{-1}\\
&  \quad\times\left.  \frac{1}{h^{2}}K\left(  \frac{a_{1}-\alpha_{1}}%
{h}\right)  K\left(  \frac{a_{2}-\alpha_{2}}{h}\right)  ^{\prime}
\left[ f\left(\alpha_{1}\right) \otimes P(x_f) \right]
da_1 da_2d\alpha_{1}d\alpha_{2}\right].
\end{align*}
Now (\ref{Bar2*b}), $\max_{\left(  x,t\right)  \in\mathcal{X\times}\left[
-1,1\right]  }\left\Vert P\left(  x,t\right)  \right\Vert =O\left(
h^{-D_{\mathcal{M}}/2}\right)  $ and Lemma \ref{Phipsi}-(iii) gives%
\[
P\left(  X_{\ell},\frac{a-\alpha}{h}\right)  \overline{\mathsf{b}}\left(
\alpha|I\right)  =B\left(  a|X_{\ell},I\right)  +o\left(
h^{s+1}\right)
\]
uniformly in $a$, $\alpha$ and $X_{\ell}$ with $\frac{a-\alpha}{h}$ in the
support of $K\left(  \cdot\right)  $, that is $\left\vert a-\alpha\right\vert \leq h$.
This gives under Assumption \ref{Riesz}%
-(ii) and by definition of $\mathbf{P}$
\begin{align}
C_{h}\left(  f,g\right) \left[P(x_f),P(x_g)\right]   &  =\int_{0}^{1}\int_{0}^{1} 
\left[g\left(  \alpha_{2}\right) \otimes P(x_g) \right]^{\prime}  \left\{  a_{1}\wedge a_{2}-a_{1}%
a_{2}\right\} 
\nonumber
\\
&  \quad\times\overline{\mathsf{R}}^{\left(  2\right)  }\left(  \alpha_{1}\right)
^{-1}\left\{  \left[  \pi\left(  \frac{a_{1}-\alpha_{1}}{h}\right)  \pi\left(
\frac{a_{2}-\alpha_{2}}{h}\right)  ^{\prime}\right]  \otimes\mathbf{P}%
\right\}  \overline{\mathsf{R}}^{\left(  2\right)  }\left(  \alpha_{2}\right)  ^{-1}
\nonumber
\\
&  \quad\times\frac{1}{h^{2}}K\left(  \frac{a_{1}-\alpha_{1}}{h}\right)
K\left(  \frac{a_{2}-\alpha_{2}}{h}\right) 
\left[f\left(  \alpha_{1}\right) \otimes P(x_f) \right] 
d\alpha_{1}d\alpha_{2}
\nonumber
\\
&  +o\left(  h^2\right)  \|P(x_f)\|\|P(x_g)\|.
\label{ChPP}
\end{align}
Set
\[
\mathfrak{P}_h
(x)
=
h^{D_{\mathcal{M}}/2}
\mathbf{P}^{1/2}P(x)
.
\]
Let $\mathcal{C} (f,g)$ be as in Lemma
\ref{Cmin}.
The expansion (\ref{InvR2}) of $\overline{\mathsf{R}}^{\left(  2\right)  }\left(  \alpha\right)
^{-1}$ then gives, under Assumption \ref{Riesz}-(i),
\begin{align*}
&
C_{h}\left(  f,g\right) \left[h^{D_{\mathcal{M}}/2}P(x_f),h^{D_{\mathcal{M}}/2}P(x_g)\right]
\\
&
\quad
= 
\mathcal{C}_{h}
\Big(  
\Omega_h(\alpha)^{-1}f(\alpha)\mathbf{1}_N^{\prime}\mathbf{P}_0 (\alpha )^{-1}\mathfrak{P}_h(x_f),
\Omega_h(\alpha)^{-1}g(\alpha)
\mathbf{1}_N^{\prime}\mathbf{P}_0 (\alpha )^{-1}\mathfrak{P}_h(x_g)
\Big) 
\\
& 
\quad
+
h
\mathcal{C}_{h}
\Big(  
\Omega_h(\alpha)^{-1}
\Omega_{1h} (\alpha)
\Omega_h(\alpha)^{-1}
f(\alpha)\mathbf{1}_N^{\prime}
\mathbf{P}_0 (\alpha )^{-1}
\mathbf{P}_{1} (\alpha )
\mathbf{P}_0 (\alpha )^{-1}
\mathfrak{P}_h(x_f),
\\
& 
\quad\quad\quad\quad\quad\quad\quad\quad\quad\quad\quad\quad
\quad\quad\quad\quad\quad\quad\quad\quad\quad\quad\quad\quad
\Omega_h(\alpha)^{-1}g(\alpha)
\mathbf{1}_N^{\prime}
\mathbf{P}_0 (\alpha )^{-1}
\mathfrak{P}_h(x_g)
\Big) 
\\
& 
\quad
+
h
\mathcal{C}_{h}
\Big(  
\Omega_h(\alpha)^{-1}
\Omega_{1h} (\alpha)
\Omega_h(\alpha)^{-1}
g(\alpha)\mathbf{1}_N^{\prime}
\mathbf{P}_0 (\alpha )^{-1}
\mathbf{P}_{1} (\alpha )
\mathbf{P}_0 (\alpha )^{-1}
\mathfrak{P}_h(x_g),
\\
& 
\quad\quad\quad\quad\quad\quad\quad\quad\quad\quad\quad\quad
\quad\quad\quad\quad\quad\quad\quad\quad\quad\quad\quad\quad
\Omega_h(\alpha)^{-1}f(\alpha)
\mathbf{1}_N^{\prime}
\mathbf{P}_0 (\alpha )^{-1}
\mathfrak{P}_h(x_f)
\Big)
\\
& 
\quad
+
h^2
\mathcal{C}_{h}
\Big(  
\Omega_h(\alpha)^{-1}
\Omega_{1h} (\alpha)
\Omega_h(\alpha)^{-1}
f(\alpha)
\mathbf{1}_N^{\prime}
\mathbf{P}_0 (\alpha )^{-1}
\mathbf{P}_{1} (\alpha )
\mathbf{P}_0 (\alpha )^{-1}
\mathfrak{P}_h(x_f),
\\
& 
\quad\quad\quad\quad\quad\quad
\Omega_h(\alpha)^{-1}
\Omega_{1h} (\alpha)
\Omega_h(\alpha)^{-1}
g(\alpha)
\mathbf{1}_N^{\prime}
\mathbf{P}_0 (\alpha )^{-1}
\mathbf{P}_{1} (\alpha )
\mathbf{P}_0 (\alpha )^{-1}
\mathfrak{P}_h(x_g)
\Big)
\\
&
\quad\quad\quad
+
o\left(  h^2\right)  . 
\end{align*}
As $s \geq 2$, Proposition \ref{SeriesB2}-(i), Lemma \ref{Invband} and the disjoint support condition in Assumption \ref{Riesz}-(ii) ensure that $\alpha \in [0,1] \mapsto \mathbf{1}_N^{\prime}\mathbf{P}_0 (\alpha )^{-1}\mathfrak{P}_h(x)$, $\mathbf{1}_N^{\prime}
\mathbf{P}_0 (\alpha )^{-1}
\mathbf{P}_{1} (\alpha )
\mathbf{P}_0 (\alpha )^{-1}
\mathfrak{P}_h(x)$ are continuously differentiable with bounded derivatives independently of $h$.
$\Omega_h(\alpha)^{-1}
$ and 
$\Omega_h(\alpha)^{-1}
\Omega_{1h} (\alpha)
\Omega_h(\alpha)^{-1}$
are constant over $[h,1-h]$ and have $O(1/h)$ derivatives outside this interval. 

Now, abbreviating $\varphi_0(\alpha|x)$ and $\varphi_1(\alpha|x)$ by removing $x$, recall
\begin{align*}
&\operatorname*{Var}\left(  \sqrt{LIh^{D_{\mathcal{M}}}}\widehat{\boldsymbol{I}%
}_{\varphi}\left(  x|I\right)  \right)
\\
&
\quad
=
C_h
\left(\varphi_{0}\left(  \alpha|x\right)  S_{0}^{\prime}+\varphi_{1}\left(
\alpha|x\right)  \frac{S_{1}^{\prime}}{h},\varphi_{0}\left(  \alpha|x\right)  S_{0}^{\prime}+\varphi_{1}\left(
\alpha|x\right)  \frac{S_{1}^{\prime}}{h}\right)
\left[h^{D_{\mathcal{M}}/2}P(x),h^{D_{\mathcal{M}}/2}P(x)\right]
\end{align*}
and then
\begin{align*}
&\operatorname*{Var}\left(  \sqrt{LIh^{D_{\mathcal{M}}}}\widehat{\boldsymbol{I}%
}_{\varphi}\left(  x|I\right)  \right)
\\
&
\quad 
=
\frac{1}{h^2}
\mathcal{C}_{h}
\Big(  
\Omega_h(\alpha)^{-1}
S_{1}^{\prime}\varphi_{1}\left(  \alpha\right) 
\mathbf{1}_N^{\prime}\mathbf{P}_0 (\alpha )^{-1}\mathfrak{P}_h(x),
\Omega_h(\alpha)^{-1}
S_{1}^{\prime}\varphi_{1}\left(  \alpha\right) 
\mathbf{1}_N^{\prime}\mathbf{P}_0 (\alpha )^{-1}\mathfrak{P}_h(x)
\Big)
\\
&
\quad
+
\frac{2}{h}
\mathcal{C}_{h}
\Big(  
\Omega_h(\alpha)^{-1}
S_{0}^{\prime}\varphi_{0}\left(  \alpha\right) 
\mathbf{1}_N^{\prime}\mathbf{P}_0 (\alpha )^{-1}\mathfrak{P}_h(x),
\Omega_h(\alpha)^{-1}
S_{1}^{\prime}\varphi_{1}\left(  \alpha\right) 
\mathbf{1}_N^{\prime}\mathbf{P}_0 (\alpha )^{-1}\mathfrak{P}_h(x)
\Big)
\\
& 
\quad
-
\frac{2}{h}
\mathcal{C}_{h}
\Big(  
\Omega_h(\alpha)^{-1}
\Omega_{1h} (\alpha)
\Omega_h(\alpha)^{-1}
S_{1}^{\prime}\varphi_{1}\left(  \alpha\right) 
\mathbf{1}_N^{\prime}
\mathbf{P}_0 (\alpha )^{-1}
\mathbf{P}_{1} (\alpha )
\mathbf{P}_0 (\alpha )^{-1}
\mathfrak{P}_h(x),
\\
& 
\quad\quad\quad\quad\quad\quad\quad\quad\quad\quad\quad\quad
\quad\quad\quad\quad\quad\quad
\Omega_h(\alpha)^{-1}
S_{1}^{\prime}\varphi_{1}\left(\alpha\right)
\mathbf{1}_N^{\prime}
\mathbf{P}_0 (\alpha )^{-1}
\mathfrak{P}_h(x)
\Big)
\\
&
\quad
+
\mathcal{C}_{h}
\Big(  
\Omega_h(\alpha)^{-1}
S_{0}^{\prime}\varphi_{0}\left(  \alpha\right) 
\mathbf{1}_N^{\prime}\mathbf{P}_0 (\alpha )^{-1}\mathfrak{P}_h(x),
\Omega_h(\alpha)^{-1}
S_{0}^{\prime}\varphi_{0}\left(  \alpha\right) 
\mathbf{1}_N^{\prime}\mathbf{P}_0 (\alpha )^{-1}\mathfrak{P}_h(x)
\Big)
\\
& 
\quad
-
2
\mathcal{C}_{h}
\Big(  
\Omega_h(\alpha)^{-1}
\Omega_{1h} (\alpha)
\Omega_h(\alpha)^{-1}
S_{0}^{\prime}\varphi_{0}\left(  \alpha\right) 
\mathbf{1}_N^{\prime}
\mathbf{P}_0 (\alpha )^{-1}
\mathbf{P}_{1} (\alpha )
\mathbf{P}_0 (\alpha )^{-1}
\mathfrak{P}_h(x),
\\
& 
\quad\quad\quad\quad\quad\quad\quad\quad\quad\quad\quad\quad
\quad\quad\quad\quad\quad\quad
\Omega_h(\alpha)^{-1}
S_{1}^{\prime}\varphi_{1}\left(\alpha\right)
\mathbf{1}_N^{\prime}
\mathbf{P}_0 (\alpha )^{-1}
\mathfrak{P}_h(x)
\Big)
\\
& 
\quad
-
2
\mathcal{C}_{h}
\Big(  
\Omega_h(\alpha)^{-1}
\Omega_{1h} (\alpha)
\Omega_h(\alpha)^{-1}
S_{1}^{\prime}\varphi_{1}\left(\alpha\right)
\mathbf{1}_N^{\prime}
\mathbf{P}_0 (\alpha )^{-1}
\mathbf{P}_{1} (\alpha )
\mathbf{P}_0 (\alpha )^{-1}
\mathfrak{P}_h(x),
\\
& 
\quad\quad\quad\quad\quad\quad\quad\quad\quad\quad\quad\quad
\quad\quad\quad\quad\quad\quad
\Omega_h(\alpha)^{-1}
S_{0}^{\prime}\varphi_{0}\left(  \alpha\right) 
\mathbf{1}_N^{\prime}
\mathbf{P}_0 (\alpha )^{-1}
\mathfrak{P}_h(x)
\Big)
\\
& 
\quad
+
\mathcal{C}_{h}
\Big(  
\Omega_h(\alpha)^{-1}
\Omega_{1h} (\alpha)
\Omega_h(\alpha)^{-1}
S_{1}^{\prime}\varphi_{1}\left(\alpha\right)
\mathbf{1}_N^{\prime}
\mathbf{P}_0 (\alpha )^{-1}
\mathbf{P}_{1} (\alpha )
\mathbf{P}_0 (\alpha )^{-1}
\mathfrak{P}_h(x),
\\
& 
\quad\quad\quad\quad\quad\quad
\Omega_h(\alpha)^{-1}
\Omega_{1h} (\alpha)
\Omega_h(\alpha)^{-1}
S_{1}^{\prime}\varphi_{1}\left(\alpha\right)
\mathbf{1}_N^{\prime}
\mathbf{P}_0 (\alpha )^{-1}
\mathbf{P}_{1} (\alpha )
\mathbf{P}_0 (\alpha )^{-1}
\mathfrak{P}_h(x)
\Big)
\\
&
\quad
+
o(1).
\end{align*}
Lemma \ref{Cmin} gives, $A$ being $\mathcal{U}_{[0,1]}$,
\begin{align*}
&
\frac{1}{h^2}
\mathcal{C}_{h}
\Big(  
\Omega_h(\alpha)^{-1}
S_{1}^{\prime}\varphi_{1}\left(  \alpha\right) 
\mathbf{1}_N^{\prime}\mathbf{P}_0 (\alpha )^{-1}\mathfrak{P}_h(x),
\Omega_h(\alpha)^{-1}
S_{1}^{\prime}\varphi_{1}\left(  \alpha\right) 
\mathbf{1}_N^{\prime}\mathbf{P}_0 (\alpha )^{-1}\mathfrak{P}_h(x)
\Big)
\\
&
\quad
=
\mathrm{Var}
\left(
\varphi_{1}\left(  A\right) 
\mathbf{1}_N^{\prime}\mathbf{P}_0 (A )^{-1}\mathfrak{P}_h(x)
\right)+o(1),
\end{align*}
\begin{align*}
&
\frac{1}{h}
\mathcal{C}_{h}
\Big(  
\Omega_h(\alpha)^{-1}
S_{0}^{\prime}\varphi_{0}\left(  \alpha\right) 
\mathbf{1}_N^{\prime}\mathbf{P}_0 (\alpha )^{-1}\mathfrak{P}_h(x),
\Omega_h(\alpha)^{-1}
S_{1}^{\prime}\varphi_{1}\left(  \alpha\right) 
\mathbf{1}_N^{\prime}\mathbf{P}_0 (\alpha )^{-1}\mathfrak{P}_h(x)
\Big)
\\
&
+
\frac{1}{h}
\mathcal{C}_{h}
\Big(  
\Omega_h(\alpha)^{-1}
S_{1}^{\prime}\varphi_{1}\left(  \alpha\right) 
\mathbf{1}_N^{\prime}\mathbf{P}_0 (\alpha )^{-1}\mathfrak{P}_h(x),
\Omega_h(\alpha)^{-1}
S_{0}^{\prime}\varphi_{0}\left(  \alpha\right) 
\mathbf{1}_N^{\prime}\mathbf{P}_0 (\alpha )^{-1}\mathfrak{P}_h(x)
\Big)
\\
&
\quad
=
\mathrm{Cov}
\left(
\int_{A}^1
\varphi_{0}\left(  \alpha\right) 
\mathbf{1}_N^{\prime}\mathbf{P}_0 (\alpha )^{-1}
\mathfrak{P}_h(x)d\alpha,
\varphi_{1}\left(  A\right) 
\mathbf{1}_N^{\prime}\mathbf{P}_0 (A )^{-1}\mathfrak{P}_h(x)
\right)
\\
&
\quad
+
\mathrm{Cov}
\left(
\int_{A}^1
\varphi_{1}\left(  \alpha\right) 
\mathbf{1}_N^{\prime}\mathbf{P}_0 (\alpha )^{-1}
\mathfrak{P}_h(x)d\alpha,
\varphi_{0}\left(  A\right) 
\mathbf{1}_N^{\prime}\mathbf{P}_0 (A )^{-1}\mathfrak{P}_h(x)
\right)
+
o(1),
\end{align*}
\begin{align*}
&
\mathcal{C}_{h}
\Big(  
\Omega_h(\alpha)^{-1}
S_{0}^{\prime}\varphi_{0}\left(  \alpha\right) 
\mathbf{1}_N^{\prime}\mathbf{P}_0 (\alpha )^{-1}\mathfrak{P}_h(x),
\Omega_h(\alpha)^{-1}
S_{0}^{\prime}\varphi_{0}\left(  \alpha\right) 
\mathbf{1}_N^{\prime}\mathbf{P}_0 (\alpha )^{-1}\mathfrak{P}_h(x)
\Big)
\\
&
\quad
=
\mathrm{Var}
\left(
\int_{A}^1
\varphi_{0}\left(  \alpha\right) 
\mathbf{1}_N^{\prime}\mathbf{P}_0 (\alpha )^{-1}\mathfrak{P}_h(x)
d\alpha
\right)
+o(1),
\end{align*}
and by (\ref{S1omegah1h}) which gives
$
S_1
\Omega_h(\alpha)^{-1}
\Omega_{1h} (\alpha)
S_0^{\prime}=1
$
and
$
S_0
\Omega_h(\alpha)^{-1}
\Omega_{1h} (\alpha)
S_0^{\prime}=0
$,
\begin{align*}
&
\frac{1}{h}
\mathcal{C}_{h}
\Big(  
\Omega_h(\alpha)^{-1}
\Omega_{1h} (\alpha)
\Omega_h(\alpha)^{-1}
S_{1}^{\prime}
\varphi_{1}\left(  \alpha\right) 
\mathbf{1}_N^{\prime}
\mathbf{P}_0 (\alpha )^{-1}
\mathbf{P}_{1} (\alpha )
\mathbf{P}_0 (\alpha )^{-1}
\mathfrak{P}_h(x),
\\
& 
\quad\quad\quad\quad\quad\quad\quad\quad\quad\quad\quad\quad
\quad\quad\quad\quad\quad\quad
\Omega_h(\alpha)^{-1}
S_{1}^{\prime}\varphi_{1}\left(\alpha\right)
\mathbf{1}_N^{\prime}
\mathbf{P}_0 (\alpha )^{-1}
\mathfrak{P}_h(x)
\Big)
\\
&
\quad
=
\mathrm{Cov}
\Bigg(
\int_{A}^{1}
S_1
\Omega_h(\alpha)^{-1}
\Omega_{1h} (\alpha)
S_0^{\prime}
\varphi_{1}\left(  \alpha\right) 
\mathbf{1}_N^{\prime}
\mathbf{P}_0 (\alpha )^{-1}
\mathbf{P}_{1} (\alpha )
\mathbf{P}_0 (\alpha )^{-1}
\mathfrak{P}_h(x)
d\alpha,
\\
& 
\quad\quad\quad\quad\quad\quad\quad\quad\quad\quad\quad\quad
\quad\quad\quad\quad\quad\quad\quad\quad\quad\quad
\varphi_{1}\left(A\right)
\mathbf{1}_N^{\prime}
\mathbf{P}_0 (A )^{-1}
\mathfrak{P}_h(x)
\Bigg)+o(1)
\\
&
\quad
=
\mathrm{Cov}
\Bigg(
\int_{A}^{1}
\varphi_{1}\left(  \alpha\right) 
\mathbf{1}_N^{\prime}
\mathbf{P}_0 (\alpha )^{-1}
\mathbf{P}_{1} (\alpha )
\mathbf{P}_0 (\alpha )^{-1}
\mathfrak{P}_h(x)
d\alpha,
\varphi_{1}\left(A\right)
\mathbf{1}_N^{\prime}
\mathbf{P}_0 (A )^{-1}
\mathfrak{P}_h(x)
\Bigg)+o(1),
\end{align*}
\begin{align*}
&
\mathcal{C}_{h}
\Big(  
\Omega_h(\alpha)^{-1}
\Omega_{1h} (\alpha)
\Omega_h(\alpha)^{-1}
S_{0}^{\prime}\varphi_{0}\left(  \alpha\right) 
\mathbf{1}_N^{\prime}
\mathbf{P}_0 (\alpha )^{-1}
\mathbf{P}_{1} (\alpha )
\mathbf{P}_0 (\alpha )^{-1}
\mathfrak{P}_h(x),
\\
& 
\quad\quad\quad\quad\quad\quad\quad\quad\quad\quad\quad\quad
\quad\quad\quad\quad\quad\quad
\Omega_h(\alpha)^{-1}
S_{1}^{\prime}\varphi_{1}\left(\alpha\right)
\mathbf{1}_N^{\prime}
\mathbf{P}_0 (\alpha )^{-1}
\mathfrak{P}_h(x)
\Big)=o(1),
\end{align*}
\begin{align*}
&
\mathcal{C}_{h}
\Big(  
\Omega_h(\alpha)^{-1}
\Omega_{1h} (\alpha)
\Omega_h(\alpha)^{-1}
S_{1}^{\prime}\varphi_{1}\left(\alpha\right)
\mathbf{1}_N^{\prime}
\mathbf{P}_0 (\alpha )^{-1}
\mathbf{P}_{1} (\alpha )
\mathbf{P}_0 (\alpha )^{-1}
\mathfrak{P}_h(x),
\\
& 
\quad\quad\quad\quad\quad\quad\quad\quad\quad\quad\quad\quad
\quad\quad\quad\quad\quad\quad
\Omega_h(\alpha)^{-1}
S_{0}^{\prime}\varphi_{0}\left(  \alpha\right) 
\mathbf{1}_N^{\prime}
\mathbf{P}_0 (\alpha )^{-1}
\mathfrak{P}_h(x)
\Big)
\\
&
\quad
=
\mathrm{Cov}
\Bigg(
\int_{A}^{1}
\varphi_{1}\left(  \alpha\right) 
\mathbf{1}_N^{\prime}
\mathbf{P}_0 (\alpha )^{-1}
\mathbf{P}_{1} (\alpha )
\mathbf{P}_0 (\alpha )^{-1}
\mathfrak{P}_h(x)
d\alpha,
\int_{A}^{1}
\varphi_{0}\left(  \alpha\right) 
\mathbf{1}_N^{\prime}
\mathbf{P}_0 (\alpha )^{-1}
\mathfrak{P}_h(x)
d\alpha
\Bigg)+o(1),
\end{align*}
\begin{align*}
&
\mathcal{C}_{h}
\Big(  
\Omega_h(\alpha)^{-1}
\Omega_{1h} (\alpha)
\Omega_h(\alpha)^{-1}
S_{1}^{\prime}\varphi_{1}\left(\alpha\right)
\mathbf{1}_N^{\prime}
\mathbf{P}_0 (\alpha )^{-1}
\mathbf{P}_{1} (\alpha )
\mathbf{P}_0 (\alpha )^{-1}
\mathfrak{P}_h(x),
\\
& 
\quad\quad\quad\quad\quad\quad
\Omega_h(\alpha)^{-1}
\Omega_{1h} (\alpha)
\Omega_h(\alpha)^{-1}
S_{1}^{\prime}\varphi_{1}\left(\alpha\right)
\mathbf{1}_N^{\prime}
\mathbf{P}_0 (\alpha )^{-1}
\mathbf{P}_{1} (\alpha )
\mathbf{P}_0 (\alpha )^{-1}
\mathfrak{P}_h(x)
\Big)
\\
& 
\quad
=
\mathrm{Var}
\Bigg(
\int_{A}^{1}
\varphi_{1}\left(  \alpha\right) 
\mathbf{1}_N^{\prime}
\mathbf{P}_0 (\alpha )^{-1}
\mathbf{P}_{1} (\alpha )
\mathbf{P}_0 (\alpha )^{-1}
\mathfrak{P}_h(x)
d\alpha
\Bigg)
+o(1).
\end{align*}
Collecting the items then gives
\begin{align*}
&
\operatorname*{Var}\left(  \sqrt{LIh^{D_{\mathcal{M}}}}\widehat{\boldsymbol{I}%
}_{\varphi}\left(  x|I\right)  \right)
=
\mathrm{Var}
\left(
\varphi_{1}\left(  A\right) 
\mathbf{1}_N^{\prime}\mathbf{P}_0 (A )^{-1}\mathfrak{P}_h(x)
\right)
\\
& 
\quad
+
2
\mathrm{Cov}
\left(
\int_{A}^1
\varphi_{0}\left(  \alpha\right) 
\mathbf{1}_N^{\prime}\mathbf{P}_0 (\alpha )^{-1}
\mathfrak{P}_h(x)d\alpha,
\varphi_{1}\left(  A\right) 
\mathbf{1}_N^{\prime}\mathbf{P}_0 (A )^{-1}\mathfrak{P}_h(x)
\right)
\\
&
\quad
-
2\mathrm{Cov}
\Bigg(
\int_{A}^{1}
\varphi_{1}\left(  \alpha\right) 
\mathbf{1}_N^{\prime}
\mathbf{P}_0 (\alpha )^{-1}
\mathbf{P}_{1} (\alpha )
\mathbf{P}_0 (\alpha )^{-1}
\mathfrak{P}_h(x)
d\alpha,
\varphi_{1}\left(A\right)
\mathbf{1}_N^{\prime}
\mathbf{P}_0 (A )^{-1}
\mathfrak{P}_h(x)
\Bigg)
\\
&
\quad
+
\mathrm{Var}
\left(
\int_{A}^1
\varphi_{0}\left(  \alpha\right) 
\mathbf{1}_N^{\prime}\mathbf{P}_0 (\alpha )^{-1}\mathfrak{P}_h(x)
d\alpha
\right)
\\
&
\quad
-
2
\mathrm{Cov}
\Bigg(
\int_{A}^{1}
\varphi_{1}\left(  \alpha\right) 
\mathbf{1}_N^{\prime}
\mathbf{P}_0 (\alpha )^{-1}
\mathbf{P}_{1} (\alpha )
\mathbf{P}_0 (\alpha )^{-1}
\mathfrak{P}_h(x)
d\alpha,
\int_{A}^{1}
\varphi_{0}\left(  \alpha\right) 
\mathbf{1}_N^{\prime}
\mathbf{P}_0 (\alpha )^{-1}
\mathfrak{P}_h(x)
d\alpha
\Bigg)
\\
&
\quad
+
\mathrm{Var}
\Bigg(
\int_{A}^{1}
\varphi_{1}\left(  \alpha\right) 
\mathbf{1}_N^{\prime}
\mathbf{P}_0 (\alpha )^{-1}
\mathbf{P}_{1} (\alpha )
\mathbf{P}_0 (\alpha )^{-1}
\mathfrak{P}_h(x)
d\alpha
\Bigg)
+o(1).
\end{align*}
This gives
\begin{align*}
&\operatorname*{Var}\left(  \sqrt{LIh^{D_{\mathcal{M}}}}\widehat{\boldsymbol{I}%
}_{\varphi}\left(  x|I\right)  \right)
\\
&
\quad
=
\mathrm{Var}
\Bigg[
\int_{A}^1
\Big(
\varphi_{0}\left(  \alpha\right) 
\mathbf{1}_N^{\prime}\mathbf{P}_0 (\alpha )^{-1}
-
\varphi_{1}\left(  \alpha\right) 
\mathbf{1}_N^{\prime}
\mathbf{P}_0 (\alpha )^{-1}
\mathbf{P}_{1} (\alpha )
\mathbf{P}_0 (\alpha )^{-1}
\Big)
\mathfrak{P}_h(x)
d\alpha
\\
&
\quad\quad\quad\quad\quad\quad\quad\quad\quad\quad\quad\quad
\quad\quad\quad\quad\quad\quad\quad\quad\quad\quad\quad\quad
+
\varphi_{1}\left(  A\right) 
\mathbf{1}_N^{\prime}\mathbf{P}_0 (A )^{-1}\mathfrak{P}_h(x)
\Bigg]
+
o(1).
\end{align*}
Observe now that
\[
\partial_{\alpha}\left[  \varphi_{1}\left(  \alpha\right)
\mathbf{P}_{0}\left(  \alpha\right)  ^{-1}\right]  
=
\left[
\partial_{\alpha} \varphi_{1}\left(  \alpha\right)  
\right]
\mathbf{P}_{0}
\left(\alpha\right)^{-1}
-
\varphi_{1}\left(  \alpha\right)  
\mathbf{P}_{0}\left(\alpha\right)^{-1}
\mathbf{P}_{1}\left(  \alpha\right)  
\mathbf{P}_{0}\left(\alpha\right)^{-1}
\]
so that
\begin{align*}
& 
- 
\int_{A}^{1}
\varphi_{1}\left(  \alpha\right)
\mathbf{P}_{0}\left(  \alpha\right)^{-1}
\mathbf{P}_{1}\left(  \alpha\right)
\mathbf{P}_{0}\left(  \alpha\right)^{-1}  
d\alpha
+
\varphi_{1}\left(A\right)  \mathbf{P}_{0}\left(  A\right)  ^{-1}\\
&
\quad
=
\int_{A}^{1}
\left(
\left[  \partial_{\alpha} \varphi_{1}\left(  \alpha\right)
\mathbf{P}_{0}\left(  \alpha\right)  ^{-1}\right]  
-
\left[
\partial_{\alpha} \varphi_{1}\left(  \alpha\right)
\right]  
\mathbf{P}_{0}
\left(\alpha\right)^{-1}
\right)
d\alpha
+
\varphi_{1}\left(A\right)  \mathbf{P}_{0}\left(  A\right)  ^{-1}
\\
&  
\quad
=
-
\int_{A}^{1} 
\left[
\partial_{\alpha}
\varphi_{1}\left(  \alpha\right)  
\right]
\mathbf{P}_{0}\left(  \alpha\right)  ^{-1}d\alpha
+\varphi_{1}\left(
1\right)  \mathbf{P}_{0}\left(  1\right)  ^{-1}
\\
& 
\quad
=
\int_{0}^{A} 
\left[
\partial_{\alpha}
\varphi_{1}\left(  \alpha\right)  
\right]
\mathbf{P}_{0}\left(  \alpha\right)^{-1}
d\alpha
-
\int_{0}^{1} 
\left[
\partial_{\alpha}
\varphi_{1}\left(  \alpha\right)  
\right]
\mathbf{P}_{0}\left(  \alpha\right)^{-1}
d\alpha
+
\varphi_{1}\left(
1\right)  \mathbf{P}_{0}\left(  1\right)^{-1}
.
\end{align*}
It then follows
\begin{align*}
&\operatorname*{Var}\left(  \sqrt{LIh^{D_{\mathcal{M}}}}\widehat{\boldsymbol{I}%
}_{\varphi}\left(  x|I\right)  \right)
\\
&
\quad
=
\mathrm{Var}
\Bigg[
\int_{0}^{A}
\Big(
\varphi_{0}\left(  \alpha|x\right) 
-
\partial_{\alpha}
\varphi_{1}\left(  \alpha|x\right) 
\Big)
\mathbf{1}_N^{\prime}\mathbf{P}_0 (\alpha |I)^{-1}
\mathbf{P} (I)^{1/2}
h^{D_{\mathcal{M}}/2} P(x)
d\alpha
\Bigg]+o(1)
\end{align*}
which is equal to
$
\sigma_L^2(x)+o(1)
$
as stated in the Lemma. The study of 
$
\operatorname*{Var}\left(  \sqrt{LI}
\int_{\mathcal{X}}
\widehat{\boldsymbol{I}%
}_{\varphi}\left(  x|I\right) dx  \right)
$
is similar, changing $h^{D_{\mathcal{M}}/2} P(x)$ into $P(x)$ and allowing for vector functions $f(\alpha|x_f)$ and $g(\alpha|x_g)$ and integrating with respect to $(x_f,x_g)$ in the key equation 
(\ref{ChPP}), which would then hold with a remainder term
$o(h^2) \left\| \int_{\mathcal{X}} |P(x)| dx \right\|_2^2 = o(h^2)$ under Assumption \ref{Riesz}-(i).
\hfill$\Box$

Consider two real valued continuous functions $\mathcal{F}_{0}\left(
b_{0},b_{1}\right)  $ and $\mathcal{F}_{1}\left(  b_{0},b_{1}\right)  $.
Define%
\begin{align*}
\varphi_{0}\left(  \alpha|x,I\right)   &  =\mathcal{F}_{0}\left(  B\left(
\alpha|x,I\right)  ,B^{\left(  1\right)  }\left(  \alpha|x,I\right)  \right)
,\quad\varphi_{1}\left(  \alpha|x,I\right)  =\mathcal{F}_{1}\left(  B\left(
\alpha|x,I\right)  ,B^{\left(  1\right)  }\left(  \alpha|x,I\right)  \right)
,\\
\widehat{\boldsymbol{I}}_{\mathcal{F}}\left(  x|I\right)   &  =\int_{0}%
^{1}\left[  \varphi_{0}\left(  \alpha|x,I\right)  S_{0}+\varphi
_{1}\left(  \alpha|x,I\right)  \frac{S_{1}}{h}\right]  \otimes
P^{\prime}\left(  x\right) \\
&  \quad\quad\quad\times\left[  \overline{\mathsf{R}}^{\left(  2\right)
}\left(  \overline{\mathsf{b}}\left(  \alpha|I\right)  ;\alpha,I\right)
\right]  ^{-1}\widehat{\mathsf{R}}^{\left(  1\right)  }\left(  \overline
{\mathsf{b}}\left(  \alpha|I\right)  ;\alpha,I\right)  d\alpha.
\end{align*}
A condition ensuring that the variances $\sigma_{L}^{2}\left(  x|I\right)  $
and $\sigma_{L}^{2}\left(  I\right)  $ of Lemma \ref{VarI} do not vanish is
(\ref{Condvar}), that is for all $x$ in $\mathcal{X}$,
\[
\varphi_{0}\left(  \alpha|x,I\right)  -\partial_{\alpha} \varphi_{1}\left(
\alpha|x,I\right)  \neq0.
\]

\begin{proposition}
\label{CLTint}Suppose  that Assumptions
\ref{Auct.A}, \ref{Kernel.A}, \ref{Spec.A} and \ref{Riesz} hold. Assume
that $\varphi_{0}\left(  \alpha|x\right)  $, $\varphi_{1}\left(
\alpha|x\right)  $ and $\partial_{\alpha} \varphi_{1}\left(  \alpha|x\right)
$are continuous functions in $\left(  \alpha,x\right)
\in\left[  0,1\right]  \times\mathcal{X}$. Let $\sigma_{L}\left(  x|I\right)
$ and $\sigma_{L}\left(  I\right)  $ be as in Lemma \ref{VarI}.

Then if (\ref{Condvar}) holds for some $\alpha$ in $\left[  0,1\right]  $ and
if $Lh^{D_{\mathcal{M}}+2}$ diverges, $\sqrt{LIh^{D_{\mathcal{M}}}%
}\widehat{\boldsymbol{I}}_{\mathcal{F}}\left(  x|I\right)  /\sigma_{L}\left(
x|I\right)  $ converges in distribution to a standard normal. If
(\ref{Condvar}) holds for some $\left(  \alpha,X\right)  $ of $\left[
0,1\right]  \times\mathcal{X}$ and $Lh^{2}$ diverges, $\sqrt{LI}%
\int_{\mathcal{X}}\widehat{\boldsymbol{I}}_{\mathcal{F}}\left(  x|I\right)
dx/\sigma_{L}$ converges in distribution to a standard normal.
\end{proposition}

\paragraph{Proof of Proposition \ref{CLTint}.}

The eigenvalues of $\mathbf{P}_{0}\left(  \alpha\right)  ^{-1}$,
$\mathbf{P}_{1}\left(  \alpha\right)  $ and $\mathbf{P}$ are bounded uniformly
in $h$ and $\alpha$ by Assumptions \ref{Riesz} and \ref{Spec.A}, and
$\left\Vert h^{D_{\mathcal{M}}/2}P\left(  x\right)  \right\Vert $ is bounded
away from $0$ and infinity by Assumptions \ref{Riesz} and \ref{Kernel.A}. Then
if (\ref{Condvar}) holds for some $\alpha$, $\sigma_{L}^{2}\left(  x|I\right)
$ is bounded away from $0$ and infinity and the exact order of
$\operatorname*{Var}\left(  \widehat{\boldsymbol{I}}_{\mathcal{F}}\left(
x|I\right)  \right)  $ is $1/LIh^{D_{\mathcal{M}}}$. We now check the
Lyapounov condition. Write $\widehat{\mathsf{R}}^{\left(  1\right)  }\left(
\alpha\right)  =\frac{1}{LI}\sum_{\ell=1}^{L}\mathbb{I}\left[  I_{\ell
}=I\right]  r_{\ell}\left(  \alpha\right)  $, with%
\[
r_{\ell}\left(  \alpha\right)  =\sum_{i=1}^{I_{\ell}}\mathsf{\int%
_{-\frac{\alpha}{h}}^{\frac{1-\alpha}{h}}}\left\{  \mathbb{I}\left(  B_{i\ell
}\leq P\left(  X_{\ell},t\right)  ^{\prime}\overline{\mathsf{b}}\left(
\alpha|I\right)  \right)  -\left(  \alpha+ht\right)  \right\}  \pi\left(
t\right)  \otimes P\left(  X_{\ell}\right)  K\left(  t\right)  dt.
\]
This gives, since the eigenvalues of $\overline{\mathsf{R}}^{\left(  2\right)
}\left(  \alpha\right)  $ are asymptotically bounded from $0$ by Lemma
\ref{R2} and (\ref{Bar2*b}),
\begin{align*}
&  \mathbb{E}\left[  \left\vert \int_{0}^{1}\left[  \varphi_{0}\left(
\alpha|x,I\right)  S_{0}+\varphi_{1}\left(  \alpha|x,I\right)
\frac{S_{1}}{h}\right]  \otimes P^{\prime}\left(  x\right)  \left[
\overline{\mathsf{R}}^{\left(  2\right)  }\left(  \alpha\right)  \right]
^{-1}\frac{r_{\ell}\left(  \alpha\right)  -\mathbb{E}\left[  r_{\ell}\left(
\alpha\right)  \right]  }{LI}d\alpha\right\vert ^{3}\right] \\
&  \quad\leq C\frac{h^{-1}\max_{x\in\mathcal{X}}\left\Vert P\left(  x\right)
\right\Vert ^{2}}{\left(  LI\right)  ^{3}}LI\operatorname*{Var}\left(
\widehat{\boldsymbol{I}}_{\mathcal{F}}\left(  x|I\right)  \right)  =\frac
{C}{L^{2}h^{D_{\mathcal{M}}+1}}\operatorname*{Var}\left(
\widehat{\boldsymbol{I}}\left(  x|I\right)  \right)  .
\end{align*}
$Lh^{D_{\mathcal{M}}+2}\rightarrow\infty$ implies that the Lyapounov condition
holds since%
\[
\frac{C}{Lh^{D_{\mathcal{M}}+1}\operatorname*{Var}^{3/2}\left(
\widehat{\boldsymbol{I}}_{\mathcal{F}}\left(  x|I\right)  \right)
}\operatorname*{Var}\left(  \widehat{\boldsymbol{I}}_{\mathcal{F}}\left(
x|I\right)  \right)  =O\left(  \frac{1}{\left(  Lh^{D_{\mathcal{M}}+2}\right)
^{1/2}}\right)  \rightarrow0
\]
This implies that $\widehat{\boldsymbol{I}}_{\mathcal{F}}\left(  x|I\right)
/\operatorname*{Var}^{1/2}\left(  \widehat{\boldsymbol{I}}_{\mathcal{F}%
}\left(  x|I\right)  \right)  $ is asymptotically $\mathcal{N}\left(
0,1\right)  $, and then the stated asymptotic normality.

For $\sqrt{LI}\int_{\mathcal{X}}\widehat{\boldsymbol{I}}_{\mathcal{F}}\left(
x|I\right)  dx$, recall that $\left\Vert \int\left\vert P\left(  x\right)
\right\vert dx\right\Vert =O\left(  1\right)  $ by Assumption \ref{Riesz}.
This also gives%
\begin{align*}
&  \mathbb{E}\left[  \left\vert \int_{\mathcal{X}}\left[  \int_{0}^{1}\left(
\varphi_{0}\left(  \alpha|X,I\right)  S_{0}+\varphi_{1}\left(
\alpha|X,I\right)  \frac{S_{1}}{h}\right)  \otimes P^{\prime}\left(
x\right)  \right]  \left[  \overline{\mathsf{R}}^{\left(  2\right)  }\left(
\alpha\right)  \right]  ^{-1}\frac{r_{\ell}\left(  \alpha\right)
-\mathbb{E}\left[  r_{\ell}\left(  \alpha\right)  \right]  }{LI}%
d\alpha\right\vert ^{3}\right] \\
&  \quad\leq C\frac{h^{-1}}{\left(  LI\right)  ^{3}}LI\operatorname*{Var}%
\left(  \int_{\mathcal{X}}\widehat{\boldsymbol{I}}_{\mathcal{F}}\left(
x|I\right)  dx\right)  =\frac{C}{L^{2}h}\operatorname*{Var}\left(
\int_{\mathcal{X}}\widehat{\boldsymbol{I}}_{\mathcal{F}}\left(  x|I\right)
dx\right)  .
\end{align*}
Therefore the Lyapounov condition holds since $Lh^{2}$ diverges, because
\[
\frac{C}{Lh\operatorname*{Var}^{3/2}\left(  \int_{\mathcal{X}}%
\widehat{\boldsymbol{I}}_{\mathcal{F}}\left(  x|I\right)  dx\right)
}\operatorname*{Var}\left(  \int_{\mathcal{X}}\widehat{\boldsymbol{I}%
}_{\mathcal{F}}\left(  x|I\right)  dx\right)  =\frac{C}{\left(  Lh^{2}\right)
^{1/2}}\rightarrow0
\]
The rest of the proof is as above.$\hfill\square$

\paragraph{Proof of Theorems \ref{FuncCLT} and \ref{A.FuncCLT}.}

Let $\widehat{\mathsf{d}}\left(  \alpha|I\right)  $ and $\widehat{\mathsf{e}%
}\left(  \alpha|I\right)  $ be as in (\ref{Let}) and (\ref{Llt}),%

\begin{align*}
\widehat{\mathsf{e}}\left(  \alpha|I\right)   &  =-\left(  \overline
{\mathsf{R}}^{\left(  2\right)  }\left(  \overline{\mathsf{b}}\left(
\alpha|I\right)  ;\alpha,I\right)  \right)  ^{-1}\widehat{\mathsf{R}}^{\left(
1\right)  }\left(  \overline{\mathsf{b}}\left(  \alpha|I\right)
;\alpha,I\right)  ,\\
\widehat{\mathsf{d}}\left(  \alpha|I\right)   &  =\widehat{\mathsf{b}}\left(
\alpha|I\right)  -\overline{\mathsf{b}}\left(  \alpha|I\right)
-\widehat{\mathsf{e}}\left(  \alpha|I\right)  .
\end{align*}
Let $\widehat{\boldsymbol{I}}_{\mathcal{F}}\left(  x|I\right)  $ be as above,
replacing $\varphi_{j}\left(  \cdot\right)  $ with $\varphi_{jI}\left(
\cdot\right)  $, $j=0,1$. Then the second-order Taylor inequality gives%
\begin{align*}
&  \widehat{\theta}\left(  x\right)  -\theta\left(  x\right) \\
&  \quad=\sum_{I\in\mathcal{I}}\int_{0}^{1}\left[  \varphi_{0I}\left(
\alpha,x\right)  \left(  \overline{B}\left(  \alpha|x,I\right)  -B\left(
\alpha|x,I\right)  \right)  +\varphi_{1I}\left(  \alpha,x\right)  \left(
\overline{B}^{\left(  1\right)  }\left(  \alpha|x,I\right)  -B^{\left(
1\right)  }\left(  \alpha|x,I\right)  \right)  \right]  d\alpha\\
&  \quad+\sum_{I\in\mathcal{I}}\widehat{\boldsymbol{I}}_{\mathcal{F}}\left(
x|I\right) \\
&  \quad+\sum_{I\in\mathcal{I}}\int_{0}^{1}\left[  \left(  \varphi_{0I}\left(
\alpha,x\right)  S_{0}+\varphi_{1I}\left(  \alpha,x\right)
\frac{S_{1}}{h}\right)  \otimes P^{\prime}\left(  x\right)  \right]
\widehat{\mathsf{d}}\left(  \alpha|I\right)  d\alpha\\
&  \quad+O\left(  1\right)  \sup_{\left(  \alpha,x,I\right)  \in\left[
0,1\right]  \times\mathcal{X\times I}}\left[  \left(  \overline{B}\left(
\alpha|x,I\right)  -B\left(  \alpha|x,I\right)  \right)  ^{2}+\left(
\overline{B}^{\left(  1\right)  }\left(  \alpha|x,I\right)  -B^{\left(
1\right)  }\left(  \alpha|x,I\right)  \right)  ^{2}\right] \\
&  \quad O\left(  1\right)  \sup_{\left(  \alpha,x,I\right)  \in\left[
0,1\right]  \times\mathcal{X\times I}}\left[  \left(  \left[  S_{0}\otimes P^{\prime}\left(  x\right)  \right]  \widehat{\mathsf{e}}\left(
\alpha|I\right)  \right)  ^{2}+\left(  \left[  \frac{S_{1}}{h}\otimes
P^{\prime}\left(  x\right)  \right]  \widehat{\mathsf{e}}\left(
\alpha|I\right)  \right)  ^{2}\right] \\
&  \quad O\left(  1\right)  \sup_{\left(  \alpha,x,I\right)  \in\left[
0,1\right]  \times\mathcal{X\times I}}\left[  \left(  \left[  S_{0}\otimes P^{\prime}\left(  x\right)  \right]  \widehat{\mathsf{d}}\left(
\alpha|I\right)  \right)  ^{2}+\left(  \left[  \frac{S_{1}}{h}\otimes
P^{\prime}\left(  x\right)  \right]  \widehat{\mathsf{d}}\left(
\alpha|I\right)  \right)  ^{2}\right]  .
\end{align*}
The second-order Taylor inequality, Theorems \ref{Bias} and \ref{Baha}, Lemma \ref{Leadterm} give, 
\begin{align*}
\widehat{\theta}\left(  x\right)  -\theta\left(  x\right)   &  =o\left(
h^{s}\right)  +\sum_{I\in\mathcal{I}}\widehat{\boldsymbol{I}}_{\mathcal{F}%
}\left(  x|I\right) \\
&  +\frac{1}{\left(  Lh^{D_{\mathcal{M}}+1}\right)  ^{1/2}}O_{\mathbb{P}}\left(
\frac{\log L}{\left(  Lh^{3D_{\mathcal{M}}+2  }\right)  ^{1/2}}+\frac{\log L}{\left(  Lh^{D_{\mathcal{M}+!}%
+2}\right)  ^{1/2}}\right) \\
&  =o\left(  h^{s}\right)  +\sum_{I\in\mathcal{I}}\widehat{\boldsymbol{I}%
}_{\mathcal{F}}\left(  x|I\right)  +o_{\mathbb{P}}\left(  \frac{1}{\left(
Lh^{D_{\mathcal{M}}}\right)  ^{1/2}}\right)  
\end{align*}
as $\frac{\log^2 L}{L h^{3D_{\mathcal{M}}+3  } }=o(1)$.

Proposition \ref{CLTint} then gives the result since the
$\widehat{\boldsymbol{I}}_{\mathcal{F}}\left(  x|I\right)  $ are independent.
The asymptotic normality of $\widehat{\theta}$ similarly follows from
Assumption \ref{Riesz}, which gives $\left\Vert \int_{\mathcal{X}}\left\vert
P\left(  x\right)  \right\vert dx\right\Vert =O\left(  1\right)  $, and
Theorem \ref{Baha} which implies%
\begin{align*}
\widehat{\theta}-\theta &  =o\left(  h^{s}\right)  +\sum_{I\in\mathcal{I}}%
\int_{\mathcal{X}}\widehat{\boldsymbol{I}}_{\mathcal{F}}\left(  x|I\right)
dx\\
&  +O\left(  \frac{\sup_{\alpha\in\left[  0,1\right]  }\left\Vert
\widehat{\mathsf{d}}\left(  \alpha|I\right)  \right\Vert }{h}\right)
+\frac{1}{L^{1/2}}O_{\mathbb{P}}\left(  \frac{\log L}{\left(
Lh^{4D_{\mathcal{M}}+3}\right)  ^{1/2}}\right) \\
&  =o\left(  h^{s}\right)  +\sum_{I\in\mathcal{I}}\int_{\mathcal{X}%
}\widehat{\boldsymbol{I}}_{\mathcal{F}}\left(  x|I\right)  dx+o_{\mathbb{P}%
}\left(  \frac{1}{L^{1/2}}\right)   .
\end{align*}
Theorems \ref{FuncCLT} and \ref{A.FuncCLT} follow from Proposition \ref{CLTint}. \hfill $\Box$

\pagebreak

\renewcommand{\thesection}{Appendix \Alph{section}}\setcounter{section}{5}

\section{\!\!\!\!\!\! - Proofs of preliminary results \label{App:Proofsprelim}}

\renewcommand{\thesection}{F.\arabic{section}}
\renewcommand{\thetheorem}{F.\arabic{theorem}}
\renewcommand{\thefootnote}{F.\arabic{footnote}}
\setcounter{footnote}{0}
\renewcommand{\theequation}{F.\arabic{equation}}\setcounter{section}{0}
\setcounter{equation}{0} \setcounter{theorem}{0} 

\section{Proposition \ref{SeriesB2} and Lemma \ref{Invband}}

These two results make use of Theorem 2.2 in Demko (1977):

\begin{theorem}[Demko (1977)]
	\label{Inv}
	Let $A=[a_{ij}, 1 \leq i,j \leq n]$ be a $n\times n$ $c$-band matrix, $a_{ij}=0$ if $|i-j| \geq c/2$.
	Suppose that $\|A\|$ and $\|A^{-1}\|$ are both less than $\kappa<\infty$. Then there are constants $0<\varrho<1$ and $C$, which only depend upon $c$ and $\kappa$ such that the entries $a^{ij}$ of $A^{-1}$ satisfy
	\[
	\left| a^{ij} \right| \leq C \cdot \varrho^{|i-j|}
	\text{ for all $1\leq i,j \leq n$.}
	\]
\end{theorem}

As using Lemma \ref{Invband} shortens the proof of Proposition \ref{SeriesB2}, this lemma is established in the first place.

\subsection{Proof of Lemma \ref{Invband}}
Note that the rows and the columns of the permutation matrix $S$ and $S^{-1}$ have a unique entry equal to $1$, all other being set to $0$, so that  for $k=2,\infty$, $\| S^{-1} M(\alpha) S^{-1} \|_{k} = \|  M(\alpha)  \|_{k}$ and $\| S M(\alpha)^{-1} S \|_{k} = \|  M(\alpha)^{-1} \|_{k}$. Hence one can assume without loss of generality that the $M(\alpha)$'s are $c$-band matrix. Let $m^{ij} (\alpha)$ be the entries of $M(\alpha)^{-1}$. Then Theorem \ref{Inv} gives that there are a $C$ and a $\varrho \in (0,1)$, which only depend upon $c$ and $C_0$, such that $|m^{ij} (\alpha)| \leq C \varrho^{|i-j|}$. Recall now that
\[
\left\|
M(\alpha)^{-1}
\right\|
=
\max_{i} \sum_{j} \left| m^{ij} (\alpha)\right|.
\]
But $\sum_{j} \left| m^{ij} (\alpha)\right| \leq C\left(1+2\sum_{k=1}^{\infty} \varrho^k \right)=C \left(1+\frac{2\varrho}{1-\varrho}\right)$, and the Lemma is proven. \hfill $\Box$

\subsection{Proof of Proposition \ref{SeriesB2}}

\subparagraph{Proof of (\ref{Vsqr}).}
Note that $\gamma (\cdot)$ is differentiable with, by the Lebesgue Dominated Convergence Theorem
\[
\gamma^{(p)} (\alpha|I)
=
\left(
\mathbb{E}
\left[
\left.
P(X )
P(X)^{\prime}
\right|
I
\right]
\right)^{-1}
\mathbb{E}
\left[
\left.
P(X )
V^{(p)}(\alpha|X,I)
\right|
I
\right]
,
\quad p=0,\ldots,s+1,
\]
so that $\gamma^{(s+1)} (\cdot|I)$ is also continuous.

By the Approximation Property S, there exists some $N \times 1$ $\gamma_p (\cdot)$ such that 
$r_p (\alpha|x) = P(x)^{\prime} \gamma_p (\alpha)-V^{(p)} (\alpha|x,I) $  satisfies\footnote{Using for instance a suitable expansion of $V(\cdot|\cdot,I)$ over $[0,1] \times \mathcal{X}_{\epsilon}$.}, for $p=0,\ldots,s+1$,
\[
\sup_{\alpha \in [0,1]}
\max_{x \in [0,1]^D}
\left|
r_p (\alpha|x)
\right|
=
h^{s+1-p}
O
\left(
\sum_{\mathbf{j}}
\sup_{\alpha \in [0,1]} \textrm{mc}_{s+1-p} \left(V^{(p)}_{\mathbf{j}}(\alpha|x,I);h \right)
\right)
=
o(h^{s+1-p}).
\]
Hence, for all $x$ of $\mathcal{X}$ and by the sieve disjoint support property,
\begin{eqnarray*}
	\lefteqn{
		\left|
		P(x)^{\prime} \gamma^{(p)} (\alpha|I)
		-
		P(x)^{\prime} \gamma_p (\alpha)
		\right|
	}
	\\
	& = &
	\left|
	P(x)^{\prime} 
	\left(
	\mathbb{E}
	\left[
	\left.
	P(X )
	P(X)^{\prime}
	\right|
	I
	\right]
	\right)^{-1}
	\mathbb{E}
	\left[
	\left.
	P(X )
	\left( P(X)^{\prime} \gamma_p (\alpha) + r_p (\alpha|X) \right)
	\right|
	I
	\right]
	-
	P(x)^{\prime} \gamma_p (\alpha)
	\right|
	\\
	& = &
	\left|
	P(x)^{\prime} 
	\left(
	\mathbb{E}
	\left[
	\left.
	P(X )
	P(X)^{\prime}
	\right|
	I
	\right]
	\right)^{-1}
	\mathbb{E}
	\left[
	\left.
	P(X )
	r_p (\alpha|X) 
	\right|
	I
	\right]
	\right|
	\\
	& \leq &
	Ch^{-D_{\mathcal{M}}/2}
	\left\| \left(
	\mathbb{E}
	\left[
	\left.
	P(X )
	P(X)^{\prime}
	\right|
	I
	\right]
	\right)^{-1}
	\mathbb{E}
	\left[
	\left.
	P(X )
	r_p (\alpha|X) 
	\right|
	I
	\right]
	\right\|_{\infty}
	\\
	& \leq &
	Ch^{-D_{\mathcal{M}}/2}
	\left\| \left(
	\mathbb{E}
	\left[
	\left.
	P(X )
	P(X)^{\prime}
	\right|
	I
	\right]
	\right)^{-1}
	\right\|_{\infty}
	\left\|
	\mathbb{E}
	\left[
	\left.
	P(X )
	r_p (\alpha|X) 
	\right|
	I
	\right]
	\right\|_{\infty}
	.
\end{eqnarray*}
As the eigenvalues of $\int_{\mathcal{X}} P(x)P(x)^{\prime}dx$ are bounded away from $0$ and $\infty$ by Assumption \ref{Riesz}-(i), so are the ones of the $c/2$ band matrix
$
\mathbb{E}
\left[
\left.
P(X )
P(X)^{\prime}
\right|
I
\right]
$
by Assumption \ref{Auct.A}, and Lemma \ref{Invband} gives
$
\left\| \left(
\mathbb{E}
\left[
\left.
P(X )
P(X)^{\prime}
\right|
I
\right]
\right)^{-1}
\right\|_{\infty}
\leq C
$.
As 
$
\left\|
\mathbb{E}
\left[
\left.
P(X )
r_p (\alpha|X) 
\right|
I
\right]
\right\|_{\infty}
\leq
o \left(h^{s+1-p}\right)
\sup_{1 \leq n \leq N} \int_{\mathcal{X}} |P_n(x)| dx$
with
$\sup_{1 \leq n \leq N} \int_{\mathcal{X}} |P_n(x)| dx \leq C h^{D_{\mathcal{M}}/2}$ by Assumption \ref{Riesz}-(i), (\ref{Vsqr}) follows from:
\[
\max_{(\alpha,x)\in [0,1] \times \mathcal{X}}
\left|
P(x)^{\prime} \gamma^{(p)} (\alpha|I)
-
P(x)^{\prime} \gamma_p (\alpha)
\right|
\leq
Ch^{-D_{\mathcal{M}}/2} h^{D_{\mathcal{M}}/2}
o \left(h^{s+1-p}\right)
=
o \left(h^{s+1-p}\right).
\]

\subparagraph{Proof of (i).}
By (\ref{V2B}), $B\left(  \alpha|x,I\right)  =\left(  I-1\right)  \int_{0}%
^{1}u^{I-2}V\left(  \alpha u|x,I\right)  du$, so that $B^{\left(  1\right)
}\left(  \alpha|x,I\right)  =\left(  I-1\right)  \int_{0}^{1}u^{I-1}V^{\left(
	1\right)  }\left(  \alpha u|x,I\right)  du$ which implies the two first
statements in (i) about lower and upper bounds for $B^{\left(  1\right)
}\left(  \alpha|x,I\right)  $ and that $B\left(  \cdot|\cdot,I\right)  $ is
$\left(  s+1\right)  $th continuously differentiable. That $B\left(
\cdot|x,I\right)  $ is $(s+2)$th continuously differentiable over $\left(
0,1\right]  $ follows from its integral expression (\ref{V2B}). Observe now
that for $p=1,\ldots,s+2$%
\[
\partial_{\alpha}^{p}
\left[  \alpha B\left(  \alpha|x,I\right)  \right]
=\alpha B^{\left(  p\right)  }\left(  \alpha|x,I\right)
+pB^{\left(  p-1\right)  }\left(  \alpha|x,I\right)
\]
with, for $p=1,\ldots,s+1$
\begin{align*}
B^{\left(  p\right)  }\left(  \alpha|x,I\right)   &  =\left(  I-1\right)
\int_{0}^{1}u^{I-2+p}V^{\left(  p\right)  }\left(  \alpha u|x,I\right)
du=\frac{I-1}{\alpha^{I-1+p}}\int_{0}^{\alpha}t^{I-2+p}V^{\left(  p\right)
}\left(  t|x,I\right)  dt\\
B^{\left(  p+1\right)  }\left(  \alpha|x,I\right)   &  =-\frac{\left(
	I-1\right)  \left(  I-1+p\right)  }{\alpha^{I+p}}\int_{0}^{\alpha}%
t^{I-2+p}V^{\left(  p\right)  }\left(  t|x,I\right)  dt+\frac{\left(
	I-1\right)  V^{\left(  p\right)  }\left(  \alpha|x,I\right)  }{\alpha}\\
&  =-\frac{I-1+p}{\alpha}B^{\left(  p\right)  }\left(  \alpha|x,I\right)
+\frac{\left(  I-1\right)  V^{\left(  p\right)  }\left(  \alpha|x,I\right)
}{\alpha}.
\end{align*}
Hence, when $\alpha$ goes to $0$%
\begin{align*}
&  \alpha B^{\left(  s+2\right)  }\left(  \alpha|x,I\right)  =-\left(
I+s\right)  B^{\left(  s+1\right)  }\left(  0|x,I\right)  +\left(  I-1\right)
V^{\left(  s+1\right)  }\left(  0|x,I\right)  +o\left(  1\right) \\
&  \quad=-\left(  I+s\right)  \left(  I-1\right)  \int_{0}^{1}u^{I+s-1}%
V^{\left(  s+1\right)  }\left(  0|x,I\right)  du+\left(  I-1\right)
V^{\left(  s+1\right)  }\left(  0|x,I\right)  +o\left(  1\right) \\
&  \quad=o\left(  1\right)
\end{align*}
uniformly with respect to $x$. As $\partial^{s+2} [\alpha B(\alpha|x,I)] = \alpha B^{(s+2)} (\alpha|x,I) + (s+2)
B^{(s+1)} (\alpha|x,I)$, it follows that $\alpha \in [0,1] \mapsto \alpha B(\alpha|x,I)$ is $(s+2)$ times continuously differentiable.

\subparagraph{Proof of (ii).}  For $\beta(\cdot|I)$ as in (\ref{Bsqr})%
\[
\beta^{\left(  p\right)  }\left(  \alpha|I\right)  =\left(  I-1\right)
\int_{0}^{1}u^{I+p-2}\gamma^{\left(  p\right)  }\left(  \alpha u|I\right)
du,\quad p=0,\ldots,s+1
\]
and%
\begin{align*}
&  
\sup_{\left(  \alpha,x\right)  \in\left[  0,1\right]  \times\mathcal{X}%
}\left\vert 
P(x)^{\prime} \beta^{(p)} (\alpha|I)
-
B^{\left(  p\right)  }\left(  \alpha|x,I\right)   \right\vert \\
&  
\quad
=
\sup_{\left(  \alpha,x\right)  \in\left[  0,1\right]  \times \mathcal{X}}
\left\vert 
\left(  I-1\right)  
\int_{0}^{1}
u^{I+p-2}
\left(
P(x)^{\prime}
\gamma^{(p)} \left( \alpha u |I \right)
-
V^{\left(  p\right)  }\left(  \alpha u|x,I\right)  
\right)  du\right\vert \\
&  
\quad
\leq
\sup_{\left(  \alpha,x\right)  \in\left[  0,1\right]\times\mathcal{X}}
\left\vert
P(x)^{\prime}
\gamma^{(p)} (\alpha|I)
- 
V^{\left(  p\right)  }\left(  \alpha|x,I\right)
\right\vert
=
o
\left(h^{s+1-p}\right)
\end{align*}
by (\ref{Vsqr}),
which gives the sieve approximation result for $B\left(  \alpha|x,I\right)  $
in (ii).

\subparagraph{Proof of (iii).}
That $\alpha \beta^{(s+2)} (\alpha)$ is continuous over $(0,1]$ with $\lim_{\alpha \downarrow 0 } \alpha \beta^{(s+2)} (\alpha)=0$ can be established as for $\alpha B^{(s+2)} (\alpha|x,I)$.  
Observe that 
$\alpha\beta^{(1)} (\alpha|I)= (I-1) \left(\gamma(\alpha|I)-\beta(\alpha|I)\right)$
by (\ref{Bsqr}).
It follows
\begin{align*}
&  \sup_{\left(  \alpha,x\right)  \in\left[  0,1\right]  \times\mathcal{X}}
\left\vert 
\sum_{n=1}^{N}
\partial_{\alpha}^{p}
\left[  \alpha\beta_{n}^{\left(  1\right)  }\left(
\alpha|I\right)  \right]  P_{n}\left(  x\right)
-
\partial_{\alpha}^{p}
\left[  \alpha B^{\left(  1\right)  }\left(\alpha|x,I\right)  \right]  
\right\vert \\
&  
\quad
\leq
\left(  I-1\right)  
\sup_{\left(  \alpha,x\right)  \in\left[0,1\right]  \times\mathcal{X}}
\left\vert 
P(x)^{\prime} \gamma^{(p)} (\alpha|I)
-
V^{\left(  p\right)  }\left(\alpha|x,I\right)
\right\vert 
\\
&  
\quad
+
\left(  I-1\right)  
\sup_{\left(  \alpha,x\right)  \in\left[0,1\right]  \times\mathcal{X}}
\left\vert 
P(x)^{\prime} \beta^{(p)} (\alpha|I)
-
B^{\left(  p\right)  }\left(\alpha|x,I\right)    
\right\vert 
\\
&  
\quad
\leq
2\left(  I-1\right)  
\sup_{\left(  \alpha,x\right)  \in\left[0,1\right]  \times\mathcal{X}}
\left\vert 
P(x)^{\prime} \gamma^{(p)} (\alpha|I)
-
V^{\left(  p\right)  }\left(\alpha|x,I\right)  
\right\vert
= 
o
\left(h^{s+1-p}\right)
\end{align*}
by (\ref{Vsqr}),
which gives the approximation result for $\alpha B^{\left(  p+1\right)  }\left(
\alpha|x,I\right)  $, $p=0,\ldots,s+1$ in (iii) using
$\partial_{\alpha}^p\left[\alpha B^{(1)} (\alpha|x,I)\right]
=
\alpha B^{(p+1)} (\alpha|X,I)+(p+1)B^{(p)} (\alpha|x,I)
$
and (\ref{Bsqr2}). $\hfill\square$ 

\section{Lemmas \ref{Phipsi} and \ref{R2}}

\subsection{\textbf{Proof of Lemma \ref{Phipsi}}}

Consider the harder $ASQR$ case. We start with (iii). Recall that 
\[
\Psi(\alpha|x,\mathsf{b})
=
P(x,t)^{\prime} \mathsf{b}
=
P(x)^{\prime}
\sum_{p=0}^{s+1}
\frac{t^p}{p!}
\mathsf{b}_p
=
P(x)^{\prime}
\sum_{p=0}^{s+1}
\frac{\left(ht\right)^p}{p!}
\beta_p
\]
where the $1 \times N$ $\beta_p(\alpha|I)$ is equal to  $\beta^{(p)} (\alpha|I)$, $\beta (\cdot|I)$ as in (\ref{Bsqr}).

\subparagraph{Proof of (iii).}
Proposition \ref{SeriesB2}-(ii,i) gives, uniformly in $\alpha$, $t$ in $\mathcal{T}_{\alpha,h}$ and $x$ in $\mathcal{X}$
\begin{eqnarray*}
	\lefteqn{
	\Psi 
	\left(t \left| x, \mathsf{b} (\alpha|I)\right.\right)
	-
	B (\alpha+ht|x,I)
	= 
	\sum_{p=0}^{s+1}
	\frac{\left(ht\right)^p}{p!}
	P(x)^{\prime} \beta^{(p)} (\alpha) - B (\alpha+ht|x,I)}
	\\
	& = &
	\sum_{p=0}^{s+1}
	\frac{\left(ht\right)^p}{p!}
	\left(B^{(p)} (\alpha|x,I)+o\left(h^{s+1-p}\right)\right)
	- 
	B (\alpha+ht|x,I)
	\\
	& = &
	B (\alpha|x,I)
	-
	B (\alpha+ht|x,I)
	+
	\sum_{p=0}^{s+1}
	\frac{\left(ht\right)^p}{p!}
	B^{(p)} (\alpha|x,I)
	+
	o\left(h^{s+1}\right)
	=
	o(h^{s+1})
\end{eqnarray*}
by the Taylor expansion formula of order $s+1$, so that
\begin{equation}\max_{\left(  \alpha,x\right)  \in\left[  0,1\right]  \times\mathcal{X}%
}\max_{t\in\mathcal{T}_{\alpha,h}}\left\vert \Psi\left(  t|x,\mathsf{b}\left(  \alpha|I\right)  \right)  -B\left(  \alpha+ht|x,I\right)  \right\vert
=o\left(  h^{s+1}\right).
\label{B2psi*}
\end{equation}

For the next  result in (iii) and since $\alpha \mapsto \alpha B(\alpha|x,I)$ is $(s+2)$ times continuously differentiable over $[0,1]$ by Proposition \ref{SeriesB2}-(i), a Taylor expansion with integral remainder shows, observing that $\partial^{p}_{\alpha} \left[\alpha B(\alpha|x,I)\right]=\alpha B^{(p)}(\alpha|x,I) + pB^{(p-1)}(\alpha|x,I)$ 
\begin{eqnarray*}
\lefteqn{
	(\alpha+ht) B(\alpha+ht|x,I) 
	- 
	\alpha B(\alpha|x,I)
	-
	\sum_{p=1}^{s+2} \frac{(ht)^p}{p!}
	\left( \alpha B^{(p)}(\alpha|x,I) +  pB^{(p-1)}(\alpha|x,I)\right)
}
\\
& = &
\frac{(ht)^{s+2}}{(s+1)!}
\int_{0}^{1}
\left[
(\alpha+htu)
B^{(s+2)}
\left(\alpha+htu|x,I\right)
-
\alpha B^{(s+2)} (\alpha|x,I)
\right]
\left(1-u\right)^{s+1}du
\\
&&
+
\frac{(s+2)(ht)^{s+2}}{(s+1)!}
\int_{0}^{1}
\left[
B^{(s+1)} (\alpha+htu|x,I)
-
B^{(s+1)}
\left(\alpha|x,I\right)
\right]
\left(1-u\right)^{s+1}du
\end{eqnarray*}
for all $t$ in $\mathcal{T}_{\alpha,h}$. It also holds
\begin{eqnarray*}
	\lefteqn{
		ht B(\alpha+ht|x,I) - \sum_{p=1}^{s+2} \frac{(ht)^p}{p!} p B^{(p-1)} (\alpha|x,I)
	}
	\\
	& = &
	\frac{(ht)^{s+2}}{s!}
	\int_{0}^{1}
	\left[
	B^{(s+1)}
	\left(\alpha+htu|x,I\right)
	-
	B^{(s+1)}
	\left(\alpha|x,I\right)
	\right]
	\left(1-u\right)^{s}du.
\end{eqnarray*}
Combining the two Taylor expansions then gives
\begin{eqnarray*}
	\lefteqn{
		\alpha B(\alpha+ht|x,I) 
		- 
		\alpha B(\alpha|x,I)
		-
		\sum_{p=1}^{s+2} \frac{(ht)^p}{p!}
		\alpha B^{(p)}(\alpha|x,I)
	}
	\\
	& = &
	\frac{(ht)^{s+2}}{(s+1)!}
	\int_{0}^{1}
	\left[
	(\alpha+htu)
	B^{(s+2)}
	\left(\alpha+htu|x,I\right)
	-
	\alpha B^{(s+2)} (\alpha|x,I)
	\right]
	\left(1-u\right)^{s+1}du
	\\
	&&
	+
	\frac{(s+2)(ht)^{s+2}}{(s+1)!}
	\int_{0}^{1}
	\left[
	B^{(s+1)} (\alpha+htu|x,I)
	-
	B^{(s+1)}
	\left(\alpha|x,I\right)
	\right]
	\left(1-u\right)^{s+1}du
	\\
	&&
	-
	\frac{(ht)^{s+2}}{s!}
	\int_{0}^{1}
	\left[
	B^{(s+1)}
	\left(\alpha+htu|x,I\right)
	-
	B^{(s+1)}
	\left(\alpha|x,I\right)
	\right]
	\left(1-u\right)^{s}du.
\end{eqnarray*}
The Dominated Convergence Theorem then gives
\[
\max_{(\alpha,x)\in[0,1]\times\mathcal{X}}
\max_{t \in \mathcal{T}_{\alpha,h}}
\left|
\alpha B(\alpha+ht|x,I) 
- 
\alpha B(\alpha|x,I)
-
\sum_{p=1}^{s+2} \frac{(ht)^p}{p!}
\alpha B^{(p)}(\alpha|x,I)
\right|
=
o(h^{s+2}).
\]
As Proposition \ref{SeriesB2}-(iii) gives
$ P(x)^{\prime} \alpha \mathsf{b}_p (\alpha) = h^{p} \alpha B^{(p)} (\alpha|x,I) + o(h^{s+2})$ uniformly, the second result of (iii) is proven.

The third result in (iii) follows from Proposition \ref{SeriesB2}-(iii). The
fourth equality of (iii) follows from the first result of (iii), which implies
\begin{align*}
o\left(  h^{s+1}\right)   &  =\max_{\left(  \alpha,x\right)  \in\left[
	0,1\right]  \times\mathcal{X}}\max_{t\in\mathcal{T}_{\alpha,h}}\left\vert
\Psi\left(  t|x,\mathsf{b}\left(  \alpha|I\right)  \right)  -B\left(
\alpha+ht|x,I\right)  \right\vert \\
&  =\max_{\left(  \alpha,x\right)  \in\left[  0,1\right]  \times\mathcal{X}%
}\max_{u\in\Psi\left[  \mathcal{T}_{\alpha,h}|x,\mathsf{b}\left(
	\alpha|I\right)  \right]  }\left\vert \Psi\left[  \Delta\left(  u|x,\mathsf{b}\left(  \alpha|I\right)  \right)  |x,\mathsf{b}\left(
\alpha|I\right)  \right]  \right. \\
&  \quad\quad\quad\quad\quad\quad\quad\quad\quad\quad\quad\quad\quad\quad
\quad\quad\quad\quad\left.  -B\left[  \alpha+h\Delta\left(  u|x,\mathsf{b}\left(  \alpha|I\right)  \right)  |X,I\right]  \right\vert \\
&  =\max_{\left(  \alpha,x\right)  \in\left[  0,1\right]  \times\mathcal{X}%
}\max_{u\in\Psi\left[  \mathcal{T}_{\alpha,h}|x,\mathsf{b}\left(
	\alpha|I\right)  \right]  }\left\vert u-B\left[  \alpha+h\Delta\left(
u|x,\mathsf{b}\left(  \alpha|I\right)  \right)  |X,I\right]
\right\vert \\
&  =\max_{\left(  \alpha,x\right)  \in\left[  0,1\right]  \times\mathcal{X}%
}\max_{u\in\Psi\left[  \mathcal{T}_{\alpha,h}|x,\mathsf{b}\left(
	\alpha|I\right)  \right]  }\left\vert B\left[  \alpha+h\frac{G\left(
	u|X,I\right)  -\alpha}{h}|X,I\right]  \right. \\
&  \quad\quad\quad\quad\quad\quad\quad\quad\quad\quad\quad\quad\quad\quad
\quad\quad\quad\quad\left.  -B\left[  \alpha+h\Delta\left(  u|x,\mathsf{b}\left(  \alpha|I\right)  \right)  |X,I\right]  \right\vert \\
&  \geq Ch\max_{\left(  \alpha,x\right)  \in\left[  0,1\right]  \times
	\mathcal{X}}\max_{u\in\Psi\left[  \mathcal{T}_{\alpha,h}|x,\mathsf{b}\left(  \alpha|I\right)  \right]  }\left\vert \frac{G\left(  u|X,I\right)
	-\alpha}{h}-\frac{\Phi\left(  u|x,\mathsf{b}\left(  \alpha|I\right)
	\right)  -\alpha}{h}\right\vert
\end{align*}
by Proposition \ref{SeriesB2}-(i) which gives that $B^{(1)} (\cdot|\cdot)$ is bounded away from $0$.

\subparagraph{Proof of (i).}
We first show that $\mathsf{b} (\alpha|I)$ belongs to $\underline{\mathcal{BI}%
}_{\alpha,h}$ for all $\alpha$.
Observe that Proposition \ref{SeriesB2}-(ii) gives, uniformly in $\alpha$, $x$ and $t$ in $\mathcal{T}_{\alpha,h} \subset [-1,1]$
\begin{align*}
& \Psi^{(1)} \left(t|x,\mathsf{b}\left(  \alpha|I\right)\right)=\partial_{t}\left[  P\left(  x,t\right)^{\prime}
\mathsf{b}\left(  \alpha|I\right)  \right] 
=
\sum_{p=1}^{s+1}h^{p}
\frac{t^{p-1}}{\left(  p-1\right)  !}
P(x)^{\prime} \beta^{(p)} (\alpha|I) \\
&  \quad=h\left(  B^{\left(  1\right)  }\left(  \alpha|X,I\right)  +o\left(
1\right)  \right)  +h^{2}\left(  \sum_{p=2}^{s+1}h^{p-2}\frac{t^{p-1}}{\left(
	p-1\right)  !}B^{\left(  p\right)  }\left(  \alpha|X,I\right)  +o\left(
1\right)  \right) \\
&  \quad=hB^{\left(  1\right)  }\left(  \alpha|X,I\right)  +O\left(  h^2\right).
\end{align*}
Hence 
$
\min_{\alpha \in [0,1]}
\min_{(t,x) \in \mathcal{T}_{\alpha,h} \times \mathcal{X}}
\Psi^{(1)} \left(t|x,\mathsf{b}\left(  \alpha|I\right)\right)
=
h
\min_{(\alpha,x) \in [0,1] \times \mathcal{X}}
B^{\left(  1\right)  }\left(  \alpha|x,I\right)  +O\left(  h^2\right)
\geq h/\underline{f}
$
for any $\underline{f}$ large enough so that $\min_{(\alpha,x) \in [0,1] \times \mathcal{X}}
B^{\left(  1\right)  }\left(  \alpha|x,I\right) > 1/\underline{f}$ and $h$ small enough as $B^{(1)} (\cdot|\cdot)$ is bounded away from $0$ by Proposition \ref{SeriesB2}-(i).
Proposition \ref{SeriesB2}-(ii) also implies
\begin{align*}
&  \max_{p=1,\ldots,s+1}\left(  \frac{\max_{(\alpha,x)\in [0,1] \times \mathcal{X}}\left\vert
	P\left(  x\right)  ^{\prime}\mathsf{b}_{p}\left(  \alpha|I\right)
	\right\vert }{h}\right)  =\max_{p=1,\ldots,s+1}\max_{\left(  \alpha,x\right)
	\in\left[  0,1\right]  \times\mathcal{X}}h^{p-1}\left\vert B^{\left(
	p\right)  }\left(  \alpha|X,I\right)  +o\left(  1\right)  \right\vert \\
&  \quad=\max_{\left(  \alpha,x\right)  \in\left[  0,1\right]  \times
	\mathcal{X}}B^{\left(  1\right)  }\left(  \alpha|X,I\right)  +o\left(
1\right)  \leq\overline{f}%
\end{align*}
provided $\overline{f}$ is large enough, since $B^{\left(  1\right)  }\left(  \cdot|\cdot,\cdot\right)  $
is bounded away from infinity by Proposition \ref{SeriesB2}, and $h$ small enough, so that
$\mathsf{b}\left(  \alpha|I\right)  $ is in $\underline{\mathcal{BI}%
}_{\alpha,h}$ for all $\alpha$.

Suppose
now that $\left\Vert \mathsf{b}-\mathsf{b}\left(  \alpha|I\right)  \right\Vert_{\infty}
\leq C_0 h^{D_{\mathcal{M}}/2+1}$. Then since $\max_{x \in \mathcal{X}} \|P(x)\|=O (h^{-D_{\mathcal{M}}/2})$ by Assumption \ref{Riesz}-(i),
\begin{align*}
\Psi^{(1)} (t|x,\mathsf{b})    & 
=
\partial_{t}
\left[  P\left(  x,t\right)^{\prime}
\left(
\mathsf{b}\left(\alpha|I\right)
+
\mathsf{b}  
-
\mathsf{b}\left(\alpha|I\right)
\right)
\right]
\geq
\partial_{t}
\left[  P\left(  x,t\right)  ^{\prime}\mathsf{b}\left(
\alpha|I\right)  \right]   -\left\Vert \mathsf{b}-\mathsf{b}\left(
\alpha|I\right)  \right\Vert_{\infty} \left\Vert P\left(  x\right)  \right\Vert \\
&  \geq
\Psi^{(1)}
\left(
t 
\left| 
x,\mathsf{b}\left(  \alpha|I\right)  \right. \right)  
-
Ch^{-D_{\mathcal{M}}/2}
\times C_0h^{D_{\mathcal{M}}/2+1}
\geq
\Psi^{(1)}
\left(
t 
\left| 
x,\mathsf{b}\left(  \alpha|I\right)  \right. \right)  
-
C \cdot C_0h,
\\
\left\vert P\left(  x\right)  ^{\prime}\mathsf{b}_{p}\right\vert  &
\leq\left\vert P\left(  x\right)  ^{\prime}\mathsf{b}_{p}\left(
\alpha|I\right)  \right\vert +\left\Vert \mathsf{b}-\mathsf{b}\left(
\alpha|I\right)  \right\Vert_{\infty} \left\Vert P\left(  x\right)  \right\Vert \\
&  \leq\left\vert P\left(  x\right)  ^{\prime}\mathsf{b}_{p}\left(
\alpha|I\right)  \right\vert + C \cdot C_0h,\quad p=1,\ldots,s+1,
\end{align*}
and $\mathcal{B}_{\infty} \left(  \mathsf{b}\left(  \alpha|I\right)
,Ch^{D_{\mathcal{M}}/2+1}\right)  \subset\underline{\mathcal{BI}}_{\alpha,h}$ for all $\alpha$
when $h\geq 1$  provided $C_0$ is small enough. Hence (i) holds.

\subparagraph{Proof of (ii).}
(ii)
follows from the Implicit Function Theorem and the definition of
$\mathcal{BI}_{\alpha,h}$.

\subparagraph{Proof of (iv).}
The first bound follows from the Cauchy-Schwarz inequality.
This bound implies for all $u$ in $\Psi\left[  \mathcal{T}_{\alpha
	,h}|x,\mathsf{b}^{1}\right]  \cap\Psi\left[  \mathcal{T}_{\alpha
	,h}|x,\mathsf{b}^{0}\right]  $%
\begin{align*}
&\left\vert \Psi\left[  \Delta\left(  u|x,\mathsf{b}^{1}\right)
|x,\mathsf{b}^{0}\right]  -\Psi\left[  \Delta\left(  u|x,\mathsf{b}^{0}\right)  |x,\mathsf{b}^{0}\right]  \right\vert 
=\left\vert \Psi\left[  \Delta\left(  u|x,\mathsf{b}^{1}\right)
|x,\mathsf{b}^{0}\right]  -u\right\vert \\
&  \quad=\left\vert \Psi\left[  \Delta\left(  u|x,\mathsf{b}^{1}\right)
|x,\mathsf{b}^{0}\right]  -\Psi\left[  \Delta\left(  u|x,\mathsf{b}^{1}\right)  |x,\mathsf{b}^{1}\right]  \right\vert \leq Ch^{-D_{\mathcal{M}%
	}/2}\left\Vert \mathsf{b}^{1}-\mathsf{b}^{0}\right\Vert_{\infty} .
\end{align*}
By definition of $\underline{\mathcal{BI}}_{\alpha,h}$%
\begin{align*}
&  \left\vert \Psi\left[  \Delta\left(  u|x,\mathsf{b}^{1}\right)
|x,\mathsf{b}^{0}\right]  -\Psi\left[  \Delta\left(  u|x,\mathsf{b}%
_{0}\right)  |x,\mathsf{b}^{0}\right]  \right\vert \\
&  \quad\geq Ch\left\vert \Delta\left(  u|x,\mathsf{b}^{1}\right)
-\Delta\left(  u|x,\mathsf{b}^{0}\right)  \right\vert =C\left\vert \Phi\left(
u|x,\mathsf{b}^{1}\right)  -\Phi\left(  u|x,\mathsf{b}^{0}\right)
\right\vert
\end{align*}
and substituting shows that the second bound of (iv) holds.

The expression in (ii) of $\partial_u \Phi\left(u|x,\mathsf{b}
\right)  $ and the definition of $\underline{\mathcal{BI}}_{\alpha,h}$
yield the third inequality, using the fourth bound which is established now.
Note that
\begin{align*}
&  \left\vert \Psi^{(1)}\left[  \Delta\left(
u|x,\mathsf{b}^{1}\right)  |x,\mathsf{b}^{1}\right]  
-
\Psi^{(1)}
\left[  \Delta\left(  u|x,\mathsf{b}^{0}\right)  |x,\mathsf{b}^{0}\right]  \right\vert \\
&  
\quad\leq
\left\vert 
\Psi^{(1)} \left[  \Delta\left(
u|x,\mathsf{b}^{1}\right)  |x,\mathsf{b}^{1}\right]  
-
\Psi^{(1)}
\left[  \Delta\left(  u|x,\mathsf{b}^{0}\right)  |x,\mathsf{b}^{1}\right]  \right\vert \\
&  \quad
+\left\vert  \Psi^{(1)} \left[  \Delta\left(
u|x,\mathsf{b}^{0}\right)  |x,\mathsf{b}^{1}\right]  - 
\Psi^{(1)}
\left[  \Delta\left(  u|x,\mathsf{b}^{0}\right)  |x,\mathsf{b}^{0}\right]  \right\vert \\
&  \quad\leq\max_{t\in\mathcal{T}_{\alpha,h}}\left\vert \partial_{t^2}^{2}%
\Psi\left(  t|x,\mathsf{b}^{1}\right)  \right\vert
\frac{\left\vert \Phi\left(  u|x,\mathsf{b}^{1}\right)  -\Phi\left(
	u|x,\mathsf{b}^{0}\right)  \right\vert }{h}
%\\ &  \quad
+\max_{t\in\mathcal{T}_{\alpha,h}}\left\vert \partial_t P\left(
x,t\right)  \left(  \mathsf{b}^{1}-\mathsf{b}^{0}\right)
\right\vert .
\end{align*}
But, by definition of $\underline{\mathcal{BI}}_{\alpha,h}$
\[
\max_{t\in\mathcal{T}_{\alpha,h}}\left\vert \partial_{t}^{2}\Psi\left(
t|x,\mathsf{b}^{1}\right)  \right\vert \leq Ch\max
_{p=2,\ldots,s+1}\left\vert \frac{P\left(  x\right)  \mathsf{b}^{1}_{p}}%
{h}\right\vert =O\left(  h\right)
\]
so that substituting and using the bound for $\Phi\left(  u|x,\mathsf{b}^{1}\right)
-\Phi\left(  u|x,\mathsf{b}^{0}\right)  $ give, for all $\alpha$, $x$,
$u$, $\mathsf{b}^{1}$ and $\mathsf{b}^{0}$%
\[
\left\vert \Psi^{(1)}\left[  \Delta\left(  u|x,\mathsf{b}%
_{1}\right)  |x,\mathsf{b}^{1}\right]  
-
\Psi^{(1)}\left[
\Delta\left(  u|x,\mathsf{b}^{0}\right)  |x,\mathsf{b}^{0}\right]  \right\vert
\leq Ch^{-D_{\mathcal{M}}/2}\left\Vert \mathsf{b}^{1}-\mathsf{b}^{0}\right\Vert_{\infty} ,
\]
which is the fourth inequality. $\hspace*{\fill}\square$

\subsection{\textbf{Proof of Lemma \ref{R2}}}

Recall
\begin{align*}
&  \overline{\mathsf{R}}^{\left(  2\right)  }\left(  \mathsf{b};\alpha
,I\right)  =\mathbb{E}\left[  \mathbb{I}\left[  B_{i\ell}\in\Psi\left(
\mathcal{T}_{\alpha,h}|X_{\ell},\mathsf{b}\right)  ,I_{\ell}=I\right]  \right.
\\
&  \quad\quad\quad\quad\quad\quad\quad\quad\quad\left.  \frac{P\left(
X_{\ell},\Delta\left(  B_{i\ell}|X_{\ell},\mathsf{b}\right)  \right)  P\left(
X_{\ell},\Delta\left(  B_{i\ell}|X_{\ell},\mathsf{b}\right)  \right)
^{\prime}}{\Psi^{(1)} \left(  \Delta\left(  B_{i\ell}|X_{\ell},\mathsf{b}\right)
|X_{\ell},\mathsf{b}\right)  }K\left(  \Delta\left(  B_{i\ell}|X_{\ell
},\mathsf{b}\right)  \right)  \right] \\
&  \quad=\int\left[  \int_{\Psi\left(  \underline{t}_{\alpha,h}|x,\mathsf{b}%
\right)  \vee B\left(  0|x,I\right)  }^{\Psi\left(  \overline{t}_{\alpha
,h}|x,\mathsf{b}\right)  \wedge B\left(  1|x,I\right)  }\frac{P\left(
x,\Delta\left(  y|x,\mathsf{b}\right)  \right)  P\left(  x,\Delta\left(
y|x,\mathsf{b}\right)  \right)  ^{\prime}}{\Psi^{(1)}\left(  \Delta\left(
y|x,\mathsf{b}\right)  |x,\mathsf{b}\right)  }K\left(  \Delta\left(
y|x,\mathsf{b}\right)  \right)  g\left(  y,x,I\right)  dy\right]  dx,
\end{align*}
In the equation above, $\Delta (u|x,I)$ is such that $\Delta\left[  \Psi\left[  t|x,\mathsf{b}\right]  |x,\mathsf{b}\right]
=t$ for all $t$ in $\mathcal{T}_{\alpha,h}$. Let
\[
\overline{t}_{\alpha,h}\left(  x,I;\mathsf{b}\right)  =\overline{t}_{\alpha
	,h}\wedge\Delta\left[  B\left(  1|x,I\right)  |x,\mathsf{b}\right]
,\quad\underline{t}_{\alpha,h}\left(  x,I;\mathsf{b}\right)  
=
\underline{t}_{\alpha,h}\vee\Delta\left[  B\left(  0|x,I\right)  |x,\mathsf{b}\right]  .
\]
The change of variable $y=\Psi\left(  t|x,\mathsf{b}\right)  $ yields that%
\[
\overline{\mathsf{R}}^{\left(  2\right)  }\left(  \mathsf{b};\alpha,I\right)
=\int\left[  \int_{\underline{t}_{\alpha,h}\left(  x,I;\mathsf{b}\right)
}^{\overline{t}_{\alpha,h}\left(  x,I;\mathsf{b}\right)  }P\left(  x,t\right)
P\left(  x,t\right)  ^{\prime}K\left(  t\right)  g\left(  \Psi\left(
t|x,\mathsf{b}\right)  ,x,I\right)  dt\right]  dx.
\]
The Dominated Convergence Theorem, Proposition \ref{SeriesB2}
-(i) and $s\geq1$, \footnote{Recall this implies that $g\left(  \cdot,\cdot,I\right)  $ is bounded
away from $0$ and infinity and differentiable with respect to its first variable.
Hence so is 
$\mathsf{b} \mapsto g\left(  \Psi\left(
t|x,\mathsf{b}\right)  ,x,I\right)$ for $t$ in the open interval
$(\underline{t}_{\alpha,h}\left(  x,I;\mathsf{b}\right),\overline{t}_{\alpha,h}\left(  x,I;\mathsf{b}\right)$.} yield that $\overline{\mathsf{R}%
}^{\left(  2\right)  }\left(  \mathsf{\cdot};\alpha,I\right)  $ is
continuously differentiable over $\underline{\mathcal{BI}}_{\alpha,h}$ with
a differential operator $\mathsf{R}^{\left(  3\right)  }\left(  \mathsf{b};\alpha,I\right)
\left[  \cdot\right]$ which maps a vector $\mathsf{d}$ of $\mathbb{R}^{N(s+2)}$ to a 
$N(s+2) \times N(s+2)$ matrix.
By the Leibniz integral rule
\begin{align*}
\overline{\mathsf{R}}^{\left(  3\right)  }\left(  \mathsf{b};\alpha,I\right)
\left[  \mathsf{d}\right]   &  =\overline{\mathsf{R}}_{0}^{\left(  3\right)
}\left(  \mathsf{b};\alpha,I\right)  \left[  \mathsf{d}\right]  +\overline
{\mathsf{R}}_{1}^{\left(  3\right)  }\left(  \mathsf{b};\alpha,I\right)
\left[  \mathsf{d}\right]  -\overline{\mathsf{R}}_{2}^{\left(  3\right)
}\left(  \mathsf{b};\alpha,I\right)  \left[  \mathsf{d}\right]  ,\\
\overline{\mathsf{R}}_{0}^{\left(  3\right)  }\left(  \mathsf{b}%
;\alpha,I\right)  \left[  \mathsf{d}\right]   &  =\int_{\mathcal{X}}\left[
\int_{\underline{t}_{\alpha,h}\left(  x,I;\mathsf{b}\right)  }^{
	\overline{t}_{\alpha,h}\left(  x,I;\mathsf{b}\right)  }P\left(  x,t\right)  P\left(
x,t\right)  ^{\prime}K\left(  t\right)  g^{\left(  1\right)  }\left(
\Psi\left(  t|x,\mathsf{b}\right)  ,x,I\right)  \left[  \mathsf{d}^{\prime
}P\left(  x,t\right)  \right]  dt\right]  dx,\\
\overline{\mathsf{R}}_{1}^{\left(  3\right)  }\left(  \mathsf{b}%
;\alpha,I\right)  \left[  \mathsf{d}\right]   &  =\int_{\mathcal{X}}P\left(
x,\overline{t}_{\alpha,h}\left(  x,I;\mathsf{b}\right)  \right)  P\left(
x,\overline{t}_{\alpha,h}\left(  x,I;\mathsf{b}\right)  \right)  ^{\prime
}K\left(  \overline{t}_{\alpha,h}\left(  x,I;\mathsf{b}\right)  \right) \\
&  \quad\quad\quad\times g\left(  \Psi\left(  \overline{t}_{\alpha,h}\left(
x,I;\mathsf{b}\right)  |x,\mathsf{b}\right)  ,x,I\right)  \left[
\mathsf{d}^{\prime}
\partial_{\mathsf{b}}\overline{t}_{\alpha,h}\left(  x,I;\mathsf{b}%
\right)  \right]  dx,\\
\overline{\mathsf{R}}_{2}^{\left(  3\right)  }\left(  \mathsf{b}%
;\alpha,I\right)  \left[  \mathsf{d}\right]   &  =\int_{\mathcal{X}}P\left(
x,\underline{t}_{\alpha,h}\left(  x,I;\mathsf{b}\right)  \right)  P\left(
x,\underline{t}_{\alpha,h}\left(  x,I;\mathsf{b}\right)  \right)  ^{\prime
}K\left(  \underline{t}_{\alpha,h}\left(  x,I;\mathsf{b}\right)  \right) \\
&  \quad\quad\quad\times g\left(  \Psi\left(  \overline{t}_{\alpha,h}\left(
x,I;\mathsf{b}\right)  |x,\mathsf{b}\right)  ,x,I\right)  \left[
\mathsf{d}^{\prime}
\partial_{\mathsf{b}}\underline{t}_{\alpha,h}\left(
x,I;\mathsf{b}\right)  \right]  dx.
\end{align*}

Consider any $\mathsf{b}$ in $\mathcal{B}_{\infty} \left(\mathsf{b}(\alpha|I),C_0 h^{D_{\mathcal{M}}/2+1}\right)$, the inequalities below being uniform over such $\mathsf{b}$.
Proposition \ref{SeriesB2}-(i), (\ref{Matnormeq}) and Assumption \ref{Riesz}-(i), \ref{Auct.A} imply
\[
\left\Vert 
\overline{\mathsf{R}}_{0}^{\left(  3\right)  }\left(
\mathsf{b};\alpha,I\right)  \left[  \mathsf{d}\right]  \right\Vert_{\infty} \leq
C\max_{x\in\mathcal{X}}\left\Vert P\left(  x\right)  \right\Vert 
\left\Vert
\mathsf{d}\right\Vert_{\infty} \leq Ch^{-D_{\mathcal{M}}/2}\left\Vert \mathsf{d}%
\right\Vert .
\]
The operators $\overline{\mathsf{R}}_{i}^{\left(  3\right)  }\left(
\mathsf{b};\alpha,I\right)  \left[  \mathsf{d}\right]  $, $i=1,2$, can be
studied in a similar way so that only $\overline{\mathsf{R}}_{1}^{\left(  3\right)  }\left(
\mathsf{b};\alpha,I\right)  \left[  \mathsf{d}\right]  $ is considered. Observe%
\[
\partial_{\mathsf{b}}\overline{t}_{\alpha,h}\left(  x,I;\mathsf{b}\right)  =\left\{
\begin{array}
[c]{cc}%
0 & \text{if }\overline{t}_{\alpha,h}\leq\Delta\left[  B\left(  1|x,I\right)
|x,\mathsf{b}\right] \\
\partial_{\mathsf{b}} \Delta\left[  B\left(  1|x,I\right)  |x,\mathsf{b}\right]
=-\frac{P\left(  x,\Delta\left(  B\left(
1|x,I\right)  |x,\mathsf{b}\right)  \right)  }{\Psi^{\left(  1\right)
}\left(  \Delta\left(  B\left(  1|x,I\right)  |x,\mathsf{b}\right)
|x,\mathsf{b}\right)  } & \text{if }\overline{t}_{\alpha,h}>\Delta\left[
B\left(  1|x,I\right)  |x,\mathsf{b}\right]
\end{array}
\right.  .
\]
But by Lemma \ref{Phipsi}-(iii,iv) and $\left\Vert \mathsf{b}-\mathsf{b}\left(
\alpha|I\right)  \right\Vert_{\infty} \leq Ch^{ D_{\mathcal{M}}/2+1}$ , for $h$ small enough,%
\begin{align*}
&  \Delta\left[  B\left(  1|x,I\right)  |x,\mathsf{b}\right]  =\frac
{\Phi\left[  B\left(  1|x,I\right)  |x,\mathsf{b}\right]  -\alpha}{h}%
=\frac{\min\left\{  \alpha+h\overline{t}_{\alpha,h},\Phi\left[  B\left(
1|x,I\right)  |x,\mathsf{b}\right]  \right\}  -\alpha}{h}\\
&  \quad\geq\frac{\min\left\{  \alpha+h\overline{t}_{\alpha,h},\Phi\left[
B\left(  1|x,I\right)  |x,\mathsf{b}\left(  \alpha|I\right)  \right]
-Ch^{-D_{\mathcal{M}}/2}\left\Vert \mathsf{b}-\mathsf{b}\left(
\alpha|I\right)  \right\Vert_{\infty} \right\}  -\alpha}{h}\\
&  \quad\geq\frac{\min\left\{  \alpha+h\overline{t}_{\alpha,h},G\left[
B\left(  1|x,I\right)  |x,I\right]  -Ch^{s+1}-Ch\right\}  -\alpha}{h}\\
&  \quad\geq\frac{\min\left\{  \alpha+h\min\left(  \frac{1-\alpha}%
{h},1\right)  ,1-Ch\right\}  -\alpha}{h}%
\end{align*}
uniformly in $\alpha$, $x$ and $\mathsf{b}$ in $\mathcal{B}_{\infty}\left(
\mathsf{b}\left(  \alpha|I\right)  ,Ch^{D_{\mathcal{M}}/2+1}\right)  $. Hence, if $\alpha\leq1-C^{\prime}h$ with $C^{\prime
}>0$ large enough%
\[
\Delta\left[  B\left(  1|x,I\right)  |x,\mathsf{b}\right]  \geq\frac
{\min\left\{  \alpha+h,1-Ch\right\}  -\alpha}{h}\geq1\geq\overline{t}_{\alpha,h}
\]
so that 
$\partial_{\mathsf{b}}\overline{t}_{\alpha,h}\left(  x,I;\mathsf{b}\right)
=0$. Hence since $\mathcal{B}_{\infty}\left(
\mathsf{b}\left(  \alpha|I\right)  ,Ch^{D_{\mathcal{M}}/2+1}\right)
\subset\underline{\mathcal{BI}}_{\alpha,h}$, the definition of
$\underline{\mathcal{BI}}_{\alpha,h}$ which implies
$\Psi^{(1)}\left(  \Delta\left(  B\left(
1|x,I\right)  |x,\mathsf{b}\right)  |x,\mathsf{b}\right) \geq h/\underline{f}$, and using $\frac{\mathbb{I}
	\left[
	\alpha
	\geq
	1-C' h
	\right]}{h}
\leq
\frac{C}{\alpha(1-\alpha)+h}$ give for $\alpha
\geq
1-C' h
$
\begin{align*}
&  \left\Vert \overline{\mathsf{R}}_{1}^{\left(  3\right)  }\left(
\mathsf{b};\alpha,I\right)  \left[  \mathsf{d}\right]  \right\Vert_{\infty} \leq
C\mathbb{I}\left[  \alpha\geq1-C^{\prime}h\right] \\
&  \quad\quad\quad\quad
\times
\left\Vert 
\int_{\mathcal{X}}P\left(
x,\overline{t}_{\alpha,h}\left(  x,I;\mathsf{b}\right)  \right)  P\left(
x,\overline{t}_{\alpha,h}\left(  x,I;\mathsf{b}\right)  \right)  ^{\prime
}\frac{\mathsf{d}^{\prime}P\left(  x,\Delta\left(  B\left(  1|x,I\right)
|x,\mathsf{b}\right)  \right)  }{\Psi^{(1)}\left(  \Delta\left(  B\left(
1|x,I\right)  |x,\mathsf{b}\right)  |x,\mathsf{b}\right)  }dx
\right\Vert_{\infty}
\\
&  
\quad
\leq 
Ch^{-1}\mathbb{I}\left[  \alpha\geq1-C^{\prime}h\right]
\max_{x\in\mathcal{X}}\left\Vert P\left(  x\right)  \right\Vert 
\left\Vert
\mathsf{d}
\right\Vert_{\infty} 
\leq 
Ch^{-1}h^{-D_{\mathcal{M}}/2}
\left\Vert
\mathsf{d}
\right\Vert_{\infty} 
\mathbb{I}\left[  \alpha\geq1-C^{\prime}h\right] \\
&  \quad\leq C\frac{h^{-D_{\mathcal{M}}/2}}{\alpha\left(  1-\alpha\right)+h}
\left\Vert \mathsf{d}\right\Vert_{\infty} .
\end{align*}
Substituting in the expression of $\overline{\mathsf{R}}^{\left(  3\right)
}\left(  \mathsf{b};\alpha,I\right)  \left[  \mathsf{d}\right]  $ then gives
uniformly in $\mathsf{d}$
\[
\max_{\alpha\in\left[  0,1\right]  }\max_{\mathsf{b\in}\mathcal{B}_{\infty}
\left(
\mathsf{b}\left(  \alpha|I\right)  ,Ch^{D_{\mathcal{M}}/2+1}\right)
}\left(  \alpha\left(  1-\alpha\right)  +h\right)  
\left\Vert 
\overline{\mathsf{R}}^{\left(  3\right)  }\left(  \mathsf{b};\alpha,I\right)  \left[
\mathsf{d}\right]  
\right\Vert_{\infty} 
\leq 
Ch^{-D_{\mathcal{M}}/2}
\left\Vert
\mathsf{d}\right\Vert_{\infty} .
\]
The Taylor inequality then shows that the bound in (i) holds.

For (ii),  Lemma \ref{Phipsi}-(iii)
and Proposition \ref{SeriesB2}-(i) give that, uniformly in $\alpha$ and $x$%
\begin{align*}
\overline{t}_{\alpha,h}\left[  x,I;\mathsf{b}\left(  \alpha|I\right)
\right]   &  =\overline{t}_{\alpha,h}\wedge\frac{\Phi\left[  B\left(
1|x,I\right)  |x,\mathsf{b}\left(  \alpha|I\right)  \right]  -\alpha
}{h}\\
&  =\overline{t}_{\alpha,h}\wedge\frac{1+o\left(  h^{s+1}\right)  -\alpha}%
{h}=\overline{t}_{\alpha,h}+o\left(  h^{s}\right)  ,\\
\underline{t}_{\alpha,h}\left[  x,I;\mathsf{b}\left(  \alpha|I\right)
\right]   &  =\underline{t}_{\alpha,h}+o\left(  h^{s}\right)  .
\end{align*}
This gives for $\overline{\mathsf{R}}^{\left(  2\right)  }\left(  \mathsf{b}\left(
\alpha|I\right)  ;\alpha,I\right)$ the expansion
\begin{align*}
&  \overline{\mathsf{R}}^{\left(  2\right)  }\left(  \mathsf{b}\left(
\alpha|I\right)  ;\alpha,I\right)  =\int\left[  \int_{\underline{t}_{\alpha
,h}\left[  x,I;\mathsf{b}\left(  \alpha|I\right)  \right]  }
^{\overline{t}_{\alpha,h}\left[  x,I;\mathsf{b}\left(  \alpha|I\right)
\right]  }\pi\left(  t\right)  \pi\left(  t\right)  ^{\prime}K\left(
t\right)  g\left(  \Psi\left(  t|x,\mathsf{b}\left(  \alpha|I\right)
\right)  |x,I\right)  dt\right] \\
&  \quad\quad\quad\quad\quad\quad\otimes P\left(  x\right)  P\left(  x\right)
^{\prime}f\left(  x,I\right)  dx\\
&  \quad=
\int\left[  
\int_{\underline{t}_{\alpha,h}\left[  x,I;\mathsf{b}\left(  \alpha|I\right)  \right]  }
^{\overline{t}_{\alpha,h}\left[  x,I;\mathsf{b}\left(  \alpha|I\right) \right]  }
\pi\left(  t\right)
\pi\left(  t\right)  ^{\prime}K\left(  t\right)  g\left[  B\left(
\alpha+ht|x,I\right)  +o\left(  h^{s+1}\right)  |x,I\right]  dt\right] \\
&  \quad\quad\quad\quad\quad\quad\otimes P\left(  x\right)  P\left(  x\right)
^{\prime}f\left(  x,I\right)  dx\\
&  \quad=
\int\left[  
\int_{\underline{t}_{\alpha,h}}^{\overline{t}_{\alpha,h}  }
\pi\left(  t\right)
\pi\left(  t\right)  ^{\prime}K\left(  t\right)  \left(  \frac{1}{B^{\left(
		1\right)  }\left(  \alpha|x,I\right)  }-ht\frac{B^{\left(  2\right)  }\left(
	\alpha|x,I\right)  }{\left(  B^{\left(  1\right)  }\left(  \alpha|x,I\right)
	\right)  ^{2}}+o\left(  h\right)  \right)  dt\right] \\
&  \quad\quad\quad\quad\quad\quad\otimes P\left(  x\right)  P\left(  x\right)
^{\prime}f\left(  x,I\right)  dx+o\left(  h^{s}\right)\\
&  \quad=\int\Omega_{h}\left(  \alpha\right)  \otimes\frac{P\left(  x\right)
	P\left(  x\right)  ^{\prime}}{B^{\left(  1\right)  }\left(  \alpha|x,I\right)
}f\left(  x,I\right)  dx\\
&  \quad\quad\quad-h\int\Omega_{1h}\left(  \alpha\right)  \otimes
\frac{P\left(  x\right)  P\left(  x\right)  ^{\prime}B^{\left(  2\right)
	}\left(  \alpha|x,I\right)  }{\left(  B^{\left(  1\right)  }\left(
	\alpha|x,I\right)  \right)  ^{2}}f\left(  x,I\right)  dx+o\left(  h\right)
\end{align*}
where all remainder terms are with respect to the matrix norm.
This together the fact that the eigenvalues of the matrices $\Omega_{h}\left(
\alpha\right)  $ and $\int_{\mathcal{X}}P\left(  x\right)  P\left(  x\right)
^{\prime}dx$ are bounded away from $0$ and infinity, the fact that $B^{\left(
1\right)  }\left(  \alpha|X,I\right)  $ is bounded away from $0$ and infinity
shows that (ii) holds.$\hspace*{\fill}\square$

\section{Lemmas \ref{HatR2}, \ref{HatR1} and \ref{Leadterm}}

The proofs of the lemmas grouped here make use of a deviation inequality from
Massart (2007). Consider $n$ independent random variables $Z_{\ell}$ and, for
a known real function $\xi\left(  z,\theta\right)  $ separable with respect to
$\theta\in\Theta$, $Z_{\ell}\left(  \theta\right)  =\xi\left(  Z_{\ell}%
,\theta\right)  $ where $\theta$ is a parameter. Let $\underline{\xi}\left(
\cdot\right)  \leq\overline{\xi}\left(  \cdot\right)  $ be two functions. A
\textit{bracket }$\left[  \underline{\xi},\overline{\xi}\right]  $ is the set
of all functions $\xi\left(  \cdot\right)  $ such that $\underline{\xi}\left(
z\right)  \leq\xi\left(  z\right)  \leq\overline{\xi}\left(  z\right)  $ for
all $z$. The next Theorem is from Massart (2007, Theorem 6.8 and
Corollary 6.9).

\begin{theorem}
\label{Dev}Assume that $\sup_{\theta\in\Theta}\left\vert Z_{\ell}\left(
\theta\right)  \right\vert \leq M_{\infty}$, $\sup_{\theta\in\Theta
}\operatorname*{Var}\left(  Z_{\ell}\left(  \theta\right)  \right)  \leq
M_{2}^{2}$ for all $\ell$ and that for any $\epsilon>0$ there exists brackets
$\left[  \underline{\xi}_{j},\overline{\xi}_{j}\right]  \subset\left[
-b,b\right]  $, $j=1,\ldots,\exp\left(  H\left(  \epsilon\right)  \right)  $,
such that%
\[
\mathbb{E}\left[  \left(  \overline{\xi}_{j}\left(  Z_{i}\right)
-\underline{\xi}_{j}\left(  Z_{i}\right)  \right)  ^{2}\right]  \leq
\frac{\epsilon^{2}}{2}\text{ and }\left\{  \xi\left(  z,\theta\right)
,\theta\in\Theta\right\}  \subset%
%TCIMACRO{\dbigcup \limits_{j=1}^{\exp\left(  H\left(  \epsilon\right)
%\right)  }}%
%BeginExpansion
{\displaystyle\bigcup\limits_{j=1}^{\exp\left(  H\left(  \epsilon\right)
\right)  }}
%EndExpansion
\left[  \underline{\xi}_{j},\overline{\xi}_{j}\right]  .
\]
Let%
\[
\mathcal{H}_{L}=54\int_{0}^{M_{2}/2}\sqrt{\min\left(  L,H\left(
\epsilon\right)  \right)  }d\epsilon+\frac{2\left(  M_{\infty}+M_{2}\right)
H\left(  M_{2}\right)  }{L^{1/2}}.
\]
Then, for any $t\in\left[  0,10L^{1/2}M_{2}/M_{\infty}\right]  $,%
\[
\mathbb{P}\left(  \sup_{\theta\in\Theta}\left\vert \sum_{i=1}^{n}\left\{
Z_{\ell}\left(  \theta\right)  -\mathbb{E}\left[  Z_{\ell}\left(
\theta\right)  \right]  \right\}  \right\vert \geq L^{1/2}\left\{
\mathcal{H}_{L}+t\right\}  \right)  \leq2\exp\left(  -\frac{t^{2}}{25}\right)
.
\]
\end{theorem}

\subsection{\textbf{Proof of Lemma \ref{HatR2}}}

Note that $\widehat{\mathsf{R}}^{\left(  2\right)  }\left(  \mathsf{b}%
;\alpha,I\right)  \mathsf{-}\overline{\mathsf{R}}^{\left(  2\right)  }\left(
\mathsf{b};\alpha,I\right)  $ is an index permutation of a $c\left(  s+2\right)  $-band matrix, so
that the order of its matrix norm is the same than the order of its largest
entry by (\ref{Matnormeq}). The generic entry of $\widehat{\mathsf{R}}^{\left(  2\right)  }\left(
\mathsf{b};\alpha,I\right)  -\overline{\mathsf{R}}^{\left(  2\right)  }\left(
\mathsf{b};\alpha,I\right)  $ can be written as%
\[
\widehat{\mathsf{r}}^{(2)}\left(  \mathsf{b};\alpha,I\right)  =\frac{1}{LIh^{\left(
D_{\mathcal{M}}+1\right)  /2}}\sum_{\ell=1}^{L}\xi_{\ell}\left(
\mathsf{b};\alpha\right)
\]
where the $\xi_{\ell}\left(  \mathsf{b};\alpha\right)  $ are centered iid with%
\begin{align*}
&  \xi_{\ell}\left(  \mathsf{b};\alpha\right)  =\sum_{i=1}^{I_{\ell}}\left\{
\mathbb{I}\left[  B_{i\ell}\in\Psi\left(  \mathcal{T}_{\alpha,h}|X_{\ell
},\mathsf{b}\right)  ,I_{\ell}=I\right]  \xi_{i\ell}\left(  \mathsf{b}\right)
\right. \\
&  \quad\quad\quad\quad\quad\quad\quad\quad\quad\left.  -\mathbb{E}\left[
\mathbb{I}\left[  B_{i\ell}\in\Psi\left(  \mathcal{T}_{\alpha,h}|X_{\ell
},\mathsf{b}\right)  ,I_{\ell}=I\right]  \xi_{i\ell}\left(  \mathsf{b}\right)
\right]  \right\} \\
&  \quad\xi_{i\ell}\left(  \mathsf{b}\right)  =\frac{h^{D_{\mathcal{M}}/2}%
}{h^{1/2}}\frac{P_{n_{1}}\left(  X_{\ell}\right)  P_{n_{2}}\left(  X_{\ell
}\right)  }{\Psi^{\left(  1\right)  }\left(  \Delta\left(  B_{i\ell}|X_{\ell
},\mathsf{b}\right)  |X_{\ell},\mathsf{b}\right)  /h}K_{p}\left(
\Delta\left(  B_{i\ell}|X_{\ell},\mathsf{b}\right)  \right)  ,\\
&  \quad K_{p}\left(  \Delta\left(  B_{i\ell}|X_{\ell},\mathsf{b}\right)
\right)  =\frac{\Delta^{p_{1}+p_{2}}\left(  B_{i\ell}|X_{\ell},\mathsf{b}%
\right)  }{p_{1}!p_{2}!}K\left(  \Delta\left(  B_{i\ell}|X_{\ell}%
,\mathsf{b}\right)  \right)  .
\end{align*}
The proof of the Lemma follows from Theorem \ref{Dev}. Observe%
\[
\left\vert \xi_{\ell}\left(  \mathsf{b};\alpha\right)  \right\vert \leq
C\frac{h^{D_{\mathcal{M}}/2}\max_{x\in\mathcal{X}}\left\Vert P\left(
x\right)  \right\Vert ^{2}}{h^{1/2}}\leq M_{\infty}\text{ with }M_{\infty
}\asymp h^{-\left(  D_{\mathcal{M}}+1\right)  /2}.
\]
for all $\alpha$ in $\left[  0,1\right]  $ and all admissible $\mathsf{b}$.
For the variance, Lemma \ref{Phipsi}-(iii,iv) gives%
\begin{align*}
\left\vert \Delta\left(  B_{i\ell}|X_{\ell},\mathsf{b}\right)  \right\vert  &
=\left\vert \frac{\Phi\left(  B_{i\ell}|X_{\ell},\mathsf{b}\right)  -\alpha
}{h}\right\vert \\
&  \leq\left\vert \frac{G\left(  B_{i\ell}|X_{\ell},I_{\ell}\right)  -\alpha
}{h}\right\vert +\left\vert \frac{\Phi\left(  B_{i\ell}|X_{\ell}%
,\mathsf{b}\left(  \alpha|I_{\ell}\right)  \right)  -G\left(  B_{i\ell
}|X_{\ell},\mathsf{b}\right)  }{h}\right\vert \\
&  +\left\vert \frac{\Phi\left(  B_{i\ell}|X_{\ell},\mathsf{b}\right)
-\Phi\left(  B_{i\ell}|X_{\ell},\mathsf{b}\left(  \alpha|I_{\ell
}\right)  \right)  }{h}\right\vert \\
&  \leq\left\vert \frac{G\left(  B_{i\ell}|X_{\ell},I_{\ell}\right)  -\alpha
}{h}\right\vert +o\left(  h^{s}\right)  +O\left(  \frac{h^{-D_{\mathcal{M}}%
/2}\times h^{D_{\mathcal{M}}/2+1}}{h}\right) \\
&  =\left\vert \frac{G\left(  B_{i\ell}|X_{\ell},I_{\ell}\right)  -\alpha}%
{h}\right\vert +O\left(  1\right)
\end{align*}
uniformly. It follows that, $U_{i\ell}=G\left(  B_{i\ell}|X_{\ell},I_{\ell
}\right)  $ being a uniform random variable independent of $\left(  x_{\ell
},I_{\ell}\right)  $
\begin{align*}
\operatorname*{Var}\left(  \xi_{\ell}\left(  \mathsf{b};\alpha\right)
\right)   &  \leq CI^{2}h^{D_{\mathcal{M}}}\max_{x\in\mathcal{X}}\left\Vert
P\left(  x\right)  \right\Vert ^{2}\int_{\mathcal{X}}\left\vert P_{n_{1}%
}\left(  x\right)  P_{n_{2}}\left(  x\right)  \right\vert dx\int%
\mathbb{I}_{\left[  -C,C\right]  }\left(  \frac{u-\alpha}{h}\right)  \frac
{du}{h}\\
&  \leq CI^{2}h^{D_{\mathcal{M}}}\max_{x\in\mathcal{X}}\left\Vert P\left(
x\right)  \right\Vert ^{2}\left(  \int_{\mathcal{X}}P_{n_{1}}^{2}\left(
x\right)  dx\right)  ^{1/2}\left(  \int_{\mathcal{X}}P_{n_{2}}^{2}\left(
x\right)  dx\right)  ^{1/2}\\
&  \leq M_{2}^{2}\text{ with }M_{2}<\infty
\end{align*}
under Assumption \ref{Riesz}-(i), uniformly in $\mathsf{b}$ and $\alpha$.

Consider now the brackets covering. The key observation is that, due to the localized sieve construction and the definition of $\Delta(\cdot|x,\mathsf{b})=\Psi^{-1} (\cdot|x,\mathsf{b})$, $\xi_{\ell
}\left(  \mathsf{b};\alpha\right)  $ only depends on a finite dimension
subvector of $\mathsf{b}$, $\mathsf{b}^{\left(  n_{1},n_{2}\right)  }$ which
groups the entries of $\mathsf{b}$ corresponding to those $P_{n}\left(
\cdot\right)  $ such that $P_{n}\left(  \cdot\right)  P_{n_{1}}\left(
\cdot\right)  \neq0$ or $P_{n}\left(  \cdot\right)  P_{n_{2}}\left(
\cdot\right)  \neq0$, so that the dimension of $\mathsf{b}^{\left(
n_{1},n_{2}\right)  }$ is less than $2c\left(  s+2\right)  $ under Assumption
\ref{Riesz}-(ii). 
Consequently the class to be bracketed is%
\[
\mathcal{F}\mathcal{=}\left\{  \xi_{\ell}\left(  \mathsf{b}^{\left(
n_{1},n_{2}\right)  };\alpha\right)  ;\alpha\in\left[  0,1\right]
,\mathsf{b}^{\left(  n_{1},n_{2}\right)  }\mathsf{\in}\mathcal{B}_{\infty} \left(
\mathsf{b}^{\left(  n_{1},n_{2}\right)  }\left(  \alpha|I\right)
,Ch^{D_{\mathcal{M}}/2+1}\right)  \right\}  .
\]
Lemma \ref{Phipsi}-(iii), $1/\left(  Lh^{D_{\mathcal{M}}+1}\right)  =o\left(
1\right)  $, van de Geer (1999, p.20) and arguing as in Lemma B.2 of Guerre and Sabbah (2012,2014)
, Lemma \ref{Phipsi}-(iii), differentiability of $K(\cdot)$ and $h \asymp L^{-C}$ imply that $\mathcal{F}$ can be bracketed with a number of brackets%
\[
\exp\left(  H_{L}\left(  \epsilon\right)  \right)  \asymp\left(  \frac{L^{C}%
}{\epsilon}\right)  ^{C}%
\]
so that%
\[
\int_{0}^{M_{2}/2}\sqrt{\min\left(  L,H_{L}\left(  \epsilon\right)  \right)
}d\epsilon\leq\left(  \frac{M_{2}}{2}\right)  ^{1/2}\left(  \int_{0}^{M_{2}%
/2}H_{L}\left(  \epsilon\right)  d\epsilon\right)  ^{1/2}=O\left(  \log
L\right)  ^{1/2}%
\]
and for the item $\mathcal{H}_{L}$ of Theorem \ref{Dev},%
\[
\mathcal{H}_{L}=O\left(  \log L\right)  ^{1/2}+O\left(  \frac{\log
L}{Lh^{D_{\mathcal{M}}+1}}\right)  ^{1/2}=O\left(  \log L\right)  ^{1/2}%
\]
since $1/\left(  Lh^{D_{\mathcal{M}}+1}\right)  $ is bounded. Hence, by
Theorem \ref{Dev} for $t\leq10L^{1/2}M_{2}/M_{\infty}$
\begin{align*}
&  \mathbb{P}\left(  \left(  Lh^{D_{\mathcal{M}}+1}\right)  ^{1/2}\sup
_{\alpha\in\left[  0,1\right]  }\sup_{\mathsf{b\in}\mathcal{B}_{\infty}\left(
\mathsf{b}\left(  \alpha|I\right)  ,Ch^{D_{\mathcal{M}}/2+1}\right)
}\left\vert \widehat{\mathsf{r}}\left(  \mathsf{b};\alpha,I\right)
\right\vert \geq C\log^{1/2}L+t\right) \\
&  \quad\quad\quad\leq2\exp\left(  -\frac{t^{2}}{25}\right)
\end{align*}
uniformly over all the non zero entries $\widehat{\mathsf{r}}\left(
\mathsf{b};\alpha,I\right)  $ of the band matrix $\widehat{\mathsf{R}%
}^{\left(  2\right)  }\left(  \mathsf{b};\alpha,I\right)  -\overline
{\mathsf{R}}^{\left(  2\right)  }\left(  \mathsf{b};\alpha,I\right)  $. This
gives, by the Bonferroni inequality,
\begin{align*}
&  \mathbb{P}\left(  \sup_{\alpha\in\left[  0,1\right]  }\sup_{\mathsf{b\in
}\mathcal{B}\left(  \mathsf{b}\left(  \alpha|I\right)
,Ch^{D_{\mathcal{M}}/2+1}\right)  }\left| \widehat{\mathsf{R}}^{\left(
2\right)  }\left(  \mathsf{b};\alpha,I\right)  -\overline{\mathsf{R}}^{\left(
2\right)  }\left(  \mathsf{b};\alpha,I\right)  \right|_{\infty} \geq\frac
{C\log^{1/2}L+t}{\left(  Lh^{D_{\mathcal{M}}+1}\right)  ^{1/2}}\right) \\
&  \quad\quad\quad\leq C N(s+2)\exp\left(  -\frac{t^{2}}{25}\right)
\end{align*}
which, as $N \asymp h^{-D_{\mathcal{M}}} \leq L^{C}$, implies the result of the lemma by the matrix norm equivalence (\ref{Matnormeq}) and since $t\leq10L^{1/2}M_{2}/M_{\infty
}=O\left(  Lh^{D_{\mathcal{M}}+1}\right)  ^{1/2}$ can be set to $t=5\tau
\log^{1/2}L$ for an arbitrary large $\tau$ as $\log L/\left(
Lh^{D_{\mathcal{M}}+1}\right)  =o\left(  1\right)  $, so that 
$ N(s+2)\exp\left(  -\tau \log L\right)= o(1)$
.$\hspace*{\fill}\square$

\subsection{\textbf{Proof of Lemma \ref{HatR1}}}

That $\max_{\alpha \in [0,1]}\left\| \mathrm{Var} \left[ \widehat{\mathsf{R}}^{(1)} \left(\overline{\mathsf{b}} (\alpha|I);\alpha,I\right) \right] \right\|_{j}=O\left(\frac{1}{LI}\right)$, $j=2,\infty$, noting that $\mathrm{Var} \left[ \widehat{\mathsf{R}}^{(1)} \left(\overline{\mathsf{b}} (\alpha|I);\alpha,I\right) \right]$ is a band matrix up to a basis permutation
follow from the expansion (\ref{VarR1}) obtained in the next section.
The rest of the proof of Lemma \ref{HatR1} is similar to the one of Lemma \ref{HatR2}. The
generic entry of $\widehat{\mathsf{R}}^{\left(  1\right)  }\left(
\mathsf{b};\alpha,I\right)  -\overline{\mathsf{R}}^{\left(  1\right)  }\left(
\mathsf{b};\alpha,I\right)  $ writes%
\[
\widehat{\mathsf{r}}^{(1)}\left(  \mathsf{b};\alpha,I\right)  =\frac{1}{LI}\sum
_{\ell=1}^{L}\xi_{\ell}\left(  \mathsf{b};\alpha\right)
\]
where the $\xi_{\ell}\left(  \mathsf{b};\alpha\right)  $ are centered iid
with, for $K_{p}\left(  t\right)  =t^{p}K\left(  t\right)  /p!$,%
\begin{align*}
\xi_{\ell}\left(  \mathsf{b};\alpha\right)   &  =\sum_{i=1}^{I_{\ell}}\left(
\mathbb{I}\left(  I_{\ell}=I\right)  \xi_{i\ell}\left(  \mathsf{b}%
;\alpha\right)  -\mathbb{E}\left[  \mathbb{I}\left(  I_{\ell}=I\right)
\xi_{i\ell}\left(  \mathsf{b};\alpha\right)  \right]  \right)  ,\\
\xi_{i\ell}\left(  \mathsf{b};\alpha\right)   &  =P_{n}\left(  X_{\ell
}\right)  \left\{  \mathsf{\int_{\underline{t}_{\alpha,h}}^{
	\overline{t}_{\alpha,h}}}\left\{  \mathbb{I}\left[  B_{i\ell}\leq\Psi\left(  t|X_{\ell
},\mathsf{b}\right)  \right]  -\left(  \alpha+ht\right)  \right\}
K_{p}\left(  t\right)  dt\right\}  .
\end{align*}
This gives
\[
\left\vert \frac{\xi_{\ell}\left(  \mathsf{b};\alpha\right)  }{\left(
h+\alpha\left(  1-\alpha\right)  \right)  ^{1/2}}\right\vert \leq
Ch^{-1/2}\max_{x\in\mathcal{X}}\left\Vert P\left(  x\right)  \right\Vert \leq
M_{\infty}\text{ with }M_{\infty}\asymp h^{-\left(  D_{\mathcal{M}}+1\right)
/2}.
\]
For the computation of the variance, Lemma \ref{Phipsi}-(iii,iv) and
Proposition \ref{SeriesB2}-(i) give uniformly in $\alpha$, $t$ in
$\mathcal{T}_{\alpha,h}$ the admissible $\mathsf{b}$ and $X_{\ell}$, and for
the uniform $U_{i\ell}=G\left(  B_{i\ell}|X_{\ell},I_{\ell}\right)  $,%
\begin{align*}
\mathbb{I}\left[  B_{i\ell}\leq\Psi\left(  t|X_{\ell},\mathsf{b}\right)
\right]   &  =\mathbb{I}\left[  B_{i\ell}\leq\Psi\left(  t|X_{\ell}%
,\mathsf{b}\left(  \alpha|I\right)  \right)  +O\left(  h\right)
\right] \\
&  =\mathbb{I}\left[  B\left(  U_{i\ell}|X_{\ell},I_{\ell}\right)  \leq
B\left(  \alpha+ht|x_{\ell},I_{\ell}\right)  +O\left(  h\right)  \right] \\
&  =\mathbb{I}\left[  U_{i\ell}\leq G\left[  B\left(  \alpha+ht|x_{\ell
},I_{\ell}\right)  +O\left(  h\right)  |X_{\ell},I_{\ell}\right]  \right] \\
&  =\mathbb{I}\left[  U_{i\ell}\leq\alpha+ht+O\left(  h\right)  \right]  .
\end{align*}
It then follows, since $U_{i\ell}\sim \mathcal{U}_{[0,1]}$ is independent of $\left(  X_{\ell}%
,I_{\ell}\right)  $%
\begin{align*}
&  \mathbb{E}\left[  \xi_{i\ell}^{2}\left(  \mathsf{b};\alpha\right)
|I_{\ell}\right] \\
&  \quad\leq\mathbb{E}\left[  P_{n}^{2}\left(  X_{\ell}\right)  \int
_{\underline{t}_{\alpha,h}}^{\overline{t}_{\alpha,h}}\int_{
	\underline{t}_{\alpha,h}}^{\overline{t}_{\alpha,h}}
\mathbb{I}\left[  U_{i\ell}\leq
\alpha+h\left(  t_{1}\wedge t_{2}\right)  +O\left(  h\right)  \right]
K_{p}\left(  t_{1}\right)  K_{p}\left(  t_{2}\right)  dt_{1}dt_{2}|I_{\ell
}\right] \\
&  \quad\quad\quad-2\mathbb{E}\left[  P_{n}^{2}\left(  X_{\ell}\right)
\int_{\underline{t}_{\alpha,h}}^{\overline{t}_{\alpha,h}}%
\int_{\underline{t}_{\alpha,h}}^{\overline{t}_{\alpha,h}}
\mathbb{I}\left[
U_{i\ell}\leq\alpha+ht_{1}+O\left(  h\right)  \right]  \left(  \alpha
+ht_{2}\right)  K_{p}\left(  t_{1}\right)  K_{p}\left(  t_{2}\right)
dt_{1}dt_{2}|I_{\ell}\right] \\
&  \quad\quad\quad+\mathbb{E}\left[  P_{n}^{2}\left(  X_{\ell}\right)
|I_{\ell}\right]  
\int_{\underline{t}_{\alpha,h}}^{\overline{t}_{\alpha,h}}\int_{\underline{t}_{\alpha,h}}^{\overline{t}_{\alpha,h}}%
\left(  \alpha+ht_{1}\right)  \left(  \alpha+ht_{2}\right)  K_{p}\left(
t_{1}\right)  K_{p}\left(  t_{2}\right)  dt_{1}dt_{2}\\
&  \quad
\leq \mathbb{E}\left[  P_{n}^{2}\left(  X_{\ell}\right)  |I_{\ell}\right]
\int_{\underline{t}_{\alpha,h}}^{\overline{t}_{\alpha,h}}%
\int_{\underline{t}_{\alpha,h}}^{\overline{t}_{\alpha,h}}\left\{
\alpha+\left| O\left(  h\right) \right|  -\alpha^{2}\right\}  
\left| K_{p}\left(  t_{1}\right)K_{p}\left(  t_{2}\right) \right|  
dt_{1}dt_{2}\leq C\left(  h+\alpha\left(
1-\alpha\right)  \right)
\end{align*}
uniformly in $\alpha$ and $\mathsf{b}$. Hence, uniformly in $\alpha$ and
$\mathsf{b}$%
\[
\operatorname*{Var}\left(  \frac{\xi_{\ell}\left(  \mathsf{b};\alpha\right)
}{\left(  h+\alpha\left(  1-\alpha\right)  \right)  ^{1/2}}\right)  \leq
M_{2}^{2}\text{ with }M_{2}<\infty.
\]

The bracketing part of the proof is similar to the one in Lemma \ref{HatR2} noticing that $\zeta_{\ell} (\mathsf{b};\alpha)$ only depends upon a subvector $\mathsf{b}^{\left(  n_{1},n_{2}\right)  }$ of $\mathsf{b}$ of finite dimension, and similar
to Guerre and Sabbah (2012,2014, Lemma B.2). This gives, for $\mathcal{H}_{L}$ as in Theorem \ref{Dev}
\[
\mathcal{H}_{L}=O\left(  \log L\right)  ^{1/2}+O\left(  \frac{\log
L}{Lh^{D_{\mathcal{M}}+1}}\right)  ^{1/2}=O\left(  \log L\right)  ^{1/2}.
\]
Arguing with Theorem \ref{Dev} then shows that the order of the largest
entry in $\widehat{\mathsf{R}}^{\left(  1\right)  }\left(  \mathsf{b}%
;\alpha,I\right)  -\overline{\mathsf{R}}^{\left(  1\right)  }\left(
\mathsf{b};\alpha,I\right)  $ is $O_{\mathbb{P}}\left(  \log L/L\right)
^{1/2}$, which gives
\[
\max_{\alpha\in\left[  0,1\right]  }
\max_{\mathsf{b\in}\mathcal{B}_{\infty} \left(
	\mathsf{b}\left(  \alpha|I\right)  ,Ch^{D_{\mathcal{M}}/2+1}\right)}
\frac{
\left\Vert \widehat{\mathsf{R}}^{\left(  1\right)  }\left(  \mathsf{b}%
;\alpha,I\right)  -\overline{\mathsf{R}}^{\left(  1\right)  }\left(
\mathsf{b};\alpha,I\right)  
\right\Vert_{\infty}
}{
\left(
h
+
\alpha(1-\alpha)
\right)^{1/2}
}
=O_{\mathbb{P}}\left(
\frac{\log L}{L}\right)^{1/2},
\]
which also gives the result for the $\|\cdot\|_{2}$ by (\ref{Vecnormeq}),
$
\mathcal{B} \left(
\mathsf{b}  ,\epsilon \right)
\subset
\mathcal{B}_{\infty} \left(
\mathsf{b}  ,\epsilon \right)
$
and $N \asymp h^{-D_{\mathcal{M}}}$.$\hspace*{\fill}\square$

\subsection{\textbf{Proof of Lemma \ref{Leadterm}}}

For (i), we first give a convenient expression for $v_h^2 (\alpha)$. 
	Recall that $S_0$ and $S_1$ are $1\times (s+2)$ vectors with
	$S_0 = [1,0,\ldots,0]$ and $S_1 = [0,1,0,\ldots,0]$, see (\ref{Selmat}).
Abbreviate $\Omega_{h}\left(  \alpha\right)  $, $\Omega_{1h}\left(
\alpha\right)  $ in $\Omega$, $\Omega_{1}$.
Define
\begin{align*}
\omega_{0}  &  =\int_{\underline{t}_{\alpha,h}}^{\overline{t}_{\alpha,h}}%
\pi\left(  t\right)  K\left(  t\right)  dt= \Omega S_0^{\prime},\quad\omega_{1}=\int_{
	\underline{t}_{\alpha,h}}^{\overline{t}_{\alpha,h}}t \pi\left(  t\right)  K\left(  t\right)
dt= \Omega S_1^{\prime}\\
\mathbf{\Pi}_{m}  &  \mathbf{=}\int_{\underline{t}_{\alpha,h}}^{
	\overline{t}_{\alpha,h}}\int_{\underline{t}_{\alpha,h}}^{\overline{t}_{\alpha,h}}
\min\left(  t_{1},t_{2}\right)  \pi\left(  t_{1}\right)  \pi\left(
t_{2}\right)  ^{\prime}K\left(  t_{1}\right)  K\left(  t_{2}\right)  dt.
\end{align*}
Note that $v_{h}^{2}\left(  \alpha\right)  =S_{1}\Omega^{-1}\mathbf{\Pi}%
_{m}\Omega^{-1}S_{1}^{\prime}$. Let $\mathbb{W}_0 (\cdot)$ be a Brownian motion over the straight line and $\mathbb{W} (\cdot)$ a Brownian motion over $[0,\infty)$, and set
\[
\mathbf{v}_h^2 (\alpha)
=
\mathrm{Var}
\left[
\int_{\underline{t}_{\alpha,h}}^{\overline{t}_{\alpha,h}} 
\mathbb{W}_0 (t)
S_1
\Omega^{-1}
\pi(t) 
K(t) 
dt
\right].
\]
As
$S_1 \Omega^{-1} \int_{\underline{t}_{\alpha,h}}^{\overline{t}_{\alpha,h}} \pi(t) K(t) dt = S_1 \Omega^{-1} \omega_0 =S_{1} S_{0}^{\prime} =0$, it follows
\begin{align*}
\mathbf{v}_h^2 (\alpha)
& =
\mathrm{Var}
\left[
\int_{\underline{t}_{\alpha,h}}^{\overline{t}_{\alpha,h}} 
\frac{\mathbb{W} (\alpha+ht) - \mathbb{W} (\alpha) }{\sqrt{h}} 
S_1
\Omega^{-1}
\pi(t) 
K(t) 
dt
\right]
\\
&
=
\mathrm{Var}
\left[
\int_{\underline{t}_{\alpha,h}}^{\overline{t}_{\alpha,h}} 
\frac{\mathbb{W} (\alpha+ht)  }{\sqrt{h}} 
S_1
\Omega^{-1}
\pi(t) 
K(t) 
dt
\right]
\\
& =
S_1 \Omega^{-1}
\frac{1}{h}
\int_{\underline{t}_{\alpha,h}}^{\overline{t}_{\alpha,h}} 
\int_{\underline{t}_{\alpha,h}}^{\overline{t}_{\alpha,h}} 
\mathrm{Cov}
\left(
\mathbb{W} (\alpha+ht_1),
\mathbb{W} (\alpha+ht_2)
\right)
\pi (t_1)
\pi (t_2)^{\prime}
K(t_1) K(t_2)
dt_1 dt_2
\Omega^{-1}
S_1^{\prime}
\\
& =
S_1 \Omega^{-1}
\frac{1}{h}
\int_{\underline{t}_{\alpha,h}}^{\overline{t}_{\alpha,h}} 
\int_{\underline{t}_{\alpha,h}}^{\overline{t}_{\alpha,h}} 
\left(
\alpha+ht_1 \wedge t_2
\right)
\pi (t_1)
\pi (t_2)^{\prime}
K(t_1) K(t_2)
dt_1 dt_2
\Omega^{-1}
S_1^{\prime}
\\
& 
=
S_1 \Omega^{-1}
\frac{1}{h}
\left(
\omega_{0} \omega_{0}^{\prime}
+
h\mathbf{\Pi}_{m}
\right)
\Omega^{-1}
S_1^{\prime}
=
\frac{1}{h}
\left(
\left(S_1 S_0^{\prime}\right)^2
+
h
S_1 \Omega^{-1}
\mathbf{\Pi}_{m}
\Omega^{-1}
S_1^{\prime}
\right)
\\
& = v_h^2 (\alpha).
\end{align*}
As
$\overline{t}_{\alpha,h} - \underline{t}_{\alpha,h} \geq 1 $ for $h$ small enough, it follows that
$\inf_{\alpha \in [0,1]} v_{h}^2 (\alpha) \geq 1/C$ while $\sup_{\alpha \in [0,1]} v_{h}^2 (\alpha) \leq C$ since the eigenvalues of $\Omega = \Omega_h (\alpha)$ stay bounded away from $0$ and infinity when $h$ goes to $0$, uniformly in $\alpha \in [0,1]$.

Abbreviate $\mathbf{P}(I)$, $\mathbf{P}_0(\alpha|I)$ and $\mathbf{P}_1 (\alpha|I)$ as
\begin{align*}
\mathbf{P}  &  =\mathbb{E}\left[  \mathbb{I}\left(  I_{\ell}=I\right)
P\left(  X_{\ell}\right)  P\left(  X_{\ell}\right)  ^{\prime}\right]  ,\\
\mathbf{P}_{0}  &  =\mathbb{E}\left[  \frac{\mathbb{I}\left(  I_{\ell
}=I\right)  P\left(  X_{\ell}\right)  P\left(  X_{\ell}\right)  ^{\prime}%
}{B^{\left(  1\right)  }\left(  \alpha|X_{\ell},I_{\ell}\right)  }\right]  ,\\
\mathbf{P}_{1}  &  =\mathbb{E}\left[  \frac{\mathbb{I}\left(  I_{\ell
}=I\right)  B^{\left(  2\right)  }\left(  \alpha|X_{\ell},I_{\ell}\right)
P\left(  X_{\ell}\right)  P\left(  X_{\ell}\right)  ^{\prime}}{\left(
B^{\left(  1\right)  }\left(  \alpha|X_{\ell},I_{\ell}\right)  \right)  ^{2}%
}\right]  ,
\end{align*}
and abbreviate $\Omega_{h}\left(  \alpha\right)  $, $\Omega_{1h}\left(
\alpha\right)  $ in $\Omega$, $\Omega_{1}$. It holds%
\[
\operatorname*{Var}\left(  \widehat{\mathsf{e}}\left(  \alpha|I\right)
\right)  =\left[  \overline{\mathsf{R}}^{\left(  2\right)  }\left(
\overline{\mathsf{b}}\left(  \alpha|I\right)  ;\alpha,I\right)  \right]
^{-1}\operatorname*{Var}\left[  \widehat{\mathsf{R}}^{\left(  1\right)
}\left(  \overline{\mathsf{b}}\left(  \alpha|I\right)  ;\alpha,I\right)
\right]  \left[  \overline{\mathsf{R}}^{\left(  2\right)  }\left(
\overline{\mathsf{b}}\left(  \alpha|I\right)  ;\alpha,I\right)  \right]  ^{-1}%
\]
with by Lemma \ref{R2}%
\begin{align*}
&  \left[  \overline{\mathsf{R}}^{\left(  2\right)  }\left(  \overline
{\mathsf{b}}\left(  \alpha|I\right)  ;\alpha,I\right)  \right]  ^{-1}=\left[
\Omega\otimes\mathbf{P}_{0}-h\Omega_{1}\otimes\mathbf{P}_{1}+o\left(
h\right)  \right]  ^{-1}\\
&  \quad=\left[  \operatorname*{Id}-h\left(  \Omega^{-1}\Omega_{1}\right)
\otimes\left(  \mathbf{P}_{0}^{-1}\mathbf{P}_{1}\right)  +o\left(  h\right)
\right]  ^{-1}\Omega^{-1}\otimes\mathbf{P}_{0}^{-1}\\
&  \quad=\Omega^{-1}\otimes\mathbf{P}_{0}^{-1}+h\left(  \Omega^{-1}\Omega
_{1}\Omega^{-1}\right)  \otimes\left(  \mathbf{P}_{0}^{-1}\mathbf{P}%
_{1}\mathbf{P}_{0}^{-1}\right)  +o\left(  h\right)
\end{align*}
uniformly in $\alpha$ where the remainder term $o\left(  h\right)  $ is with
respect to the matrix norm. 
Now (\ref{Bar2*b}) in the proof of Theorem \ref{Bias} and Lemma \ref{Phipsi}%
-(iii,iv) show that $\left(  LI\right)  \operatorname*{Var}\left[
\widehat{\mathsf{R}}^{\left(  1\right)  }\left(  \overline{\mathsf{b}}\left(
\alpha|I\right)  ;\alpha,I\right)  \right]  $ admits the expansion, with
uniform remainder terms,%
\begin{align}
&  \mathbb{E}\left[  \int_{\underline{t}_{\alpha,h}}^{\overline{t}_{\alpha,h}%
}\int_{\underline{t}_{\alpha,h}}^{\overline{t}_{\alpha,h}}\left\{  G\left[
B\left(  \alpha+ht_{1}|X_{\ell},I_{\ell}\right)  \wedge B\left(  \alpha
+ht_{2}|X_{\ell},I_{\ell}\right)  +o\left(  h\right)  |X_{\ell},I_{\ell
}\right]  \right.  \right. 
\nonumber \\
&  \quad\quad\quad-G\left[  B\left(  \alpha+ht_{1}|X_{\ell},I_{\ell}\right)
+o\left(  h\right)  |X_{\ell},I_{\ell}\right]  \left(  \alpha+ht_{2}\right)
-G\left[  B\left(  \alpha+ht_{2}|X_{\ell},I_{\ell}\right)  +o\left(  h\right)
|X_{\ell},I_{\ell}\right]  \left(  \alpha+ht_{1}\right) \nonumber \\
&  \quad\quad\quad\left.  \left.  +\left(  \alpha+ht_{1}\right)  \left(
\alpha+ht_{2}\right)  \right\}  \pi\left(  t_{1}\right)  \pi\left(
t_{2}\right)  ^{\prime}K\left(  t_{1}\right)  K\left(  t_{2}\right)
dt_{1}dt_{2}\otimes\mathbb{I}\left(  I_{\ell}=I\right)  P\left(  x_{\ell
}\right)  P\left(  X_{\ell}\right)  ^{\prime}\right] \nonumber \\
&  \quad=\int_{\underline{t}_{\alpha,h}}^{\overline{t}_{\alpha,h}}%
\int_{\underline{t}_{\alpha,h}}^{\overline{t}_{\alpha,h}}\left\{
\alpha+h\left(  t_{1}\wedge t_{2}\right)  -\alpha^{2}-h\alpha\left(
t_{1}+t_{2}\right)  \right\}  \pi\left(  t_{1}\right)  \pi\left(
t_{2}\right)  ^{\prime}K\left(  t_{1}\right)  K\left(  t_{2}\right)
dt_{1}dt_{2}+o\left(  h\right) \nonumber \\
&  \quad=\alpha\left(  1-\alpha\right)  \omega_{0}\omega_{0}^{\prime}%
\otimes\mathbf{P}+h\left\{  \mathbf{\Pi}_{m}-\alpha\left(  \omega_{0}%
\omega_{1}^{\prime}+\omega_{1}\omega_{0}^{\prime}\right)  \right\}
\otimes\mathbf{P}+o\left(  h\right)  .
\label{VarR1}
\end{align}
Hence an elementary expansion gives, uniformly in $\alpha\in\left[
0,1\right]  $, $\operatorname*{Var}\left(  \widehat{\mathsf{e}}\left(
\alpha|I\right)  \right)  =\mathsf{V}_{e}/\left(  LI\right)  +o\left(
h\right)  $ with%
\begin{align*}
\mathsf{V}_{e}  &  =\alpha\left(  1-\alpha\right)  \left[  \Omega^{-1}%
\omega_{0}\omega_{0}^{\prime}\Omega^{-1}\right]  \otimes\left[  \mathbf{P}%
_{0}^{-1}\mathbf{PP}_{0}^{-1}\right] \\
&  +h\alpha\left(  1-\alpha\right)  \left[  \Omega^{-1}\Omega_{1}\Omega
^{-1}\omega_{0}\omega_{0}^{\prime}\Omega^{-1}\right]  \otimes\left[
\mathbf{P}_{0}^{-1}\mathbf{P}_{1}\mathbf{P}_{0}^{-1}\mathbf{PP}_{0}%
^{-1}\right] \\
&  +h\alpha\left(  1-\alpha\right)  \left[  \Omega^{-1}\omega_{0}\omega
_{0}^{\prime}\Omega^{-1}\Omega_{1}\Omega^{-1}\right]  \otimes\left[
\mathbf{P}_{0}^{-1}\mathbf{PP}_{0}^{-1}\mathbf{P}_{1}\mathbf{P}_{0}%
^{-1}\right] \\
&  +h\left[  \Omega^{-1}\left(  \mathbf{\Pi}_{m}-\left(  \omega_{1}\omega
_{0}^{\prime}+\omega_{0}\omega_{1}^{\prime}\right)  \right)  \Omega
^{-1}\right]  \otimes\left[  \mathbf{P}_{0}^{-1}\mathbf{PP}_{0}^{-1}\right]  .
\end{align*}
Observe now that $\Omega^{-1}\omega_{0}=S_{0}^{\prime}$, $\Omega^{-1}\omega_{1}=S_{1}^{\prime}$
and $\Omega^{-1}\Omega_{1}\Omega^{-1}\omega_{0}=\Omega^{-1}\Omega_{1}%
S_{0}^{\prime}=\Omega^{-1}\omega_{1}=S_{1}^{\prime}$. This gives%
\begin{align*}
\mathsf{V}_{e}  &  =\alpha\left(  1-\alpha\right)  \left[  S_{0}^{\prime}
S_{0}\right]  \otimes\left[  \mathbf{P}_{0}^{-1}\mathbf{PP}_{0}^{-1}\right] \\
&  +h\alpha\left(  1-\alpha\right)  \left[  S_{1}^{\prime}S_{0}\right]
\otimes\left[  \mathbf{P}_{0}^{-1}\mathbf{P}_{1}\mathbf{P}_{0}^{-1}%
\mathbf{PP}_{0}^{-1}\right] \\
&  +h\alpha\left(  1-\alpha\right)  \left[  S_{0}^{\prime}S_{1}\right]
\otimes\left[  \mathbf{P}_{0}^{-1}\mathbf{PP}_{0}^{-1}\mathbf{P}_{1}%
\mathbf{P}_{0}^{-1}\right] \\
&  +h\left[  \Omega^{-1}\mathbf{\Pi}_{m}\Omega^{-1}-\left(  S_{1}S_{0}%
^{\prime}+S_{0}S_{1}^{\prime}\right)  \right]  \otimes\left[  \mathbf{P}%
_{0}^{-1}\mathbf{PP}_{0}^{-1}\right]  .
\end{align*}
Since the eigenvalues of $\mathbf{P}_{0}^{-1}$, $\mathbf{P}$, $\mathbf{P}_{1}%
$, $\Omega^{-1}$ and 
$\Omega_{1}$ are bounded away from infinity uniformly in
$\alpha$ and 
$
\widehat{\mathsf{e}}_{0}\left(  \alpha|I\right)
=
\mathsf{S}_0
\widehat{\mathsf{e}}\left(  \alpha|I\right)
$, 
it follows that $\max_{\alpha\in\left[  0,1\right]  }\left\Vert
\operatorname*{Var}\left(  \widehat{\mathsf{e}}_{0}\left(  \alpha|I\right)
\right)  \right\Vert =O\left(  1/L\right)  $ and then uniformly in $\alpha$ and $x$
\[
\operatorname*{Var}\left(  P\left(  x\right)  ^{\prime}\widehat{\mathsf{e}%
}_{0}\left(  \alpha|I\right)  \right)  
\asymp
\left\| P(x) \right\|^2
\asymp \frac{1}{h^{D_{\mathcal{M}}}}
\]
by Assumption \ref{Riesz}-(i).
For $\operatorname*{Var}\left(  \widehat{\mathsf{e}}_{1}\left(  \alpha
|I\right)  /h\right)  $, recall that $\widehat{\mathsf{e}}_{1}\left(
\alpha|I\right)  =\mathsf{S}_{1}\widehat{\mathsf{e}}\left(  \alpha|I\right)  $, see (\ref{Selmat}).
It gives that the leading term of $L \operatorname*{Var}\left( \widehat{\mathsf{e}}_{1}\left(  \alpha|I\right)  \right)  $ is
\begin{align*}
\mathsf{S}_{1}\mathsf{V}_{e}\mathsf{S}_{1}^{\prime}  &  =h\left(  S_{1}\Omega
^{-1}\mathbf{\Pi}_{m}\Omega^{-1}S_{1}^{\prime}\right)  \left(  \mathbf{P}_{0}%
^{-1}\mathbf{PP}_{0}^{-1}\right) \\
&  =hv_{h}^{2}\left(  \alpha\right)  \mathbb{E}^{-1}\left[  \frac
{\mathbb{I}\left(  I_{\ell}=I\right)  P\left(  X_{\ell}\right)  P\left(
X_{\ell}\right)  ^{\prime}}{B^{\left(  1\right)  }\left(  \alpha|X_{\ell
},I_{\ell}\right)  }\right] \\
&  \quad\quad\quad\times\mathbb{E}\left[  \mathbb{I}\left(  I_{\ell}=I\right)
P\left(  X_{\ell}\right)  P\left(  X_{\ell}\right)  ^{\prime}\right]
\mathbb{E}^{-1}\left[  \frac{\mathbb{I}\left(  I_{\ell}=I\right)  P\left(
X_{\ell}\right)  P\left(  X_{\ell}\right)  ^{\prime}}{B^{\left(  1\right)
}\left(  \alpha|X_{\ell},I_{\ell}\right)  }\right]
\end{align*}
as $v_{h}^{2}\left(  \alpha\right)  =S_{1}\Omega^{-1}\mathbf{\Pi}%
_{m}\Omega^{-1}S_{1}^{\prime}$. 
This also gives the expansion for 
$Lh \operatorname*{Var}\left( \widehat{\mathsf{e}}_{1}\left(  \alpha|I\right)  /h\right)  $ and the bounds for the eigenvalues of $\operatorname*{Var}\left(  P\left(  x\right)  ^{\prime}\widehat{\mathsf{e}}_{1}\left(  \alpha|I\right)  /h\right)  $ since $ \| P(x)\|\asymp h^{-D_{\mathcal{M}}/2}$ uniformly in $x$ under Assumption \ref{Riesz}-(i), as $v_h^2 (\alpha)$ stays bounded away from $0$ and infinity uniformly in $\alpha$.

For (ii), we just show that $\max_{\left(  \alpha,x\right)  \in\left[
0,1\right]  \times\mathcal{X}}\left\vert P\left(  x\right)  ^{\prime
}\widehat{\mathsf{e}}_{1}\left(  \alpha|I\right)  /h\right\vert =O_{\mathbb{P}%
}\left(  \left(  \log L/Lh^{D_{\mathcal{M}}+1}\right)  ^{1/2}\right)  $, the other result being proved in a similar way. Since
$\max_{x\in\left[  0,1\right]  }\left\Vert P\left(  x\right)  \right\Vert
=O\left(  h^{-D_{\mathcal{M}}/2}\right)  $ and%
\begin{align*}
\max_{\left(  \alpha,x\right)  \in\left[  0,1\right]  \times\mathcal{X}%
}\left\vert \frac{P\left(  x\right)  ^{\prime}\widehat{\mathsf{e}}_{1}\left(
\alpha|I\right)  }{h}\right\vert 
& \leq
\left(  \max_{\left(  \alpha,x\right)
\in\left[  0,1\right]  \times\mathcal{X}}\left\vert \frac{P\left(  x\right)
^{\prime}\widehat{\mathsf{e}}_{1}\left(  \alpha|I\right)  }{h^{1/2}\left(
1+\left\Vert P\left(  x\right)  \right\Vert \right)  }\right\vert \right)
\times h^{-1/2}\left(  1+\max_{x\in\left[  0,1\right]  }\left\Vert P\left(
x\right)  \right\Vert \right) \\
& =
O \left( \frac{1}{\sqrt{h^{D_{\mathcal{M}}+1}}}\right)
\max_{\left(  \alpha,x\right)
	\in\left[  0,1\right]  \times\mathcal{X}}\left\vert \frac{P\left(  x\right)
	^{\prime}\widehat{\mathsf{e}}_{1}\left(  \alpha|I\right)  }{h^{1/2}\left(
	1+\left\Vert P\left(  x\right)  \right\Vert \right)  }\right\vert,
\end{align*}
it is sufficient to show%
\begin{equation}
\max_{\left(  \alpha,x\right)  \in\left[  0,1\right]  \times\mathcal{X}%
}\left\vert \frac{P\left(  x\right)  ^{\prime}\widehat{\mathsf{e}}_{1}\left(
\alpha|I\right)  }{h^{1/2}\left(  1+\left\Vert P\left(  x\right)  \right\Vert
\right)  }\right\vert =O_{\mathbb{P}}\left(  \left(  \frac{\log L}{L}\right)
^{1/2}\right)  . \label{TBP_Lead}%
\end{equation}
Write%
\[
\frac{P\left(  x\right)  ^{\prime}\widehat{\mathsf{e}}_{1}\left(
\alpha|I\right)  }{h^{1/2}\left(  1+\left\Vert P\left(  x\right)  \right\Vert
\right)  }=\frac{1}{L}\sum_{\ell=1}^{L}\xi_{\ell}\left(  \alpha,x\right)
\]
with%
\begin{align*}
\xi_{\ell}\left(  \alpha,x\right)   &  =\sum_{i=1}^{I_{\ell}}\left(
\mathbb{I}\left(  I_{\ell}=I\right)  \xi_{i\ell}\left(  \alpha,x\right)
-\mathbb{E}\left[  \mathbb{I}\left(  I_{\ell}=I\right)  \xi_{i\ell}\left(
\alpha,x\right)  \right]  \right)  ,\\
\xi_{i\ell}\left(  \alpha,x\right)   &  =\frac{P\left(  x\right)  ^{\prime
}\mathsf{S}_{1}\left[  \overline{\mathsf{R}}^{\left(  2\right)  }\left(  \overline
{\mathsf{b}}\left(  \alpha|I\right)  ;\alpha,I\right)  \right]  ^{-1}P\left(
X_{\ell}\right)  }{h^{1/2}\left(  1+\left\Vert P\left(  x\right)  \right\Vert
\right)  }\\
&  \times\left\{  \int_{\underline{t}_{\alpha,h}}^{\overline{t}_{\alpha,h}%
}\left\{  \mathbb{I}\left[  B_{i\ell}\leq\Psi\left(  t|X_{\ell},\overline
{\mathsf{b}}\left(  \alpha|I\right)  \right)  \right]  -\left(  \alpha
+ht\right)  \right\}  K\left(  t\right)  dt\right\}  .
\end{align*}
This gives, uniformly in $\left(  \alpha,x\right)  \in\left[  0,1\right] \times \mathcal{X} $, using the bound of
$Lh \operatorname*{Var}\left( \widehat{\mathsf{e}}_{1}\left(  \alpha|I\right)  /h\right)  $ for the second
\begin{align*}
\left\vert \xi_{\ell}\left(  \alpha,x\right)  \right\vert  &  \leq
Ch^{-1/2}\frac{\left(  \max_{x\in\mathcal{X}}\left\Vert P\left(  x\right)
\right\Vert \right)  ^{2}}{1+\max_{x\in\mathcal{X}}\left\Vert P\left(
x\right)  \right\Vert }\leq M_{\infty}\text{ with }M_{\infty}\asymp
h^{-\left(  D_{\mathcal{M}}+1\right)  /2},\\
\operatorname*{Var}\left(  \xi_{\ell}\left(  \alpha,x\right)  \right)   &
\leq C\frac{\left(  \max_{x\in\mathcal{X}}\left\Vert P\left(  x\right)
\right\Vert \right)  ^{2}}{\left(  1+\max_{x\in\mathcal{X}}\left\Vert P\left(
x\right)  \right\Vert \right)  ^{2}}\leq M_{2}\text{ with }M_{2}\asymp1.
\end{align*}
The Implicit Function Theorem and the FOC $\overline{\mathsf{R}}^{\left(
1\right)  }\left(  \overline{\mathsf{b}}\left(  \alpha|I\right)
;\alpha,I\right)  =0$, Lemma \ref{R2} with (\ref{Bar2*b}) and $s\geq
1$ give that $\alpha\mapsto\overline{\mathsf{b}}\left(
\alpha|I\right)  $ is $\left\Vert \cdot\right\Vert $-Lipshitz with a Lipshitz
constant of order $L^{C}$, as $\alpha\mapsto\left[  \overline{\mathsf{R}%
}^{\left(  2\right)  }\left(  \overline{\mathsf{b}}\left(  \alpha|I\right)
;\alpha,I\right)  \right]  ^{-1}$ and $x\mapsto P\left(  x\right)  /\left(
1+\left\Vert P\left(  x\right)  \right\Vert \right)  $ is H\"{o}lder in a similar sense, see Assumption \ref{Riesz}-(iii). Lemma \ref{Phipsi}%
-(iii), $h \asymp L^{-C}$, arguing as Guerre and Sabbah (2012, 2014, Lemma B.2) imply that
$\left\{  \xi_{\ell}\left(  \alpha,X\right)  ;\left(  \alpha,X\right)
\in\left[  0,1\right]  \times\mathcal{X}\right\}  $ can be bracketed with a
number of brackets%
\[
\exp\left(  H_{L}\left(  \epsilon\right)  \right)  \asymp\left(  \frac{L^{C}%
}{\epsilon}\right)  ^{C}\text{.}%
\]
Arguing as in the proof of Lemma \ref{HatR2} gives, for the item
$\mathcal{H}_{L}$ of Theorem \ref{Dev},
\[
\mathcal{H}_{L}=O\left(  \log L\right)  ^{1/2}+O\left(  \frac{\log
L}{Lh^{D_{\mathcal{M}}+1}}\right)  ^{1/2}=O\left(  \log L\right)  ^{1/2}%
\]
and then (\ref{TBP_Lead}) holds.$\hspace*{\fill}\square$

\section{Proof of Lemma \ref{Cmin}}

The proof of Lemma \ref{Cmin} is based on the following lemma.

\begin{lemma}
\label{Cov} Let $k_{1}\left(  \cdot\right)  $ with primitive $K_{1}\left(
\cdot\right)  $ and $k_{2}\left(  \cdot\right)
$ be two functions over $\left[  0,1\right]  $. Then, if $A$ is a random
variable with a uniform distribution over $\left[  0,1\right]  $ and for any
choice of $K_{1}\left(  \cdot\right)  $,
\begin{align*}
&  \int_{0}^{1}\int_{0}^{1}k_{1}\left(  a_{1}\right)  k_{2}\left(
a_{2}\right)  \left[  a_{1}\wedge a_{2}-a_{1}a_{2}\right]  da_{1}da_{2}\\
&  \quad=-\int_{0}^{1}k_{2}\left(  a_{2}\right)  \left\{  \int_{0}^{a_{2}%
}\left(  K_{1}\left(  a_{1}\right)  -\mathbb{E}\left[  K_{1}\left(  A\right)
\right]  \right)  da_{1}\right\}  da_{2}.
\end{align*}

\end{lemma}

\paragraph{Proof of Lemma \ref{Cov}.}

Observe that%
\begin{align*}
&  \int_{0}^{1}\int_{0}^{1}k_{1}\left(  a_{1}\right)  k_{2}\left(
a_{2}\right)  \left[  a_{1}\wedge a_{2}-a_{1}a_{2}\right]  da_{1}da_{2}\\
&  \quad=\mathbb{E}\left[  \int_{0}^{1}k_{1}\left(  a_{1}\right)
\mathbb{I}\left[  A\leq a_{1}\right]  da_{1}\int_{0}^{1}k_{2}\left(
a_{2}\right)  \mathbb{I}\left[  A\leq a_{2}\right]  da_{1}\right] \\
&  \quad\quad\quad-\mathbb{E}\left[  \int_{0}^{1}k_{1}\left(  a_{1}\right)
\mathbb{I}\left[  A\leq a_{1}\right]  da_{1}\right]  \mathbb{E}\left[
\int_{0}^{1}k_{2}\left(  a_{2}\right)  \mathbb{I}\left[  A\leq a_{2}\right]
da_{1}\right] \\
&  \quad=\operatorname*{Cov}\left(  \int_{0}^{A}k_{1}\left(  a\right)
da,\int_{0}^{A}k_{2}\left(  a\right)  da\right)  =\operatorname*{Cov}\left(
K_{1}\left(  A\right)  ,K_{2}\left(  A\right)  \right)
\end{align*}
which does not depend upon the choice of the primitives. Integrating by parts
now gives%
\begin{align*}
\operatorname*{Cov}\left(  K_{1}\left(  A\right)  ,K_{2}\left(  A\right)
\right) & =\int_{0}^{1}K_{2}\left(  a_{2}\right)  \left(  K_{1}\left(
a_{2}\right)  -\mathbb{E}\left[  K_{2}\left(  A\right)  \right]  \right)
da_{2}\\
&  =\int_{0}^{1}K_{2}\left(  a_{2}\right)  d\left[  \int_{0}^{a_{2}%
}\left(  K_{1}\left(  a_{1}\right)  -\mathbb{E}\left[  K_{2}\left(  A\right)
\right]  \right)  da_{1}\right] \\
&  =-\int_{0}^{1}k_{2}\left(  a_{2}\right)  \left\{  \int_{0}^{a_{2}%
}\left(  K_{1}\left(  a_{1}\right)  -\mathbb{E}\left[  K_{2}\left(  A\right)
\right]  \right)  da_{1}\right\}  da_{2}%
\end{align*}
since $\int_{0}^{a_{2}}\left(  K_{1}\left(  a_{1}\right)  -\mathbb{E}\left[
K_{2}\left(  A\right)  \right]  \right)  da_{1}$ vanishes for $a_{2}=0$ and
$a_{2}=1$.$\hfill\square$

\bigskip

\paragraph{Proof of Lemma \ref{Cmin}.}
Abbreviate $\mathcal{C}_{h} (f,g)$ into $\mathcal{C}_h$.
Set $k_f \left(\alpha,t\right)= f^{\prime} (\alpha) \pi (t) K(t)$
and $K_f (\alpha,t) = \int_{-\infty}^{t} k_f(\alpha,u)du$, so that
\[
\int_{-\infty}^{a} \frac{1}{h} k_f \left(\alpha,\frac{u-\alpha}{h}\right) du
\stackrel{u=\alpha+ht}{=}
\int_{-\infty}^{\frac{a-\alpha}{h}}  k_f \left(\alpha,t\right) dt
=
K_f \left(\alpha,\frac{a-\alpha}{h}\right).
\]
It follows from Lemma \ref{Cov} that%
\begin{align*}
\mathcal{C}_{h}
&= 
\int_{0}^{1}
\int_{0}^{1}
\left\{
\int_{0}^{1}
\int_{0}^{1}
\frac{1}{h^2}
k_f \left(\alpha_1,\frac{a_1-\alpha_1}{h}\right)
k_g \left(\alpha_2,\frac{a_2-\alpha_2}{h}\right)
\left[  a_{1}\wedge a_{2}-a_{1}a_{2}\right]
da_1da_2
\right\}
d\alpha_1 d\alpha_2
\\
& =
-
\int_{0}^{1}
\int_{0}^{1}
\left\{
\int_{0}^{1}
\frac{1}{h}
k_g \left(\alpha_2,\frac{a_2-\alpha_2}{h}\right)
\right.\\
& \quad
\times
\left.
\left[
\int_{0}^{a_2}
\left(
K_f \left(\alpha_1,\frac{a_1-\alpha_1}{h}\right)
-
\mathbb{E}
\left[
K_f \left(\alpha_1,\frac{A-\alpha_1}{h}\right)
\right]
\right)
da_1
\right]
da_2
\right\}
d\alpha_1
d\alpha_2
\\
&  =-\mathcal{T}_{h}+\mathcal{J}_{h}
\end{align*}
with
\begin{align*}
\mathcal{T}_{h}
&
=
\int_{0}^{1}
\int_{0}^{1}
\int_{0}^{1}
\frac{1}{h}
k_g \left(\alpha_2,\frac{a_2-\alpha_2}{h}\right)
%\right.\\
%& \quad
%\times
%\left.
\left[
\int_{0}^{a_2}
K_f \left(\alpha_1,\frac{a_1-\alpha_1}{h}\right)
da_1
\right]
da_2
d\alpha_1
d\alpha_2,
\\
\mathcal{J}_{h}
&=
\int_{0}^{1}
\int_{0}^{1}
\left\{
\int_{0}^{1}
\frac{a_2}{h}
k_g \left(\alpha_2,\frac{a_2-\alpha_2}{h}\right)
%\right.\\
%& \quad
%\times
%\left.
\mathbb{E}
\left[
K_f \left(\alpha_1,\frac{A-\alpha_1}{h}\right)
\right]
da_2
\right\}
d\alpha_1
d\alpha_2.
\end{align*}

Consider first $\mathcal{J}_{h}$, which satisfies
\[
\mathcal{J}_{h}
=
\int_{0}^{1}
\mathbb{E}
\left[
K_f \left(\alpha_1,\frac{A-\alpha_1}{h}\right)
\right]
d\alpha_{1}
\int_{0}^{1}
\left\{
\int_{0}^{1}
\frac{a_2}{h}
k_g \left(\alpha_2,\frac{a_2-\alpha_2}{h}\right)
da_2
\right\}
d\alpha_2.
\]
The change of variable $a_{2}=\alpha_{2}+ht$
and the definition of $\Omega_{h}\left(  \alpha_{2}\right)  $ give
\begin{align*}
&
\int_{0}^{1}
\left\{
\int_{0}^{1}
\frac{a_2}{h}
k_g \left(\alpha_2,\frac{a_2-\alpha_2}{h}\right)
da_2
\right\}
d\alpha_2
=
\int_{0}^{1}
\left\{
\int_{-\frac{\alpha_2}{h}}^{\frac{1-\alpha_2}{h}}
(\alpha_2+ht)
k_g \left(\alpha_2,t\right)
da_2
\right\}
d\alpha_2
\\
&
\quad
=
\int_{0}^{1}
g(\alpha_2)^{\prime}
\left\{
\int_{-\frac{\alpha_2}{h}}^{\frac{1-\alpha_2}{h}}
\left(\alpha_2+ht\right)
\pi (t) K(t)
dt
\right\}
d\alpha_2
\\
&
\quad
=
\int_{0}^{1}
g(\alpha_2)^{\prime}
\left\{
\alpha_2 \Omega_h (\alpha_2) S_0^{\prime}
+
h
\Omega_h (\alpha_2) S_1^{\prime}
\right\}
d\alpha_2
\\
& \quad
=
\int_{0}^{1}
\alpha
g(\alpha)^{\prime}
\Omega_h (\alpha) S_0^{\prime}
d\alpha
+
h
\int_{0}^{1}
g(\alpha)^{\prime}
\Omega_h (\alpha) S_1^{\prime}
d\alpha.
\end{align*}
For the first item in $\mathcal{J}_h$, it holds first, integrating by parts
\begin{align*}
\mathbb{E}
\left[
K_f \left(\alpha_1,\frac{A-\alpha_1}{h}\right)
\right]
& =
\int_{0}^{1}
\int_{-\infty}^{a} \frac{1}{h} k_f \left(\alpha,\frac{u-\alpha}{h}\right) du
da
\\
& = 
\int_{-\infty}^{1} \frac{1}{h} k_f \left(\alpha,\frac{u-\alpha}{h}\right) du
-
\int_{0}^{1}\frac{a}{h} k_f \left(\alpha,\frac{a-\alpha}{h}\right) da
\end{align*}
with
\begin{align*}
\int_{0}^{1} \frac{1}{h} k_f \left(\alpha,\frac{u-\alpha}{h}\right) du
& =
f(\alpha)^{\prime}
\int_{-\frac{\alpha}{h}}^{\frac{1-\alpha}{h}} \pi (t) K(t) dt
=
f(\alpha)^{\prime}
\Omega_h (\alpha) S_0^{\prime},
\\
\int_{-\infty}^{0} \frac{1}{h} k_f \left(\alpha,\frac{u-\alpha}{h}\right) du
& =
f(\alpha)^{\prime}
\int_{-\infty}^{-\frac{\alpha}{h}} \pi (t) K(t) dt.
\end{align*} 
It follows
\begin{align*}
&
\int_{0}^{1}
\mathbb{E}
\left[
K_f \left(\alpha_1,\frac{A-\alpha_1}{h}\right)
\right]
d\alpha_{1}
-
\int_{0}^{1}
f(\alpha_1)^{\prime}
\int_{-\infty}^{-\frac{\alpha_1}{h}} \pi (t) K(t) dt
d \alpha_1
\\
& \quad 
= 
\int_{0}^{1}
f(\alpha)^{\prime}
\Omega_h (\alpha) S_0^{\prime}
d\alpha
-
\int_0^1
f(\alpha)^{\prime}
\left\{
\int_{-\frac{\alpha}{h}}^{\frac{1-\alpha}{h}}
(\alpha+ht)
\pi (t)
K(t)
dt
\right\}
d\alpha
\\
&
\quad 
= 
\int_{0}^{1}
f(\alpha)^{\prime}
(1-\alpha)
\Omega_h (\alpha) S_0^{\prime}
d\alpha
-
h
\int_{0}^{1}
f(\alpha)^{\prime}
\Omega_h (\alpha) S_{1}^{\prime}
d\alpha.
\end{align*}
Hence
\begin{align}
&  
\mathcal{J}_{h}=
\left[  \int_{0}^{1}\alpha g\left(  \alpha\right)^{\prime}
\Omega_{h}\left(  \alpha\right)  d\alpha\right]  
S_{0}^{\prime} S_{0}
\left[
\int_{0}^{1}  \left(  1-\alpha\right)  \Omega_{h}\left(
\alpha\right)  f\left(  \alpha\right) d\alpha\right] 
\nonumber \\
&  
\quad\quad\quad
+h
\left[  \int_{0}^{1}g\left(  \alpha\right)^{\prime}  \Omega
_{h}\left(  \alpha\right)  d\alpha\right]  S_{1}^{\prime}S_{0}\left[  \int%
_{0}^{1}  \left(  1-\alpha\right)  \Omega_{h}\left(
\alpha\right) f\left(  \alpha\right) d\alpha\right] 
\nonumber \\
&  \quad\quad\quad
-
h\left[  \int_{0}^{1}\alpha g\left(  \alpha\right)^{\prime}
\Omega_{h}\left(  \alpha\right)  d\alpha\right]  S_{0}^{\prime}S_{1}\left[
\int_{0}^{1}  \Omega_{h}\left(  \alpha\right) f\left(  \alpha\right)
d\alpha\right] 
\nonumber \\
&  \quad\quad\quad
-h^{2}\left[  \int_{0}^{1}g\left(  \alpha\right)^{\prime}  \Omega
_{h}\left(  \alpha\right)  d\alpha\right]  S_{1}^{\prime}S_{1}\left[  \int%
_{0}^{1}  \Omega_{h}f\left(  \alpha\right)\left(  \alpha\right)
d\alpha\right] 
\nonumber \\
& 
\quad\quad\quad
+
\left[
\int_{0}^{1}
g(\alpha)^{\prime}
\Omega_h (\alpha) 
\left(\alpha S_0^{\prime} + h S_1^{\prime} \right)
d\alpha
\right]
\left[
\int_{0}^{1}
\int_{-\infty}^{-\frac{\alpha_1}{h}} \pi (t)^{\prime} K(t) dt
d\alpha
f(\alpha)
\right].
\label{Jh}
\end{align}

Consider now $\mathcal{T}_{h}$, which satisfies
\begin{align*}
	\mathcal{T}_{h}
	&
	=
	\int_{0}^{1}
	\int_{0}^{1}
	\left\{
	\int_{0}^{1}
	\frac{1}{h}
	k_g \left(\alpha_2,\frac{a_2-\alpha_2}{h}\right)
	%\right.\\
	%& \quad
	%\times
	%\left.
	\left[
	\int_{0}^{a_2}
	K_f \left(\alpha_1,\frac{a_1-\alpha_1}{h}\right)
	da_1
	\right]
	da_2
	\right\}
	d\alpha_1
	d\alpha_2
	\\
	& 
	\stackrel{a_2=\alpha_2+ht_2}{=}
	\int_{0}^{1}
	\int_{0}^{1}
	\Bigg\{
	\int_{-\frac{\alpha_2}{h}}^{\frac{1-\alpha_2}{h}}
	k_g \left(\alpha_2,t_2\right)
	\\
	&
	\quad\quad\quad\quad\quad\quad\quad\quad\quad
	\times
	\left[
	\int_{0}^{\alpha_2+ht_2}
	\left(
	\int_{-\infty}^{a_1}
	\frac{1}{h}
	k_f \left(\alpha_1,\frac{a-\alpha_1}{h}\right)
	da
	\right)
	da_1
	\right]
	dt_2
	\Bigg\}
	d\alpha_1
	d\alpha_2
	\\
	& 
	\stackrel{a_1=\alpha_1+ht}{=}
	\int_{0}^{1}
	\int_{0}^{1}
	\left\{
	\int_{-\frac{\alpha_2}{h}}^{\frac{1-\alpha_2}{h}}
	k_g \left(\alpha_2,t_2\right)
	\left[
	\int_{0}^{\alpha_2+ht_2}
	\left(
	\int_{-\infty}^{\frac{a_1-\alpha_1}{h}}
	k_f \left(\alpha_1,t\right)
	dt
	\right)
	da_1
	\right]
	\right\}
	d\alpha_1
	d\alpha_2.
\end{align*}
For the item between squared brackets, integrating by parts gives
\begin{align*}
&
\int_{0}^{\alpha_2+ht_2}
\left(
\int_{-\infty}^{\frac{a_1-\alpha_1}{h}}
k_f \left(\alpha_1,t\right)
dt
\right)
da_1
= 
\int_{0}^{\alpha_2}
\left(
\int_{-\infty}^{\frac{a_1-\alpha_1}{h}}
k_f \left(\alpha_1,t\right)
dt
\right)
da_1
\\
& 
\quad\quad\quad\quad\quad\quad\quad\quad\quad
+
\int_{\alpha_2}^{\alpha_2+ht_2}
\left(
\int_{-\infty}^{\frac{a_1-\alpha_1}{h}}
k_f \left(\alpha_1,t\right)
dt
\right)
d\left[a_1 - \alpha_2 - ht_2\right]
\\
& 
\quad
= 
\int_{0}^{\alpha_2}
\left(
\int_{-\infty}^{\frac{a_1-\alpha_1}{h}}
k_f \left(\alpha_1,t\right)
dt
\right)
da_1
+
h t_2
\int_{-\infty}^{\frac{\alpha_2-\alpha_1}{h}}
k_f \left(\alpha_1,t\right)
dt
\\
&
\quad\quad\quad
-
\int_{\alpha_2}^{\alpha_2+ht_2}
\left(
a_1 - \alpha_2 - ht_2
\right)
\frac{1}{h}
k_f \left(\alpha_1,\frac{a_1-\alpha_1}{h}\right)
d a_1
\\
& 
=
\int_{0}^{\alpha_2}
\left(
\int_{-\infty}^{\frac{a_1-\alpha_1}{h}}
k_f \left(\alpha_1,t\right)
dt
\right)
da_1
+
h t_2
\int_{-\infty}^{\frac{\alpha_2-\alpha_1}{h}}
k_f \left(\alpha_1,t\right)
dt
\\
&
\quad\quad\quad
+
\left(
ht_2
\right)^2
\int_{0}^{1}
\left(
1-u
\right)
\frac{1}{h}
k_f \left(\alpha_1,\frac{\alpha_2+u\cdot ht_2-\alpha_1}{h}\right)
d u
.
\end{align*}
It follows that 
\[
\mathcal{T}_{h}=\mathcal{T}_{0}+h\mathcal{T}_{1}%
+h^{2}\mathcal{T}_{2}\] 
with
\begin{align*}
\mathcal{T}_{0}  
&  =
\int_{0}^{1}
\int_{0}^{1}
\left\{
\int_{-\frac{\alpha_2}{h}}^{\frac{1-\alpha_2}{h}}
k_g \left(\alpha_2,t_2\right)
dt_2
\left[
\int_{0}^{\alpha_2}
\left(
\int_{-\infty}^{\frac{a_1-\alpha_1}{h}}
k_f \left(\alpha_1,t\right)
dt
\right)
da_1
\right]
\right\}
d\alpha_1
d\alpha_2
\\
& =
\int_{0}^{1}
\int_{0}^{1}
\left\{
g (\alpha_2)^{\prime}
\Omega_h (\alpha_2)
S_0^{\prime}
\left[
\int_{0}^{\alpha_2}
\left(
\int_{-\infty}^{\frac{a_1-\alpha_1}{h}}
k_f \left(\alpha_1,t\right)
dt
\right)
da_1
\right]
\right\}
d\alpha_1
d\alpha_2
,
\\
\mathcal{T}_{1}  
&  =
\int_{0}^{1}
\int_{0}^{1}
\left\{
\int_{-\frac{\alpha_2}{h}}^{\frac{1-\alpha_2}{h}}
k_g \left(\alpha_2,t_2\right)
\left[
t_2
\int_{-\infty}^{\frac{\alpha_2-\alpha_1}{h}}
k_f \left(\alpha_1,t\right)
dt
\right]
d t_2
\right\}
d\alpha_1
d\alpha_2
\\
&  =
\int_{0}^{1}
\int_{0}^{1}
\left\{
g(\alpha_2)^{\prime}
\Omega_h  (\alpha_2)
S_1^{\prime}
\left[
\int_{-\infty}^{\frac{\alpha_2-\alpha_1}{h}}
k_f \left(\alpha_1,t\right)
dt
\right]
\right\}
d\alpha_1
d\alpha_2
,
\\
\mathcal{T}_{2}  
&  =
	\int_{0}^{1}
\int_{0}^{1}
\Bigg\{
\int_{-\frac{\alpha_2}{h}}^{\frac{1-\alpha_2}{h}}
k_g \left(\alpha_2,t_2\right)
t_{2}^{2}
\\
&
\quad\quad\quad
\times
\left[
\int_{0}^{1}
\left(
1-u
\right)
\frac{1}{h}
k_f \left(\alpha_1,\frac{\alpha_2+u\cdot ht_2-\alpha_1}{h}\right)
d u
\right]
dt_2
\Bigg\}
d\alpha_1
d\alpha_2.
\end{align*}

Consider first $\mathcal{T}_0$. A first integration by parts gives
\begin{align*}
\mathcal{T}_{0}  
& =
-
\int_{0}^{1}
\left\{
\int_{0}^{1}
\left[
\int_{0}^{\alpha_2}
\left(
\int_{-\infty}^{\frac{a_1-\alpha_1}{h}}
k_f \left(\alpha_1,t\right)
dt
\right)
da_1
\right]
d
\left[
\int_{\alpha_2}^{1}
g (a_2)^{\prime}
\Omega_h (a_2)
S_0^{\prime}
da_2
\right]
\right\}
d\alpha_1
\\
& =
\int_{0}^{1}
\int_{0}^{1}
\left(
\int_{\alpha_2}^{1}
g (a_2)^{\prime}
\Omega_h (a_2)
S_0^{\prime}
da_2
\right)
\left(
\int_{-\infty}^{\frac{\alpha_2-\alpha_1}{h}}
k_f \left(\alpha_1,t\right)
dt
\right)
d\alpha_1
d\alpha_2.
\end{align*}
Integrating by parts again then gives
\begin{align*}
\mathcal{T}_{0}  
& =
-
\int_{0}^{1}
\left\{
\int_{0}^{1}
\left(
\int_{-\infty}^{\frac{\alpha_2-\alpha_1}{h}}
k_f \left(\alpha_1,t\right)
dt
\right)
d
\left[
\int_{\alpha_2}^{1}
\int_{a_2}^{1}
g (\alpha)^{\prime}
\Omega_h (\alpha)
S_0^{\prime}
d\alpha
da_2
\right]
\right\}
d\alpha_1
\\
& = 
\int_{0}^{1}
\int_{a_2}^{1}
g (\alpha)^{\prime}
\Omega_h (\alpha)
S_0^{\prime}
d\alpha
da_2
\times
\int_{0}^{1}
\left(
\int_{-\infty}^{-\frac{\alpha_1}{h}}
k_f \left(\alpha_1,t\right)
dt
\right)
d\alpha_1
\\
&
\quad
+
\int_{0}^{1}
\int_{0}^{1}
\left(
\int_{\alpha_2}^{1}
\int_{a_2}^{1}
g (\alpha)^{\prime}
\Omega_h (\alpha)
S_0^{\prime}
d\alpha
da_2
\right)
\frac{1}{h}
k_f \left(\alpha_1,\frac{\alpha_2-\alpha_1}{h}\right)
d\alpha_1
d \alpha_2
\\
& = 
\int_{0}^{1}
a_2
g (a_2)^{\prime}
\Omega_h (a_2)
S_0^{\prime}
da_2
\times
\int_{0}^{1}
\left(
\int_{-\infty}^{-\frac{\alpha_1}{h}}
k_f \left(\alpha_1,t\right)
dt
\right)
d\alpha_1
\\
&
\quad
+
\int_{0}^{1}
\int_{0}^{1}
\left(
\int_{\alpha_2}^{1}
\int_{a_2}^{1}
g (\alpha)^{\prime}
\Omega_h (\alpha)
S_0^{\prime}
d\alpha
da_2
\right)
\frac{1}{h}
k_f \left(\alpha_1,\frac{\alpha_2-\alpha_1}{h}\right)
d\alpha_1
d \alpha_2,
\end{align*}
as 
$
\int_{0}^{1}
\int_{a_2}^{1}
g (\alpha)^{\prime}
\Omega_h (\alpha)
S_0^{\prime}
d\alpha
da_2
=
\int_{0}^{1}
a_2
g (a_2)^{\prime}
\Omega_h (a_2)
S_0^{\prime}
da_2
$.
As the first item is left unchanged in the final expression of $\mathcal{T}_0$, we focus now on the second integral. It holds 
\begin{align*}
&
\int_{0}^{1}
\int_{0}^{1}
\left(
\int_{\alpha_2}^{1}
\int_{a_2}^{1}
g (\alpha)^{\prime}
\Omega_h (\alpha)
S_0^{\prime}
d\alpha
da_2
\right)
\frac{1}{h}
k_f \left(\alpha_1,\frac{\alpha_2-\alpha_1}{h}\right)
d\alpha_1
d \alpha_2
\\
&
\quad
\stackrel{\alpha_2=\alpha_1+ht}{=}
\int_{0}^{1}
\left\{
\int_{-\frac{\alpha_1}{h}}^{\frac{1-\alpha_1}{h}}
\left[
\int_{\alpha_1+ht}^{1}
\int_{a_2}^{1}
g (\alpha)^{\prime}
\Omega_h (\alpha)
S_0^{\prime}
d\alpha
da_2
\right]
k_f \left(\alpha_1,t\right)
dt
\right\}
d\alpha_1
.
\end{align*}
Now a second-order Taylor expansion with integral remainder gives
\begin{align*}
&
\int_{\alpha_1+ht}^{1}
\int_{a_2}^{1}
g (\alpha)^{\prime}
\Omega_h (\alpha)
S_0^{\prime}
d\alpha
da_2
=
\int_{\alpha_1}^{1}
\int_{a_2}^{1}
g (\alpha)^{\prime}
\Omega_h (\alpha)
S_0^{\prime}
d\alpha
da_2
-
ht
\int_{\alpha_1}^{1}
g (\alpha)^{\prime}
\Omega_h (\alpha)
S_0^{\prime}
d\alpha
\\
&
\quad\quad\quad\quad\quad\quad\quad\quad\quad\quad\quad\quad
\quad\quad\quad
+
\frac{(ht)^2}{2}
\left(
g (\alpha_1)^{\prime}
\Omega_h (\alpha_1)
S_0^{\prime}
+
\epsilon (t;\alpha_1)
\right)
\text{ where }
\\
&
\epsilon (t;\alpha_1)
=
\int_0^{1}
\left[
g (\alpha_1+htu)^{\prime}
\Omega_h (\alpha_1+htu)
S_0^{\prime}
-
g (\alpha_1)^{\prime}
\Omega_h (\alpha_1)
S_0^{\prime}
\right]
(1-u)du.
\end{align*}
Note that $|\epsilon (t;\alpha_1)| \leq C\left(h \mathbb{I} (h\leq\alpha_1\leq 1-h) + 1-\mathbb{I} (h\leq\alpha_1\leq 1-h)\right)$. It then follows, recalling
$k_f \left(\alpha_1,t\right) = f^{\prime} (\alpha_1) \pi(t) K(t)$,
\begin{align*}
&
\int_{0}^{1}
\int_{0}^{1}
\left(
\int_{\alpha_2}^{1}
\int_{a_2}^{1}
g (\alpha)^{\prime}
\Omega_h (\alpha)
S_0^{\prime}
d\alpha
da_2
\right)
\frac{1}{h}
k_f \left(\alpha_1,\frac{\alpha_2-\alpha_1}{h}\right)
d\alpha_1
d \alpha_2
\\
& \quad 
=
\int_{0}^{1}
\left\{
\left[
\int_{\alpha_1}^{1}
\int_{a_2}^{1}
g (\alpha)^{\prime}
\Omega_h (\alpha)
S_0^{\prime}
d\alpha
da_2
\right]
\int_{-\frac{\alpha_1}{h}}^{\frac{1-\alpha_1}{h}}
k_f \left(\alpha_1,t\right)
dt
\right\}
d\alpha_1
\\
&
\quad\quad
-
h
\int_{0}^{1}
\left\{
\left[
\int_{\alpha_1}^{1}
g (\alpha)^{\prime}
\Omega_h (\alpha)
S_0^{\prime}
d\alpha
\right]
\int_{-\frac{\alpha_1}{h}}^{\frac{1-\alpha_1}{h}}
t
k_f \left(\alpha_1,t\right)
dt
\right\}
d\alpha_1
\\
& \quad\quad
+
\frac{h^2}{2}
\int_{0}^{1}
\left\{
\left[
g (\alpha_1)^{\prime}
\Omega_h (\alpha_1)
S_0^{\prime}
\right]
\int_{-\frac{\alpha_1}{h}}^{\frac{1-\alpha_1}{h}}
t^2
k_f \left(\alpha_1,t\right)
dt
\right\}
d\alpha_1
+
o(h^2)
\\
& \quad 
=
\int_{0}^{1}
\left[
\int_{\alpha_1}^{1}
\int_{a_2}^{1}
g (\alpha)^{\prime}
\Omega_h (\alpha)
S_0^{\prime}
d\alpha
da_2
\right]
f(\alpha_1)^{\prime}
\Omega_h (\alpha_1) S_0^{\prime}
d\alpha_1
\\
& \quad\quad
-
h
\int_{0}^{1}
\left[
\int_{\alpha_1}^{1}
g (\alpha)^{\prime}
\Omega_h (\alpha)
S_0^{\prime}
d\alpha
\right]
f(\alpha_1)^{\prime}
\Omega_h (\alpha_1) S_1^{\prime}
d\alpha_1
\\
& \quad\quad
+
\frac{h^2}{2}
\int_{0}^{1}
g (\alpha_1)^{\prime}
\Omega_h (\alpha_1)
S_0^{\prime}
d\alpha
f(\alpha_1)^{\prime}
\Omega_h (\alpha_1) S_2^{\prime}
d\alpha_1
+o (h^2).
\end{align*}
As
\begin{align*}
&
\int_{0}^{1}
\left[
\int_{\alpha_1}^{1}
\int_{a_2}^{1}
g (\alpha)^{\prime}
\Omega_h (\alpha)
d\alpha
da_2
\right]
S_0^{\prime}
S_0
\Omega_h (\alpha_1) f(\alpha_1)
d\alpha_1
\\
&
\quad
=
\int_{0}^{1}
\left[
\int_{\alpha_1}^{1}
g (\alpha)^{\prime}
\Omega_h (\alpha)
d\alpha
\right]
S_0^{\prime}
S_0
\left[
\int_{0}^{\alpha_1}
\Omega_h (\alpha) f(\alpha)
d\alpha
\right]
d\alpha_1,
\end{align*}
it gives for $\mathcal{T}_0$,
\begin{align*}
\mathcal{T}_0
& =
\int_{0}^{1}
\left[
\int_{\alpha_1}^{1}
g (\alpha)^{\prime}
\Omega_h (\alpha)
d\alpha
\right]
S_0^{\prime}
S_0
\left[
\int_{0}^{\alpha_1}
\Omega_h (\alpha) f(\alpha)
d\alpha
\right]
d\alpha_1
\\
& \quad
-
h
\int_{0}^{1}
\left[
\int_{\alpha_1}^{1}
g (\alpha)^{\prime}
\Omega_h (\alpha)
d\alpha
\right]
S_0^{\prime} S_1
\Omega_h (\alpha_1) f(\alpha_1)
d\alpha_1
\\
& \quad
+
\frac{h^2}{2}
\int_{0}^{1}
g (\alpha_1)^{\prime}
\Omega_h (\alpha_1)
S_0^{\prime} S_2
\Omega_h (\alpha_1) f(\alpha_1)
d\alpha_1
\\
& \quad
+
\int_{0}^{1}
a_2
g (a_2)^{\prime}
\Omega_h (a_2)
S_0^{\prime}
da_2
\times
\int_{0}^{1}
\left(
\int_{-\infty}^{-\frac{\alpha_1}{h}}
k_f \left(\alpha_1,t\right)
dt
\right)
d\alpha_1
+
o(h^2).
\end{align*}

Let us now turn to $\mathcal{T}_1$. Integrating by parts  gives
\begin{align*}
\mathcal{T}_1
& =
-
\int_{0}^{1}
\left\{
\int_{0}^{1}
\left[
\int_{-\infty}^{\frac{\alpha_2-\alpha_1}{h}}
k_f \left(\alpha_1,t\right)
dt
\right]
d
\left[
\int^{1}_{\alpha_2}
g(\alpha)^{\prime}
\Omega_h  (\alpha)
S_1^{\prime}
d\alpha
\right]
\right\}
d\alpha_1
\\
& = 
\int_{0}^{1}
\left[
\int^{1}_{0}
g(\alpha)^{\prime}
\Omega_h  (\alpha)
S_1^{\prime}
d\alpha
\right]
\left[
\int_{-\infty}^{-\frac{\alpha_1}{h}}
k_f \left(\alpha_1,t\right)
dt
\right]
d \alpha_1
\\
&
\quad
+
\int_{0}^{1}
\left\{
\int_{0}^{1}
\left[
\int^{1}_{\alpha_2}
g(\alpha)^{\prime}
\Omega_h  (\alpha)
S_1^{\prime}
d\alpha
\right]
\frac{1}{h}
k_f \left(\alpha_1,\frac{\alpha_2-\alpha_1}{h}\right)
d \alpha_2
\right\}
d \alpha_1
\end{align*}
where
\begin{align*}
&
\int_{0}^{1}
\left\{
\int_{0}^{1}
\left[
\int^{1}_{\alpha_2}
g(\alpha)^{\prime}
\Omega_h  (\alpha)
S_1^{\prime}
d\alpha
\right]
\frac{1}{h}
k_f \left(\alpha_1,\frac{\alpha_2-\alpha_1}{h}\right)
d \alpha_2
\right\}
d \alpha_1
\\
&
\quad
\stackrel{\alpha_2=\alpha_1+ht}{=}
\int_{0}^{1}
\left\{
\int_{-\frac{\alpha_1}{h}}^{\frac{1-\alpha_1}{h}}
\left[
\int^{1}_{\alpha_1+ht}
g(\alpha)^{\prime}
\Omega_h  (\alpha)
S_1^{\prime}
d\alpha
\right]
k_f \left(\alpha_1,t\right)
d t
\right\}
d \alpha_1
\\
& 
\quad
= 
\int_{0}^{1}
\left\{
\int_{-\frac{\alpha_1}{h}}^{\frac{1-\alpha_1}{h}}
\left[
\int^{1}_{\alpha_1}
g(\alpha)^{\prime}
\Omega_h  (\alpha)
S_1^{\prime}
d\alpha
-ht
g(\alpha_1)^{\prime}
\Omega_h  (\alpha_1)
S_1^{\prime}
\right]
k_f \left(\alpha_1,t\right)
d t
\right\}
d \alpha_1
\\
&  
\quad\quad
+
\int_{0}^{1}
\Bigg\{
\int_{-\frac{\alpha_1}{h}}^{\frac{1-\alpha_1}{h}}
\Bigg[
ht
\int_0^1
\Big(
g(\alpha_1+uht)^{\prime}
\Omega_h  (\alpha_1+uht)
\\
&
\quad\quad\quad\quad\quad\quad\quad\quad\quad\quad\quad\quad
-
g(\alpha_1)^{\prime}
\Omega_h  (\alpha_1)
\Big)
S_1^{\prime}
(1-u)du
\Bigg]
k_f \left(\alpha_1,t\right)
d t
\Bigg\}
d \alpha_1
\\
& =
\int_{0}^{1}
\left[
\int^{1}_{\alpha_1}
g(\alpha)^{\prime}
\Omega_h  (\alpha)
d\alpha
\right]
S_1^{\prime}
S_0
\Omega_h  (\alpha_1)
f(\alpha_1)
d\alpha_1
\\
&
\quad\quad\quad
-
h
\int_{0}^{1}
g(\alpha_1)^{\prime}
\Omega_h  (\alpha_1)
S_1^{\prime}
S_1
\Omega_h  (\alpha_1)
f(\alpha_1)
d\alpha_1
+
o(h).
\end{align*}
It follows that
\begin{align*}
\mathcal{T}_1
& =
\int_{0}^{1}
\left[
\int^{1}_{\alpha_1}
g(\alpha)^{\prime}
\Omega_h  (\alpha)
d\alpha
\right]
S_1^{\prime}
S_0
\Omega_h  (\alpha_1)
f(\alpha_1)
d\alpha_1
\\
& \quad\quad
-
h
\int_{0}^{1}
g(\alpha_1)^{\prime}
\Omega_h  (\alpha_1)
S_1^{\prime}
S_1
\Omega_h  (\alpha_1)
f(\alpha_1)
d\alpha_1
\\
&
\quad\quad
+
\left[
\int^{1}_{0}
g(\alpha)^{\prime}
\Omega_h  (\alpha)
d\alpha
\right]
S_1^{\prime}
\int_{0}^{1}
\left[
\int_{-\infty}^{-\frac{\alpha_1}{h}}
k_f \left(\alpha_1,t\right)
dt
\right]
d \alpha_1
+
o(h).
\end{align*}

For $\mathcal{T}_2$, it holds
\begin{align*}
\mathcal{T}_{2}  
&  =
\int_{0}^{1}
\big(
1-u
\big)
\Bigg[
\int_{0}^{1}
\int_{0}^{1}
\Bigg\{
\int_{-\frac{\alpha_2}{h}}^{\frac{1-\alpha_2}{h}}
k_g \left(\alpha_2,t_2\right)
t_{2}^{2}
\frac{1}{h}
k_f \left(\alpha_1,\frac{\alpha_2+u\cdot ht_2-\alpha_1}{h}\right)
dt_2
\Bigg\}
d\alpha_1
d\alpha_2
\Bigg]
du
\\
& 
\stackrel{a_1=\alpha_1-uht_2}{=}
\int_{0}^{1}
\big(
1-u
\big)
\Bigg[
\int_{0}^{1}
\Bigg\{
\int_{-\frac{\alpha_2}{h}}^{\frac{1-\alpha_2}{h}}
k_g \left(\alpha_2,t_2\right)
t_{2}^{2}
\\
&
\quad\quad\quad
\times
\Bigg(
\int_{-uht_2}^{1-uht_2}
f \left(a_1+uht_2\right)^{\prime}
\frac{1}{h}
\pi
\left(\frac{\alpha_2-a_1}{h}\right)
K
\left(\frac{\alpha_2-a_1}{h}\right)
da_1
\Bigg)
dt_2
\Bigg\}
d\alpha_2
\Bigg]
du
\\
& 
=
\int_{0}^{1}
\big(
1-u
\big)
\Bigg[
\int_{0}^{1}
\Bigg\{
\int_{-\frac{\alpha_2}{h}}^{\frac{1-\alpha_2}{h}}
k_g \left(\alpha_2,t_2\right)
t_{2}^{2}
\\
&
\quad\quad\quad
\times
\Bigg(
\int_{-uht_2}^{1-uht_2}
f \left(a_1 \right)^{\prime}
\frac{1}{h}
\pi
\left(\frac{\alpha_2-a_1}{h}\right)
K
\left(\frac{\alpha_2-a_1}{h}\right)
da_1
\Bigg)
dt_2
\Bigg\}
d\alpha_2
\Bigg]
du
+
o(1)
\\
&
\stackrel{a_1=\alpha_2+ht_1}{=}
\int_{0}^{1}
\big(
1-u
\big)
\Bigg[
\int_{0}^{1}
\Bigg\{
\int_{-\frac{\alpha_2}{h}}^{\frac{1-\alpha_2}{h}}
g(\alpha_2)^{\prime} t_2^2 \pi(t_2)K(t_2)
\\
&
\quad\quad\quad\quad\quad\quad\quad\quad\quad
\times
\Bigg(
\int_{-\frac{\alpha_2+uht_2}{h}}^{\frac{1-\alpha_2-uht_2}{h}}
f \left(\alpha_2 \right)^{\prime}
\pi
\left(t_1\right)
K
\left(t_1\right)
dt_1
\Bigg)
dt_2
\Bigg\}
d\alpha_2
\Bigg]
du
+
o(1)
\\
& =
\int_{0}^{1}
\big(
1-u
\big)
\Bigg[
\int_{0}^{1}
\Bigg\{
\int_{-\frac{\alpha_2}{h}}^{\frac{1-\alpha_2}{h}}
g(\alpha_2)^{\prime} t_2^2 \pi(t_2)K(t_2)
S_0 \Omega_h (\alpha_2+uht_2) f(\alpha_2) dt_2
\Bigg\}
d\alpha_2
\Bigg]
du
+
o(1)
\\
& =
\int_{0}^{1}
\big(
1-u
\big)
\Bigg[
\int_{0}^{1}
\Bigg\{
\int_{-\frac{\alpha_2}{h}}^{\frac{1-\alpha_2}{h}}
g(\alpha_2)^{\prime} t_2^2 \pi(t_2)K(t_2)
S_0 \Omega_h (\alpha_2) f(\alpha_2) dt_2
\Bigg\}
d\alpha_2
\Bigg]
du
+
o(1),
\end{align*}
and then
\[
\mathcal{T}_{2}
=
\frac{1}{2}
\int_{0}^{1}
g(\alpha)^{\prime}
\Omega_h (\alpha)
S_2^{\prime}
S_0
\Omega_h (\alpha) f(\alpha)
d\alpha
+o(1).
\]

As $\mathcal{T}_{h}=\mathcal{T}_{0}+h\mathcal{T}_{1}%
	+h^{2}\mathcal{T}_{2}$, collecting all the items gives
\begin{align}
\mathcal{T}_{h}
&
=
\int_{0}^{1}
\left[
\int_{\alpha}^{1}
g (a)^{\prime}
\Omega_h (a)
da
\right]
S_0^{\prime}
S_0
\left[
\int_{0}^{\alpha}
\Omega_h (a) f(a)
da
\right]
d\alpha
\nonumber
\\
& \quad\quad
-
h
\int_{0}^{1}
\left[
\int_{\alpha}^{1}
g (\alpha)^{\prime}
\Omega_h (a)
da
\right]
\left(
S_0^{\prime} S_1
+
S_1^{\prime} S_0
\right)
\Omega_h (\alpha) f(\alpha)
d\alpha
\nonumber
\\
& \quad\quad
-
h^2
\int_{0}^{1}
g(\alpha)^{\prime}
\Omega_h  (\alpha)
S_1^{\prime}
S_1
\Omega_h  (\alpha)
f(\alpha)
d\alpha
\nonumber
\\
& \quad\quad
+
\frac{h^2}{2}
\int_{0}^{1}
g (\alpha)^{\prime}
\Omega_h (\alpha)
S_0^{\prime} S_2
\Omega_h (\alpha) f(\alpha)
d\alpha
\nonumber
\\
& \quad
+
\left[
\int_{0}^{1}
g (\alpha)^{\prime}
\Omega_h (\alpha)
\left(
\alpha
S_0^{\prime}
+
h
S_1^{\prime}
\right)
d\alpha
\right]
\times
\int_{0}^{1}
\left(
\int_{-\infty}^{-\frac{\alpha}{h}}
k_f \left(\alpha,t\right)
dt
\right)
d\alpha
+
o(h^2).
\label{Th}
\end{align}

We now prepare to compute the expansion of $\mathcal{J}_{h}-\mathcal{T}_{h}$ from
	(\ref{Jh}) and (\ref{Th}).
Observe $\int_{0}^{1}\left[  \int_{\alpha}^{1}g\left(  a\right)^{\prime}  \Omega
	_{h}\left(  a\right) S da\right]  d\alpha=\int_{0}^{1}\alpha g\left(
	\alpha\right)^{\prime}  \Omega_{h}\left(  \alpha\right) S d\alpha$, so that
\begin{align*}
&  \left[  \int_{0}^{1}\alpha g\left(  \alpha\right)^{\prime}  \Omega_{h}\left(
\alpha\right)  d\alpha\right]  S_{0}^{\prime}S_{0}\left[  \int_{0}^{1}  \left(  1-\alpha\right)  \Omega_{h}\left(  \alpha\right) f\left(
\alpha\right)
d\alpha\right] \\
&  \quad\quad\quad-\int_{0}^{1}\left[  \int_{\alpha}^{1}g\left(  a\right)^{\prime}
\Omega_{h}\left(  a\right)  da\right]  S_{0}^{\prime}S_{0}\left[  \int%
_{0}^{\alpha}\Omega_{h}\left(  a\right)  f\left(  a\right)  da\right]
d\alpha\\
&  \quad=-\left[  \int_{0}^{1}\alpha g\left(  \alpha\right)^{\prime}  \Omega_{h}\left(
\alpha\right)  d\alpha\right]  S_{0}^{\prime}S_{0}\left[  \int_{0}^{1}\alpha
  \Omega_{h}\left(  \alpha\right) f\left(  \alpha\right) d\alpha\right] \\
&  \quad\quad\quad+\left[  \int_{0}^{1}\alpha g\left(  \alpha\right)^{\prime}
\Omega_{h}\left(  \alpha\right)  d\alpha\right]  S_{0}^{\prime}S_{0}\left[
\int_{0}^{1}  \Omega_{h}\left(  \alpha\right)f\left(  \alpha\right)
d\alpha\right] \\
&  \quad\quad\quad-\int_{0}^{1}\left[  \int_{\alpha}^{1}g\left(  a\right)^{\prime}
\Omega_{h}\left(  a\right)  da\right]  S_{0}^{\prime}S_{0}\left[  \int_{0}%
^{1}\Omega_{h}\left(  a\right)  f\left(  a\right)  da\right]  d\alpha\\
&  \quad\quad\quad+\int_{0}^{1}\left[  \int_{\alpha}^{1}g\left(  a\right)^{\prime}
\Omega_{h}\left(  a\right)  da\right]  S_{0}^{\prime}S_{0}\left[  \int%
_{\alpha}^{1}\Omega_{h}\left(  a\right)  f\left(  a\right)  da\right]
d\alpha\\
&  \quad=\int_{0}^{1}\left[  \int_{\alpha}^{1}g\left(  a\right)^{\prime}  \Omega
_{h}\left(  a\right)  da\right]  S_{0}^{\prime}S_{0}\left[  \int_{\alpha}%
^{1}\Omega_{h}\left(  a\right)  f\left(  a\right)  da\right]  d\alpha\\
&  -\left[  \int_{0}^{1}\alpha g\left(  \alpha\right)^{\prime}  \Omega_{h}\left(
\alpha\right)  d\alpha\right]  S_{0}^{\prime}S_{0}\left[  \int_{0}^{1}
\Omega_{h}\left(  \alpha\right)
f\left(\alpha\right)  \alpha  d\alpha\right]  ,\\
&  \quad=\operatorname*{Cov}\left(  \int_{A}^{1}g\left(  a\right)^{\prime}  \Omega
_{h}\left(  a\right)  da S_{0}^{\prime},S_{0}\int_{A}^{1}\Omega_{h}\left(
a\right) f\left(  a\right)    da\right)  .
\end{align*}
Similarly, $\int_{0}^{1}\left[  \int_{0}^{\alpha}\Omega_{h}\left(  a\right) f\left(  \alpha\right)
  da\right]  d\alpha=\int_{0}^{1}  \left(  1-\alpha\right)  \Omega_{h}\left(  \alpha\right)f\left(
  \alpha\right)
d\alpha$ gives, after an integration by parts,%
\begin{align*}
&  \left[  \int_{0}^{1}g\left(  \alpha\right)^{\prime}  \Omega_{h}\left(
\alpha\right)  d\alpha\right]  S_{1}^{\prime}S_{0}\left[  \int_{0}^{1}  \left(  1-\alpha\right)  \Omega_{h}\left(  \alpha\right) f\left(
\alpha\right)
d\alpha\right] \\
&  \quad\quad\quad-\int_{0}^{1}  \left[  \int_{\alpha
}^{1}g\left(  a\right)^{\prime}  \Omega_{h}\left(  a\right)  da\right]  S_{1}^{\prime}
S_{0}\Omega_{h}\left(  \alpha\right)  f\left(  \alpha\right) d\alpha\\
&  \quad=\left[  \int_{0}^{1}g\left(  \alpha\right)^{\prime}  \Omega_{h}\left(
\alpha\right)  d\alpha\right]  S_{1}^{\prime}S_{0}\left[  \int_{0}^{1}\left(
\int_{0}^{\alpha}\Omega_{h}\left(  a\right)  f\left(  a\right)  da\right)
d\alpha\right] \\
&  \quad\quad\quad-\int_{0}^{1}g\left(  \alpha\right)^{\prime}  \Omega_{h}\left(
\alpha\right)  S_{1}^{\prime}S_{0}\left[  \int_{0}^{\alpha}\Omega_{h}\left(
a\right)  f\left(  a\right)  da\right]  d\alpha\\
&  \quad=-\operatorname*{Cov}\left(  g\left(  A\right)^{\prime}  \Omega_{h}\left(
A\right)  S_{1}^{\prime}, S_{0}\int_{0}^{A}  \Omega_{h}\left(
a\right) f\left(  a\right) da  \right) \\
&  \quad=\operatorname*{Cov}\left(  g\left(  A\right)^{\prime}  \Omega_{h}\left(
A\right)  S_{1}^{\prime},  S_{0}\int_{A}^{1}  \Omega_{h}\left(
a\right)  f\left(  a\right)da  \right)  ,\\
&  \int_{0}^{1}  \left[  \int_{\alpha}^{1}g\left(
a\right)  \Omega_{h}\left(  a\right)  da\right]  S_{0}^{\prime}S_{1}\Omega
_{h}\left(  \alpha\right)  f\left(  \alpha\right) d\alpha\\
&  \quad\quad\quad-\left[  \int_{0}^{1}\alpha g\left(  \alpha\right)
\Omega_{h}\left(  \alpha\right)  d\alpha\right]  S_{0}^{\prime}S_{1}\left[
\int_{0}^{1}  \Omega_{h}\left(  \alpha\right) f\left(  \alpha\right)
d\alpha\right] \\
&  \quad=\operatorname*{Cov}\left(  \left[  \int_{A}^{1}g\left(  a\right)^{\prime}
\Omega_{h}\left(  a\right)  da\right]  S_{0}^{\prime},S_{1}  \Omega
_{h}\left(  A\right) f\left(  A\right) \right)  ,
\end{align*}
and, for any conformable deterministic $S_f$ and $S_g$,%
\begin{align*}
&  \int_{0}^{1}  g\left(  \alpha\right)^{\prime}  \Omega
_{h}\left(  \alpha\right)  S_g^{\prime} S_f  \Omega_{h}\left(
\alpha\right)  f\left(  \alpha\right) d\alpha
%\\
%&  \quad\quad\quad
-\left[  \int_{0}^{1}g\left(  \alpha\right)^{\prime}  \Omega
_{h}\left(  \alpha\right)  d\alpha\right]  S_g^{\prime} S_{f}  \left[
\int_{0}^{1}  \Omega_{h}\left(  \alpha\right) f\left(  \alpha\right)
d\alpha\right] \\
&  \quad=\operatorname*{Cov}\left(  g\left(  A\right)^{\prime}  \Omega_{h}\left(
A\right)  S_{g}^{\prime}, S_{f} \Omega_{h}\left(  A\right)  f\left(  A\right)\right)  .
\end{align*}
Collecting these items gives the expansion of $\mathcal{C}_{h}$ stated in the
Lemma.$\hfill\square$

\end{document}